\documentclass[traditabstract]{aa}

\usepackage{graphicx}
\usepackage{txfonts}
\usepackage{natbib}
\usepackage{subfigure}
\usepackage{url}
\usepackage{longtable}
\usepackage{supertabular}
\usepackage{rotating}
\usepackage{multirow}
\usepackage{gensymb} 
\usepackage{pifont}

\def\ujy{\,$\mu$Jy}
\def\mum{\,$\mu$m}
\def\nodata{...}
\def\p20ghz{P$^{\rm 20\,GHz}_{\rm core}$}
\def\500mhz{500\,MHz}
\def\3ghz{3\,GHz}
\def\20ghz{20\,GHz}

\def\spitzer{{\it Spitzer\,}}

\begin{document}

\title{Jet and Torus Orientations in High Redshift Radio Galaxies}

\author{G. Drouart
          \inst{1,2,3},
        C. De Breuck
          \inst{1},
        J. Vernet
          \inst{1},
        R. A. Laing
          \inst{1},
        N. Seymour
          \inst{3},
        D. Stern
          \inst{4},
        M. Haas
          \inst{5},
        E. A. Pier
          \inst{6},
        B. Rocca-Volmerange
          \inst{2}
}

\institute{\inst{1}European Southern Observatory, Karl Schwarzschild Stra\ss e 2, 
                   85748 Garching bei M\"unchen, Germany \\
           \inst{2}Institut d'Astrophysique de Paris, 98bis boulevard Arago, 75014 Paris, France \\
           \inst{3}CSIRO Astronomy \& Space Science, PO Box 76, Epping, NSW 1710, Australia \\
           \inst{4}Jet Propulsion Laboratory, California Institute of Technology, 
                   Mail Stop 169-221, Pasadena, CA 91109, USA \\
           \inst{5}Astronomisches Institut, Ruhr-Universit\"at Bochum, Universit\"atstr. 150, 44801 Bochum, Germany \\
           \inst{6}Oceanit Laboratories, 828 Fort Street Mall, Suite 600, Honolulu, HI 96813, USA \\
}

\date{resubmitted, 30th August 2012}

\abstract{ {

We examine the relative orientation of radio jets and 
dusty tori surrounding the AGN in powerful radio galaxies at
$z>1$. The radio core dominance
$R = P_{\rm core}^{\rm 20\,GHz}/P_{\rm extended}^{\rm 500\,MHz}$ serves as an orientation
indicator, measuring the ratio between the anisotropic Doppler-beamed core emission 
and the isotropic lobe emission. Assuming a fixed cylindrical geometry for the hot, dusty
torus, we derive its inclination $i$ by fitting optically-thick radiative transfer models
to spectral energy distributions obtained with the
{\textit Spitzer Space Telescope}.  We find a highly significant
anti-correlation ($p<0.0001$) between $R$ and $i$ in our sample of 35 type~2
AGN combined with a sample of 18 $z\sim 1$ 3CR sources containing both
type 1 and 2 AGN. This analysis provides observational evidence both for the
Unified scheme of AGN and for the common assumption that radio jets
are in general perpendicular to the plane of the torus. The use of 
inclinations derived from mid-infrared photometry breaks several 
degeneracies which have been problematic in earlier analyses. We illustrate this by deriving the core Lorentz factor
$\Gamma$ from the $R$-$i$ anti-correlation, finding $\Gamma \ga 1.3$.

}
{}{}{}{}
}


\titlerunning{AGN configuration in HzRG}
\authorrunning{G. Drouart et al.}
\maketitle


\section{Introduction} \label{sec:intro}

Radio galaxies are among the most luminous objects in the Universe
over the entire electromagnetic spectrum. Their powerful radio
emission betrays the presence of a central massive black hole
\citep[][]{Blandford1982} up to a few billion $M_\odot$ in mass
\citep{McLure2006,Nesvadba2011}. Early in the development of the
subject, local radio galaxies were associated with massive elliptical
galaxies \citep{Matthews1964}. It is now well established that
powerful radio sources are also hosted by massive galaxies at higher
redshift \citep[e.g.][]{DeBreuck2010}, as expected from the
bulge--black hole mass relation \citep{Ferrarese2000}. Radio galaxies
appear to be a singular stage in the evolution of massive galaxies,
observed during a peak of activity. This phase presents a unique chance
to test models of galaxy formation and the interaction between the AGN
and their host galaxies \citep[e.g.][]{Nesvadba2008}.

According to current understanding, an active galactic
  nucleus (AGN) consists of an accretion disk around a supermassive black hole
\citep[SMBH;][]{Rees1984}. About 10\% of AGN show radio jets
\citep[e.g.][]{Best2005}; these are expected to be aligned with the
black hole spin axis. A dusty torus has been hypothesized to explain
the observed dichotomy between unobscured (type 1) AGN where the
observer can see the region close to the black hole directly and
obscured (type 2) AGN with a more edge-on view \citep[for a review,
  see][]{Antonucci1993}. As a result of this geometry, the AGN
emission is anisotropic at most wavelengths. The accretion disk is a
powerful source of soft and hard X-rays which are attenuated when
passing through the torus. Assuming a universal torus shape, the
dominating factor determining the amount of obscuration along the line
of sight is the orientation of the torus with respect to the
observer. In particular, soft X-rays are sensitive to the hydrogen
column density ($N_H$) which varies from $10^{20}$\,cm$^{-2}$ in type 1
AGN to $>10^{24}$\,cm$^{-2}$ type 2 AGN \citep[e.g.][]{Ibar2007}.

In the optical domain, the effects of inclination can explain why the Broad Line
Regions (BLR) are observed directly in type 1 AGN, while in type 2's,
they are mostly obscured or seen in scattered emission
\citep{Antonucci1985,Tran1992}. In both types, Narrow Line Regions
(NLR) are observed further out from the torus. In the case of type 2's,
the torus acts as a coronograph blocking the much brighter central
emission, and allows the NLR to be traced out to
$\sim$100\,kpc \citep{Reuland2003,VillarMartin2003}.  Such detailed
studies have shown that the NLR is frequently aligned with the
radio jets \citep{McCarthy1987}. Moreover, optical polarimetry of type
2 AGN reveals a continuum polarization angle mostly perpendicular to
the radio axis, implying that the light passing through torus opening
is scattered by dust clouds along the radio jets
\citep{diSerego1989,diSerego1993,Cimatti1993,Hines1994,Vernet2001}. Both
observations imply that the radio jets are aligned orthogonally to the
equatorial plane of the torus.

The torus surrounding the SMBH re-processes a significant fraction of
the AGN radiation (X-rays, UV and optical) into mid-IR thermal dust
emission. This establishes a radial temperature gradient within the
torus ranging from the sublimation temperature ($\sim$1500\,K) at the
inner surface of the torus to a few hundred Kelvin in the outer
parts. The torus geometry creates a strong anisotropy in the hot dust
emission because the amount of extinction towards the innermost
(hottest) parts is very sensitive to the orientation with respect to
the observer \citep[][ hereafter PK92]{Pier1992}.  It should therefore
be possible to derive the inclination by modelling of the mid-IR
SED. The magnitude of the variation with inclination is illustrated by
the difference between the mean type 1 and type 2 mid-IR SEDs
  of an isotropically selected AGN sample where the type 2 AGN SED
can be reproduced by simple reddening of the type 1 distribution
\citep{Haas2008}.

The most isotropic emission from AGN comes from the radio lobes which
mark the interaction between the jets and the surrounding
intergalactic medium. This can happen on scales as large as several
Mpc and produces steep-spectrum synchrotron emission. In contrast, the
radio cores generally have flatter spectral indices and are
anisotropic as they are subject to Doppler beaming effects. The radio
cores will look brighter if the jet axis is observed closer to the
line of sight (e.g.\ in type 1 AGN). The ratio of the core to
  total radio emission (core dominance) is a proxy for orientation
  \citep[e.g.][]{Scheuer1979,Kapahi1982}.

In summary, previous observations suggest that the radio jets
  are orthogonal to the equatorial plane of the torus. Assuming a
  generic torus geometry, an inclination can be derived by fitting the
  mid-IR SED; the core dominance can also be used to estimate the
  inclination of the radio jets. Provided that these assumptions are
correct, the two measures of inclination should be consistent. In this
paper, we make use of the unique database consisting of six-band
mid-IR data for 70 radio galaxies spanning $z$=1 to $z$=5.2
\citep[][S07 and DB10 hereafter]{Seymour2007,DeBreuck2010}.  We derive
the inclination angle by fitting dust emission from the torus, and
compare it with the radio core dominance. We indeed
  find a significant correlation between these parameters, consistent
  with the previously mentioned observational statements which
  sample directions in the plane of the sky. We also
use this constraint on the orientation to estimate the pc-scale jet
speeds from the core/jet dominance.

This paper is organised as follows. 
Sections \ref{sec:shzrg} and \ref{sec:3CR_sample} describe the samples.
In Section \ref{sec:ir}, we describe our approach to modelling the
mid-IR emission, using both empirical correlations and a torus model for a
sub-sample with well-sampled AGN dust emission. We discuss the
implications in Section \ref{sec:disc}. Throughout this paper, we
adopt the current standard cosmological model 
($H_0= 70$\,kms$^{-1}$\,Mpc$^{-1}$, $\Omega_{\Lambda}=0.7$, $\Omega_{M}=0.3$).


\section{Spitzer High Redshift Radio Galaxy sample} \label{sec:shzrg}

Our \textit{Spitzer} High Redshift Radio Galaxy (SHzRG) sample is
selected from a compendium of the most powerful  radio galaxies
known in 2002. From this parent sample of 225 HzRGs with spectroscopic
redshifts, a subset was selected in order to sample the
redshift-radio power plane evenly out to the highest redshifts available. The
full description of this sample is presented by S07. In short, the
HzRG are distributed across $1 < z < 5.2$ and have
$P^{\rm{3\,GHz}}>10^{26}$\,WHz$^{-1}$, where $P^{\rm{3\,GHz}}$ is
the total luminosity at a rest-frame frequency of \3ghz (Table 1 of S07).

\subsection{Infrared data}

The mid-IR data presented here were obtained with the \spitzer \textit{Space
Telescope} \citep{Werner2004} during Cycles 1 and 4. All galaxies were
observed in the four Infrared Array Camera \citep[IRAC;][]{Fazio2004} channels (3.6, 4.5, 5.8,
8\mum), the 16\mum\ peak-up mode of the Infrared Spectrograph \citep[IRS;][]{Houck2004} and the
24\mum\ channel of the Multiband Infared Photometer \citep[MIPS;][]{Rieke2004}.  The 24 sources with the
lowest expected background emission were also observed with the 70 and
160\mum\ channels of MIPS. We refer the reader to S07 and DB10 for the
full description of the data reduction procedures and the full
photometric data (Table~3 of DB10).  We augment our photometry with
$K$-band magnitudes from S07.

\subsection{Radio data and core dominance}
\label{sec:calcul_core}

The radio morphologies of HzRGs are dominated by steep-spectrum
radio lobes with fainter and flatter spectrum radio cores
\citep{Carilli1997,Pentericci2000}. A variety of physical processes
contribute to the radio emission at different restframe frequencies
\citep[e.g.][]{Blundell1999}. Above $\nu \approx $1\,GHz, Doppler boosting will
introduce an orientation bias in the radio cores. In addition, the
radio lobes are affected by synchrotron and inverse Compton losses. At
very low frequencies $\nu\la $100\,MHz, synchrotron self-absorption,
free-free absorption and any low-energy cutoff of the relativistic
particles may reduce the observed synchrotron emission. We therefore
choose a rest-frame wavelength of \500mhz to obtain an accurate 
measure of the energy injected by the AGN which is also independent of orientation. Observationally, the
\500mhz luminosities P$^{\rm 500\,MHz}_{\rm extended}$ can also be determined most
uniformly over the entire sky (DB10).

Our measure of orientation for the radio jets is the ratio of core to
extended emission or {\em core dominance $R$
\citep{Scheuer1979,Kapahi1982}.} Since the flatter spectrum cores can only be
spatially resolved with interferometers at high frequencies, this is
defined as $R =
P_{\rm core}^{\rm{20\,GHz}}/P_{\rm extended}^{\rm{500\,MHz}}$, where
$P_{\rm core}^{20\rm{\,GHz}}$ is the 20\,GHz restframe core luminosity.

\begin{figure}[t] \centering
  \includegraphics[width=0.5\textwidth]{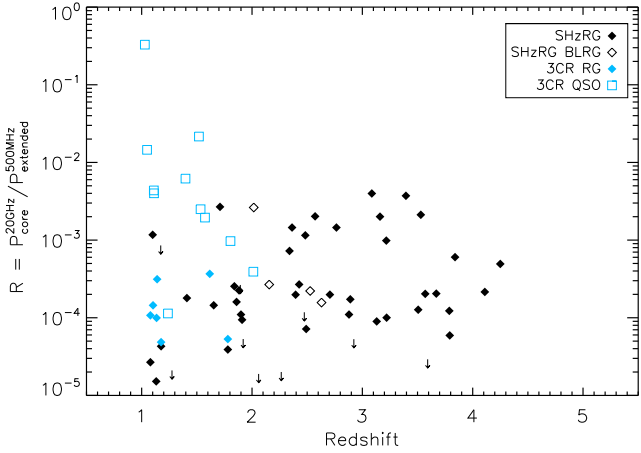}
  \caption{Core dominance $R$ versus redshift. Note the lower
    redshifts for the s3CR sample (blue diamonds and open squares). The
    open black diamonds correspond to radio galaxies with observed
    broad lines, see Section~\ref{sec:limit_obs}.}
  \label{fig:R_vs_z}
\end{figure}

Core flux densities are either taken from the literature (see
Table~\ref{tab:full_radio}), or  measured directly from radio maps 
by the following method. First, we take the {\it Spitzer}/IRAC
3.6\,$\mu$m image and overlay the radio contours. After identifying
the core with the host galaxy, we measure the integrated total and
core flux densities using the {\sc aips} verb {\sc tvstat}.
If only the lobes are identified, we derive an upper limit to 
the core dominance from the 3\,$\sigma$ sensitivity of the radio
map. Next, we calculate $P_{\rm core}^{20\rm{\,GHz}}$ using 8.4\,GHz
observations and the core spectral index
$\alpha_{\rm core}$, defined as $S \propto \nu^{\alpha}$. If no
spectral index is available, we use the median value from our sample
$\langle \alpha_{\rm core} \rangle = -0.8$. Note that given the large range of $R$, the
exact value of $\alpha_{\rm core}$ does not significantly affect our
results. At higher redshifts, inverse Compton losses can increase
significantly, which may affect $R$ values. However,
Fig.~\ref{fig:R_vs_z} does not show a significant  dependence
of $R$ on redshift, so we consider this effect to be negligible.

Table~\ref{tab:full_radio} lists the values of P$^{500\rm{\,MHz}}_{\rm extended}$
(summing the components in the case of multiple detections) and
$R$, together with the core flux densities and spectral indices
$\alpha_{\rm core}$ (see references in Table~\ref{tab:full_radio}).

\begin{table*}[ht]
\caption{Radio data for the SHzRG sample from the literature and core
  dominance calculated. ($^{*}$) mark the core flux recalculated in
  this paper, see Section~\ref{sec:calcul_core}. The letters S, D and T refer to the
  morphology of the radio source, with 1, 2 or 3 indentified
  components, respectively. Names in bold are the sSHzRG sample.}
\label{tab:full_radio}
 \begin{tabular}{lcc ccc ll}
\hline
Name & $z$ & log P$^{\rm 500\,MHz}_{\rm extended}$ & S$^{\rm 8.4\,GHz}_{\rm core}$ & $\alpha^{8.4}_{4.8}$ core & R & Morph & References \\
     &     & [W.Hz$^-1$]       & [mJy]            &                       &   &       &            \\
\hline \hline 
\textbf{6C0032+412}      & 3.670     & 28.75   &         0.25 &      \nodata &      0.00020  & T      & \cite{Blundell1998} \\              
\textbf{MRC0037-258}     & 1.100     & 27.72   &         1.74 &          0.9 &       0.0012  & T      & \cite{DeBreuck2010,Kapahi1998} \\   
\textbf{6C0058+495}      & 1.173     & 27.33   &  $<$ 0.00055 &      \nodata &  $<$ 0.00085$^{*}$  & D & \cite{Blundell1998} \\             
\textbf{MRC0114-211}     & 1.410     & 28.66   &         1.60 &      \nodata &      0.00018  & T      & \cite{DeBreuck2010,Kapahi1998} \\   
TNJ0121+1320             & 3.516     & 28.49   &      \nodata &      \nodata &      \nodata  & D      & \cite{DeBreuck2000}  \\             
6C0132+330               & 1.710     & 27.64   &         1.63 &          0.1 &       0.0027  & T      & \cite{DeBreuck2010}  \\             
6C0140+326               & 4.413     & 28.73   &      \nodata &      \nodata &      \nodata  & D      & \cite{Blundell1998}  \\             
\textbf{MRC0152-209}     & 1.920     & 28.20   &  $<$ 0.00008 &      \nodata &  $<$ 5.3e-05$^{*}$  & D & \cite{Pentericci2000} \\           
\textbf{MRC0156-252}     & 2.016     & 28.46   &         6.58 &         -0.7 &       0.0026  & T      & \cite{Carilli1997}  \\              
TNJ0205+2242             & 3.506     & 28.46   &         0.10 &         -0.6 &      0.00013$^{*}$  & T & \cite{DeBreuck2000} \\             
\textbf{MRC0211-256}     & 1.300     & 27.78   &      \nodata &      \nodata &      \nodata  & S      & \cite{DeBreuck2000} \\              
\textbf{TXS0211-122}     & 2.340     & 28.48   &         1.23 &         -1.0 &      0.00073  & T      & \cite{Carilli1997} \\               
3C65                     & 1.176     & 28.63   &         0.52 &         -0.2 &      4.3e-05  & T      & \cite{Carilli1997,Corbin1998} \\    
MRC0251-273              & 3.160     & 28.54   &         1.35 &         -1.6 &       0.0020  & T      & \cite{DeBreuck2010,Kapahi1998} \\   
MRC0316-257              & 3.130     & 28.95   &         0.28 &         -0.6 &      9.0e-05$^{*}$  & T & \cite{McCarthy1991,Athreya1997} \\ 
\textbf{MRC0324-228}     & 1.894     & 28.49   &  $<$ 0.00080 &      \nodata &  $<$ 0.00026  & T      & \cite{DeBreuck2010,McCarthy1991}  \\
\textbf{MRC0350-279}     & 1.900     & 28.25   &         0.19 &      \nodata &      0.00011$^{*}$  & T & \cite{DeBreuck2010,Kapahi1998} \\  
\textbf{MRC0406-244}     & 2.427     & 29.03   &         1.59 &         -0.8 &      0.00027  & T      & \cite{Carilli1997}  \\              
4C60.07                  & 3.788     & 29.20   &         0.21 &         -1.7 &      0.00012  & T      & \cite{Carilli1997}  \\              
\textbf{PKS0529-549}     & 2.575     & 29.16   &      \nodata &      \nodata &      \nodata  & D      & \cite{Broderick2007} \\             
WNJ0617+5012             & 3.153     & 28.02   &      \nodata &      \nodata &      \nodata       & D & \cite{DeBreuck2000} \\             
4C41.17                  & 3.792     & 29.18   &         0.27 &         -0.2 &      5.9e-05  & T      & \cite{Carilli1994}  \\              
WNJ0747+3654             & 2.992     & 28.14   &      \nodata &      \nodata &      \nodata  & S      & \cite{DeBreuck2000}  \\             
6CE0820+3642             & 1.860     & 28.28   &         0.29 &         -1.2 &      0.00016  & T      & \cite{DeBreuck2010,LawGreen1995a}  \\
5C7.269                  & 2.218     & 27.82   &      \nodata &      \nodata &      \nodata  & D & \cite{DeBreuck2010} \\             
\textbf{USS0828+193}     & 2.572     & 28.44   &         2.81 &         -0.7 &       0.0020  & T      & \cite{Carilli1997}  \\              
6CE0901+3551             & 1.910     & 28.19   &         0.14 &      \nodata &      9.4e-05  & T      & \cite{DeBreuck2010}  \\             
\textbf{B20902+34}       & 3.395     & 28.78   &         8.90 &         -0.1 &       0.0037  & T      & \cite{Carilli1994}  \\              
\textbf{6CE0905+3955}    & 1.883     & 28.17   &         0.33 &      \nodata &      0.00022  & T      & \cite{LawGreen1995b}  \\             
TNJ0924-2201             & 5.195     & 29.51   &      \nodata &      \nodata &      \nodata  & D & \cite{DeBreuck2000} \\             
6C0930+389               & 2.395     & 28.41   &         0.29 &         -0.8 &      0.00020  & T      & \cite{Pentericci2000}  \\           
USS0943-242              & 2.923     & 28.62   &  $<$ 0.00008 &      \nodata &  $<$ 5.3e-05$^{*}$  & D & \cite{Carilli1997} \\              
\textbf{3C239}           & 1.781     & 29.00   &         0.44 &      \nodata &      3.9e-05  & T      & \cite{Best1997}  \\                 
MG1019+0534              & 2.765     & 28.57   &         2.02 &         -1.0 &       0.0015  & T      & \cite{Pentericci2000}  \\           
\textbf{MRC1017-220}     & 1.768     & 27.94   &      \nodata &      \nodata &      \nodata  & S      & \cite{Pentericci2000}  \\           
WNJ1115+5016             & 2.540     & 27.82   &      \nodata &      \nodata &      \nodata  & D      & \cite{DeBreuck2000}  \\             
\textbf{3C257}           & 2.474     & 29.16   &  $<$ 0.00090 &      \nodata &  $<$ 0.00012$^{*}$  & D & \cite{vanBreugel1998}\\            
\textbf{WNJ1123+3141}    & 3.217     & 28.51   &         0.93 &      \nodata &      0.00098  & T      & \cite{White1997}  \\                
\textbf{PKS1138-262}     & 2.156     & 29.07   &         1.98 &         -1.3 &      0.00027  & T      & \cite{Carilli1997}  \\              
3C266                    & 1.275     & 28.54   &  $<$ 0.00018 &      \nodata &  $<$ 2.1e-05  & D      & \cite{Best1997}  \\                 
\textbf{6C1232+39}       & 3.220     & 28.93   &         0.35 &         -0.2 &      0.00010  & T      & \cite{Carilli1997}  \\              
USS1243+036              & 3.570     & 29.23   &         0.70 &         -1.0 &      0.00020  & T      & \cite{vanOjik1996}  \\              
TNJ1338-1942             & 4.110     & 28.71   &         0.16 &         -1.0 &      0.00021  & T      & \cite{Pentericci2000}  \\           
\textbf{4C24.28}         & 2.879     & 29.05   &         0.51 &         -0.6 &      0.00011  & T      & \cite{Carilli1997}  \\              
3C294                    & 1.786     & 28.96   &      \nodata &      \nodata &      \nodata  & T      & \cite{McCarthy1990}  \\             
\textbf{USS1410-001}     & 2.363     & 28.41   &         2.27 &         -0.7 &       0.0015  & T      & \cite{Carilli1997}  \\              
8C1435+635               & 4.250     & 29.40   &         1.31 &         -1.3 &      0.00049  & T      & \cite{Carilli1997}  \\              
\textbf{USS1558-003}     & 2.527     & 28.82   &         0.97 &         -0.1 &      0.00022  & T      & \cite{Pentericci2000}  \\           
USS1707+105              & 2.349     & 28.63   &      \nodata &      \nodata &      \nodata       & D & \cite{Pentericci2001} \\           
\textbf{LBDS53W002}      & 2.393     & 27.78   &      \nodata &      \nodata &      \nodata  & S      & \cite{Fomalont2002} \\              
LBDS53W091               & 1.552     & 27.04   &      \nodata &      \nodata &      \nodata       & D & \cite{Rigby2007} \\                
\textbf{3C356}           & 1.079     & 28.35   &         0.22 &      \nodata &      2.7e-05  & T      & \cite{Best1997}  \\                 
7C1751+6809              & 1.540     & 27.46   &      \nodata &      \nodata &      \nodata       & D & \cite{DeBreuck2000} \\             
7C1756+6520              & 1.416     & 27.40   &      \nodata &      \nodata &      \nodata       & D & \cite{Rigby2007} \\                
\textbf{3C368}           & 1.132     & 28.52   &         0.16 &         -0.5 &      1.5e-05  & T      & \cite{Best1997}  \\                 
\textbf{7C1805+6332}     & 1.840     & 27.78   &         0.15 &         -1.2 &      0.00025  & T      & \cite{DeBreuck2010}  \\             
4C40.36                  & 2.265     & 28.79   &  $<$ 0.00008 &      \nodata &  $<$ 2.0e-05  & D      & \cite{Carilli1997}  \\              
\textbf{TXSJ1908+7220}   & 3.530     & 29.12   &         4.80 &         -1.3 &       0.0021  & T      & \cite{Pentericci2000}  \\           
WNJ1911+6342             & 3.590     & 28.14   &      \nodata &      \nodata &      \nodata  & S      & \cite{DeBreuck2000}  \\             
\textbf{TNJ2007-1316}    & 3.840     & 29.13   &         1.60 &      \nodata &      0.00060  & T      & \cite{DeBreuck2010}  \\             
\textbf{MRC2025-218}     & 2.630     & 28.74   &         0.38 &         -0.9 &      0.00016  & T      & \cite{Carilli1997}  \\              
MRC2048-272              & 2.060     & 28.72   &  $<$ 0.00008 &      \nodata &  $<$ 1.9e-05  & D      & \cite{Pentericci2000}  \\           
MRC2104-242              & 2.491     & 28.84   &         0.19 &         -1.6 &      7.2e-05  & T      & \cite{Pentericci2000}  \\           
\textbf{4C23.56}         & 2.483     & 28.93   &         4.95 &         -0.9 &       0.0012  & T      & \cite{Carilli1997}  \\              
MG2144+1928              & 3.592     & 29.08   &  $<$ 0.00008 &      \nodata &  $<$ 2.9e-05$^{*}$  & D & \cite{Carilli1997} \\              
USS2202+128              & 2.706     & 28.54   &         0.14 &         -2.5 &      0.00020  & T      & \cite{Carilli1997}  \\              
\textbf{MRC2224-273}     & 1.679     & 27.52   &      \nodata &      \nodata &      \nodata  & S      & \cite{Pentericci2000}  \\           
\textbf{B3J2330+3927}    & 3.086     & 28.33   &         3.98 &         -0.1 &       0.0040  & T      & \cite{PerezTorres2005}  \\          
4C28.58                  & 2.891     & 28.91   &         0.70 &         -0.2 &      0.00017  & T      & \cite{Cai2002,Chambers1996}  \\     
\textbf{3C470}           & 1.653     & 28.79   &         1.20 &      \nodata &      0.00014  & T      & \cite{Best1997} \\                  
\hline
  \end{tabular}                   

\end{table*}

\begin{figure}[t] \centering
  \includegraphics[width=0.5\textwidth]{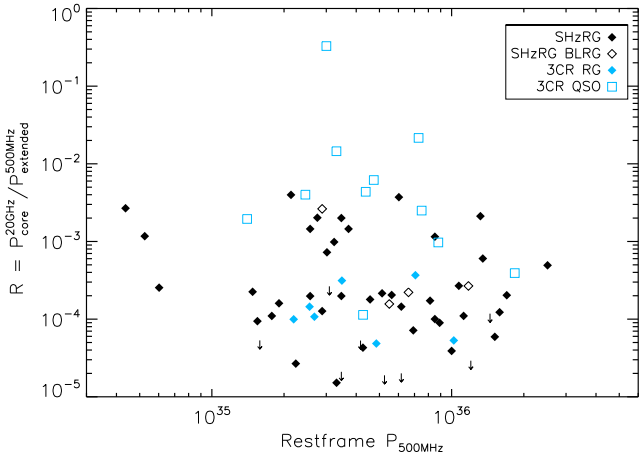}
  \caption{Core dominance $R$ versus P$^{\rm 500\,MHz}_{\rm extended}$. Note that the s3CR quasars (open blue
    squares) typically have higher values of $R$ than the galaxies from
    the same sample (filled blue diamonds).}
  \label{fig:l500_vs_R}
\end{figure}

\section{3CR sample}\label{sec:3CR_sample}

In order to compare the results of our approach for type 1 and 2
radio-loud AGN, samples of the two classes with matched selection
criteria are required.  Unfortunately, there is no type 1 (quasar)
sample in the literature which matches our radio-galaxy sample and has
both radio and \spitzer observations. The best sample available,
although significantly smaller than our SHzRG sample and with a lower
median redshift, is the 3CR sample \citep{Spinrad1985}.

\subsection{Infrared data}

A selection of 64 3CR high-redshift sources has been observed with the four
IRAC channels (3.6, 4.5, 5.8, 8\mum), IRS (16\mum) and the MIPS1 channel
(24\mum). We refer to \cite{Haas2008} for a full description of the
sample and data reduction. Table~\ref{tab:full_3CR} reports the
{\it Spitzer} photometry.

\begin{table*}[ht] \centering
\caption{3CR sample, \spitzer observations, Q = quasars, G = radio galaxies, N = undefined, Names in bold correspond to the s3CR sample.}
\label{tab:full_3CR}
\begin{tabular}{llc ccc ccc}
\hline
Name & Type & z & 3.6\mum & 4.5\mum & 5.8\mum & 8\mum & 16\mum & 24\mum  \\
     &      &   &  [\ujy] & [\ujy]  & [\ujy]  & [\ujy]& [\ujy] & [\ujy]   \\
\hline \hline 

\textbf{3C~002}   & Q    &  1.04 &   283 $\pm$    42 &   330 $\pm$    50 &   530 $\pm$    80 &   809 $\pm$   121 &  1550 $\pm$   233 &  2970 $\pm$   446 \\ 
\textbf{3C~009}   & Q &  2.01 &   884 $\pm$   133 &  1080 $\pm$   162 &  1590 $\pm$   239 &  2220 $\pm$   333 &  3330 $\pm$   500 &  3470 $\pm$   520 \\ 
\textbf{3C~013}   & G    &  1.35 &   133 $\pm$    20 &   133 $\pm$    20 &   147 $\pm$    22 &   283 $\pm$    42 &   375 $\pm$    56 &  2060 $\pm$   309 \\ 
\textbf{3C~014}   & Q    &  1.47 &  1040 $\pm$   156 &  1710 $\pm$   257 &  2740 $\pm$   411 &  4150 $\pm$   623 &  7070 $\pm$  1061 & 10300 $\pm$  1545 \\ 
\textbf{3C~036}   & G    &  1.30 &   163 $\pm$    24 &   205 $\pm$    31 &   256 $\pm$    38 &   360 $\pm$    54 &   560 $\pm$    84 &   874 $\pm$   131 \\ 
\textbf{3C~043}   & Q    &  1.47 &   193 $\pm$    29 &   270 $\pm$    41 &   356 $\pm$    53 &   445 $\pm$    67 &  1010 $\pm$   152 &  1610 $\pm$   242 \\ 
\textbf{3C~065}   & G &  1.18 &   202 $\pm$    30 &   233 $\pm$    35 &   299 $\pm$    45 &   418 $\pm$    63 &   798 $\pm$   120 &  1700 $\pm$   255 \\ 
\textbf{3C~068.1} & Q &  1.24 &   967 $\pm$   145 &  1430 $\pm$   215 &  2040 $\pm$   306 &  2780 $\pm$   417 &  3800 $\pm$   570 &  7760 $\pm$  1164 \\ 
\textbf{3C~068.2} & G    &  1.58 &   105 $\pm$    16 &   129 $\pm$    19 &   137 $\pm$    21 &   112 $\pm$    17 &  1340 $\pm$   201 &  1170 $\pm$   176 \\ 
3C~119   & G    &  1.02 &   802 $\pm$   120 &   878 $\pm$   132 &  1280 $\pm$   192 &  1850 $\pm$   278 &  4820 $\pm$   723 &  8260 $\pm$  1239 \\ 
\textbf{3C~124}   & G    &  1.08 &   144 $\pm$    22 &   120 $\pm$    18 &   188 $\pm$    28 &   310 $\pm$    47 &  1840 $\pm$   276 &  3560 $\pm$   534 \\ 
\textbf{3C~173}   & N    &  1.03 &   163 $\pm$    24 &   172 $\pm$    26 &   197 $\pm$    30 &   227 $\pm$    34 &   374 $\pm$    56 &   710 $\pm$   107 \\ 
\textbf{3C~181}   & Q    &  1.38 &   348 $\pm$    52 &   485 $\pm$    73 &   722 $\pm$   108 &  1110 $\pm$   167 &  2180 $\pm$   327 &  4260 $\pm$   639 \\ 
\textbf{3C~186}   & Q    &  1.06 &   791 $\pm$   119 &  1020 $\pm$   153 &  1410 $\pm$   212 &  1960 $\pm$   294 &  3660 $\pm$   549 &  6660 $\pm$   999 \\ 
\textbf{3C~190}   & Q    &  1.20 &   739 $\pm$   111 &   908 $\pm$   136 &  1290 $\pm$   194 &  1740 $\pm$   261 &  3310 $\pm$   497 &  6690 $\pm$  1004 \\ 
\textbf{3C~191}   & Q    &  1.96 &   333 $\pm$    50 &   399 $\pm$    60 &   655 $\pm$    98 &  1010 $\pm$   152 &  2270 $\pm$   341 &  3810 $\pm$   572 \\ 
3C~194   & G    &  1.78 &   201 $\pm$    30 &   176 $\pm$    26 &   164 $\pm$    25 &   208 $\pm$    31 &   509 $\pm$    76 &   885 $\pm$   133 \\ 
\textbf{3C~204}   & Q &  1.11 &   917 $\pm$   138 &  1250 $\pm$   188 &  1920 $\pm$   288 &  2540 $\pm$   381 &  4730 $\pm$   710 &  7360 $\pm$  1104 \\ 
\textbf{3C~205}   & Q &  1.53 &  1460 $\pm$   219 &  2080 $\pm$   312 &  2920 $\pm$   438 &  4090 $\pm$   614 &  7320 $\pm$  1098 & 12800 $\pm$  1920 \\ 
\textbf{3C~208}   & Q &  1.11 &   660 $\pm$    99 &   803 $\pm$   120 &  1160 $\pm$   174 &  1620 $\pm$   243 &  2980 $\pm$   447 &  5870 $\pm$   881 \\ 
\textbf{3C~208.1} & N    &  1.02 &   331 $\pm$    50 &   430 $\pm$    65 &   656 $\pm$    98 &   954 $\pm$   143 &  1360 $\pm$   204 &  2110 $\pm$   317 \\ 
\textbf{3C~210}   & G    &  1.17 &   256 $\pm$    38 &   336 $\pm$    50 &   489 $\pm$    73 &  1090 $\pm$   164 &  3410 $\pm$   512 &  4430 $\pm$   665 \\ 
\textbf{3C~212}   & Q &  1.05 &   925 $\pm$   139 &  1430 $\pm$   215 &  2340 $\pm$   351 &  3400 $\pm$   510 &  6710 $\pm$  1007 & 10800 $\pm$  1620 \\ 
\textbf{3C~220.2} & Q    &  1.16 &   592 $\pm$    89 &   870 $\pm$   131 &  1330 $\pm$   200 &  2000 $\pm$   300 &  4150 $\pm$   623 &  6720 $\pm$  1008 \\ 
3C~222   & G    &  1.34 &    83 $\pm$    12 &    91 $\pm$    14 &    73 $\pm$    11 &    65 $\pm$    10 &   331 $\pm$    50 &   229 $\pm$    34 \\ 
\textbf{3C~225A}  & G    &  1.56 &    47 $\pm$     7 &    49 $\pm$     7 &    71 $\pm$    11 &   108 $\pm$    16 &   321 $\pm$    48 &$<$1070            \\ 
3C~230   & G    &  1.49 &  1040 $\pm$   156 &   672 $\pm$   101 &   438 $\pm$    66 &   317 $\pm$    48 &  1150 $\pm$   173 &  1560 $\pm$   234 \\ 
3C~238   & G    &  1.40 &    65 $\pm$    10 &    77 $\pm$    12 &    84 $\pm$    12 & $<$92             &$<$283             &   266 $\pm$    40 \\ 
\textbf{3C~239}   & G &  1.78 &    96 $\pm$    14 &   111 $\pm$    17 &   130 $\pm$    20 &   142 $\pm$    21 &   651 $\pm$    98 &  1450 $\pm$   218 \\ 
\textbf{3C~241}   & G &  1.62 &    92 $\pm$    14 &   101 $\pm$    15 &   116 $\pm$    17 &   161 $\pm$    24 &   389 $\pm$    58 &   591 $\pm$    89 \\ 
\textbf{3C~245}   & Q &  1.03 &  1420 $\pm$   213 &  1900 $\pm$   285 &  3350 $\pm$   503 &  5270 $\pm$   790 & 10400 $\pm$  1560 & 20400 $\pm$  3060 \\ 
3C~249   & G    &  1.55 &    54 $\pm$     8 &    52 $\pm$     8 &    42 $\pm$     6 &    47 $\pm$     7 &   194 $\pm$    29 &$<$516             \\ 
3C~250   & G    &  1.26 &    61 $\pm$     9 &    59 $\pm$     9 &    46 $\pm$     7 &    29 $\pm$     4 &   162 $\pm$    24 &$<$147             \\ 
\textbf{3C~252}   & G &  1.10 &   225 $\pm$    34 &   382 $\pm$    57 &   787 $\pm$   118 &  1390 $\pm$   209 &  3900 $\pm$   585 &  7000 $\pm$  1050 \\ 
3C~255   & G    &  1.36 &    85 $\pm$    13 &    86 $\pm$    13 &    57 $\pm$     9 &    22 $\pm$     3 &$<$116             &$<$241             \\ 
3C~256   & G    &  1.82 &    34 $\pm$     5 &    37 $\pm$     6 &    43 $\pm$     7 &    75 $\pm$    11 &   743 $\pm$   111 &  1900 $\pm$   285 \\ 
\textbf{3C~257}   & G    &  2.47 &    85 $\pm$    13 &   111 $\pm$    17 &   194 $\pm$    29 &   322 $\pm$    48 &    \nodata        &  1360 $\pm$   204 \\ 
3C~266   & G    &  1.27 &    68 $\pm$    10 &    73 $\pm$    11 &    45 $\pm$     7 &   102 $\pm$    15 &   370 $\pm$    56 &   980 $\pm$   147 \\ 
\textbf{3C~267}   & G &  1.14 &   153 $\pm$    23 &   218 $\pm$    33 &   414 $\pm$    62 &   739 $\pm$   111 &  2370 $\pm$   356 &  3730 $\pm$   560 \\ 
\textbf{3C~268.4} & Q &  1.40 &  1060 $\pm$   159 &  1560 $\pm$   234 &  2220 $\pm$   333 &  3330 $\pm$   500 &  7580 $\pm$  1137 & 11600 $\pm$  1740 \\ 
\textbf{3C~270.1} & Q &  1.52 &   606 $\pm$    91 &   944 $\pm$   142 &  1430 $\pm$   214 &  2260 $\pm$   339 &  3910 $\pm$   587 &  5470 $\pm$   821 \\ 
\textbf{3C~280.1} & Q    &  1.66 &   378 $\pm$    57 &   512 $\pm$    77 &   777 $\pm$   116 &  1170 $\pm$   176 &  1680 $\pm$   252 &  2160 $\pm$   324 \\ 
\textbf{3C~287}   & Q    &  1.05 &   613 $\pm$    92 &   735 $\pm$   110 &  1050 $\pm$   157 &  1560 $\pm$   234 &  3430 $\pm$   515 &  5820 $\pm$   873 \\ 
3C~294   & G    &  1.79 & $<$93             &$<$103             &    68 $\pm$    10 &    67 $\pm$    10 &\nodata            &   348 $\pm$    52 \\ 
\textbf{3C~297}   & N    &  1.41 &   119 $\pm$    18 &   126 $\pm$    19 &   122 $\pm$    18 &   121 $\pm$    18 &$<$288             &   432 $\pm$    65 \\ 
\textbf{3C~298}   & Q    &  1.44 &  1600 $\pm$   240 &  2390 $\pm$   359 &  3710 $\pm$   556 &  5510 $\pm$   827 &  9160 $\pm$  1374 & 12600 $\pm$  1890 \\ 
\textbf{3C300.1} & G    &  1.16 &   148 $\pm$    22 &   158 $\pm$    24 &   133 $\pm$    20 &   220 $\pm$    33 &   751 $\pm$   113 &  1220 $\pm$   183 \\ 
\textbf{3C~305.1} & G    &  1.13 &   181 $\pm$    27 &   282 $\pm$    42 &   495 $\pm$    74 &   972 $\pm$   146 &  2410 $\pm$   362 &  2490 $\pm$   374 \\ 
\textbf{3C~318}   & Q &  1.57 &   343 $\pm$    51 &   427 $\pm$    64 &   571 $\pm$    86 &   806 $\pm$   121 &  1960 $\pm$   294 &  3400 $\pm$   510 \\ 
3C~322   & G    &  1.68 &   128 $\pm$    19 &   135 $\pm$    20 &    94 $\pm$    14 &   120 $\pm$    18 &   411 $\pm$    62 &   804 $\pm$   121 \\ 
\textbf{3C~324}   & G    &  1.21 &   165 $\pm$    25 &   160 $\pm$    24 &   178 $\pm$    27 &   450 $\pm$    68 &  2580 $\pm$   387 &  2820 $\pm$   423 \\ 
\textbf{3C~325}   & G &  1.13 &   472 $\pm$    71 &   565 $\pm$    85 &   708 $\pm$   106 &  1200 $\pm$   180 &  1990 $\pm$   299 &  3030 $\pm$   455 \\ 
3C~326.1 & G    &  1.83 &    29 $\pm$     4 &    34 $\pm$     5 &    26 $\pm$     4 &    72 $\pm$    11 &   829 $\pm$   124 &  1430 $\pm$   215 \\ 
\textbf{3C~356}   & G &  1.08 &   108 $\pm$    16 &   110 $\pm$    16 &   122 $\pm$    18 &   434 $\pm$    65 &  2270 $\pm$   341 &  4060 $\pm$   609 \\ 
\textbf{3C~368}   & G    &  1.13 &   126 $\pm$    19 &   112 $\pm$    17 &   112 $\pm$    17 &   210 $\pm$    32 &  1370 $\pm$   206 &  3250 $\pm$   488 \\ 
\textbf{3C~418}   & Q    &  1.69 &  1130 $\pm$   170 &  1630 $\pm$   245 &  2470 $\pm$   371 &  3900 $\pm$   585 &  6680 $\pm$  1002 & 13600 $\pm$  2040 \\ 
\textbf{3C~432}   & Q &  1.80 &   420 $\pm$    63 &   526 $\pm$    79 &   857 $\pm$   129 &  1490 $\pm$   224 &  2710 $\pm$   407 &  3940 $\pm$   591 \\ 
\textbf{3C~437}   & G    &  1.48 &    82 $\pm$    12 &    85 $\pm$    13 &    97 $\pm$    15 &    80 $\pm$    12 &   384 $\pm$    58 &   941 $\pm$   141 \\ 
\textbf{3C~454.1} & G    &  1.84 &    77 $\pm$    12 &    76 $\pm$    11 &   112 $\pm$    17 &   135 $\pm$    20 &   612 $\pm$    92 &  1500 $\pm$   225 \\ 
\textbf{3C~454.0} & Q    &  1.76 &   339 $\pm$    51 &   481 $\pm$    72 &   811 $\pm$   122 &  1220 $\pm$   183 &  2490 $\pm$   374 &  4150 $\pm$   623 \\ 
\textbf{3C~469.1} & G    &  1.34 &   160 $\pm$    24 &   244 $\pm$    37 &   509 $\pm$    76 &  1090 $\pm$   164 &  3270 $\pm$   491 &  1970 $\pm$   296 \\ 
\textbf{3C~470}   & G    &  1.65 &    50 $\pm$     7 &    75 $\pm$    11 &    72 $\pm$    11 &   266 $\pm$    40 &  1510 $\pm$   227 &  2650 $\pm$   398 \\ 
4C~13.66 & G    &  1.45 &    24 $\pm$     4 &    24 $\pm$     4 &    21 $\pm$     3 &    18 $\pm$     3 &$<$260             &   276 $\pm$    41 \\ 
\textbf{4C~16.49} & Q    &  1.88 &   329 $\pm$    49 &   420 $\pm$    63 &   573 $\pm$    86 &   743 $\pm$   111 &  1070 $\pm$   161 &  1830 $\pm$   275 \\ 
\hline
\end{tabular}
\end{table*}

\subsection{Radio data and core dominance}

In order to compare the 3CR and SHzRG sources, we need equivalent
radio data for both groups.  We take the relevant measurements for the
subset of 3CR sources from \citet{Hoekstra1997}. Selecting sources with both
\spitzer and radio data gives us a subset of 18 sources: 11 quasars
and 7 radio galaxies. The radio
data are reported in Table~\ref{tab:full_radio_3C}.  We recalculate
the \500mhz rest-frame luminosity from the 178\,MHz flux densities
\cite[and references therein]{Hoekstra1997}, using a spectral index
$\alpha=-1.5$, which is typical of powerful steep-spectrum radio
sources in this frequency range. The 20\,GHz rest-frame core
luminosities are calculated using the flux density at 5\,GHz assuming
$\alpha_{\rm core} = -0.8$. The derived values of $R$ are also
reported in Table~\ref{tab:full_radio_3C}.

\begin{table}[ht] \centering
\caption{Radio data for the s3CR sample. Types: Q=quasar, G=radio
  galaxy. ($^*$) Radio galaxies common with SHzRG
  sample. \label{tab:full_radio_3C} }
\begin{tabular}{l ccc cc }
\hline
Name       & Type & $z$ & log P$^{\rm 500\,MHz}$ & S$^{5\rm{\,GHz}}_{\rm core}$ & $R$ \\
           &      &     & [W.Hz$^{-1}$]     &   [mJy]     &     \\
\hline \hline
3C009      &  Q  & 2.01   &  29.26 &    0.55 &  0.00039 \\    
3C065$^*$  &  G  & 1.18   &  28.69 &    0.76 &  4.9e-05 \\ 
3C068.1    &  Q  & 1.24   &  28.63 &    0.83 &  0.00011 \\ 
3C204      &  Q  & 1.11   &  28.39 &    0.34 &   0.0040 \\ 
3C205      &  Q  & 1.53   &  28.88 &    0.67 &   0.0025 \\ 
3C208      &  Q  & 1.11   &  28.64 &    0.54 &   0.0044 \\ 
3C212      &  Q  & 1.05   &  28.52 &    0.88 &    0.015 \\ 
3C239$^*$  &  G  & 1.78   &  29.01 &    0.33 &  5.3e-05 \\ 
3C241      &  G  & 1.62   &  28.85 &    0.34 &  0.00037 \\ 
3C245      &  Q  & 1.03   &  28.48 &    1.40 &     0.33 \\ 
3C252      &  G  & 1.10   &  28.41 &    0.32 &  0.00014 \\ 
3C267      &  G  & 1.14   &  28.54 &    0.59 &  0.00031 \\ 
3C268.4    &  Q  & 1.40   &  28.68 &    0.60 &   0.0062 \\ 
3C270.1    &  Q  & 1.52   &  28.86 &    0.87 &    0.022 \\ 
3C318      &  Q  & 1.57   &  28.15 &    0.75 &   0.0019 \\ 
3C325      &  G  & 1.13   &  28.34 &    0.82 &  9.9e-05 \\ 
3C356$^*$  &  G  & 1.08   &  28.43 &    0.38 &  0.00011 \\ 
3C432      &  Q  & 1.80   &  28.94 &    0.31 &  0.00097 \\

\hline
\end{tabular}
\end{table}


\section{Modelling the torus emission} \label{sec:ir}

\subsection{Contributions to the infrared SED} \label{sec:contributions}

The near- to far-IR emission from radio galaxies and quasars consists of several
components: (i) non-thermal synchrotron emission, (ii) line emission,
(iii) starburst- and AGN-heated dust continuum, (iv) an (old) stellar
population. We now discuss the importance of these contributions in turn.

Of the type 2 objects in our sample, the source which is likely to have the highest synchrotron
contribution at 5\mum~ wavelength is B3\,J2330+3927, which
has the highest value of R and one of the flattest core spectra 
($\alpha=-0.1$).  Its extrapolated synchrotron contribution is
$<$5\,\% of the total emission at 5\mum. The synchrotron contribution can
therefore be safely ignored for all of the type 2 objects.
However in type 1 objects, this assumption may not be valid. We further discuss this point
in Section~\ref{sec:comp_disc}.

The line emission is dominated by fine structure and CO lines in the
far-IR \citep[e.g.][]{Smail2011}, and in the mid-IR by polycyclic
aromatic hydrocarbon (PAH) emission. The silicate emission/absorption 
is also seen in type 1/2 AGN, respectively. Both can affect the
broad-band photometry. Due to the large redshift range of our sources,
the sampling of the SED differs from object to object. In order to
keep our study as homogeneous as possible, we focus only on the part
of the SED below the silicate feature ($<$$10$\mum), which is 
dominated by AGN heated dust emission (see below). Mid-IR
spectroscopy with IRS \citep[][Rawlings et al., in
  prep.]{Seymour2008,Leipski2010} shows that PAH features are 
either not detected or relatively weak in HzRGs.

The contribution of the dust continuum spans three orders of magnitude
in wavelength.  The energy source of this emission (AGN and/or
starburst) has been the subject of considerable debate
\citep[e.g.][]{Sanders1988,Haas2003}. To disentangle these two
components, good sampling of the entire IR SED is essential. The
\textit{Herschel} Radio Galaxy Evolution or \textit{projet HeRG\'E}
will observe our sample between 70\mum\ and 500\mum. First results on
PKS1138$-$262 \citep{Seymour2012} show that the AGN component
dominates for $\lambda_{\rm{rest}}\lesssim 30$\,\mum.  The contribution
of starburst-heated dust may vary from source to source, but is
unlikely to contribute significantly at
$\lambda_{\rm{rest}}\lesssim$10~\mum\, as the required dust
temperature would be more than several hundred Kelvin, which cannot
be sustained throughout the host galaxy. We consequently focus only on
$\lambda_{\rm{rest}}\lesssim$10~\mum.  A crucial ingredient in
modelling the torus is the 9.7\mum\ silicate feature, which can be used
as an indicator of its intrinsic properties \citep[e.g. the
  differences between clumpy and continuous or optically thin and
  thick tori and the dust composition;][]{vanBemmel2003}.  Using this
feature to characterize the torus is beyond the scope of this paper,
since our broad-band photometry is not particularly sensitive to it.

Emission from the old stellar population peaks at
$\lambda_{\rm{rest}}\sim$1.6\,$\mu$m (S07).  In this paper, we model
this contribution assuming a formation redshift $z_{\rm form}$=$10$,
using an elliptical-galaxy template predicted by PEGASE.2
\citep{Fioc1997}.  The stellar contribution from this galaxy template
is then normalised for the bluest available band and subtracted in
each filter. Uncertainties for the subtracted data are taken to be the 
rms of the uncertainties of the individual measurement and
the stellar fit. For the type 2 3CR radio galaxies, we follow the same
method. For the type 1 3CR AGN, we do not
subtract any stellar contribution, as the transmitted non-thermal
continuum is expected to dominate over the stellar emission. The remaining flux
is considered to be a ``pure'' AGN contribution.

\subsection{Sub-samples with well-sampled AGN dust emission}

After stellar subtraction (see Section~\ref{sec:contributions}), the
signal to noise ratio (S/N) is calculated for each data-point. Then,
we create a preferred sub-sample fulfilling the criteria that: (a)
there are at least 3 points with S/N $> 2$ and (b) there is no
significant contribution from emission with
$\lambda_{\rm{rest}}>10$\,\mum\ in the passband. This in practice
restricts the wavelength range to
2\,\mum\ $<\lambda_{\rm{rest}}<8$\,\mum\ in any passband. 35 radio
galaxies from the SHzRG sample and 50 sources (22 radio galaxies, 25
quasars and 3 unidentified) from the 3CR sample meet these
criteria. We refer to these two restricted samples as `sSHzRG' and
`s3CR' respectively. Note that only 31 objects have a known core
dominance value in sSHzRG and 18 in the s3CR (11 quasars and 7 radio
galaxies).

\subsection{Empirical approach} \label{sec:emp}

\begin{figure}[t] \centering
  \includegraphics[width=0.5\textwidth]{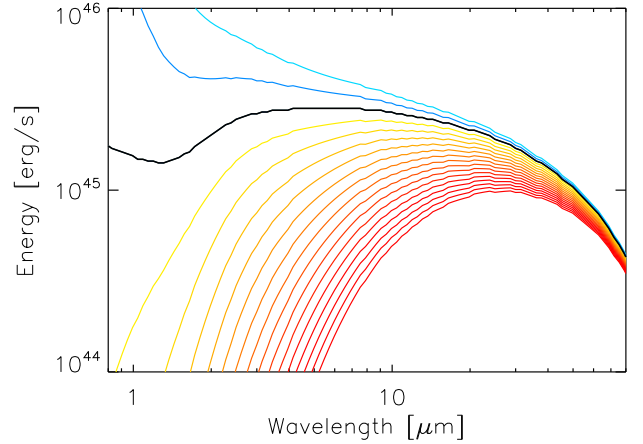}
  \caption{Template SEDs for the empirical approach. The black curve is
    the mean quasar spectrum from \cite{Richards2006}. The
    bluer/redder curves correspond to a increasing extinction (from
    -12 to 90 $A_v$ in steps of 6), using the \citet{Fitzpatrick1999}
    extinction law.}
  \label{fig:emp_torus}
\end{figure}

\begin{figure}[t] \centering
  \includegraphics[width=0.5\textwidth]{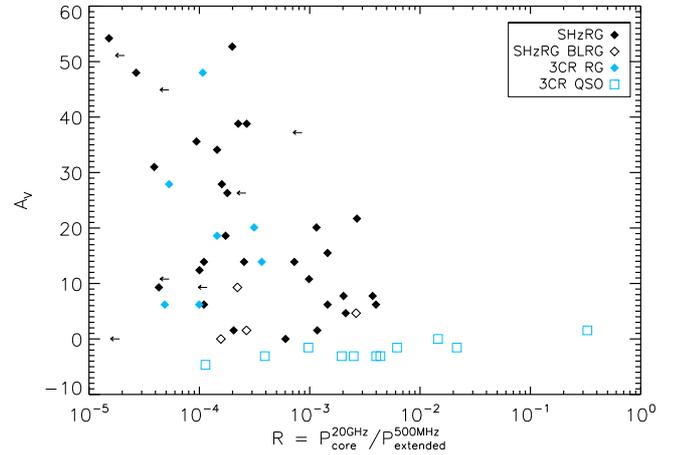}
  \caption{Extinction $A_V$ versus core dominance $R$. Note the
    absence of points in the top right part of the plot.  This figure
    is consistent with Fig. 12 of \cite{Cleary2007}.}
  \label{fig:emp_cf20}
\end{figure}

To first order, increasing the viewing angle of the torus causes it to
act as a varying dust screen: higher inclinations lead to increased
extinction of the hottest dust located in the innermost parts of the
torus.  \citet{Leipski2010} and \citet{Haas2008} have shown that the
mean radio galaxy SED can be approximated by applying an extinction
law to the mean quasar SED.

Following the same approach, we model this dust absorption using the
mean Sloan Digital Sky Survey (SDSS) quasar spectrum
\citep{Richards2006} and a \cite{Fitzpatrick1999} extinction law with
classical Galactic dust properties ($R_V=3.1$,
Fig. \ref{fig:emp_torus}). This law extrapolates Galactic dust
properties to the mid-IR without any specific treatment of the
silicate absorption and emission features around
10\,\mum. This latter approximation is still valid as we are
  focusing on the 2-8\mum\ part of the SED.

Using a standard $\chi^2$ minimization technique, we fit two
parameters (extinction $A_V$ and normalization) to the data for our
sSHzRG and s3CR samples. Tables \ref{tab:results_shzrg} and
\ref{tab:results_3CR} report the resulting values of $A_V$.
Figure \ref{fig:emp_cf20} plots $A_V$ against $R$ for the
  subset of objects with radio data available. Note the apparently
unphysical negative $A_{V}$ values for the majority of s3CR quasars,
which are bluer than the composite SDSS template.  The lack of points
in the upper right part of this plot indicates an absence of
core-dominated objects with high extinction, as expected from the
orientation-based Unified model \citep[e.g.][]{Antonucci1993}.

In order to test the correlation between $A_V$ and core dominance
taking the upper limits in $R$ into account, we use the survival
analysis package within IRAF. The generalized Spearman rank test gives
probabilities of non-correlation of $p=0.009$ and $p=0.0001$  for sSHzRG
and sSHzRG+s3CR samples, respectively. We note that \citet{Cleary2007} report a similar
distribution (their Fig.~12), although with a different core
dominance definition and using the silicate optical depth, 
$\tau_{9.7 \mu m}$ as a mesure of extinction.

\subsection{Torus models} \label{sec:model}

\begin{figure}[t] \centering
  \includegraphics[width=0.5\textwidth]{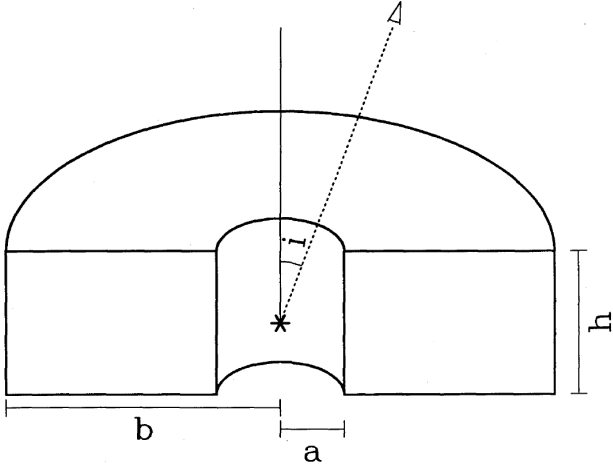}
  \caption{Sketch of the PK model \citep{Pier1992}. The inner radius,
    {\it a}, outer radius, {\it b} and height, {\it h} are
    marked, and the torus is viewed from angle {\it i}, measured as
    shown. The central nucleus is indicated by an asterisk. The
    density is assumed to be constant throughout the torus.}
  \label{fig:pk_torus}
\end{figure}

\begin{table}[t] \centering
\caption{Parameters of the torus model, $\tau_r$ : radial depth ,
  $\tau_z$ vertical depth, $a/h$ ratio of inner radius over height and
  $b/h$ : outer radius. For the score value, see Section~\ref{sec:model}.}  
\begin{tabular}{cccccc}
\hline
 & \multicolumn{4}{c}{Parameters} & \\
\cline{2-5}
model & $\tau_r$ & $\tau_z$ & $a/h$ & $b/h$ & score \\
\hline \hline
$a$  & 0.1  & 0.1  & 0.3 & 1.3  &  59   \\   
$b$  & 1.0  & 0.1  & 0.1 & 10.1 &  70   \\
$c$  & 1.0  & 1.0  & 0.3 & 1.3  &  64   \\
$d$  & 1.0  & 1.0  & 0.2 & 1.2  &  68   \\
$e$  & 1.0  & 1.0  & 0.1 & 1.1  &  54   \\
$f$  & 0.1  & 0.1  & 0.1 & 1.1  &  67   \\
$gg$ & 1.0  & 1.0  & 1.0 & 2.0  &  33   \\
$hh$ & 0.1  & 0.1  & 1.0 & 2.0  &  49   \\
$ii$ & 10.0 & 10.0 & 1.0 & 2.0  &   2   \\
$j$  & 10.0 & 1.0  & 0.1 & 10.1 &  50   \\
$o$  & 1.0  & 0.1  & 1.0 & 11.0 &  28   \\
$p$  & 10.0 & 1.0  & 1.0 & 11.0 &  14   \\
$t$  & 1.0  & 0.1  & 0.3 & 10.3 &  59   \\
$w$  & 10.0 & 1.0  & 0.3 & 10.3 &  44   \\
$y$  & 10.0 & 10.0 & 1.0 & 2.0  &   2   \\
\hline
  \end{tabular}

\label{tab:parameters}
\end{table}

We now aim to reproduce the observed range in extinction assuming a
torus geometry observed at varying angles to the line of sight. The
infrared torus SED has been modelled extensively, assuming structures
with various degrees of complexity.  Despite the evidence for a clumpy
structure \citep{Krolik1988}, the first models to solve
the radiative transfer equations used the continuous density
approximation \citep{Pier1992}.  Later, a treatment of the clumpy
structure was given by \cite{Nenkova2002}. Currently, models
with continuous and clumpy structures are available in the literature
with a range of geometries and using different computational techniques to 
solve the radiative transfer equation. Examples include: continuous models
\citep{Pier1992,Dullemond2005,Granato1994,RowanRobinson1995};
clumpy models 
\citep{vanBemmel2003,Honig2006,Schartmann2005,Fritz2006,Nenkova2008a}
and bi-phased models \citep{Stalevski2012}.

Clumpy models are closer to reality than
continuous models, but the latter require fewer
free parameters and remain valid if the inter-clump distance
is not significantly larger than the clump size. While more
sophisticated modelling may be possible for some individual objects,
our aim is to extract global trends from as many sources as possible in
our sample. Because of the small number and irregular sampling of the
data points in the mid-IR, we therefore opt for the Pier \& Krolik
continuous model (PK model, hereafter).
Trying to derive detailed information on the torus itself is beyond
the scope of this paper, and remains challenging even at lower
redshift \citep[e.g.][]{RamosAlmeida2009,Kishimoto2011}.

The PK model assumes a cylindrical geometry for the torus. The opening angle of the torus is
$\theta=tan^{-1}(2a/h)$, where $a/h$ is the aspect ratio (Fig.~\ref{fig:pk_torus}).
It models the spectrum of radiation from the central point sources as a power
law decreasing from UV to IR. As the index of this power-law $\alpha$
does not have a significant effect for our purpose, we take the
default value $\alpha=1$. We adopt an effective temperature for the
inner edge of the torus of $T_{\rm eff}$=1000\,K as this is the only
temperature available for all the geometries in the template
library from PK92 (see Table~\ref{tab:parameters}).  Indeed, a higher
$T_{\rm eff}$ corresponds to a higher contribution from the hottest
dust (i.e. 1-2\mum) but would not affect the global analysis presented
here.

The inner radius $a$ (see
Fig.~\ref{fig:pk_torus}) is then set by the central source luminosity. For
each inclination $i$, an SED is computed by solving the radiative
transfer equations. The final SED is the sum of the thermal torus
emission and the absorbed quasar component. For a full description of
the model and an example of its application, we refer the reader to
\cite{Pier1992,Pier1993}.

\begin{figure*}[ht] \centering
  \begin{tabular}{ccccc}
    \includegraphics[width=0.18\textwidth]{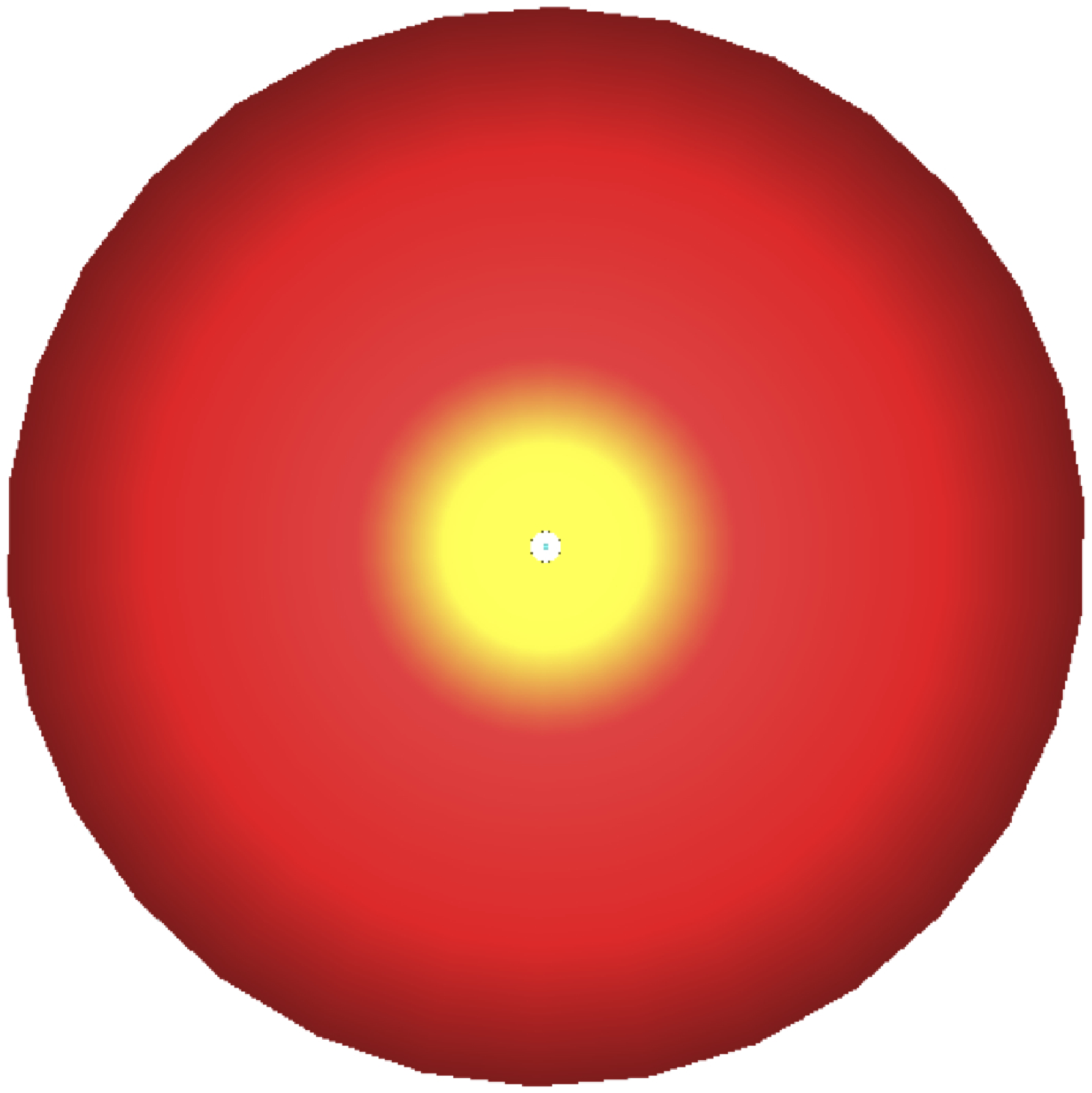} & \includegraphics[width=0.18\textwidth]{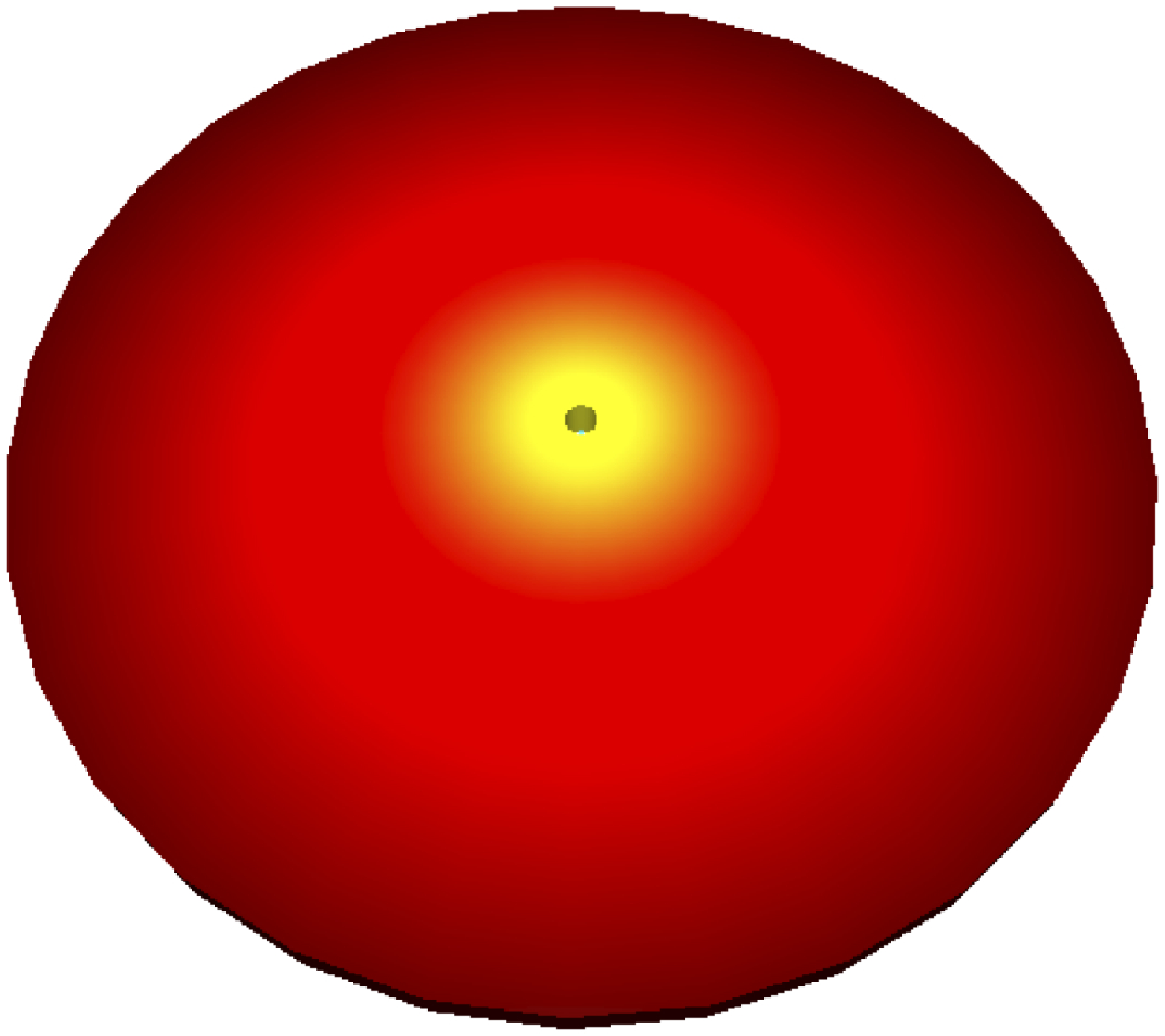} & \includegraphics[width=0.18\textwidth]{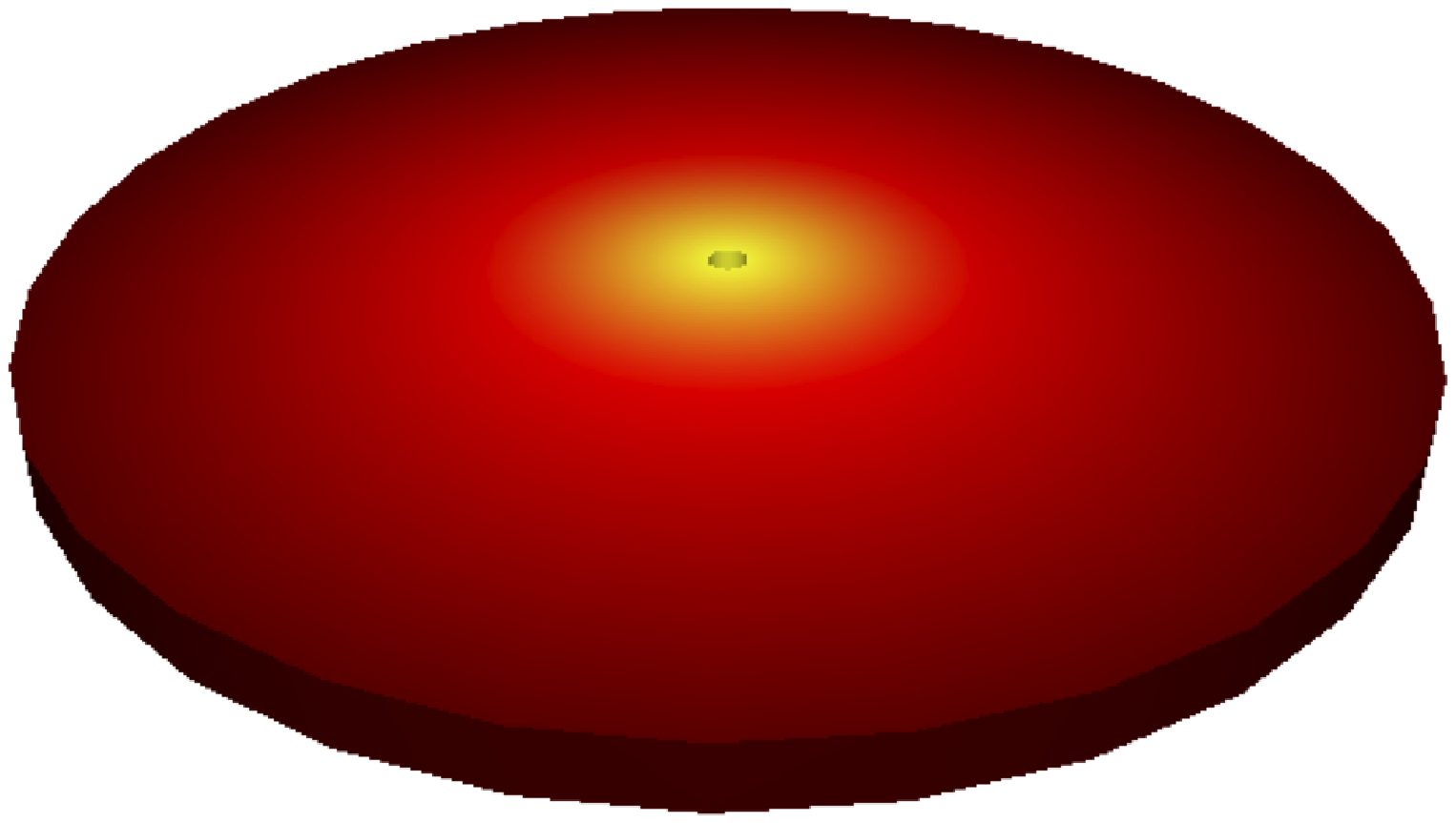} & \includegraphics[width=0.18\textwidth]{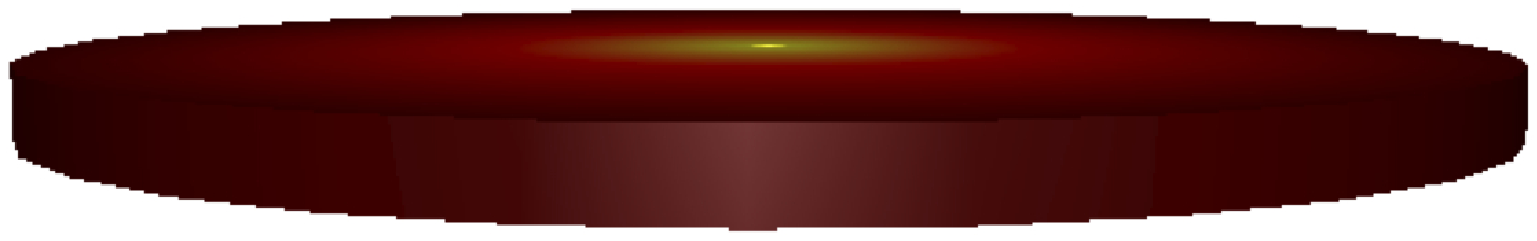} & \includegraphics[width=0.18\textwidth]{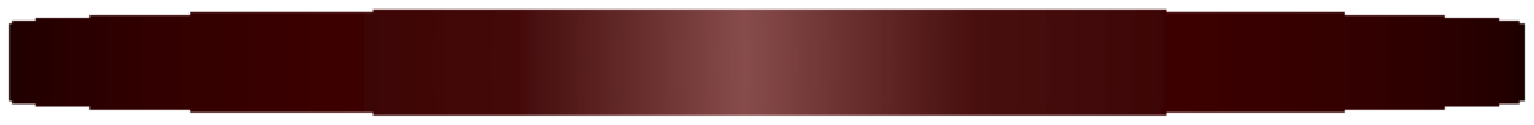} \\
0\degree & 45\degree & 60\degree & 85\degree & 90\degree \\
    \includegraphics[width=0.18\textwidth]{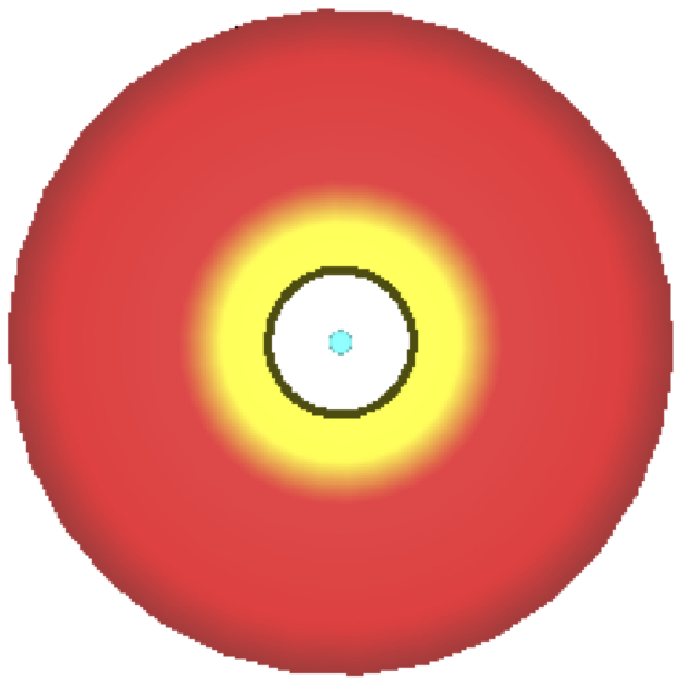}  & \includegraphics[width=0.18\textwidth]{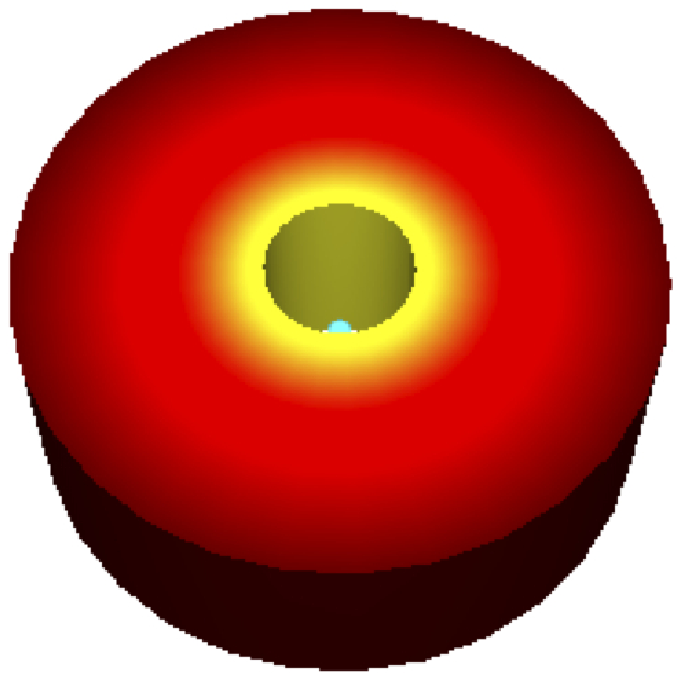}  & \includegraphics[width=0.18\textwidth]{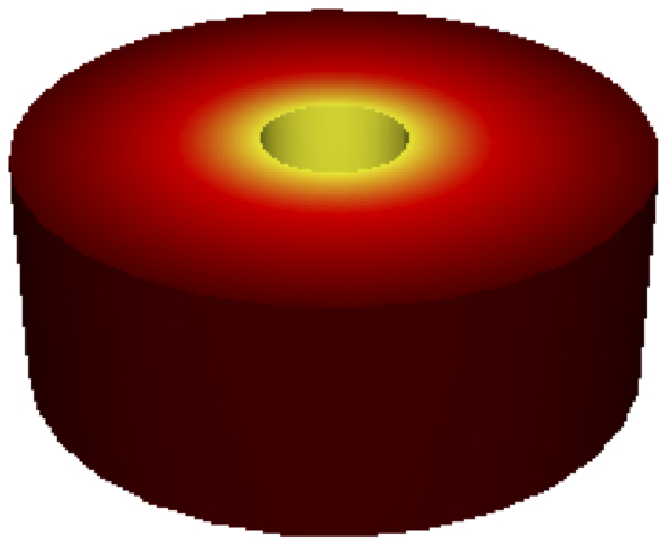} & \includegraphics[width=0.18\textwidth]{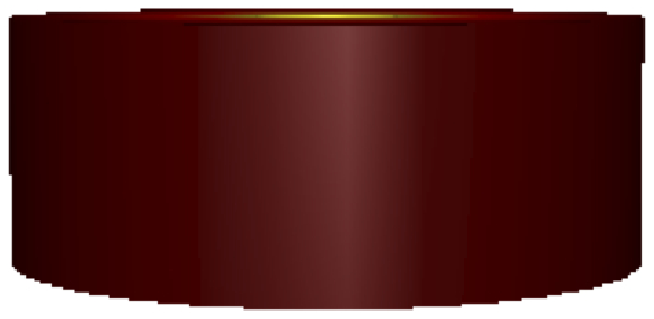} & \includegraphics[width=0.18\textwidth]{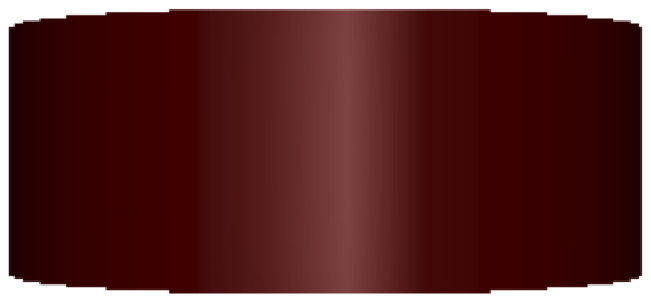} \\
  \end{tabular}
  \caption{Modeled tori for $\tau_r/\tau_z=10$ (top, model $w$, disky) and $\tau_r/\tau_z=1$ (bottom, model $c$, chunky) for increasing inclinations. Note the appearance of the central engine at 45\degree (light blue point), and the complete disappearance of the hottest part for 90\degree (yellow area) due to the cylindrical configuration. This artificially emphasises the difference of the SEDs in the range 85\degree$<$$i$$<$90\degree.
}
  \label{fig:torus}
\end{figure*}

\begin{figure}[t]
\includegraphics[width=0.5\textwidth]{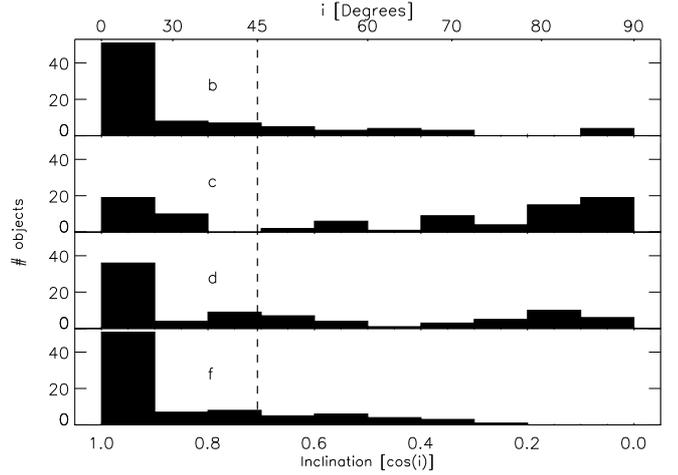} \\
    \caption{The distribution of inclination angles $i$ for the four best-scoring 
      models. The vertical dashed line indicates $i$=45\degree.}
  \label{fig:histo}
\end{figure}

The main goal of this approach is to find a physically meaningful
geometry which can give an adequate description of the entire sample,
to estimate torus inclinations and thereby to test the general configuration 
mentioned in Section~\ref{sec:intro}.  We have considered a
set of representative candidate models, as listed in Table~\ref{tab:parameters}.

Some reasonable arguments and previous observations help us to define
the average geometry and to select among the models in
Table~\ref{tab:parameters}. First, the statistical study of the
relative frequency of type 1 and 2 AGN by \citet{Barthel1989} implies
a torus opening angle $\theta \sim 45^\circ$, which corresponds to
$a/h \sim 0.3$ in the PK models. Second, X-ray observations of nearby
AGN (e.g. NGC~1068) set a lower limit on the torus opacity of $\tau
\ge 1$ \citep{Mulchaey1992}. Only the $c$ and $w$ model geometries
satisfy both criteria.  The main difference between these geometries
is illustrated in Fig. \ref{fig:torus}: they have chunky ($c$) and
disky ($w$) shapes, respectively.

A more objective approach is to give each model a score based on the
goodness of fit averaged over the whole sample.  The fit again uses
$\chi^2$ minimization with two free parameters per fit: the
normalization of each model and the inclination $i$ of the torus.
This score (rightmost column of Table~\ref{tab:parameters}) is the
number of ``reasonable'' fits, i.e.\ those with an associated
probability of $>$95\%. The average score is therefore an indicator of
how well the model fits the samples: higher scores correspond to
models which performed better at modelling the observed SEDs.
Fig. \ref{fig:histo} shows the inclination distribution for the four
geometries with the best scores for both samples (s3CR and
sSHzRG). Models with narrow opening angles (e.g. $b$ and $f$, and $d$
to a lesser extent) artificially bias the distribution towards low
inclinations. We therefore do not use the scores in
Table~\ref{tab:parameters} to simply select between all the torus
geometries; instead, the scores are used to select only between models
$c$ and $w$, which already meet criteria set by the physical arguments
and observations described above. Of these, we select $c$, which has
the higher score (Table~\ref{tab:parameters}).  The inclinations $i$
from model $c$ with their 5\,\mum\ rest frame monochromatic energies
of the torus model $\nu P^{\rm AGN}_{\nu\,(5\mu \rm{m})}$ are reported
in Table~\ref{tab:results_shzrg} and Table~\ref{tab:results_3CR} for
the sSHzRG and s3CR samples, respectively.  Their SEDs are presented
in Fig.~\ref{fig:full_SED} and Fig.~\ref{fig:full_SED_3C}.
Section~\ref{sec:disc} discusses these inclinations in more detail.

\begin{figure*}[ht] \centering
  \includegraphics[width=1\textwidth]{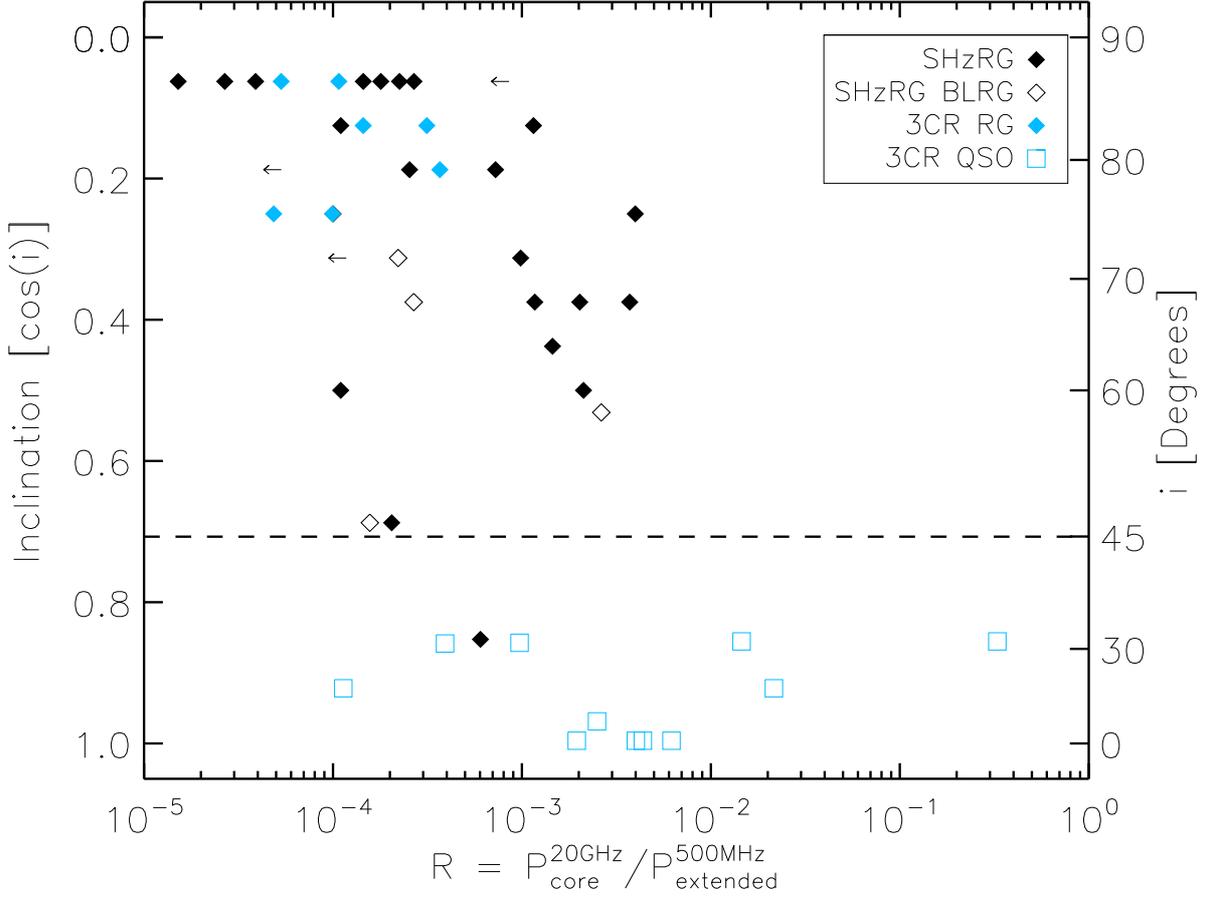}
  \caption{A plot of inclination $i$ against core dominance $R$. The dashed line at i=45\degree
corresponds to the division between radio galaxies and quasars derived by 
\citet{Barthel1989}. Filled diamonds represent the type 2 radio galaxies from
the sSHzRG (black) and s3CR (blue) samples, respectively. Open symbols
represent type 1 objects: sSHzRG broad-line radio galaxies (black
diamonds) and s3CR quasars (blue squares).}
  \label{fig:cf20_vs_inc}
\end{figure*}

Figure \ref{fig:cf20_vs_inc} shows a plot of core dominance $R$
against the inclination $i$ derived from the mid-IR observations. The
distribution of the points in the diagram suggests an anti-correlation
between $i$ and $R$, which is confirmed with a generalized Spearman
rank test for the sSHzRG sample alone ($p=0.011$).  Adding the s3CR
sample causes the correlation to become extremely significant
($p<0.0001$), but note that this is primarily because of the inclusion
of the (less obscured) s3CR quasars, which extends the ranges of both
$R$ and $i$.

We would like to add a cautionary note on using the individual
inclinations at face value. One source of uncertainty stems from the
misalignment of the radio jets with respect to the torus axis
This misalignment adds an intrinsic scatter in the $R$-$i$ relation. A
similar scatter is observed in the plane of the sky through
polarimetric measurements.  
  
The distribution of inclinations within $0^\circ < i \la 45^\circ$ is
inconsistent with the expectations for an isotropically-distributed
parent sample: far too many values are clustered around $i \approx
0$\degree or $i \approx 30$\degree.  In particular, inclinations close
to $\approx 0$ seem suspicious since they would be expected for
blazars which are absent from our sample.  

\begin{table}[t] \centering
  \caption{Results from the modelling of the SHzRG sample from
    Section~\ref{sec:emp} and \ref{sec:model}.}
  \label{tab:results_shzrg}
  \begin{tabular}{lcccc}
    \hline
    Name  & $A_v$ & $i$ & log $\nu P^{\rm AGN}_{\nu\,(5\mu \rm{m})}$ \\ 
          &        & [\degree] & [W]  \\ 
    \hline \hline
    6C~0032+412      &   1.5 &    46 &    38.87 \\ 
    MRC~0037-258     &   1.5 &    67 &    38.11 \\ 
    6C~0058+495      &  37.2 &    86 &    38.04 \\ 
    MRC~0114-211     &  26.3 &    86 &    38.40 \\ 
    MRC~0152-209     &  10.8 &    79 &    38.85 \\ 
    MRC~0156-252     &   4.7 &    57 &    39.05 \\ 
    MRC~0211-256     &   3.1 &    55 &    37.92 \\ 
    TXS~0211-122     &  13.9 &    79 &    39.13 \\ 
    MRC~0324-228     &  26.3 &    86 &    38.32 \\ 
    MRC~0350-279     &   6.2 &    59 &    37.99 \\ 
    MRC~0406-244     &  38.8 &    86 &    38.74 \\ 
    PKS~0529-549     &  34.1 &    86 &    38.59 \\ 
    USS~0828+193     &   7.8 &    67 &    39.31 \\ 
    B2~0902+34       &   7.8 &    67 &    38.61 \\ 
    6CE~0905+3955    &  38.8 &    86 &    38.64 \\ 
    3C~239           &  31.0 &    86 &    38.43 \\ 
    MRC~1017-220     &   4.7 &    53 &    38.47 \\ 
    3C~257           &   9.3 &    71 &    39.04 \\ 
    WN~J1123+3141    &  10.8 &    71 &    39.27 \\ 
    PKS~1138-262     &   1.5 &    67 &    39.32 \\ 
    6C~1232+39       &  12.4 &    75 &    38.70 \\ 
    4C~24.28         &  13.9 &    82 &    38.90 \\ 
    USS~1410-001     &   6.2 &    64 &    38.83 \\ 
    USS~1558-003     &   9.3 &    71 &    38.91 \\ 
    LBDS~53W002      &  12.4 &    79 &    38.52 \\ 
    3C~356.0         &  48.0 &    86 &    38.07 \\ 
    3C~368           &  54.2 &    86 &    37.91 \\ 
    7C~1805+6332     &  13.9 &    79 &    38.06 \\ 
    TXS~J1908+7220   &   4.7 &    59 &    39.43 \\ 
    TN~J2007-1316    &   0.0 &    31 &    38.83 \\ 
    MRC~2025-218     &   0.0 &    46 &    38.25 \\ 
    4C~23.56         &  20.1 &    82 &    39.34 \\ 
    MRC~2224-273     &   4.7 &    79 &    38.30 \\ 
    B3~J2330+3927    &   6.2 &    75 &    39.35 \\ 
    3C~470           &  34.1 &    86 &    38.49 \\ 
    \hline
  \end{tabular}                                       

\end{table}

\begin{table}[t] \centering
  \caption{Results for the s3CR sample, from Section~\ref{sec:emp} and
    \ref{sec:model}. Results for the s3CR sample. Types: Q=quasar,
    G=radio galaxy and N=unidentified.}
  \label{tab:results_3CR}
  \begin{tabular}{lccccc}
    \hline
    Name & Type & $A_v$ & $i$ & log $\nu P^{\rm AGN}_{\nu\,(5\mu \rm{m})}$ \\ 
    & & & [\degree] & [W]   \\ 
    \hline \hline
    3C~002      & Q &  -1.5 &     5 &    38.25  \\ 
    3C~009     & Q &  -3.1 &    30 &    39.36  \\ 
    3C~013      & G &   6.2 &    67 &    37.92  \\ 
    3C~014      & Q &  -1.5 &    31 &    39.27  \\ 
    3C~036      & G &   1.5 &    59 &    38.07  \\ 
    3C~043      & Q &  -3.1 &     7 &    38.45  \\ 
    3C~065     & G &   6.2 &    75 &    38.08  \\ 
    3C~068.1   & Q &  -4.7 &    22 &    38.93  \\ 
    3C~068.2    & G &  68.2 &    89 &    38.35  \\ 
    3C~124      & G &  38.8 &    86 &    38.09  \\ 
    3C~173      & N & -10.8 &    12 &    37.79  \\ 
    3C~181      & Q &  -1.5 &     7 &    38.68  \\ 
    3C~186      & Q &  -3.1 &     5 &    38.73  \\ 
    3C~190      & Q &  -4.7 &     5 &    38.81  \\ 
    3C~191      & Q &  -1.5 &    14 &    39.00  \\ 
    3C~204     & Q &  -3.1 &     5 &    38.87  \\ 
    3C~205     & Q &  -3.1 &    14 &    39.37  \\ 
    3C~208     & Q &  -3.1 &     5 &    38.68  \\ 
    3C~208.1    & N &  -1.5 &    11 &    38.26  \\ 
    3C~210      & G &  21.7 &    82 &    38.56  \\ 
    3C~212     & Q &   0.0 &    31 &    38.79  \\ 
    3C~220.2    & Q &   0.0 &     5 &    38.76  \\ 
    3C~225A     & G &  15.5 &    82 &    37.89  \\ 
    3C~239     & G &  27.9 &    86 &    38.33  \\ 
    3C~241     & G &  13.9 &    79 &    38.03  \\ 
    3C~245     & Q &   1.5 &    31 &    38.92  \\ 
    3C~252     & G &  18.6 &    82 &    38.59  \\ 
    3C~257      & G &   6.2 &    67 &    38.96  \\ 
    3C~267     & G &  20.1 &    82 &    38.38  \\ 
    3C~268.4   & Q &  -1.5 &     5 &    39.21  \\ 
    3C~270.1   & Q &  -1.5 &    22 &    39.02  \\ 
    3C~280.1    & Q &  -3.1 &    30 &    38.84  \\ 
    3C~287      & Q &  -1.5 &     5 &    38.61  \\ 
    3C~297      & N & -10.8 &     7 &    37.92  \\ 
    3C~298      & Q &  -1.5 &    31 &    39.37  \\ 
    3C~300.1    & G &  29.4 &    86 &    37.86  \\ 
    3C~305.1    & G &  17.0 &    82 &    38.44  \\ 
    3C~318     & Q &  -3.1 &     5 &    38.77  \\ 
    3C~324      & G &  41.8 &    86 &    38.30  \\ 
    3C~325     & G &   6.2 &    75 &    38.46  \\ 
    3C~356     & G &  48.0 &    86 &    38.05  \\ 
    3C~368      & G &  31.0 &    86 &    38.01  \\
    3C~418      & Q &  -1.5 &    22 &    39.39  \\ 
    3C~432     & Q &  -1.5 &    30 &    39.01  \\ 
    3C~437      & G &  38.8 &    86 &    37.79  \\ 
    3C~454.1    & G &  23.2 &    82 &    38.32  \\ 
    3C~454.0    & Q &  -1.5 &    31 &    38.97  \\ 
    3C~469.1    & G &  15.5 &    79 &    38.75  \\ 
    3C~470      & G &  37.2 &    86 &    38.39  \\ 
    4C~16.49    & Q &  -4.7 &    30 &    38.82  \\ 
    \hline                                            
  \end{tabular}                                       
  
\end{table}

\section{Discussion} \label{sec:disc}

\subsection{From extinction to inclination}

\begin{figure}[ht] \centering
  \includegraphics[width=0.5\textwidth]{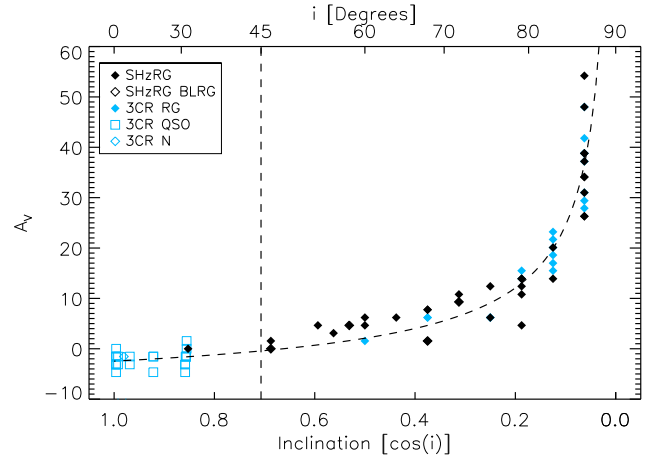}
  \caption{Inclination $i$ obtained from the torus model plotted
    against extinction $A_V$ from the empirical approach. The dashed vertical line is plotted at $i = 45^\circ$,
    and the dashed curve corresponds to Eq. \ref{eq:av_inc}.}
  \label{fig:inc_vs_av}
\end{figure}

Both the empirical approach (Section~\ref{sec:emp}) and the torus model
(Section~\ref{sec:model}) successfully fit the hot dust emission. We
now check the consistency between these two methods by comparing the
empirical extinction and the orientation from our best torus model.
As shown in Fig. \ref{fig:inc_vs_av}, these two parameters are tightly
correlated.  Their relation is well described by the following function:

\begin{equation}
A_v = 8.43 (\cos i)^{-0.62} - 10.89
\label{eq:av_inc}
\end{equation}

Since the link between extinction and inclination is essentially determined
by the geometry of the torus, the small scatter provides support for
our choice of a single torus model ($c$) for the entire sample
(sSHzRG and s3CR).

\subsection{Isolating the torus emission}

\label{sec:limit_torus}
\subsubsection{Stellar and extended warm dust emission}
 
Here, we investigate the impact of the assumptions made in
Section~\ref{sec:contributions} to isolate the hot dust component from
all other contributions in the SED. As we assume that the main stellar
contribution is dominated by an evolved population ($>$500~Myr old),
all galaxies of our sample have already formed the bulk of their
stellar mass (see S07). The SED of such a population has a
characteristic shape for $\lambda_{\rm{rest}} \ga 1.6$\,\mum, as
illustrated by the stellar SEDs plotted in
Fig.~\ref{fig:full_SED}. This shape is relatively independent of age,
since the light is dominated by low-mass stars. The assumption of a
single high formation redshift therefore has minor consequences for
the separation of stellar and hot dust components.

A more important aspect is the SED coverage, which varies systematically 
with redshift. The number of data
points measuring purely stellar emission depends on the sampling
beyond the stellar bump ($\sim$1.6\,$\mu$m) and the relative hot dust
contribution in the redmost stellar dominated IRAC channel. This may,
for instance, cause an overestimate of stellar emission for $z$$<$2 and
lead to a higher inclination and extinction estimate. However, we do not observe
such a redshift bias in the sSHzRG sample. On the other hand, the 
s3CR sample, which contains only $z$$<$2 objects, may be more affected by
this effect. We indeed find that s3CR type 2 sources mostly have a high
derived inclination ($70$\degree$<$$i$$<$$90$\degree).  Adding near-IR
photometry could solve this problem, but this is not available for most of
this sample.

Similarly, the reddest end of the SED may be affected by extended warm
dust emission. To minimize this problem, we cut at 8~$\mu$m as
explained in Section~\ref{sec:contributions}. Nevertheless, in some
cases, a rising spectrum from $\lambda_{\rm{obs}}$=16~$\mu$m
to $\lambda_{\rm{obs}}$$\ge$24\,$\mu$m (e.g.\ MRC~0152$-$209, MRC~0406$-$244) suggests that even
at $\lambda_{\rm{rest}} = 8$\,\mum, a contribution from such a
component cannot be excluded.

Indeed, we note that most of the galaxies for which the fit converged
to the highest inclination (i.e. the reddest model) may be affected
(MRC~0211$-$256, MRC~0324$-$228, 6CE~0905+3955, 3C~356, 3C~368,
7C~1805+6332, 3C~470). This could partially explain the clustering of
points at $i = 86^\circ$. To test the influence of this
contamination, we reduced by a factor of three the flux in the redmost
filter for these seven sources. The effect is small: 
the inclination only decreases in three of them, with a maximum change of 7\degree.  To quantify
this effect, a tightly sampled SED over the wavelength range 
8\,$\mu$m$<$$\lambda_{\rm{rest}}$$\lesssim$20\,$\mu$m would be required.
Alternatively, we would have to isolate the torus spatially from 
more extended emission. This has been attempted for nearby type 2 AGN
\citep{vanderWolk2010} where such an extended component has indeed 
been identified at $\lambda_{\rm{rest}}$$=$12~$\mu$m. Both types of
observation are beyond the reach of current facilities for our samples.

\subsubsection{Additional complications for type 1 AGN}\label{sec:comp_disc}

In addition to the difficulty of estimating the inclination of an
individual type 1 AGN, as described in Section \ref{sec:emp} and
\ref{sec:model}, there are two further complications: at the blue end
of the spectrum, transmitted quasar continuum emission can still
outshine both the host galaxy and the hot dust emission from the
torus. The PK model already includes the power-law emission from the
AGN. As explained in Section~\ref{sec:contributions}, we have not
included a contribution from the host galaxy in fitting type~1 AGN. As
a test, we have included this component in the same way as in the
type~2 AGN, i.e. assuming that the bluest IRAC point is 100\% stellar
emission. In all 11 s3CR quasars, we find that adding this stellar
component severely degrades the quality of the fit, but even in this
extreme case, the $R$-$i$ correlation still remains significant at the
$p$=0.002 level. It is still possible that a smaller stellar component
has a smaller contribution on the blue end. Subtracting such a partial
stellar population would slightly increase the inclinations found.

On the red end of the torus emission, some sources may also have a
contribution from core synchrotron emission \citep{Cleary2007}. This
effect is similar to the extended dust component described above, but
given the relative small core dominance in our samples
(Tables~\ref{tab:full_radio} and \ref{tab:full_radio_3C}), we do not
expect such contributions to have a significant effect.

Despite the above complications, we do find that all quasars have
$i$$<$45\degree (see Fig. \ref{fig:cf20_vs_inc}) as expected from
unified models.

\subsection{The relative orientations of jets and tori}
\label{sec:limit_obs}

Our mid-IR SED fitting provides the first estimates of AGN torus
inclinations at high redshift. The significant correlation between the
radio core dominance and the torus inclination
(Fig.~\ref{fig:cf20_vs_inc}) implies that the radio jets are indeed
generally orthogonal to the equatorial plane of the torus.
Orthogonality in projection on the plane of the sky had already been
inferred from polarimetric measurements
\citep[e.g.][]{Vernet2001}. Our result confirms this geometry 
in an orthogonal plane containing the line of sight. 
Together, these observations provide further evidence in favour of the
orientation-based unified scheme for AGN.

In the choice of our geometrical model of the torus, we have assumed a
half opening angle $\theta$=45\degree, as constrained by statistical
studies of the relative numbers of type 1 and 2 AGN in a
radio-selected sample \citep{Barthel1989}. Other studies have
suggested slightly different values of $\theta$ \citep[e.g.][ find
  $\theta$=56\degree]{Willott2000}. The torus opening angle may well
vary from object to object. This will produce a natural scatter in the
$R$-$i$ relationship. In addition, the transition between direct
and obscured view is not expected to be sharp due to the clumpiness of
the obscuring material. Two classes of radio galaxies are expected to
have inclinations near this transition region. First, radio galaxies
with observed broad H$\alpha$ emission
\citep{Humphrey2008,Nesvadba2011}, are indeed found at
45\degree$<$$i$$<$75\degree\ (open diamonds in
Fig.~\ref{fig:cf20_vs_inc}). Second, the two radio galaxies with
observed broad absorption lines \citep[MRC~2025$-$218 and
  TXS~J1908+7220; ][]{Humphrey2008,DeBreuck2001}, have $i$=46\degree
and $i$=59\degree, respectively, consistent with the expectation to
observe them near the grazing line of sight along the torus
\citep[e.g.][]{Ogle1999}.

In addition, the torus model adopted to be most representative for the
complete sample may not be the optimal choice for a given object. This
is particularly true for type~1 AGN, where we cannot
constrain the orientation within the range where we have a direct view
of the central source, i.e. $0^\circ < i \la 45^\circ$ (see Sections
\ref{sec:emp}, \ref{sec:model} and \ref{sec:comp_disc}).
Given these caveats, we deliberately do not quote uncertainties on
$i$. In Section~\ref{sec:core_prominence} (below), we use only the
inclinations for the Type 2 galaxies.

\subsection{Constraints on the core Lorentz factor}
\label{sec:core_prominence}

\begin{figure}[ht]
\includegraphics[width=8.5cm]{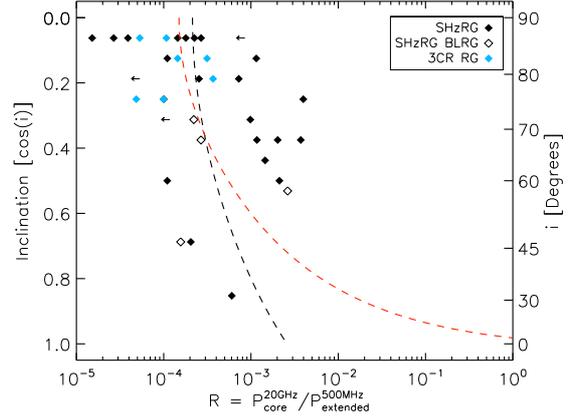}
\caption[]{Core dominance $R$ versus inclination $i$. The black dashed
  line is the best fit for all type 2 radio galaxies, $\Gamma = 1.3$. The red
  dashed line represents the case $\Gamma = 5$.}
\label{fig:lorentz_factor}
\end{figure}

Many lines of evidence lead to the conclusion that the jets in
radio-loud AGN are relativistic on pc scales, with Lorentz factors of
at least 2 and perhaps as high as 50.  The highly significant
correlation between $R$ and $i$ (Fig. \ref{fig:cf20_vs_inc}) arises
naturally if the core radio emission comes from the base of a
relativistic jet.  If the intrinsic ratio between core and extended
emission is constant, then we can use the core dominance for simple
assumptions about the jet flow.
For anti-parallel, identical jets with velocity $\beta c$ (Lorentz factor $\Gamma
= (1-\beta^2)^{-1/2}$) and spectral index
$\alpha$ at an angle $i$ to the line of sight, the predicted value of $R$
is:
\[
R = R_0 [(1-\beta\cos i)^{-(2+\alpha)} + (1+\beta\cos i)^{-(2+\alpha)}]
\]
where $2R_0$ is the value of $R$ when the jets are in the plane of the sky
($i$=90\degree). 

Fig.~\ref{fig:lorentz_factor} plots $R$ against $i$ for all of the
type 2 radio galaxies (sSHzRG and s3CR) with measured $R$ values. For a
spectral index $\alpha$=$-$0.9 (the median value of both samples), the
best-fitting Lorentz factor derived from an unweighted least-squares
fit to the relation between $\log R$ and $i$ is $\Gamma = 1.3$, with
$R_0 = 10^{-4}$. As our type 2 samples are restricted to inclinations
$i > 30^\circ$, the Lorentz factor is not well constrained (the
dependence of $R$ on $i$ is quite flat). Only very low values of
$\Gamma$ are firmly excluded, and adequate fits can be found for any
$\Gamma \ga 1.3$ (for example, see Fig. \ref{fig:lorentz_factor},
where the red dashed line represents $\Gamma = 5$). The dispersion in
$\log R$ for the best-fitting $\Gamma$ is 0.59.  A stronger constraint
on $\Gamma$ could be obtained from the $R$$-$$i$ relation for a
comparable sample of broad-lined objects (or just from the median
value of $R$ if inclinations are not available).  As noted earlier
(Section~\ref{sec:limit_obs}), we do not believe that our estimates of
inclination for the s3CR quasars are reliable enough for this purpose.

One potential source of additional scatter in the relation between
core dominance and inclination is any dependence of the intrinsic
core/extended ratio on luminosity. Such a dependence was found by
\citet{Giovannini1988} for a sample of sources covering a wide range
in luminosity. We have checked for this effect over the restricted
luminosity range of the present type~2 sample, but find no detectable
correlation between core dominance and luminosity
(Fig. \ref{fig:l500_vs_R}).

The most appropriate comparison study in the literature is a Bayesian
analysis of the core dominance distribution for a sample of powerful
FR\,II radio sources including both broad and narrow-line objects at
$z$$<$1 \citep{Mullin2009}. This gave $\Gamma = 10^{+3}_{-7}$ with a
dispersion of 0.62 in $\log R$ assuming a single value of $\Gamma$ and
a dispersion in intrinsic core dominance.  A value of $\Gamma = 10$,
would also be fully consistent with the present data.  The jet/counter-jet
ratio on kpc scales provides an independent estimate of
orientation. This can be measured adequately only for nearby,
low-luminosity radio galaxies, for which \citet{Laing1999} found
$\Gamma = 2.4$ and a dispersion of 0.45 in $\log R$.

Various lines of argument suggest that parsec-scale jets have velocity
gradients and that the single-velocity model used here is
oversimplified. If this is the case, then our analysis is sensitive to
the {\em slowest-moving} component which makes a significant
contribution to the rest-frame emission. Faster material will dominate
only at smaller angles to the line of sight.

We conclude that the relation between core dominance and inclination
for the present sample of high-redshift radio galaxies is fully
consistent with their cores being the bases of relativistic jets, as
inferred for radio-loud AGN at lower redshift, but that we can set
only a lower bound on the Lorentz factor, $\Gamma \ga 1.3$ without
additional data.


\section{Summary and conclusions}

We have examined the relative inclinations of jets and tori in
powerful radio-loud AGN at $z > 1$. To estimate the orientation of the
radio jets, we have introduced a new definition of the radio core
dominance $R$ as the ratio between high-frequency, anisotropic core
emission and low-frequency, isotropic extended emission, both measured
in the rest frame. To estimate the orientation of the dusty torus, we
fit optically-thick radiative transfer models to existing and new {\it
  Spitzer} 3.6--24\,$\mu$m photometry. To isolate the hot dust
emission in the torus from non-thermal and host galaxy contributions,
we subtract an evolved stellar population in the type 2 sources, and
restrict to the 2\,$\mu$m$<$$\lambda_{\rm rest}$$<$8\,$\mu$m region.

Using this method, we derive radio core dominance and torus
inclination values for 42 type~2 and 11 type~1 AGN. This mid-IR
determination of the inclination allows us to draw the following
conclusions:

\begin{enumerate}
\item The significant correlation between $R$ and $i$ implies that 
  radio jets are indeed approximately orthogonal to the equatorial plane of the
  torus as predicted by the orientation-based AGN unified
  scheme. A similar result in the plane of the sky has been reported
  using polarimetric measurements, but our results provide additional
  evidence in the complementary direction, i.e. along the line of
  sight.
\item The $R$--$i$ correlation is consistent with the radio cores
  being the bases of relativistic jets with Lorentz factors
  $\Gamma \ga 1.3$.
\item Assuming our torus geometry is representative, we can estimate
  inclinations for larger samples of type~2 AGN using the relation
  (eq.~\ref{eq:av_inc}) between inclination and the obscuration $A_V$
  as derived from a simple reddening of a mean type~1 template.
\end{enumerate}

The present study characterises the anisotropic component of the dust
IR SED in type~2 AGN. The remaining more isotropic dust emission,
dominating at longer wavelenghts, is powered by a combination of AGN
and starbursts. Our new \textit{Projet HeRG\'E} aims to disentangle
these two contributions by adding {\it Herschel} 70--500\,$\mu$m
photometry \citep{Seymour2012} to our sample of 71 radio galaxies at
1$<$$z$$<$5.2.

\begin{acknowledgements} 

We thank the referee for their careful reading of the manuscript and
their constructive comments. The authors thank K. Blundell for useful
comments and her participation in the improvement of this paper and
C. Leipski for reanalysing some of the 3CR photometry. Nick Seymour is
the recipient of an Australian Research Council Future Fellowship.
This work is based on observations made with the \textit{Spitzer Space
  Telescope}, which is operated by the Jet Propulsion Laboratory,
California Institute of Technology under contract with NASA.

\end{acknowledgements}

\bibliographystyle{aa.bst}
\bibliography{Library}

\newpage

\begin{figure*}[ht]
  \begin{center}
  \begin{tabular}{r@{}c@{}l} 
\includegraphics[height=42mm,trim= 0 39 15  0,clip=true]{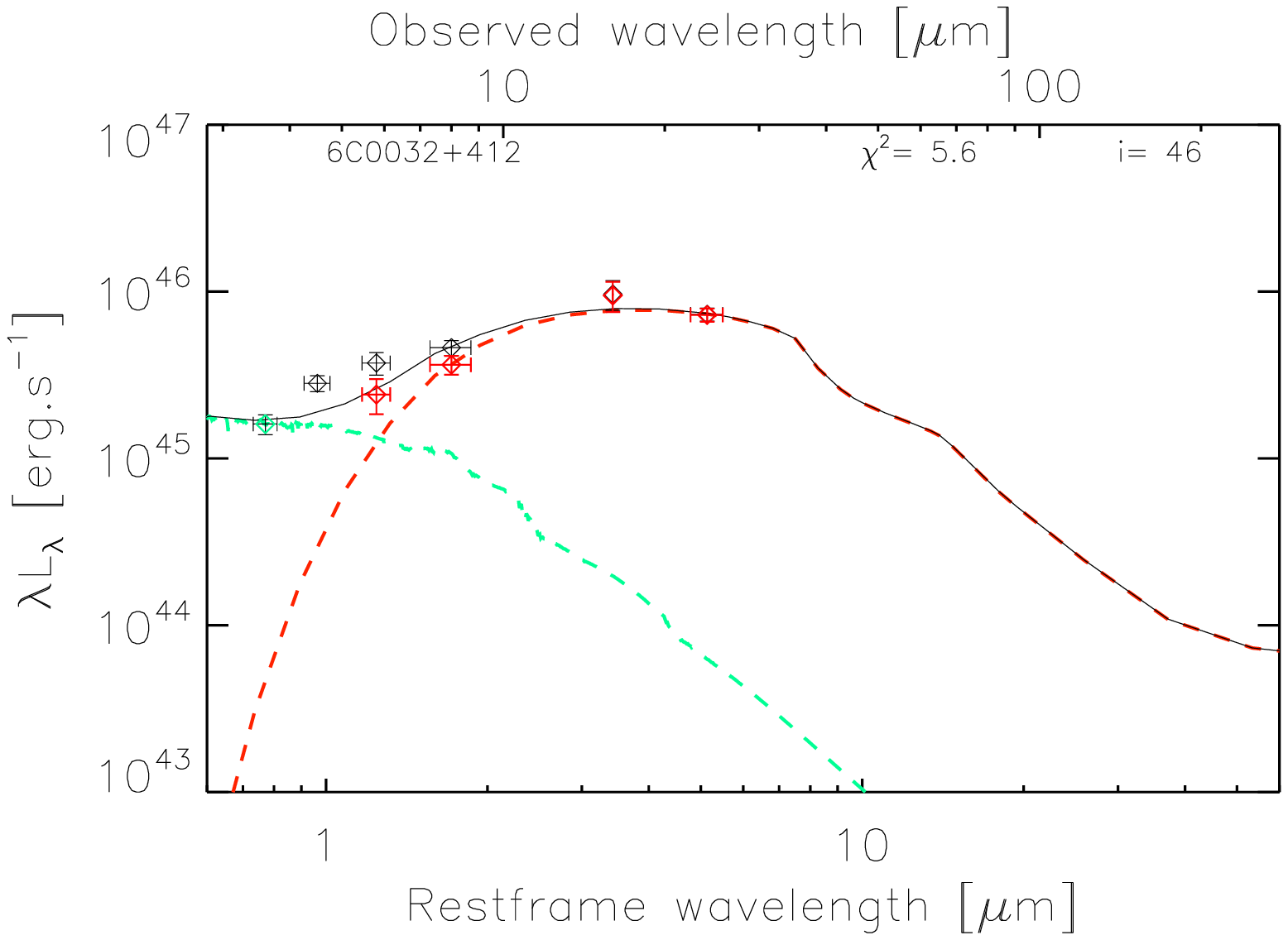}  &
\includegraphics[height=42mm,trim=95 39 15  0,clip=true]{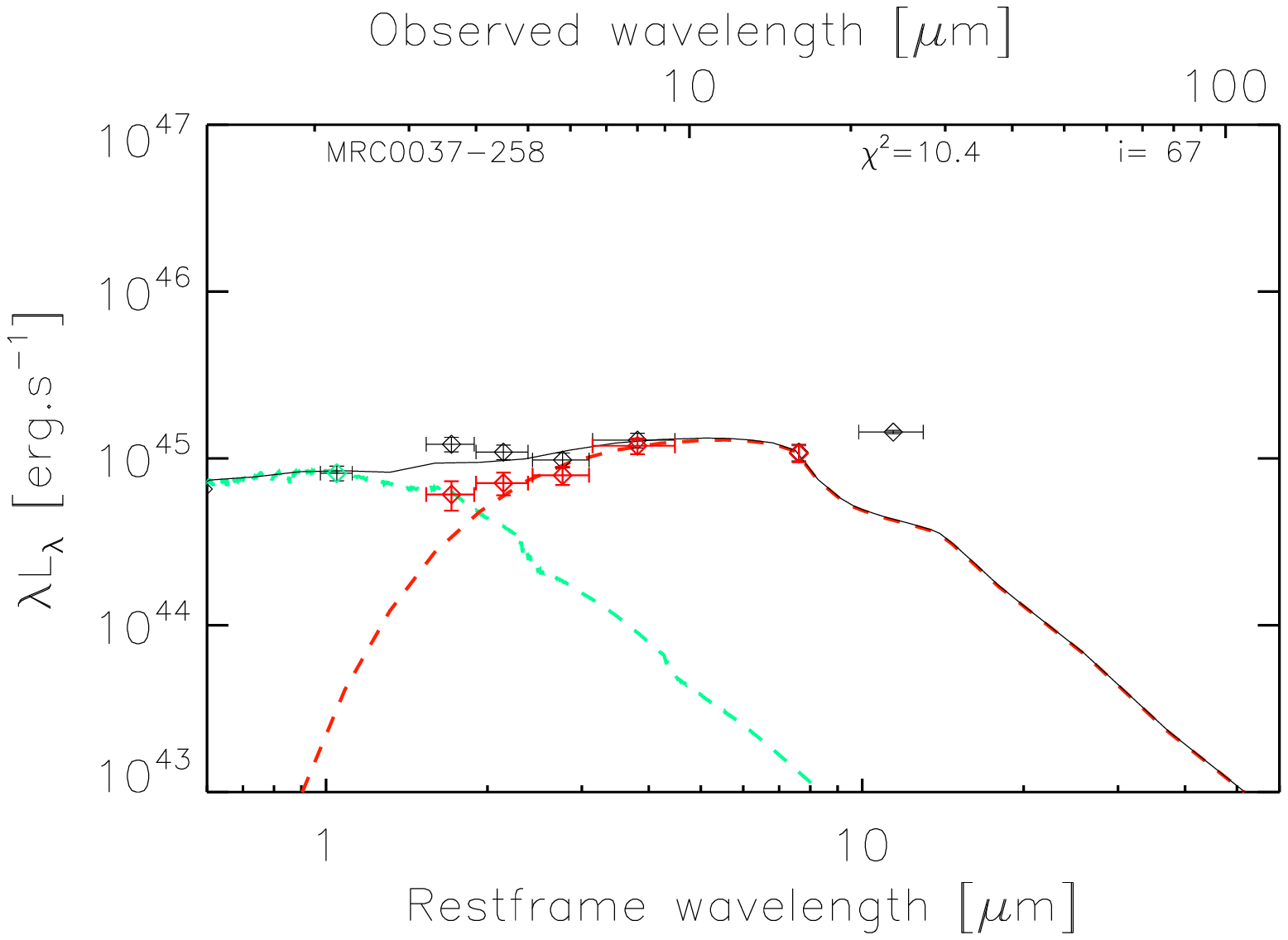} &
\includegraphics[height=42mm,trim=95 39 15  0,clip=true]{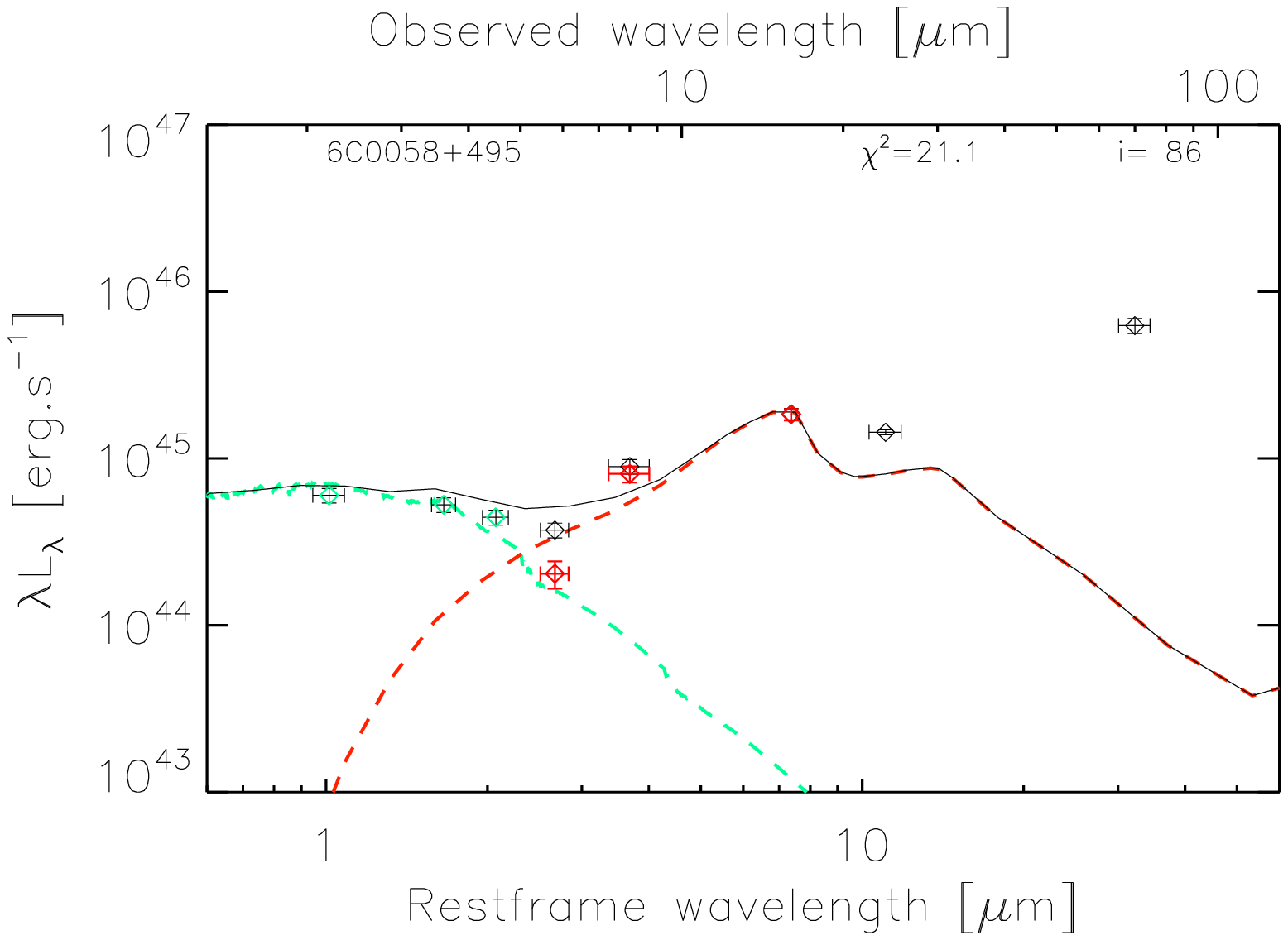} \\
\includegraphics[height=37mm,trim= 0 39 15 41,clip=true]{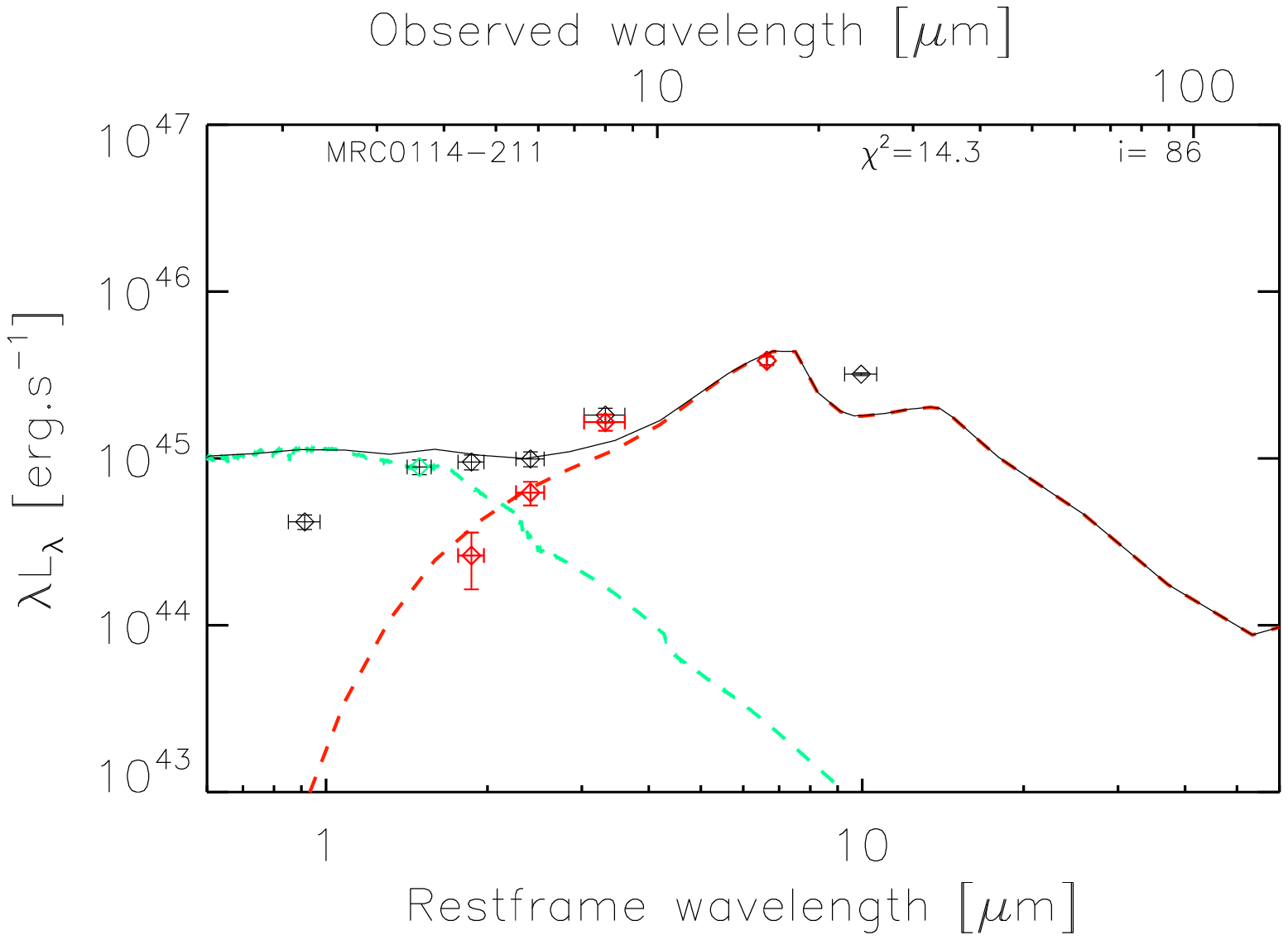} &
\includegraphics[height=37mm,trim=95 39 15 41,clip=true]{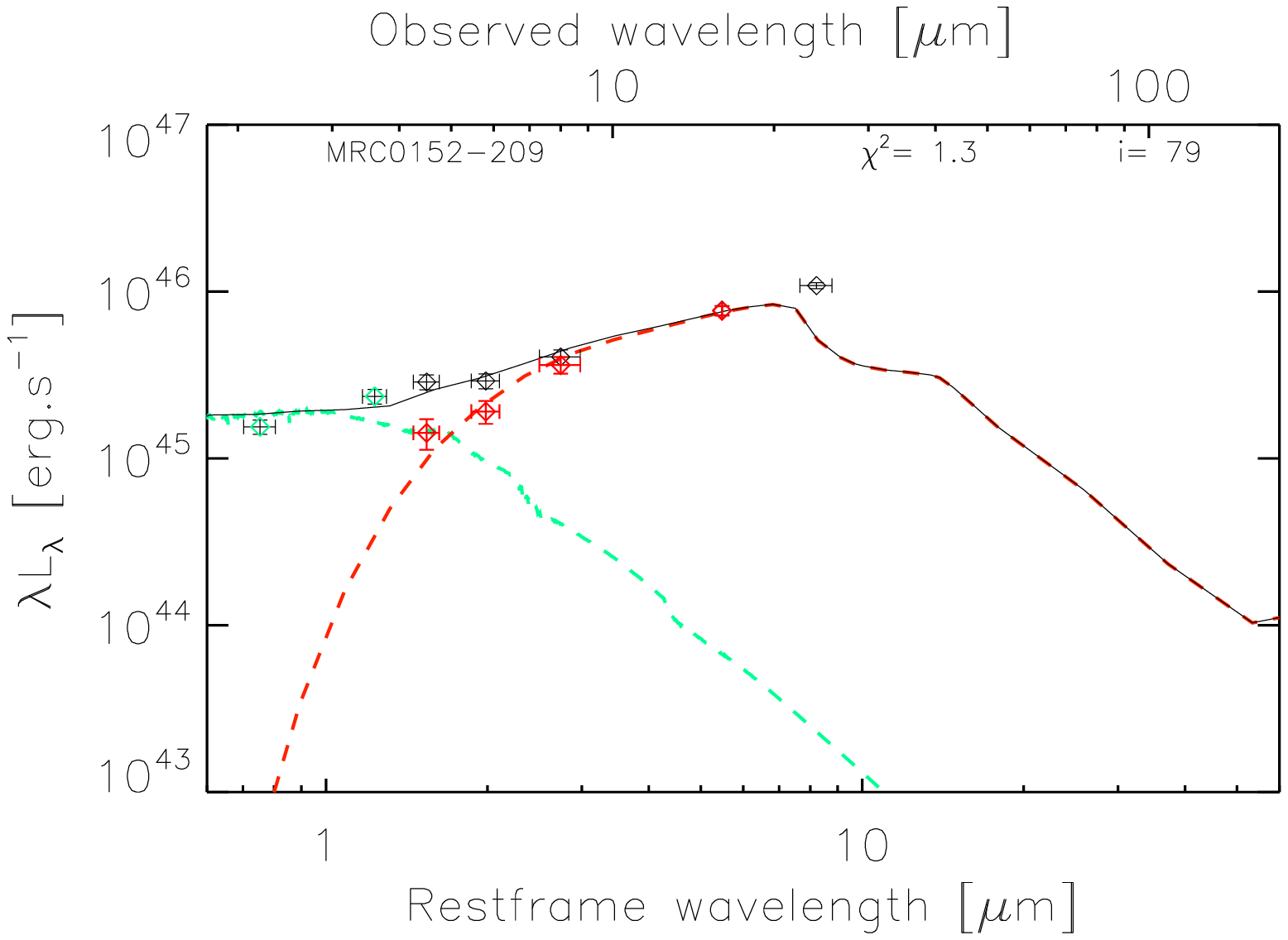} &
\includegraphics[height=37mm,trim=95 39 15 41,clip=true]{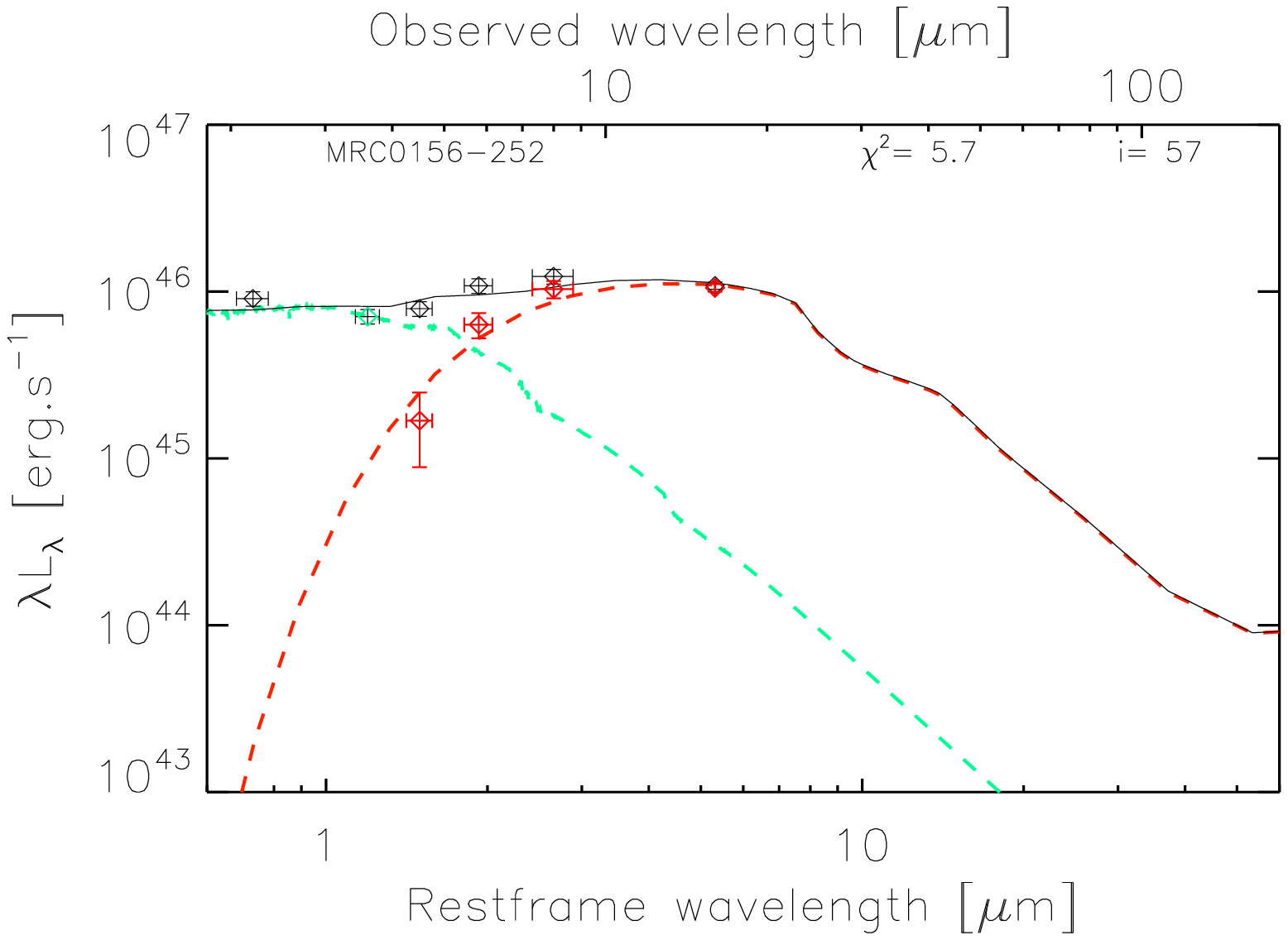} \\
\includegraphics[height=37mm,trim= 0 39 15 41,clip=true]{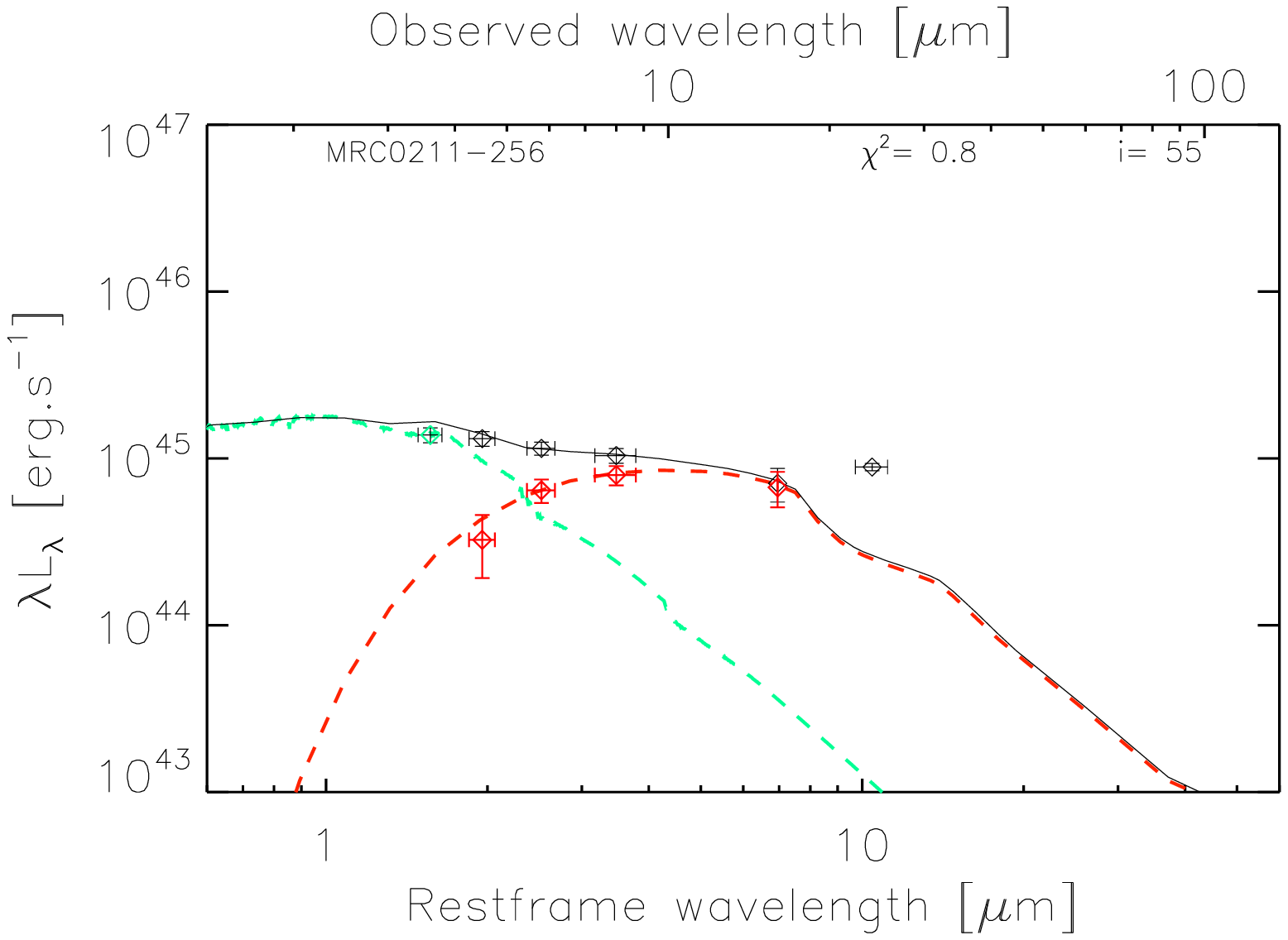} &
\includegraphics[height=37mm,trim=95 39 15 41,clip=true]{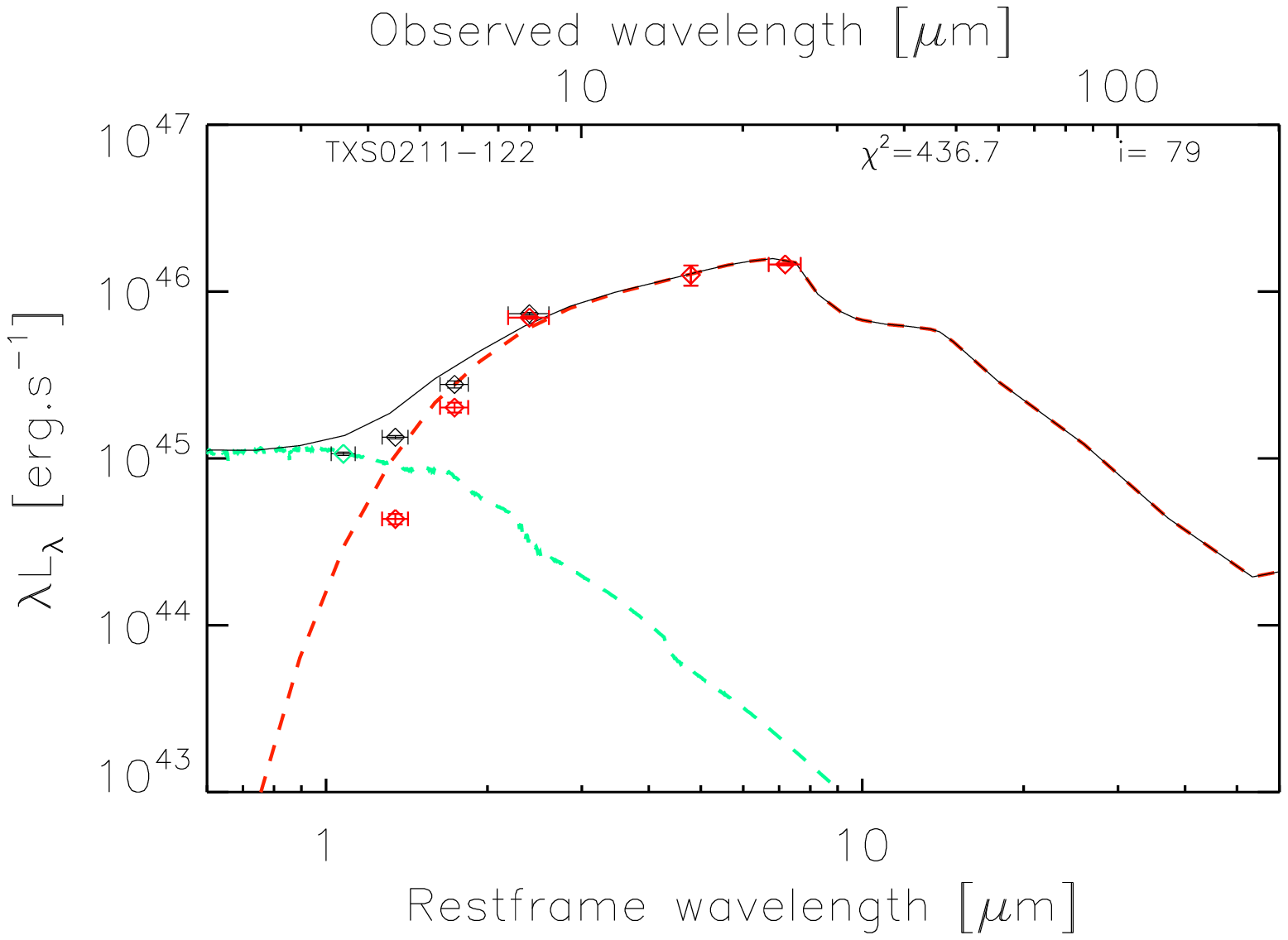} &
\includegraphics[height=37mm,trim=95 39 15 41,clip=true]{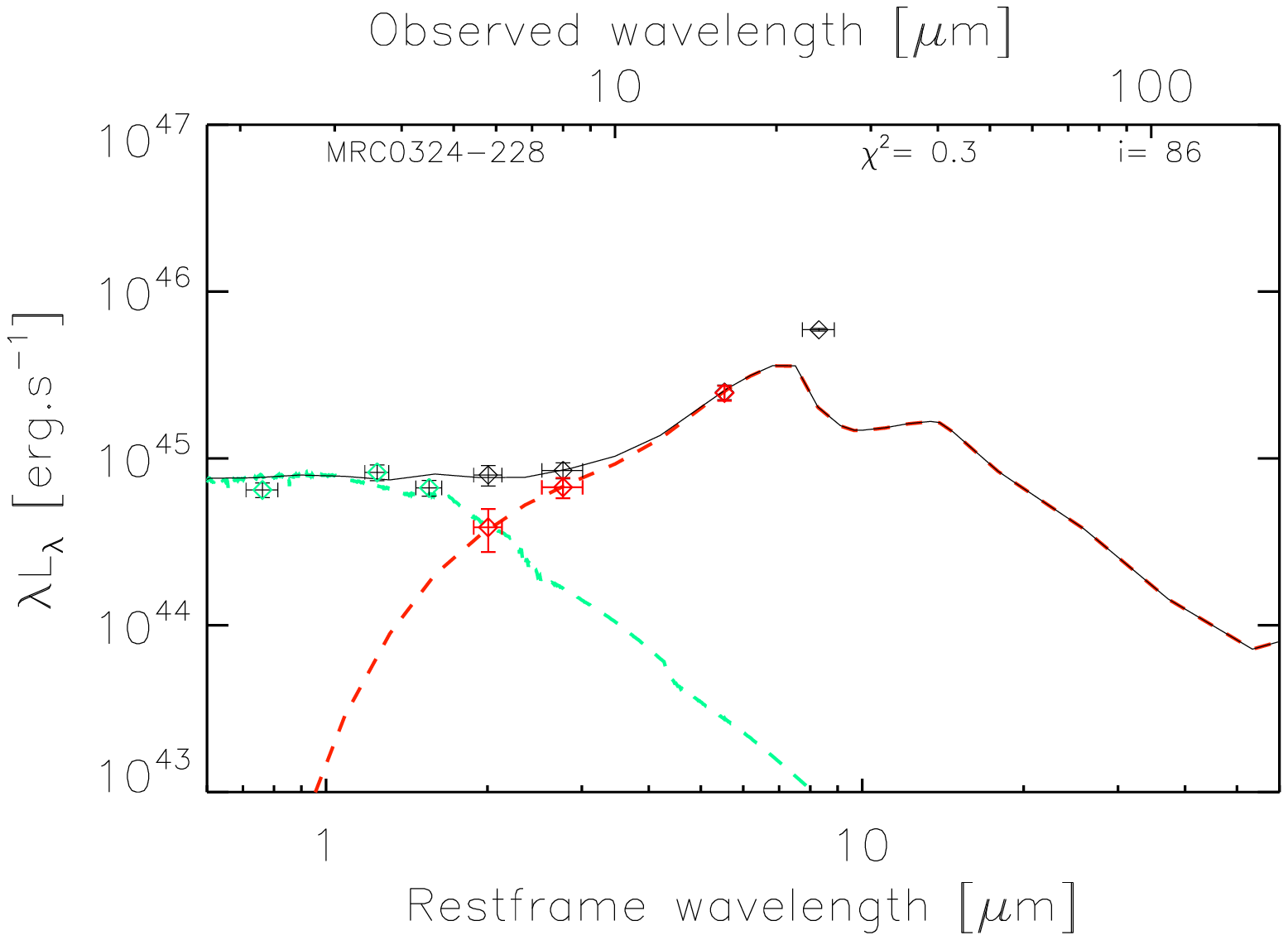} \\
\includegraphics[height=37mm,trim= 0 39 15 41,clip=true]{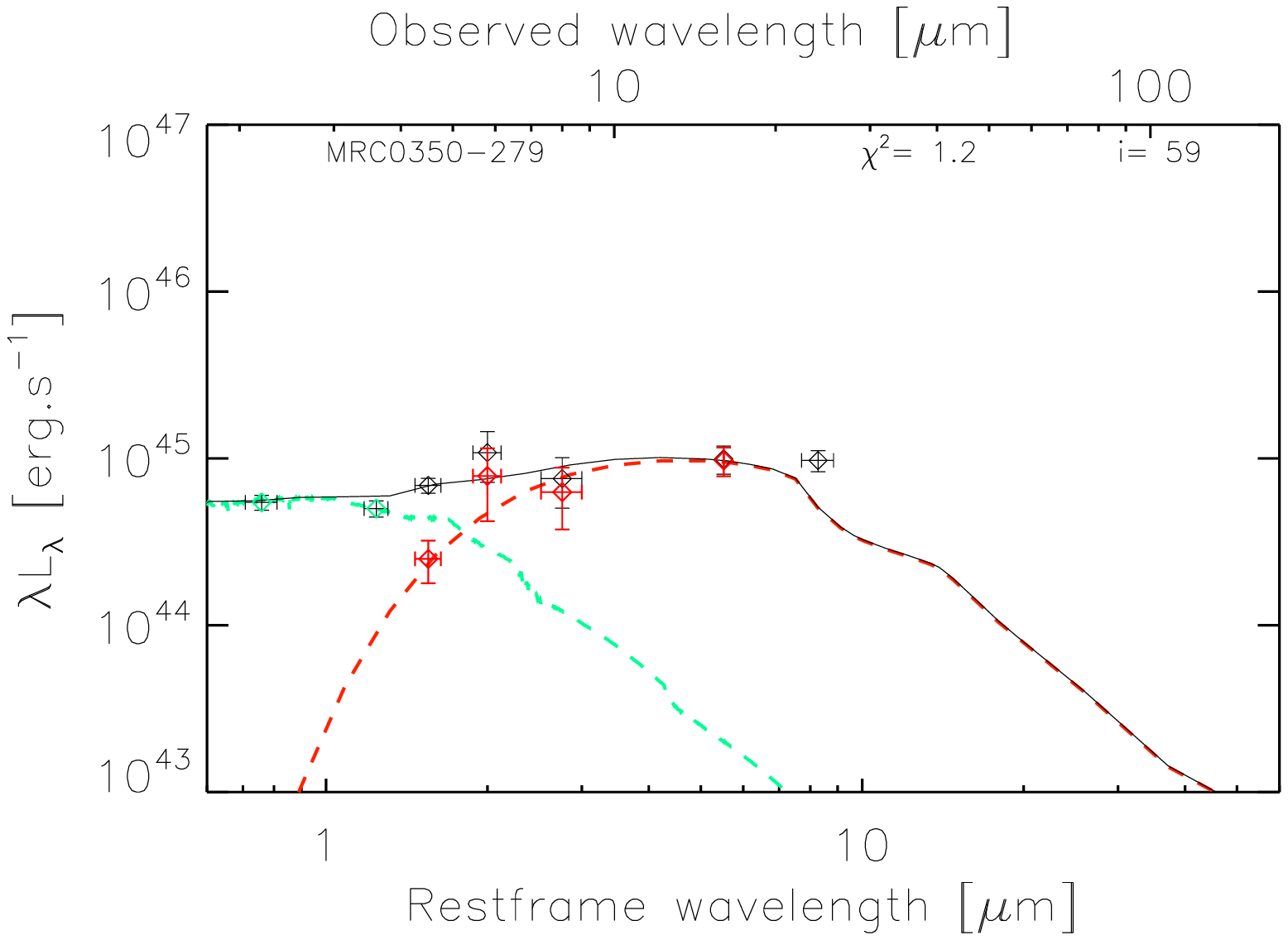} &
\includegraphics[height=37mm,trim=95 39 15 41,clip=true]{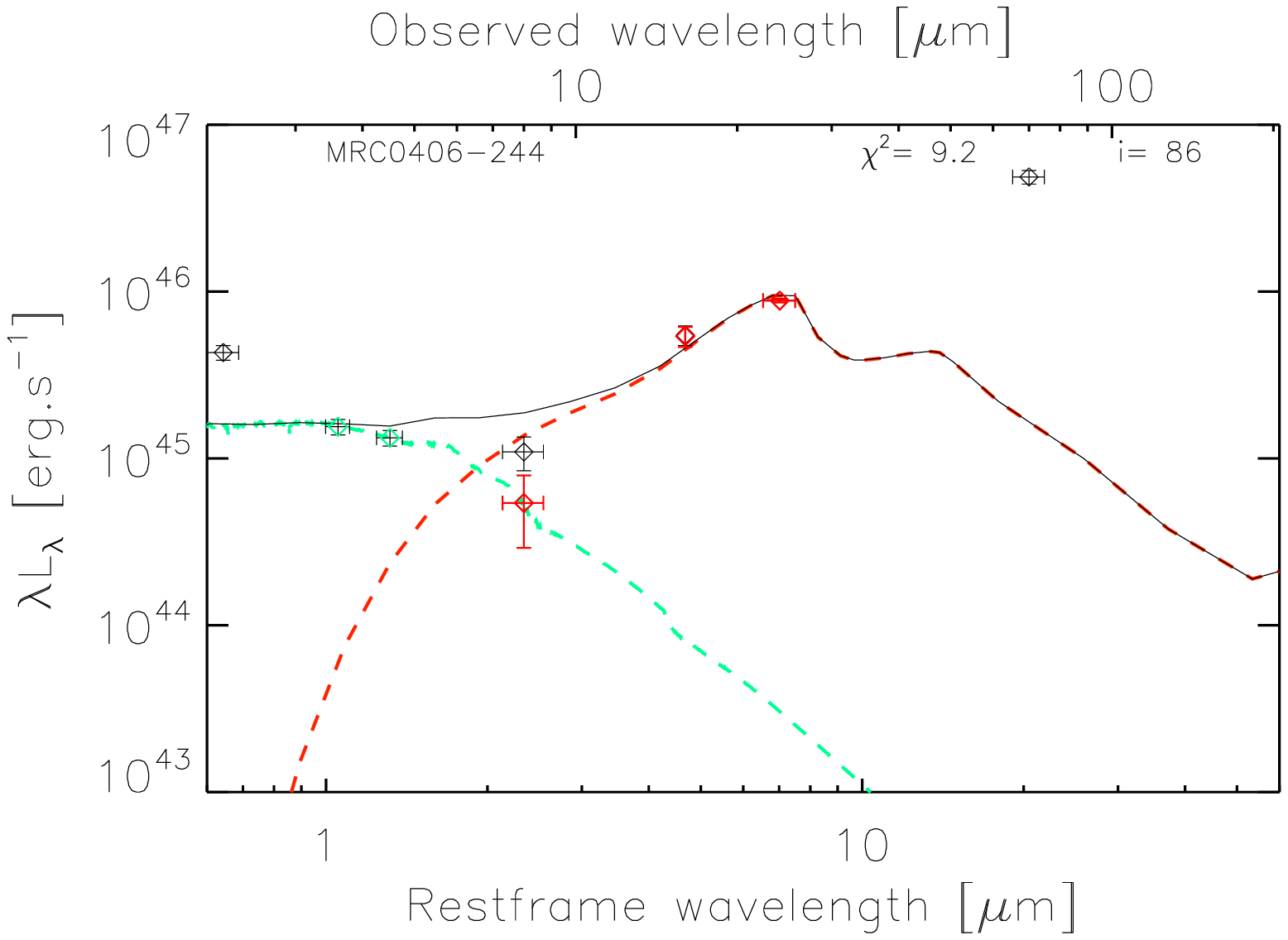} &
\includegraphics[height=37mm,trim=95 39 15 41,clip=true]{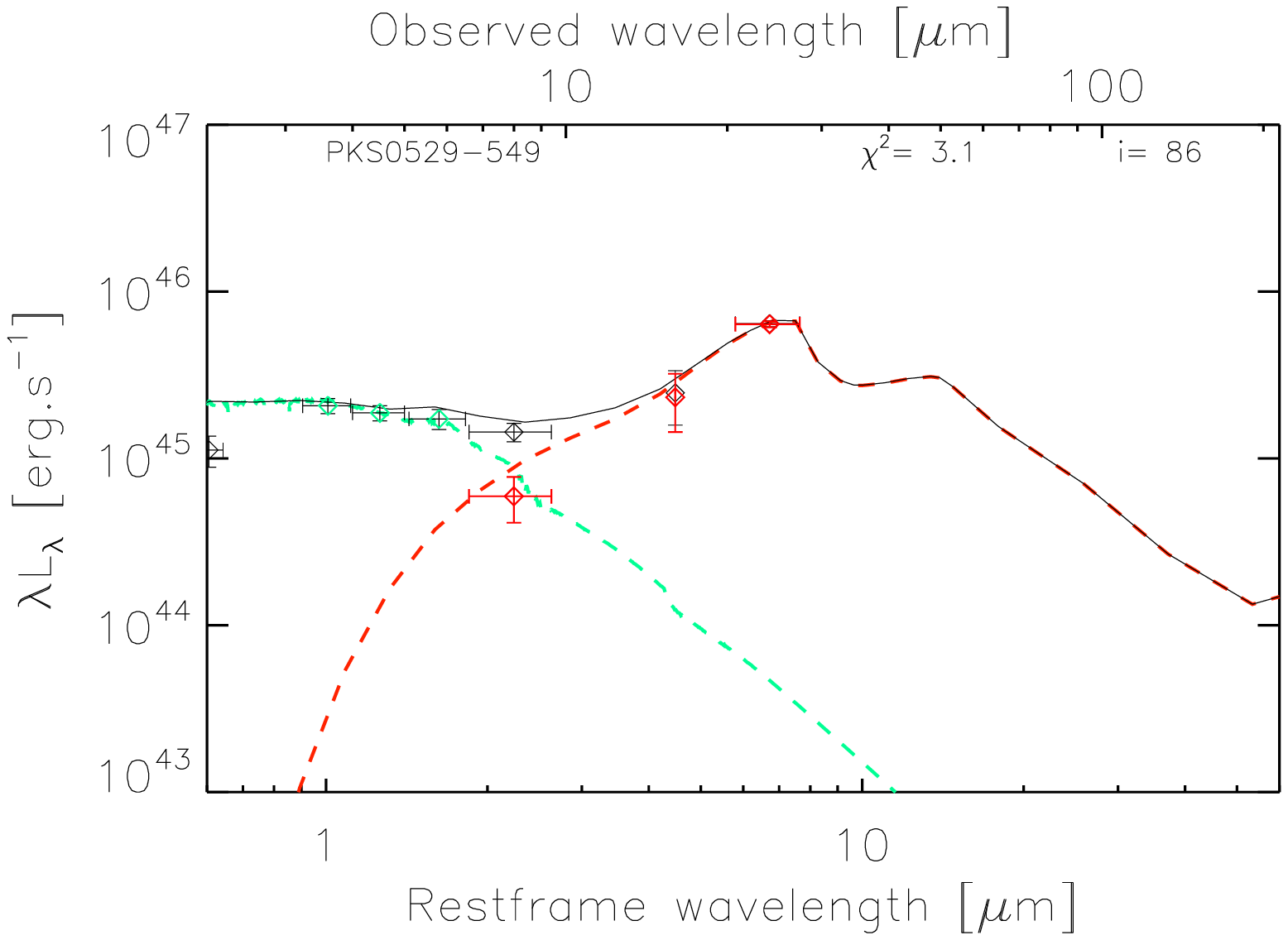} \\
\includegraphics[height=37mm,trim= 0 39 15 41,clip=true]{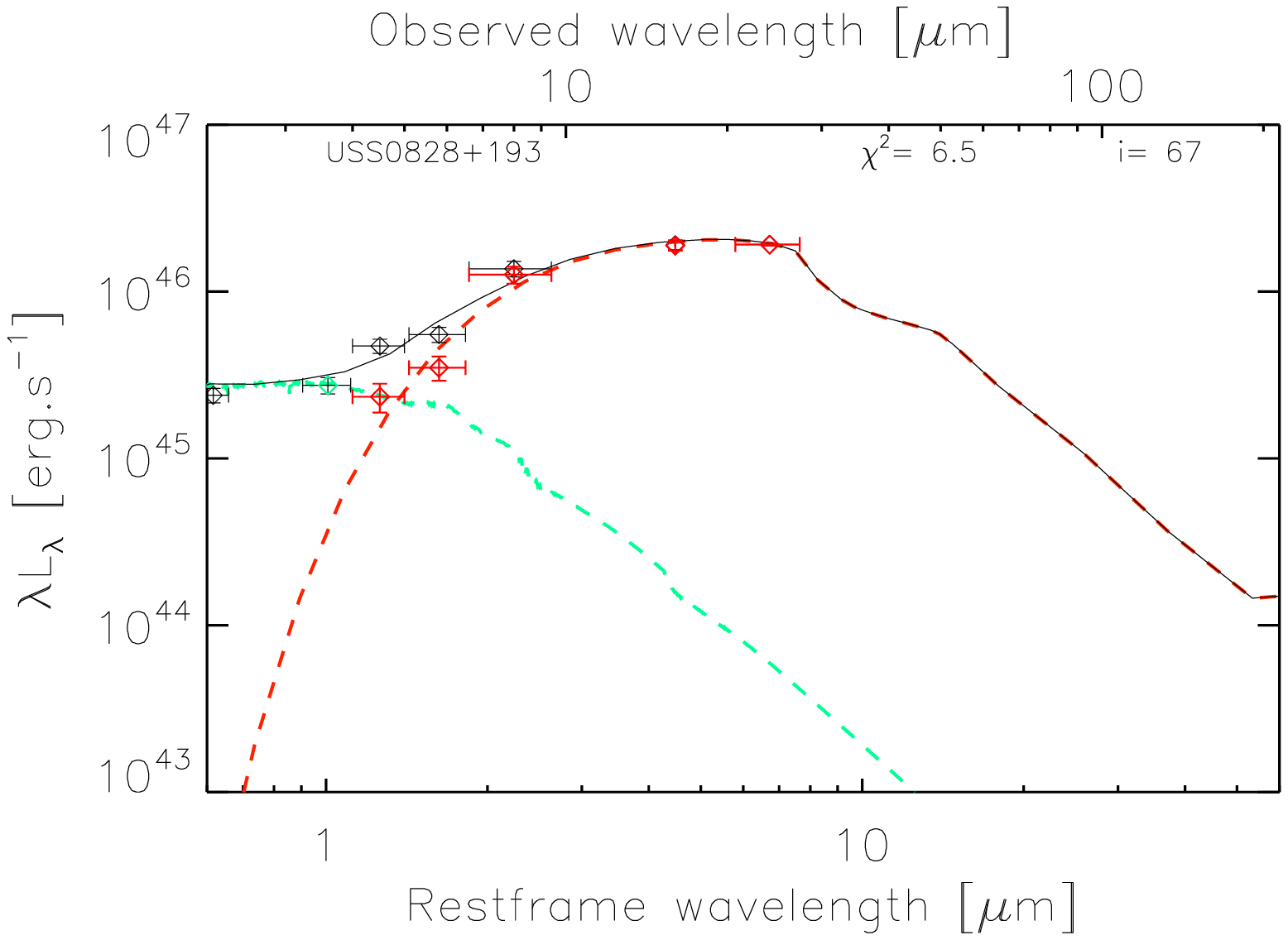} &
\includegraphics[height=37mm,trim=95 39 15 41,clip=true]{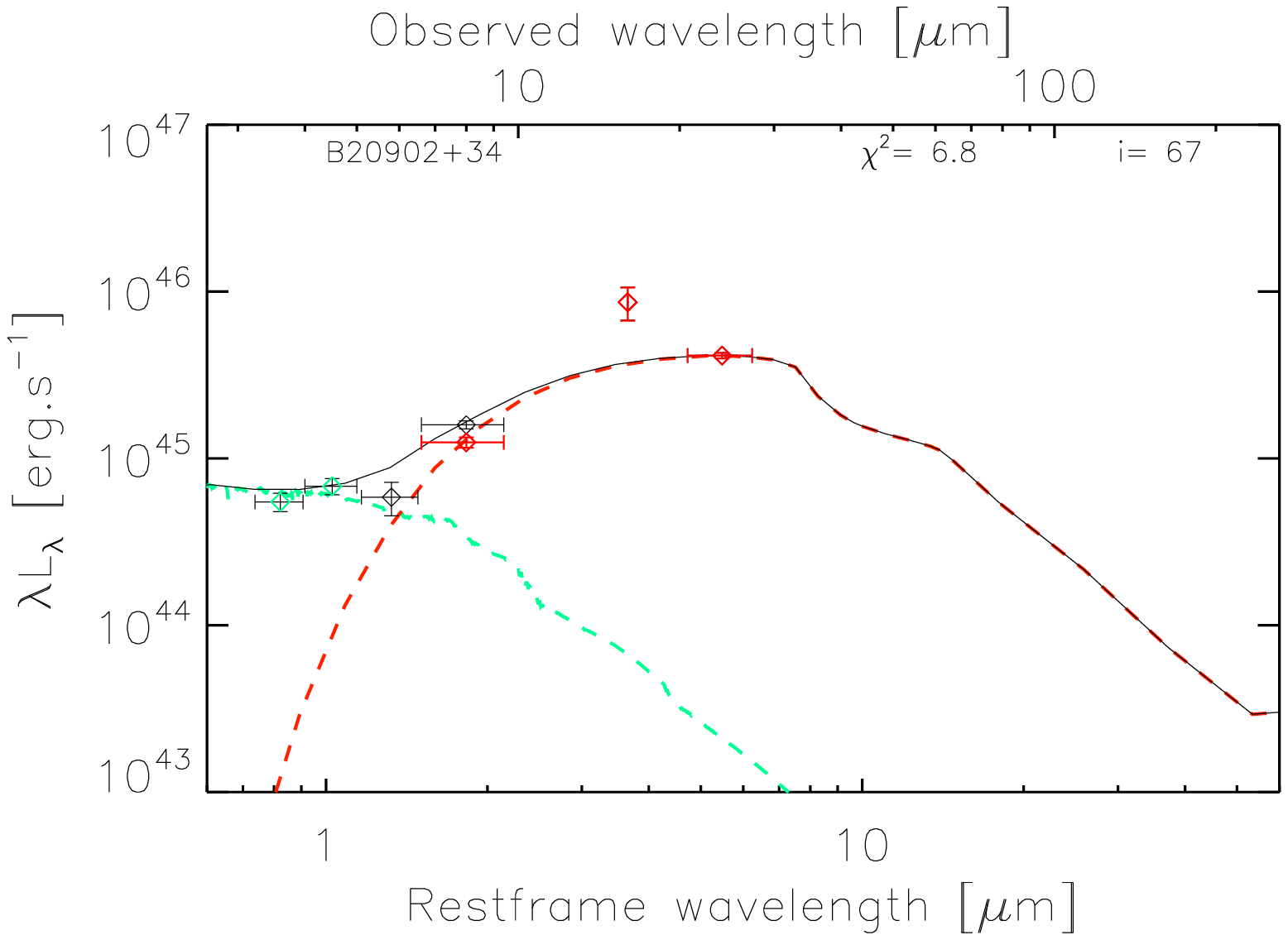}   &
\includegraphics[height=37mm,trim=95 39 15 41,clip=true]{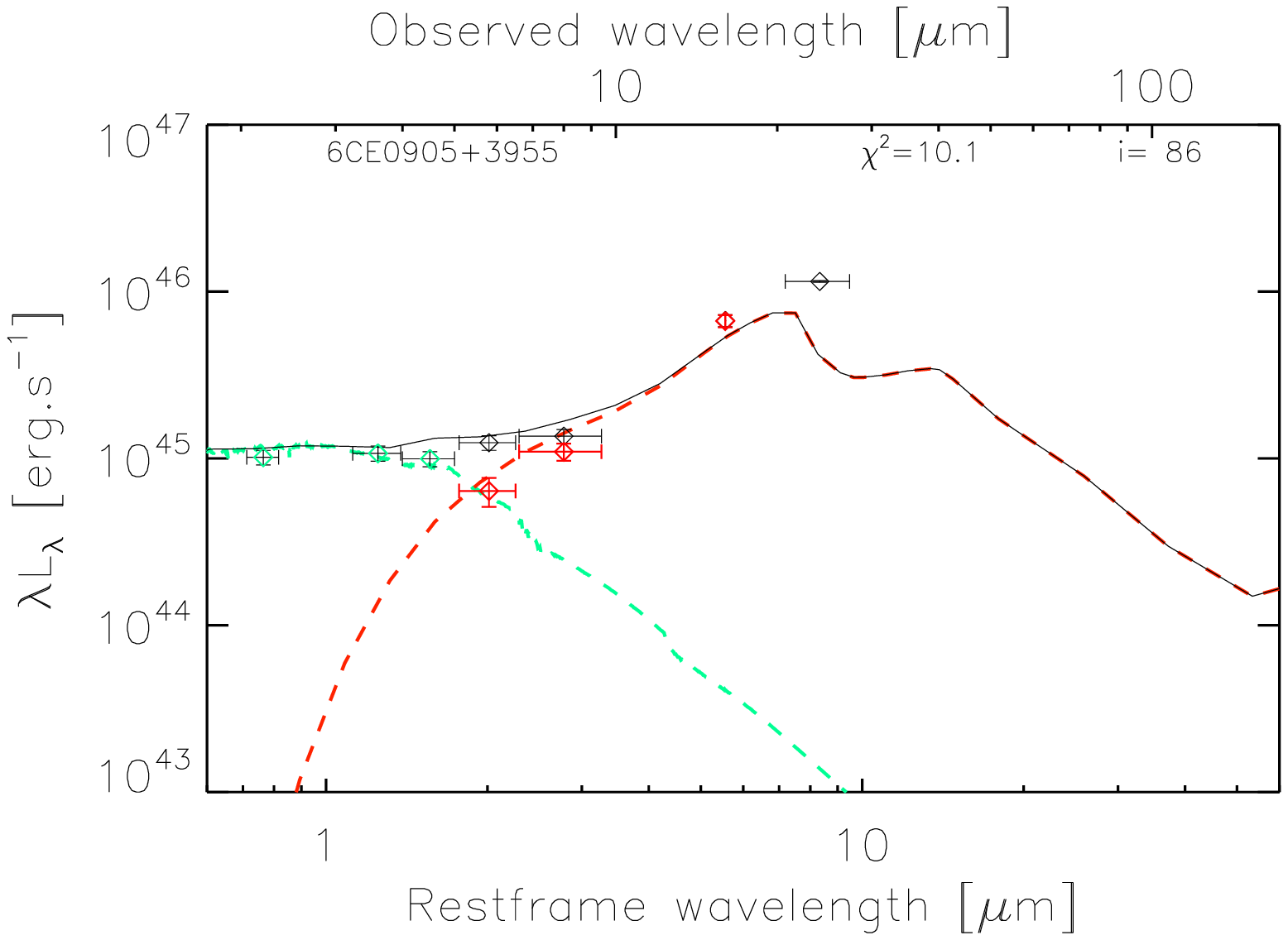}\\
\includegraphics[height=42mm,trim= 0  0 15 41,clip=true]{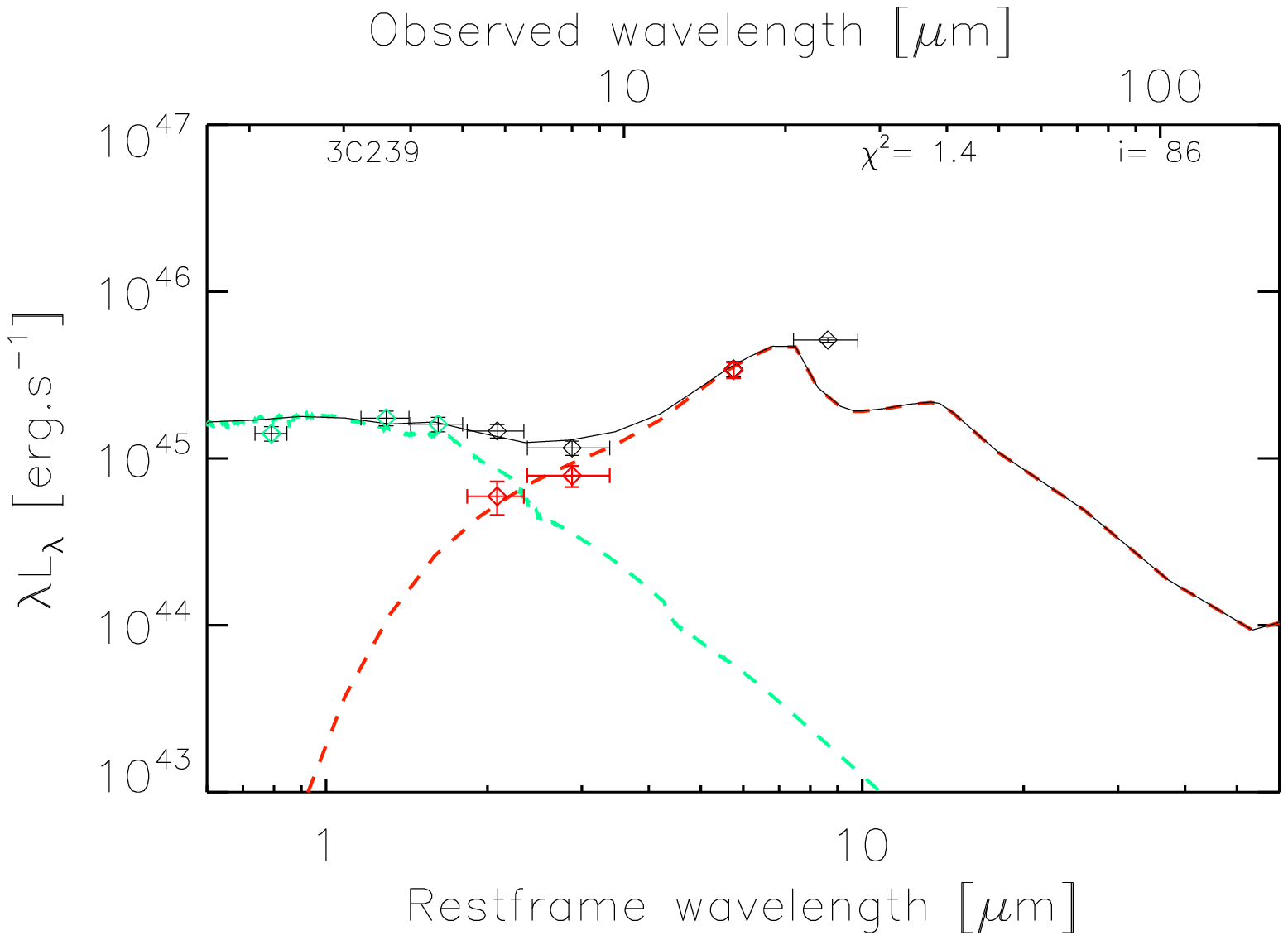}       &
\includegraphics[height=42mm,trim=95  0 15 41,clip=true]{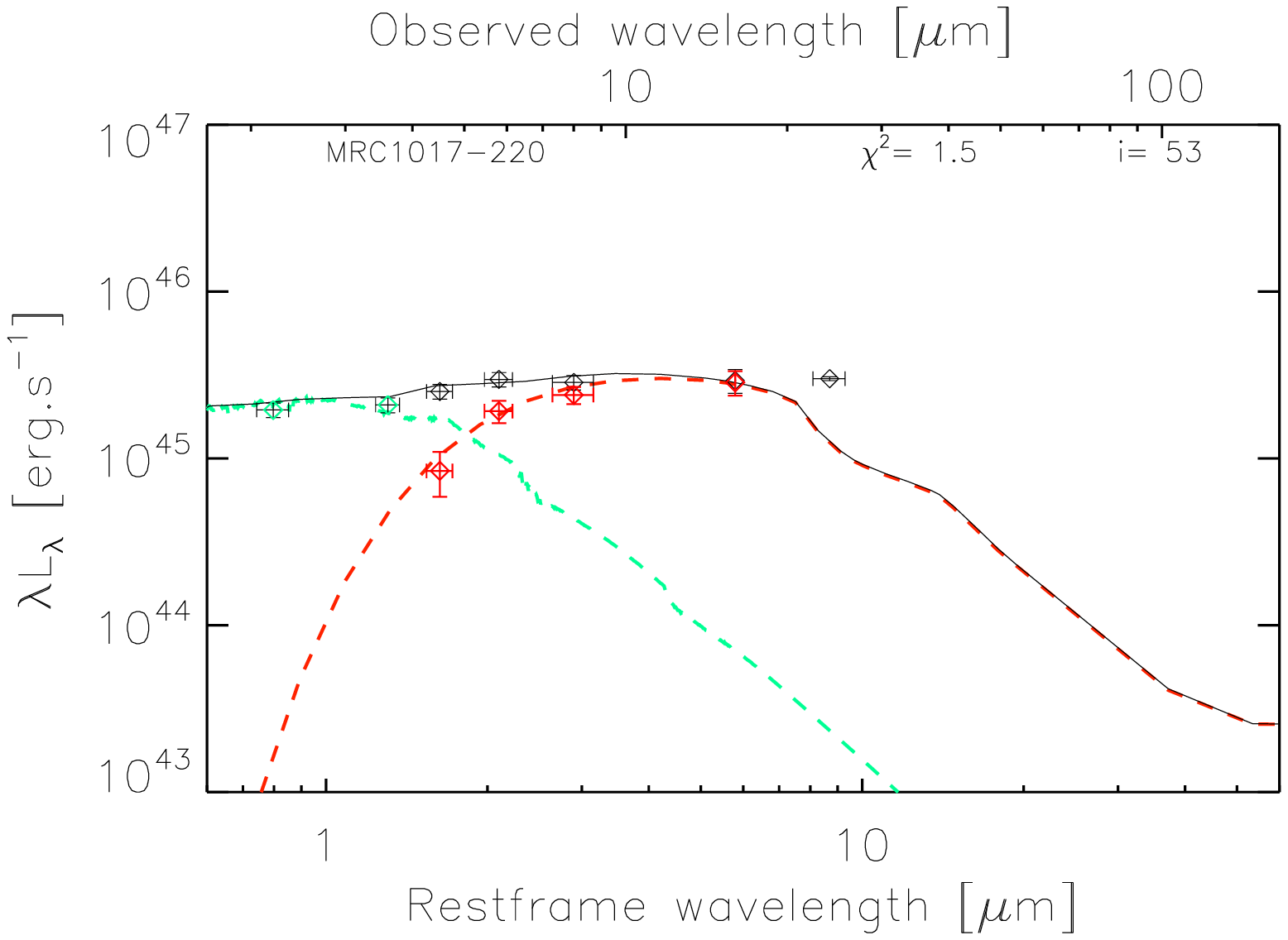} &
\includegraphics[height=42mm,trim=95  0 15 41,clip=true]{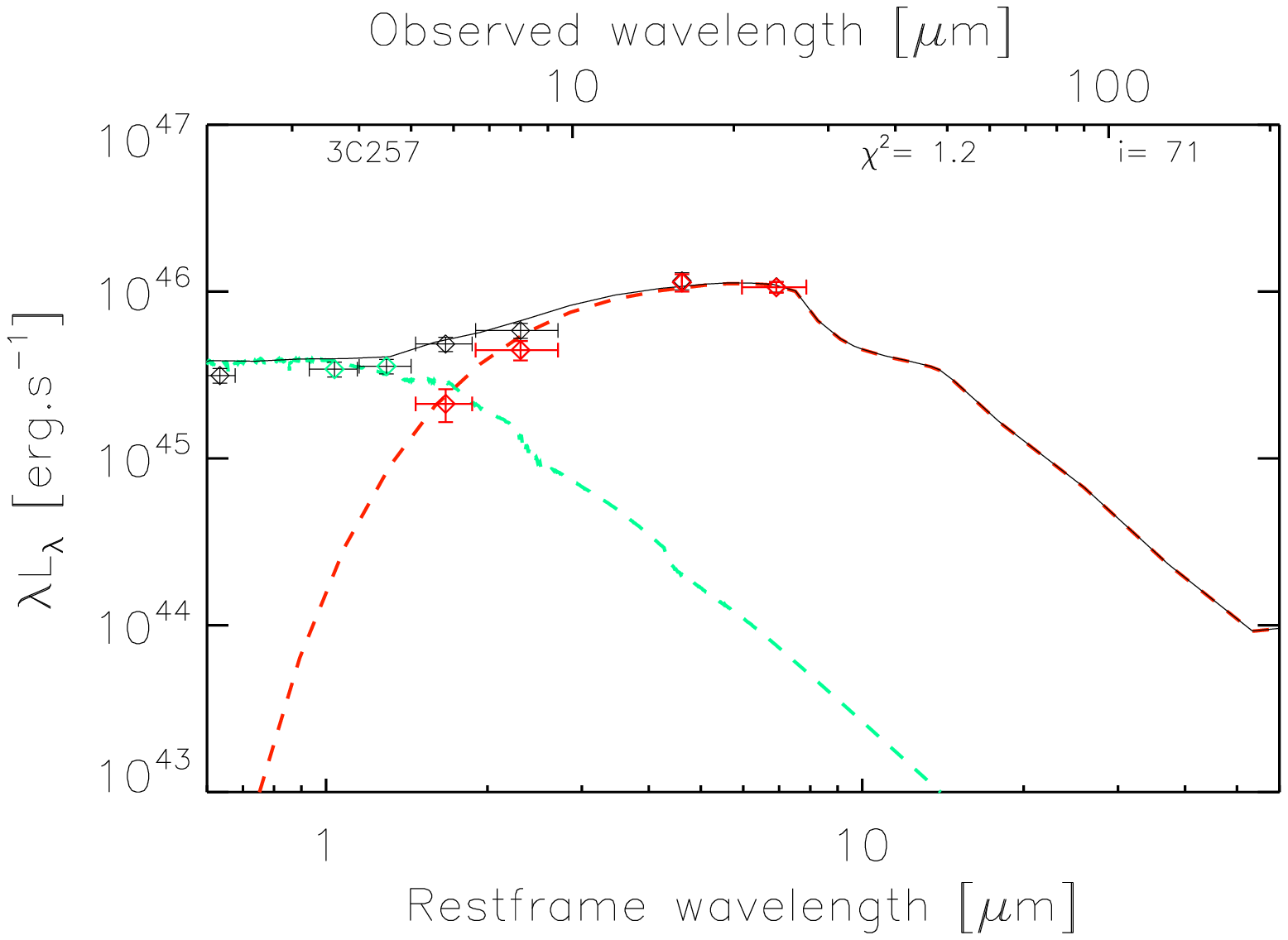}       \\
  \end{tabular}
\end{center}
\end{figure*}

\newpage

\begin{figure*}[ht]
  \begin{center}
  \begin{tabular}{r@{}c@{}l }
\includegraphics[height=42mm,trim= 0 39 15  0,clip=true]{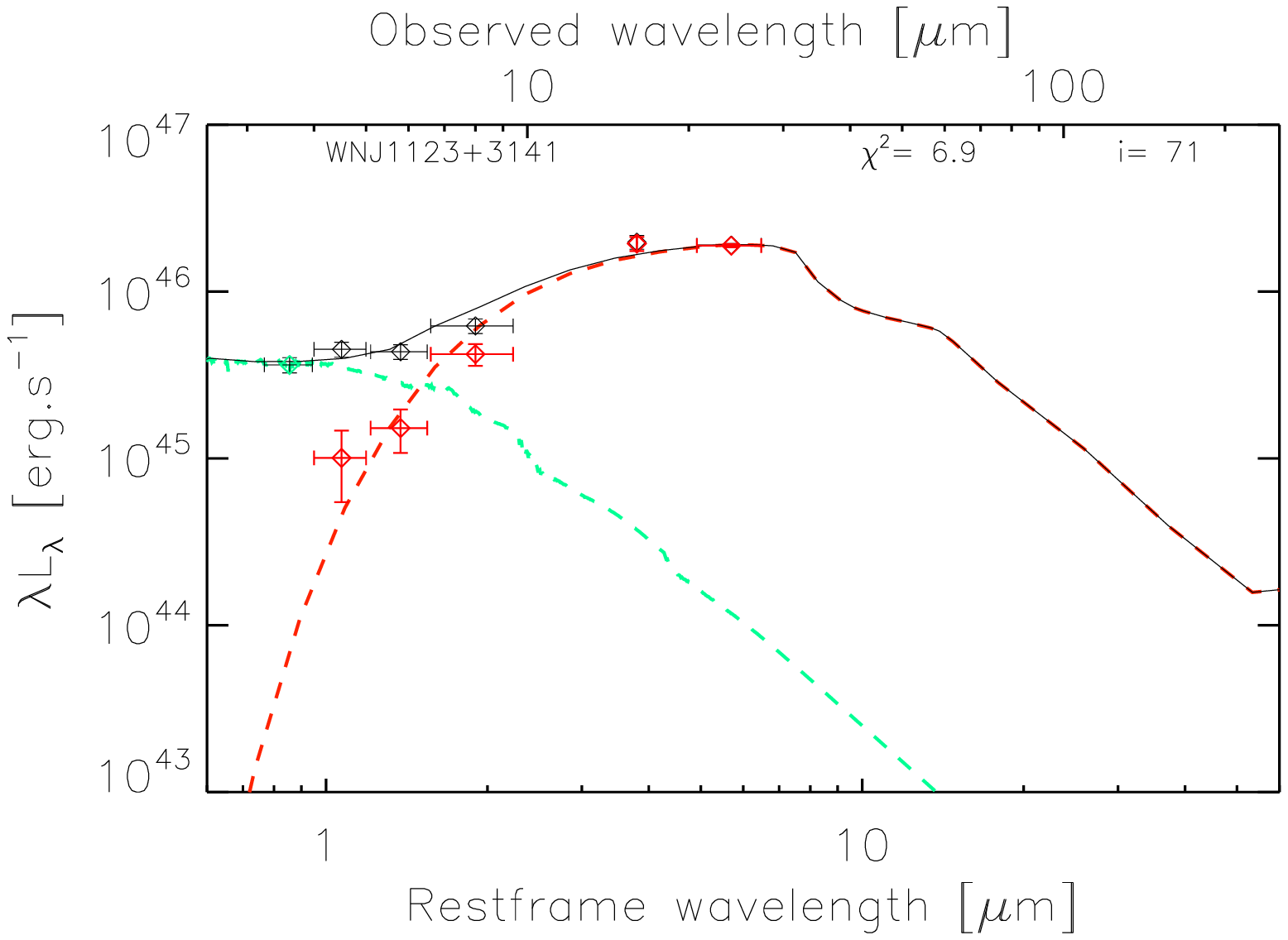} & 
\includegraphics[height=42mm,trim=95 39 15  0,clip=true]{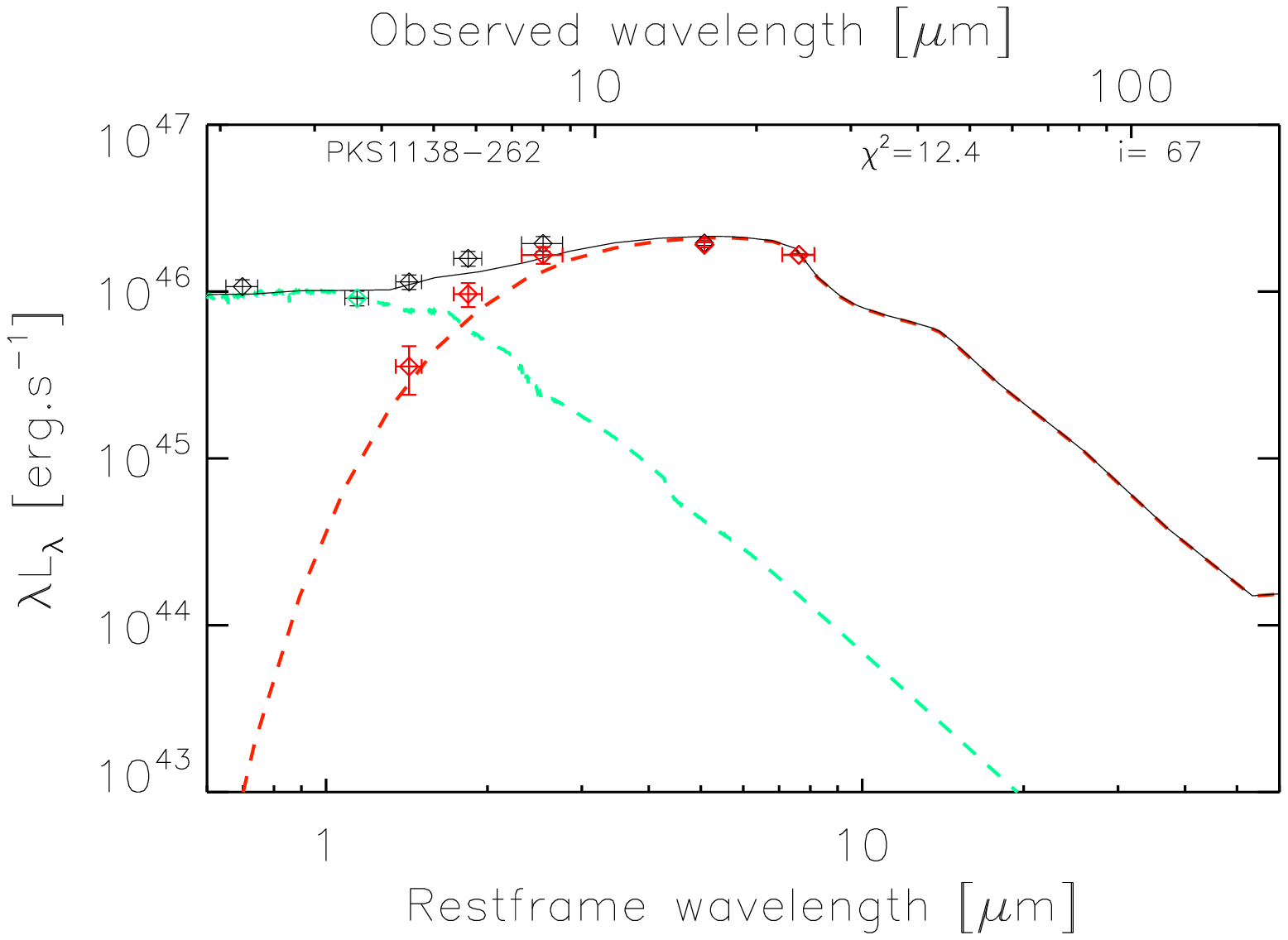}  &
\includegraphics[height=42mm,trim=95 39 15  0,clip=true]{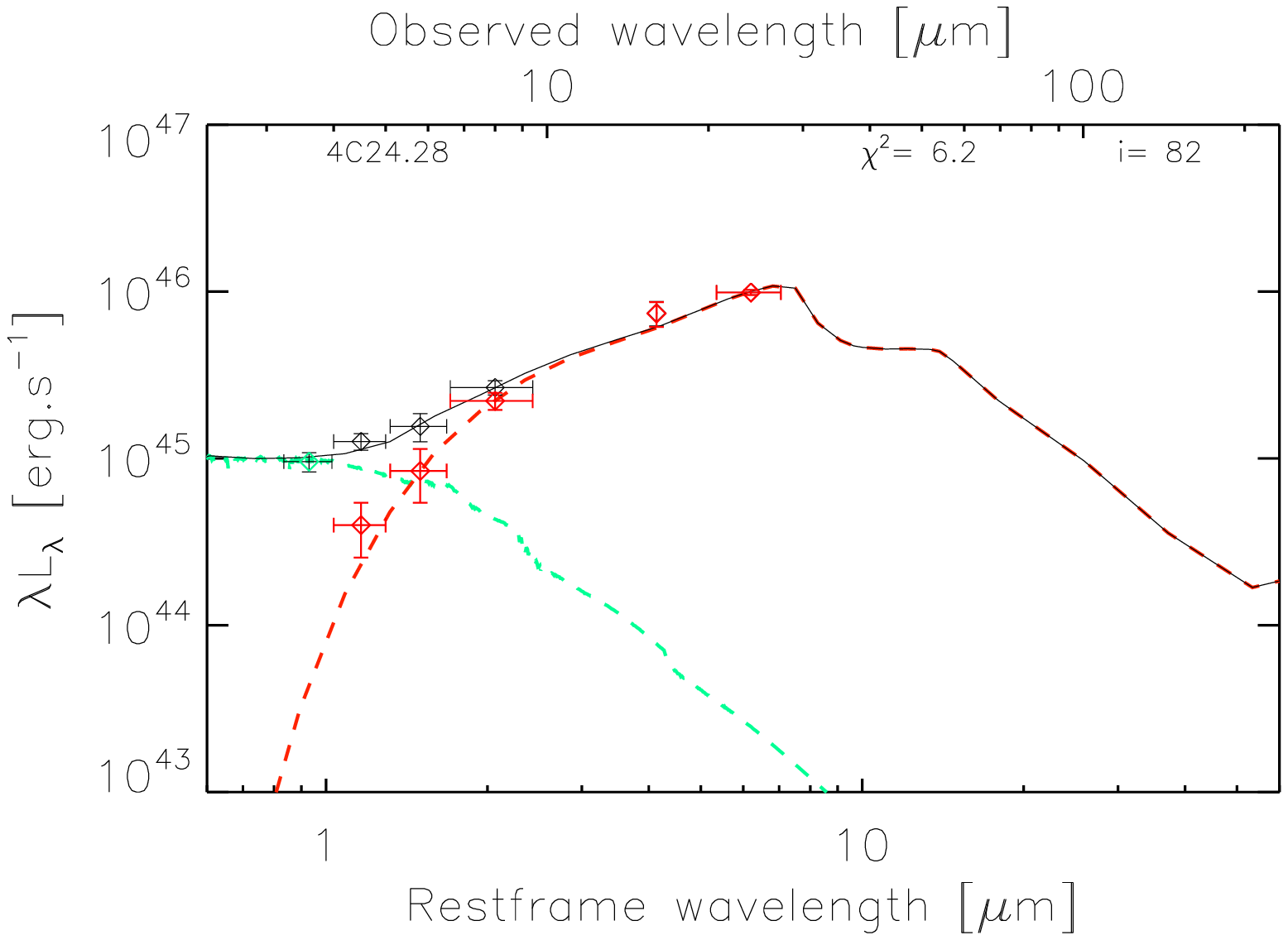}      \\
\includegraphics[height=37mm,trim= 0 39 15 41,clip=true]{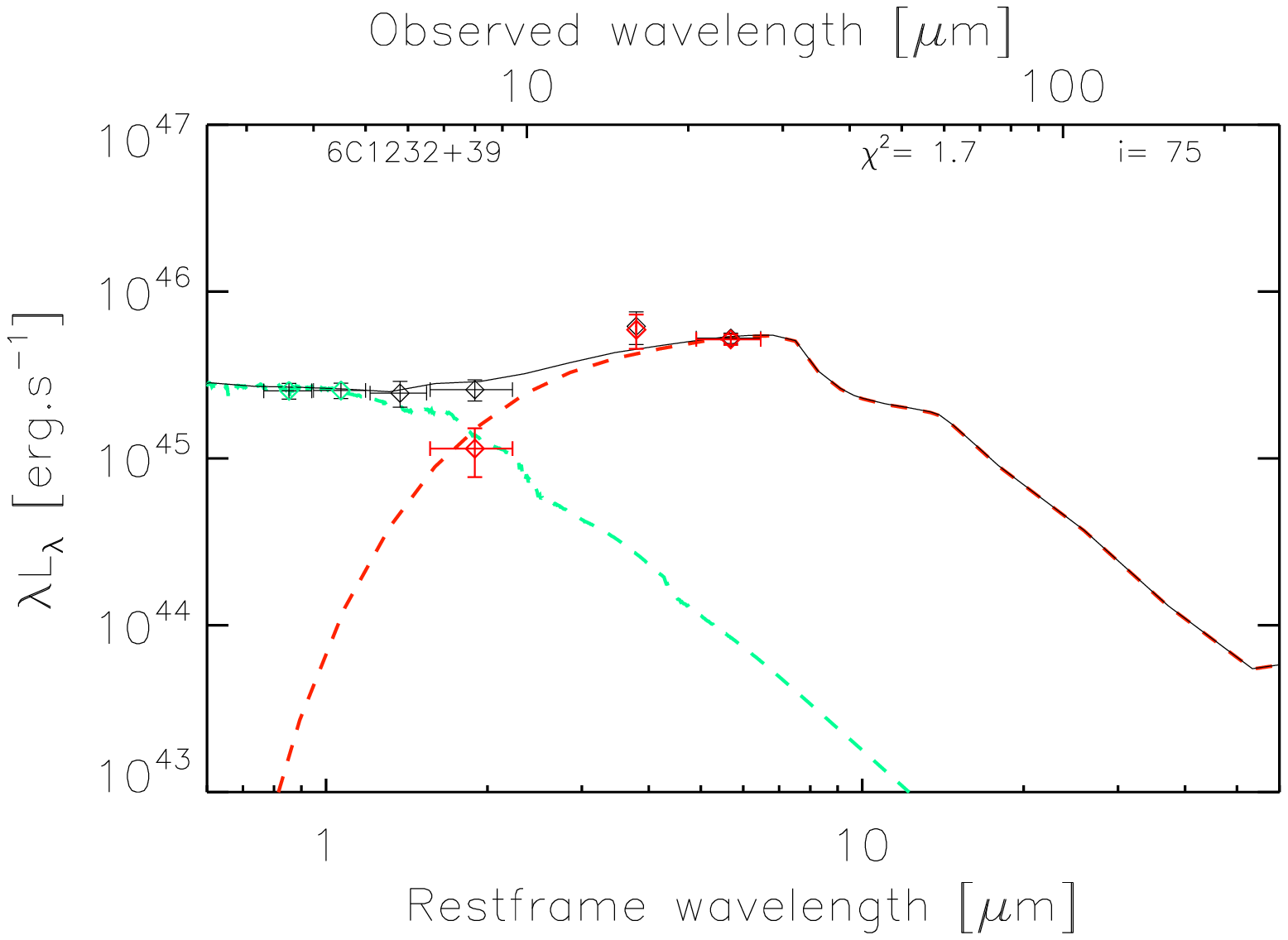}    &
\includegraphics[height=37mm,trim=95 39 15 41,clip=true]{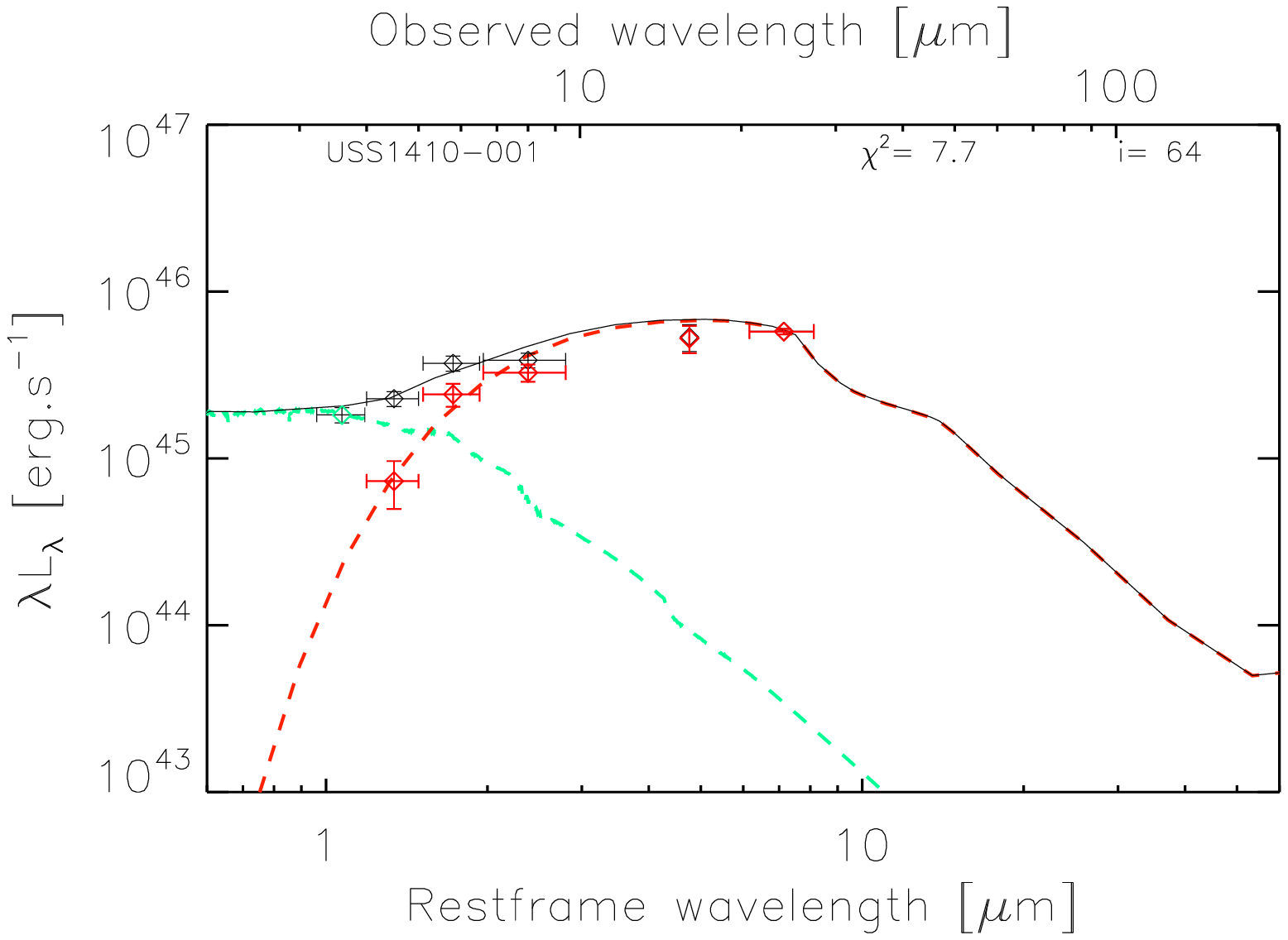}  &
\includegraphics[height=37mm,trim=95 39 15 41,clip=true]{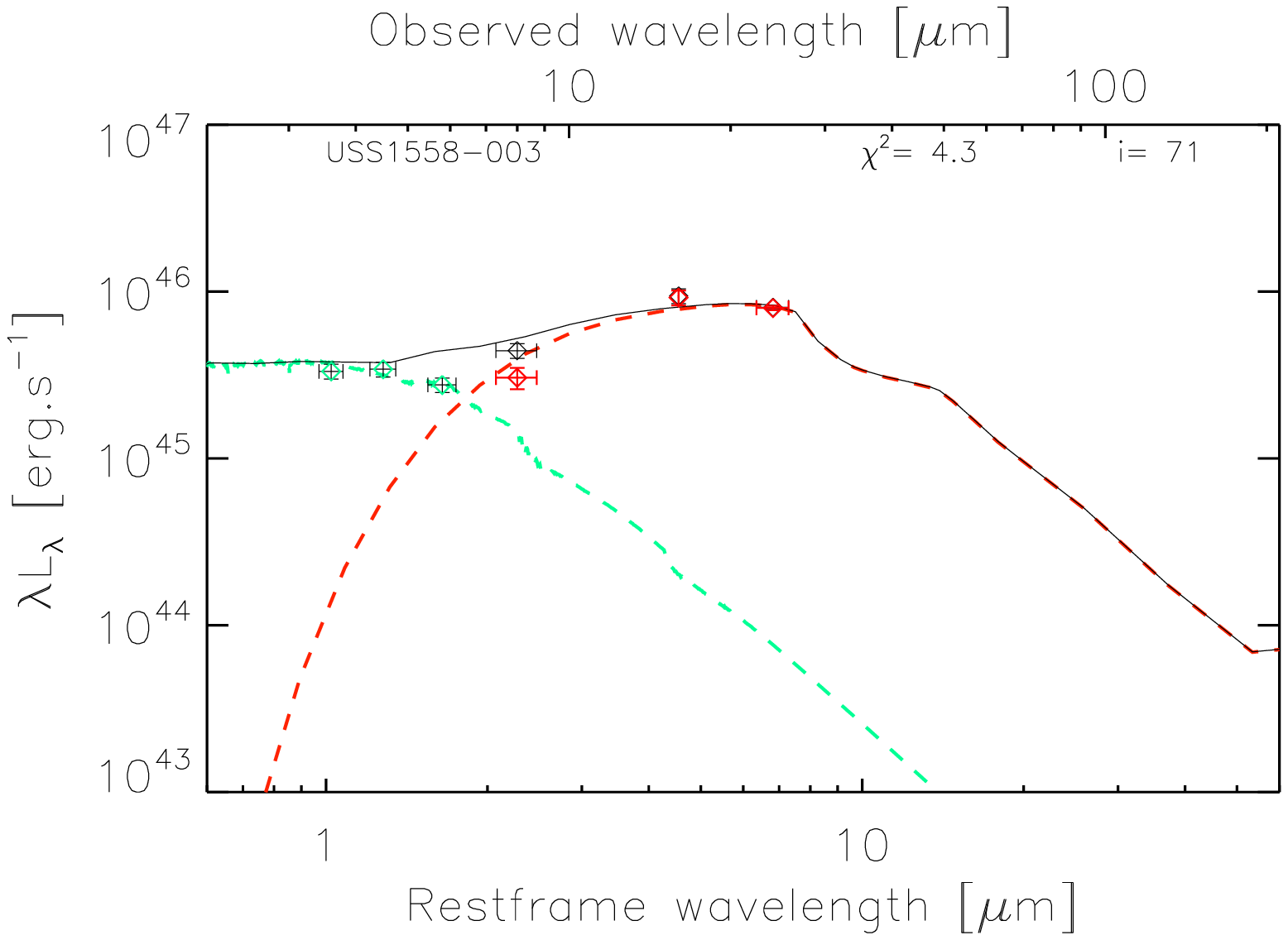}  \\
\includegraphics[height=37mm,trim= 0 39 15 41,clip=true]{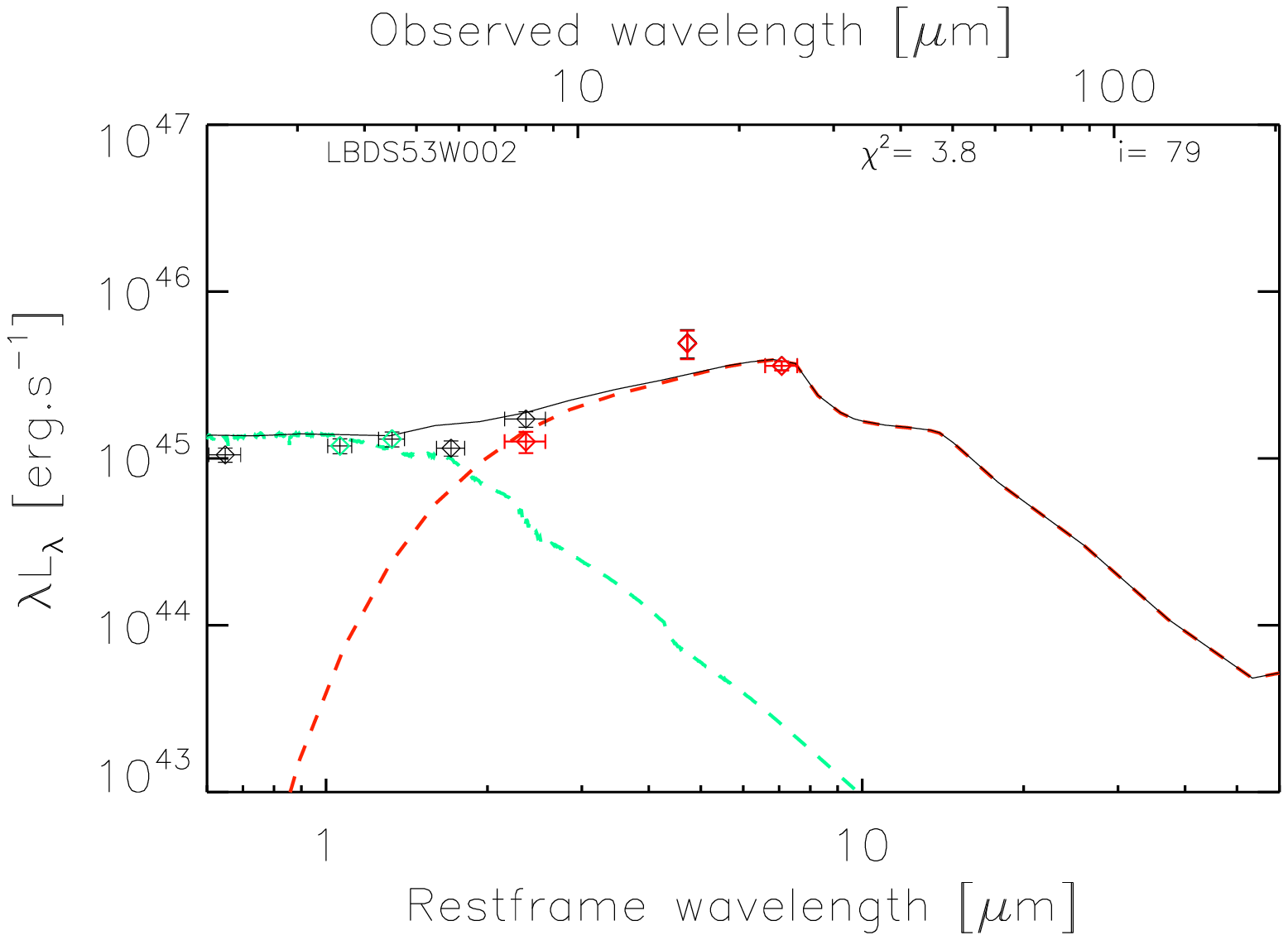}   &
\includegraphics[height=37mm,trim=95 39 15 41,clip=true]{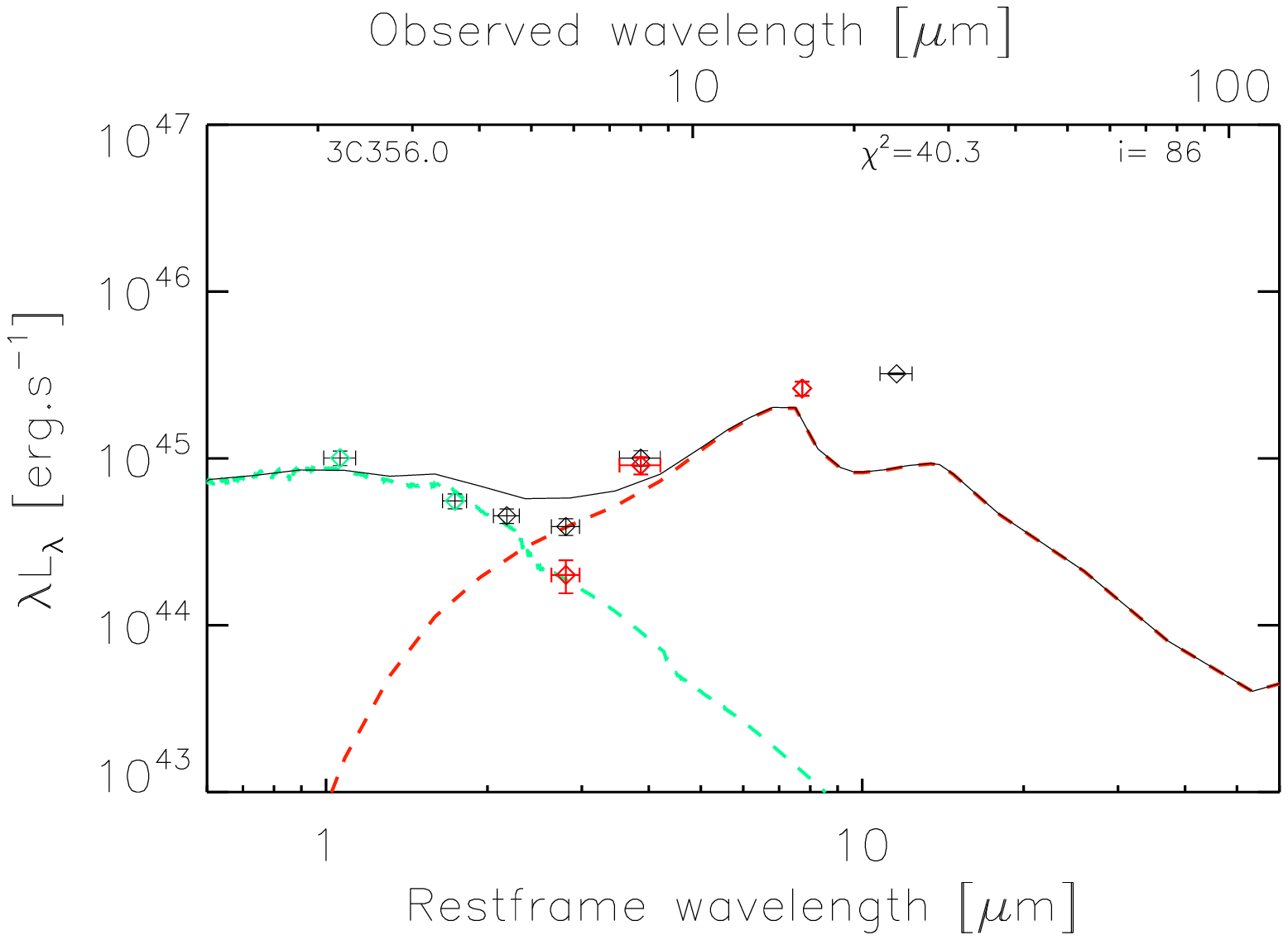}      &
\includegraphics[height=37mm,trim=95 39 15 41,clip=true]{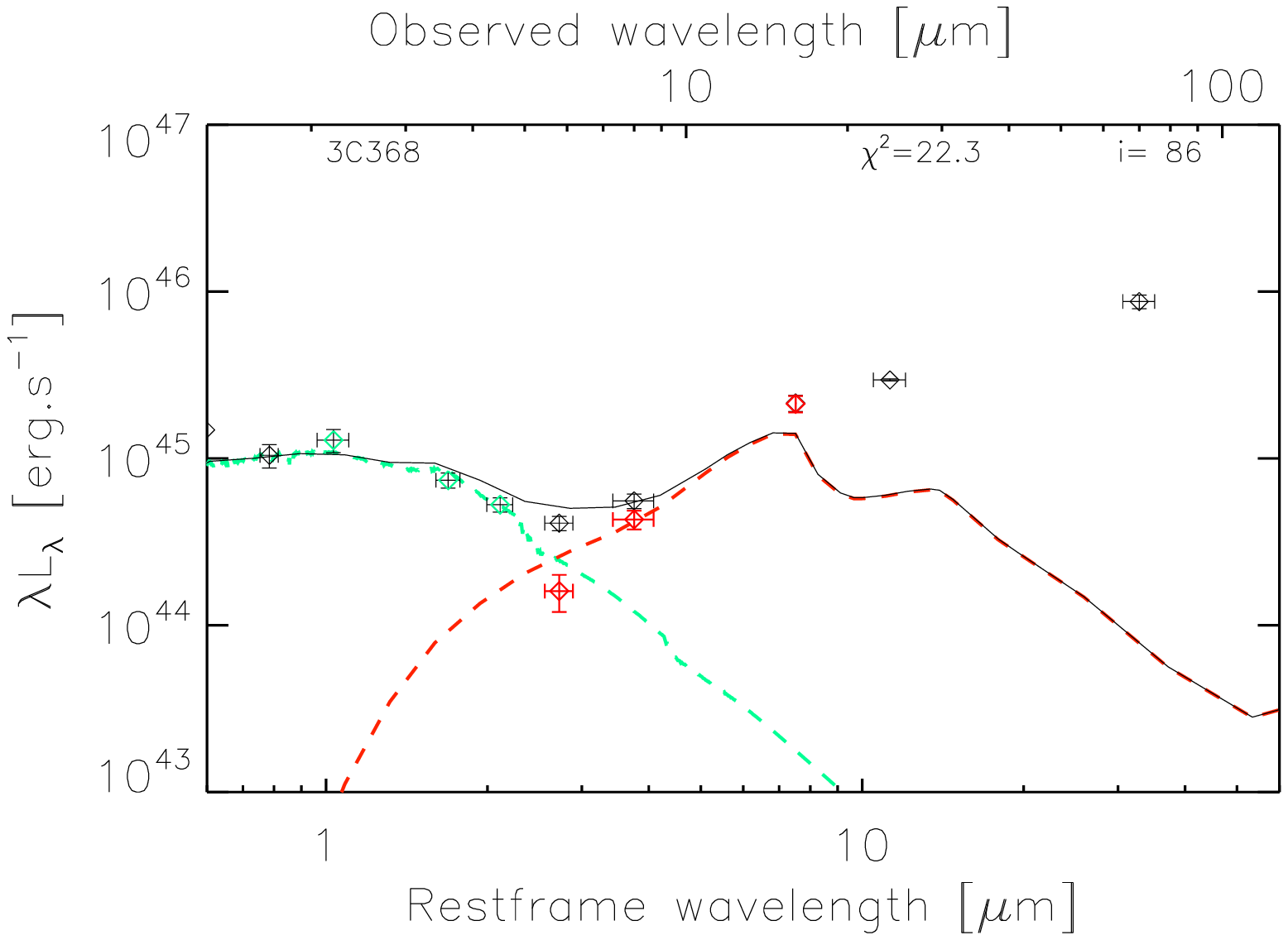}        \\
\includegraphics[height=37mm,trim= 0 39 15 41,clip=true]{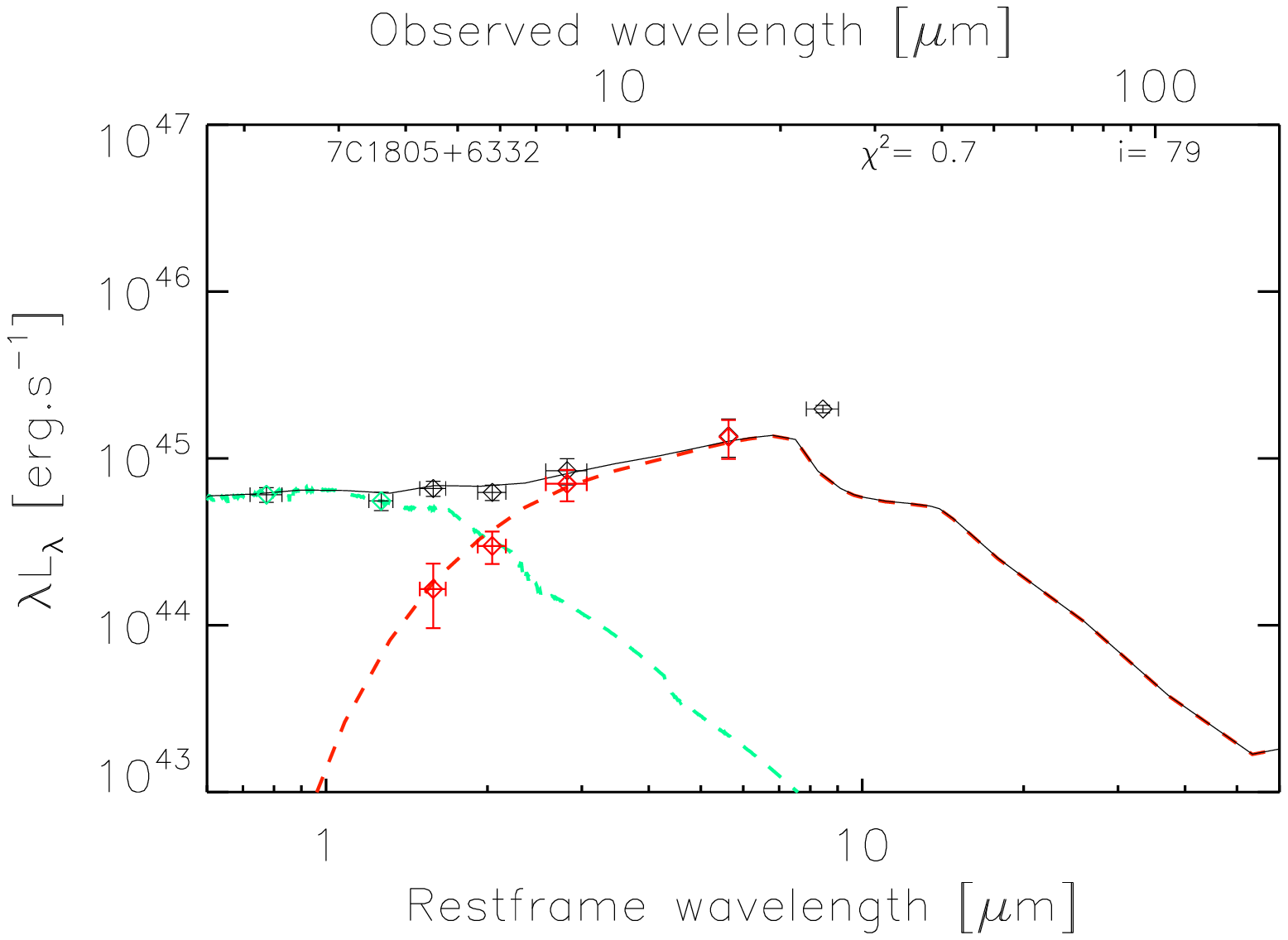}  &
\includegraphics[height=37mm,trim=95 39 15 41,clip=true]{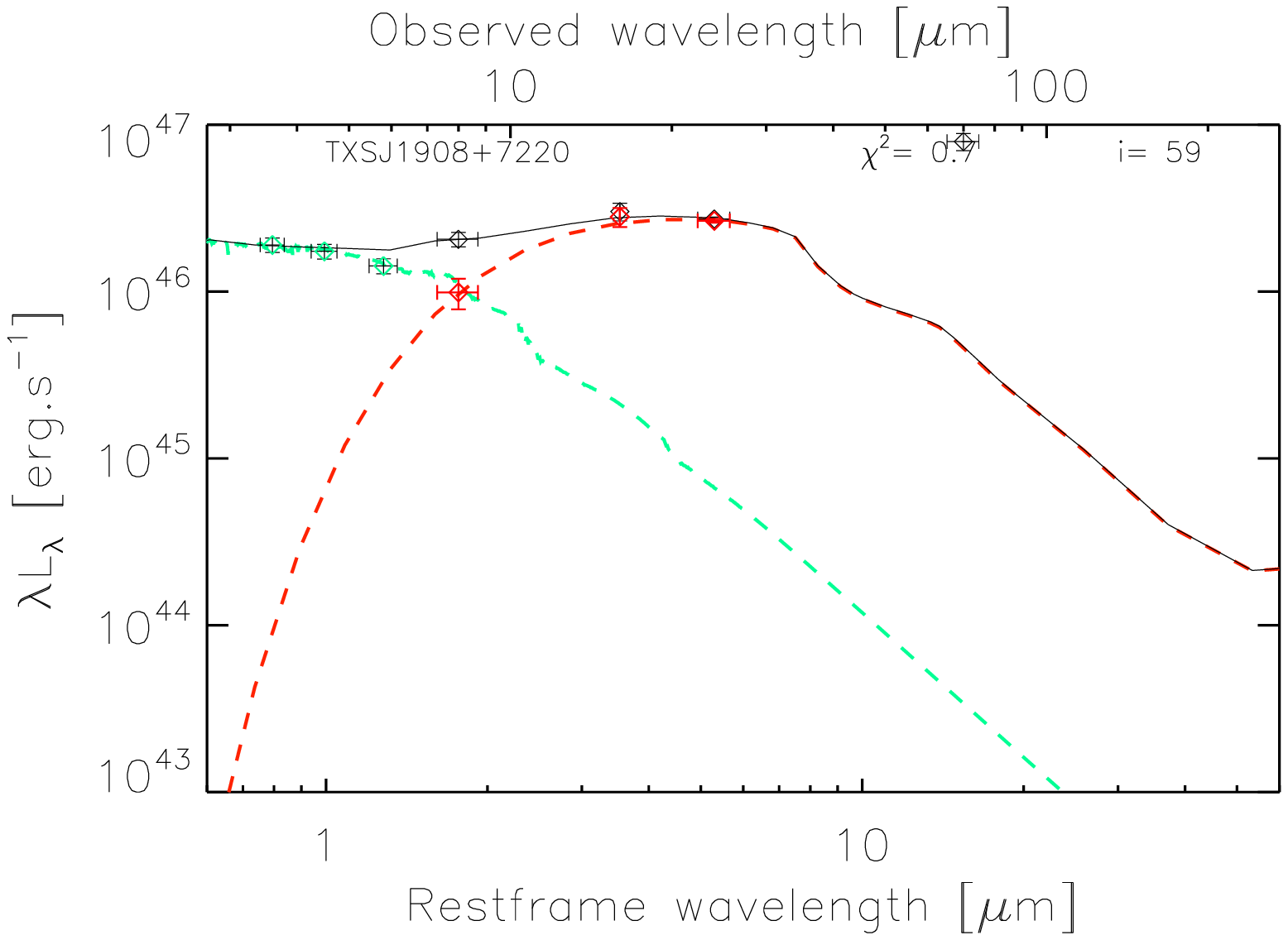}&
\includegraphics[height=37mm,trim=95 39 15 41,clip=true]{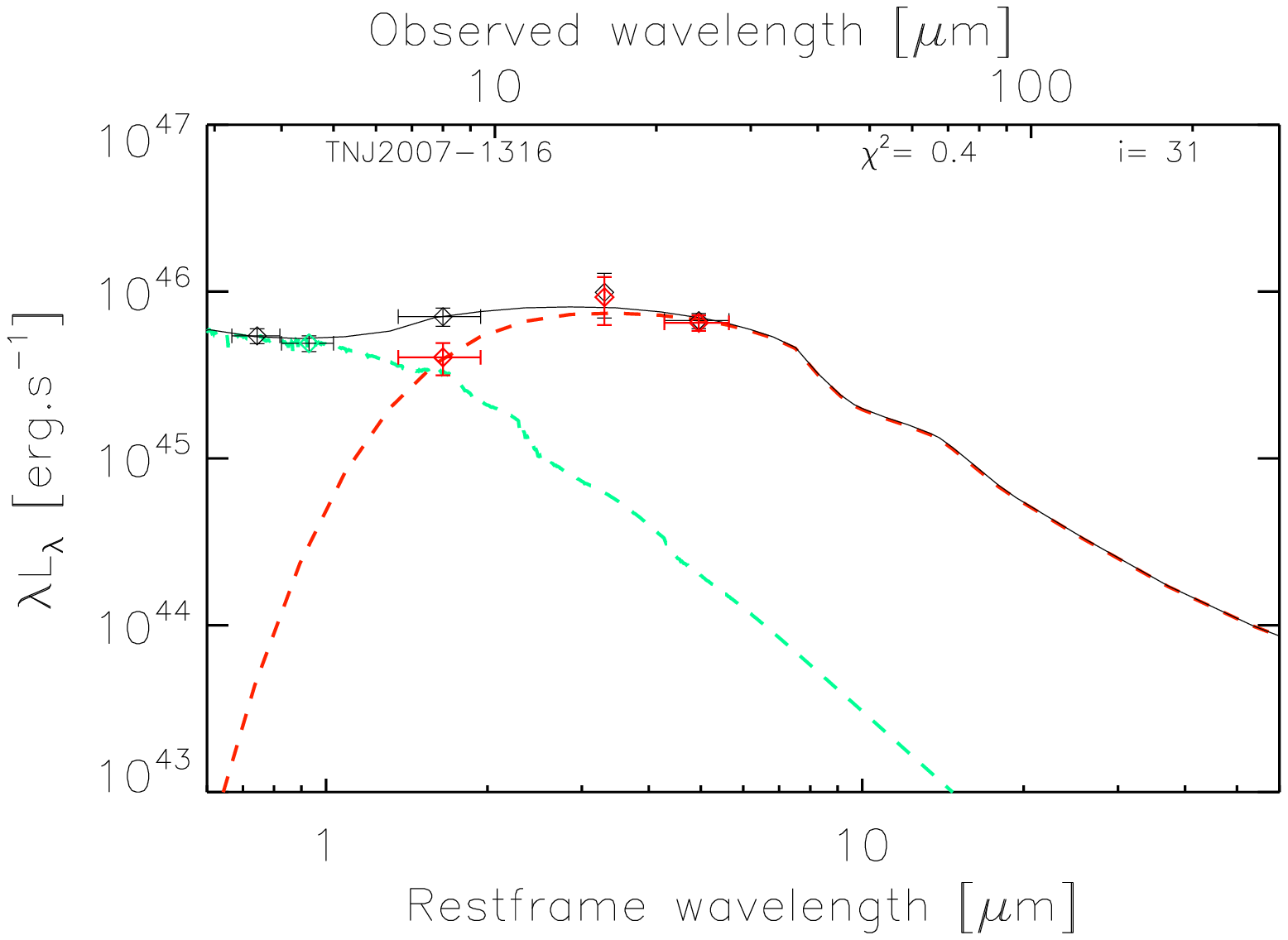} \\
\includegraphics[height=37mm,trim= 0 39 15 41,clip=true]{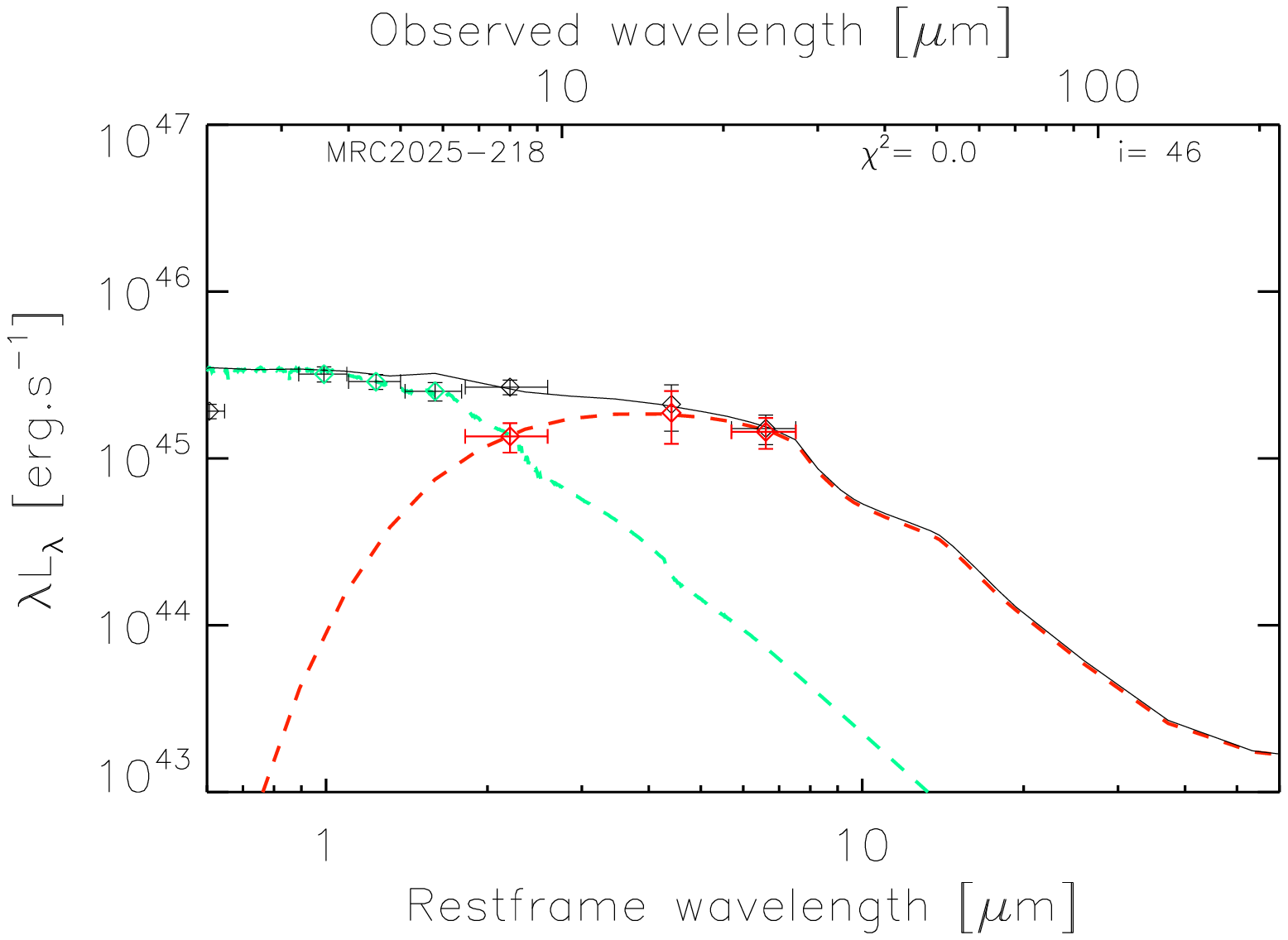}  & 
\includegraphics[height=37mm,trim=95 39 15 41,clip=true]{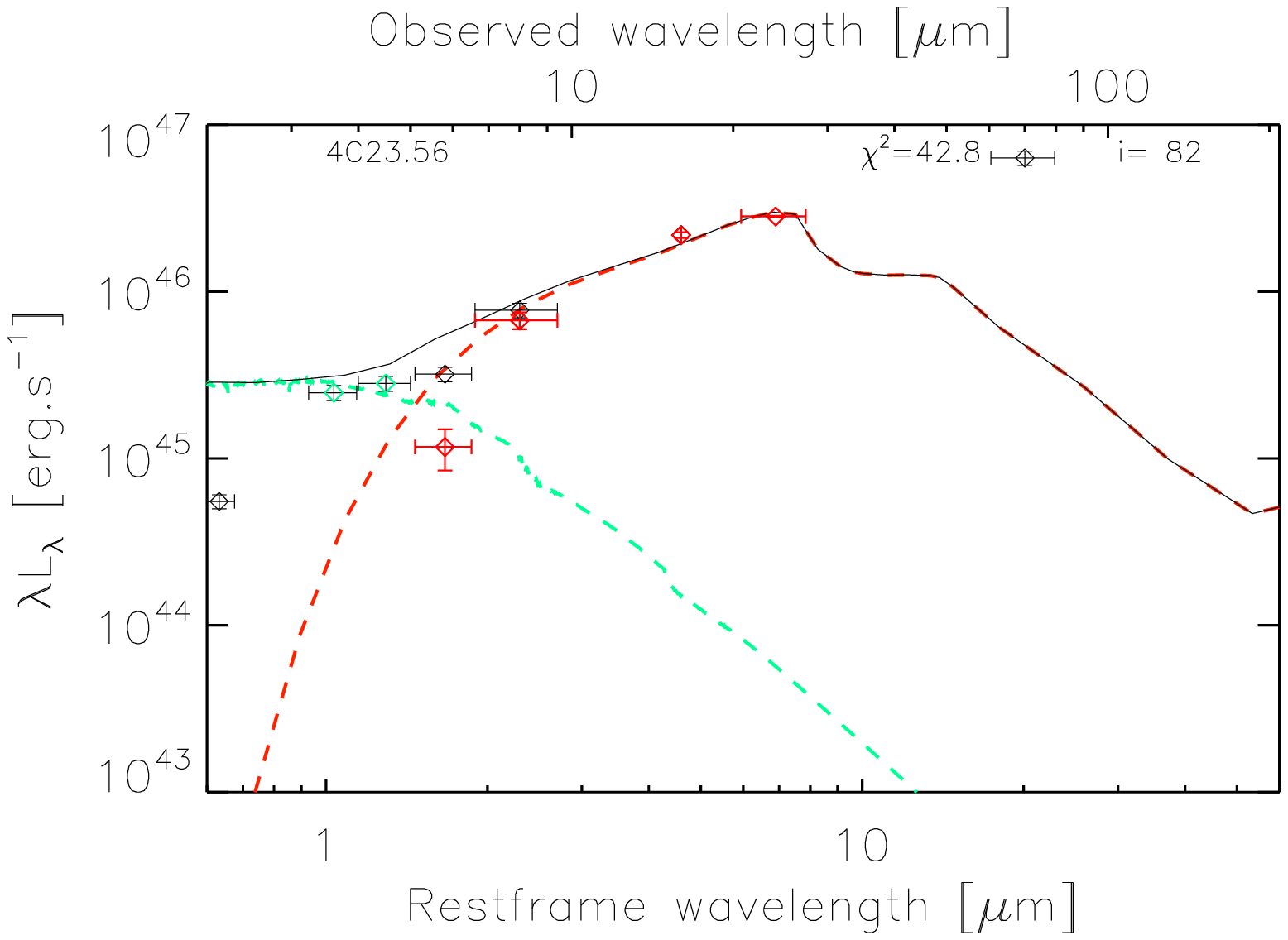}      &
\includegraphics[height=37mm,trim=95 39 15 41,clip=true]{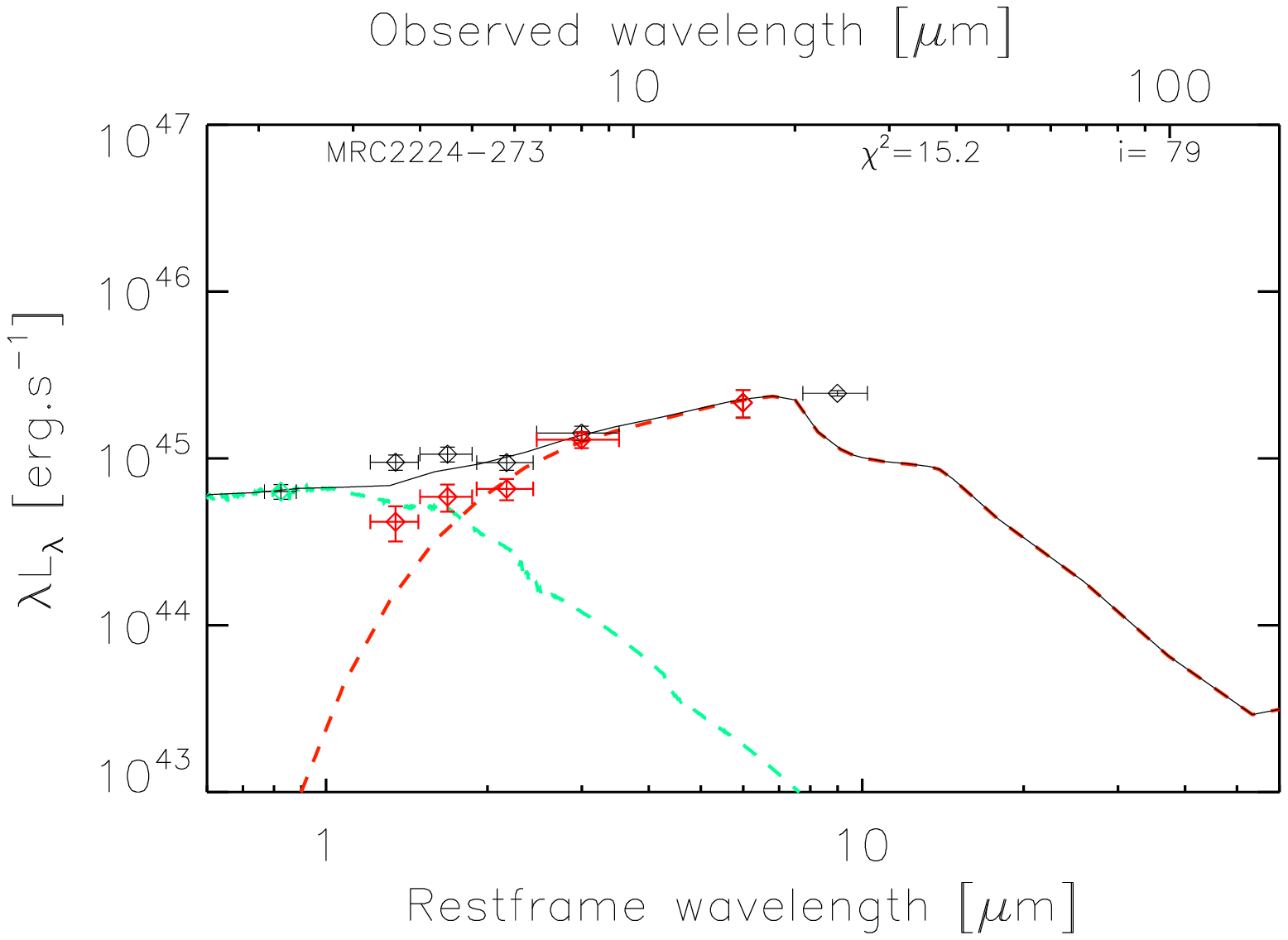}  \\
\includegraphics[height=42mm,trim= 0  0 15 41,clip=true]{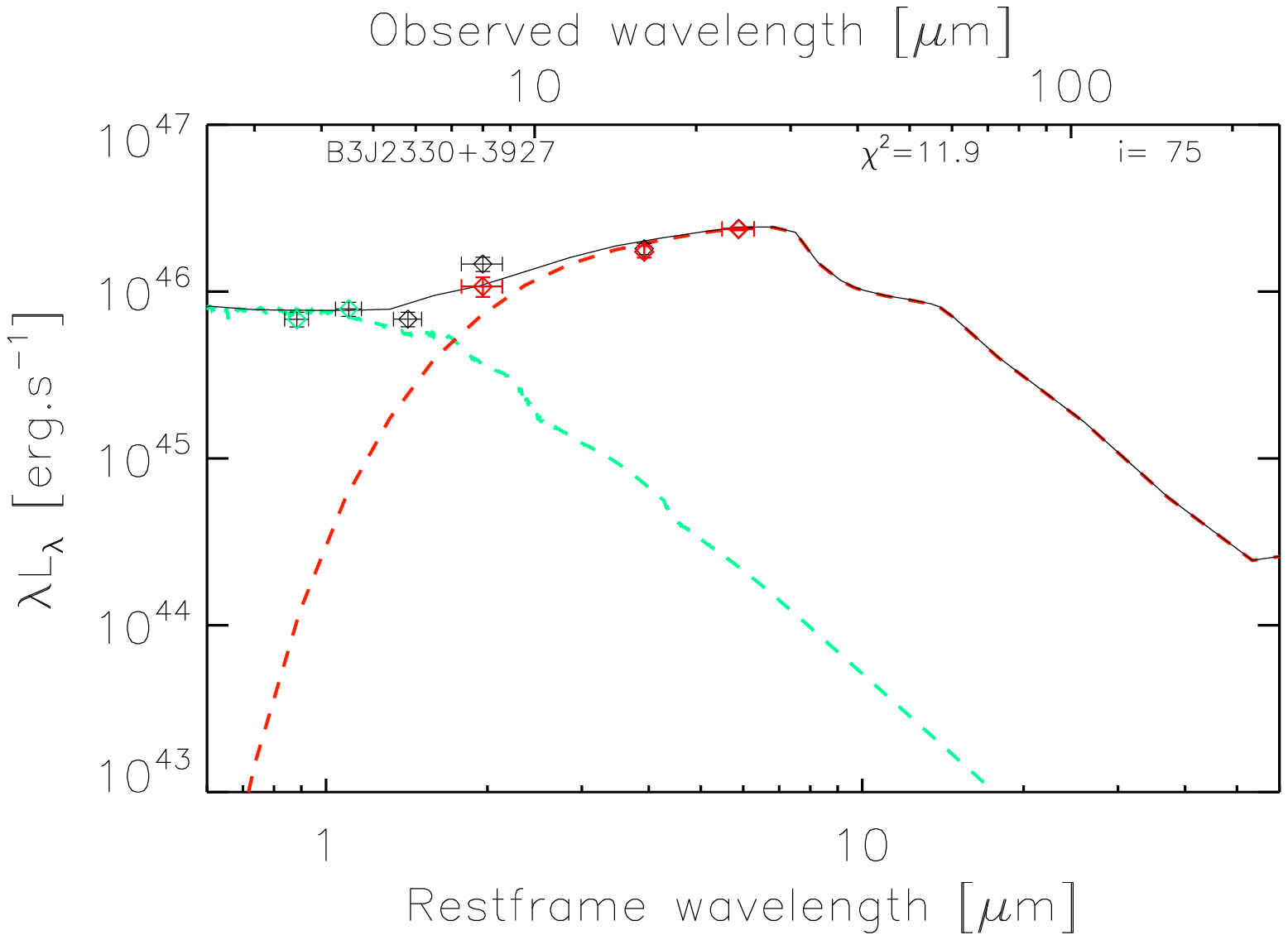} &
\includegraphics[height=42mm,trim=95  0 15 41,clip=true]{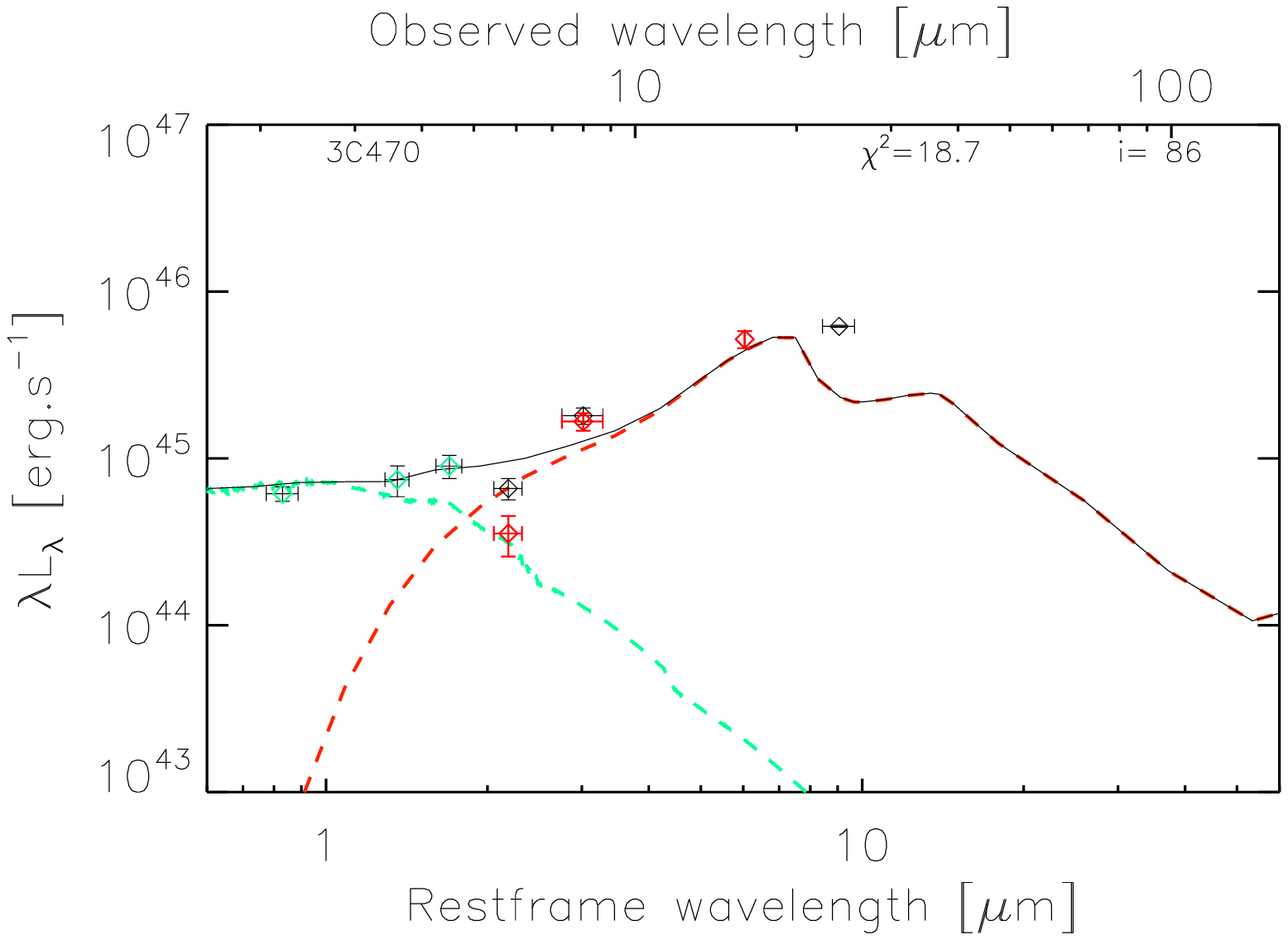}        & \\
  \end{tabular}              
  \caption{Mid-IR SEDs for the sSHzRG sample. Black points: photometric data.  Green points:
    measurements used to normalise the contribution from an old
    stellar population. Red points: results after subtraction of the
    stellar contribution.  Dashed green line: model stellar
    emission. Dashed red line: best-fitting torus model. Full black
    line: sum of stellar and torus models.}
  \label{fig:full_SED}
\end{center}
\end{figure*}

\newpage

\begin{figure*}[ht]
  \begin{center}
  \begin{tabular}{r@{}c@{}l} 
\includegraphics[height=42mm,trim= 0 39 15  0,clip=true]{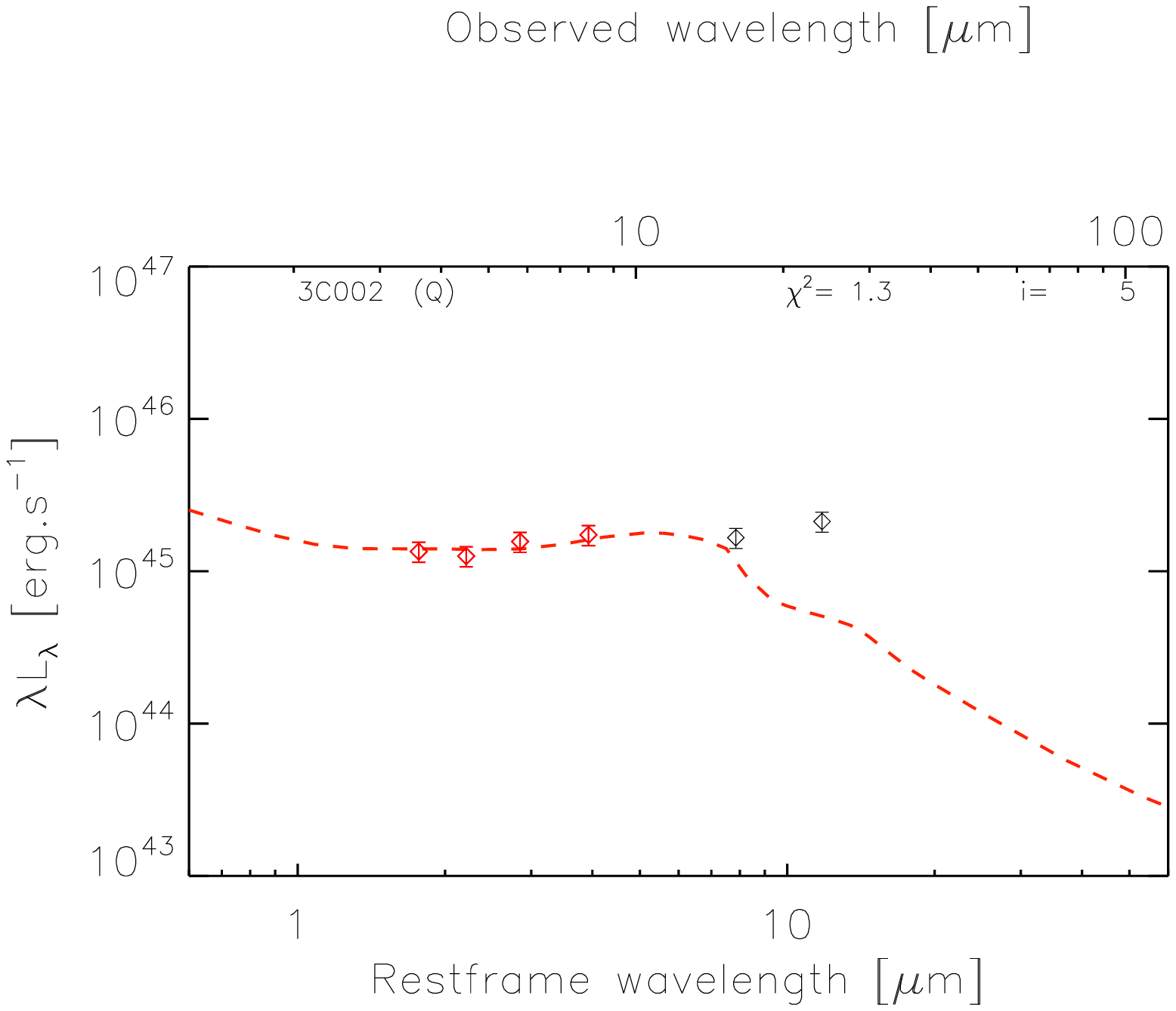} &
\includegraphics[height=42mm,trim=95 39 15  0,clip=true]{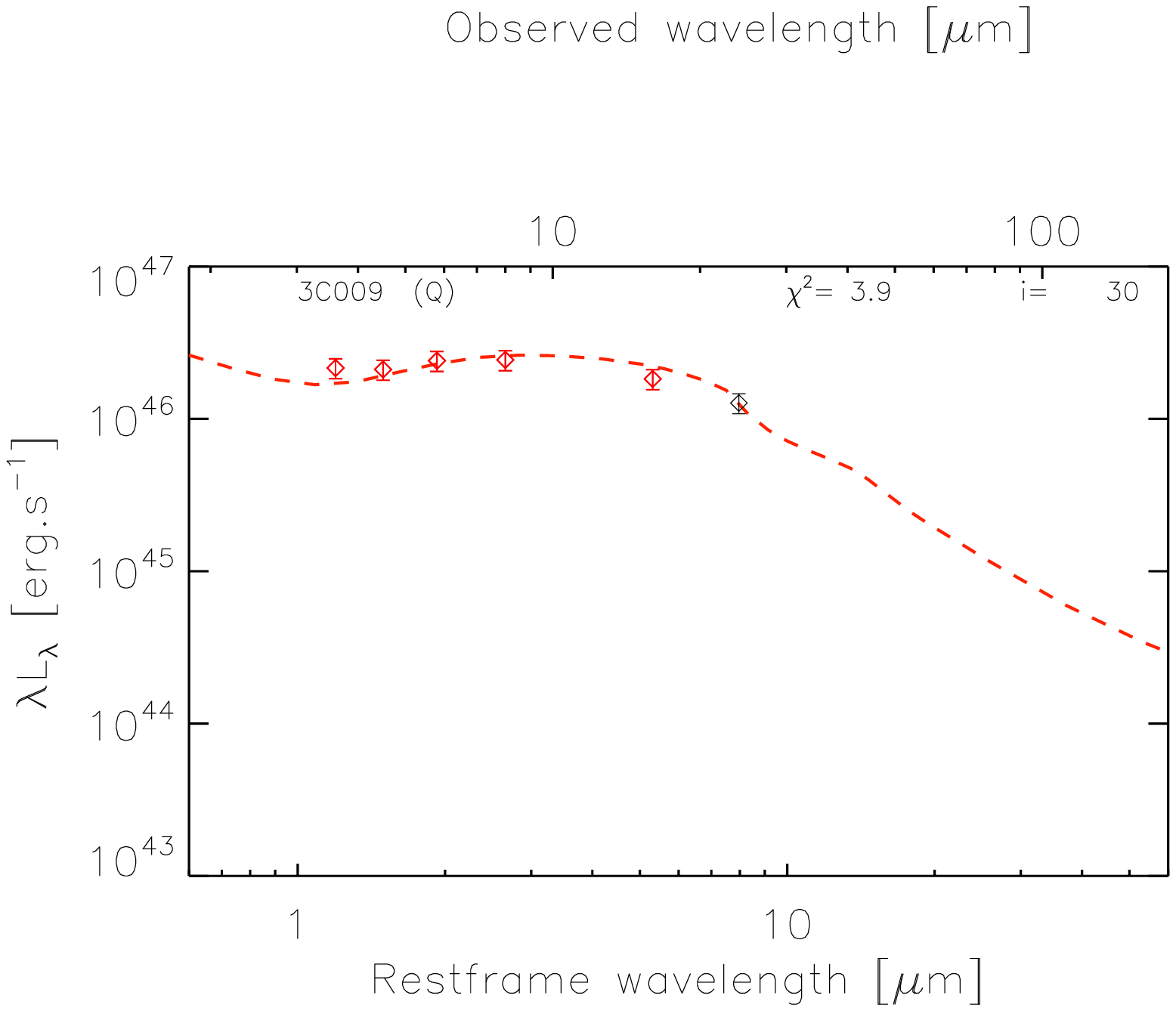} &
\includegraphics[height=42mm,trim=95 39 15  0,clip=true]{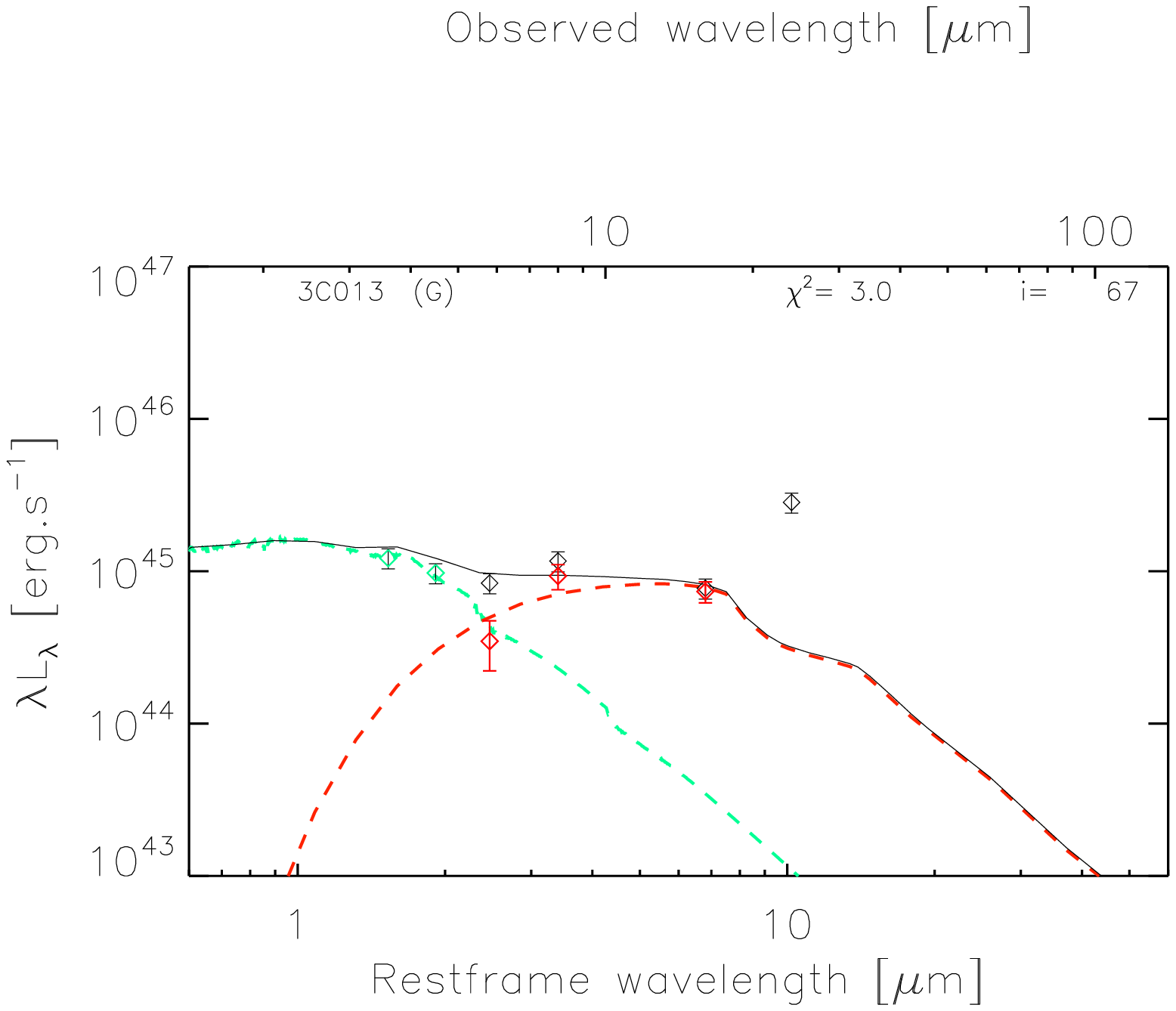} \\
\includegraphics[height=37mm,trim= 0 39 15 41,clip=true]{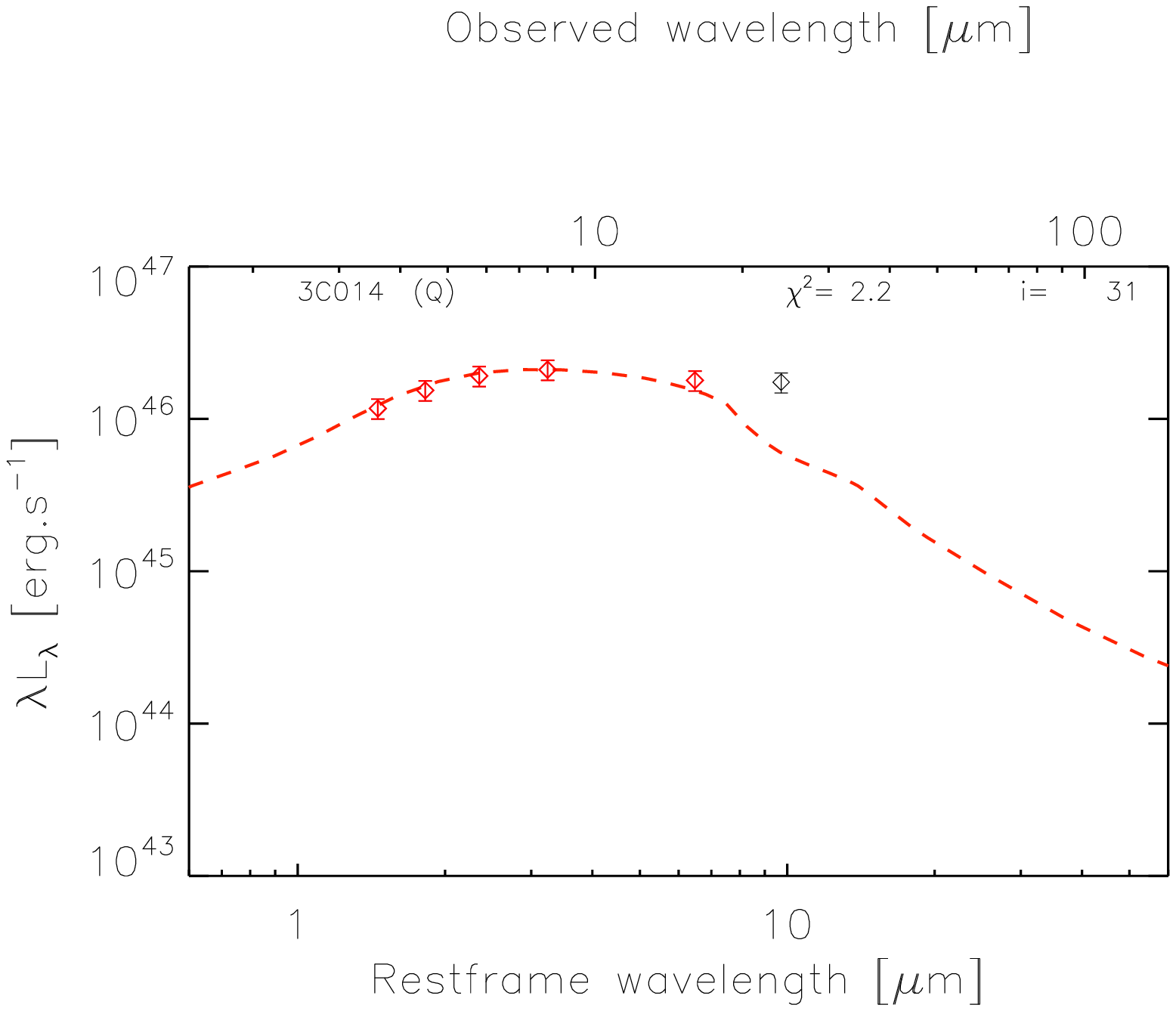} &
\includegraphics[height=37mm,trim=95 39 15 41,clip=true]{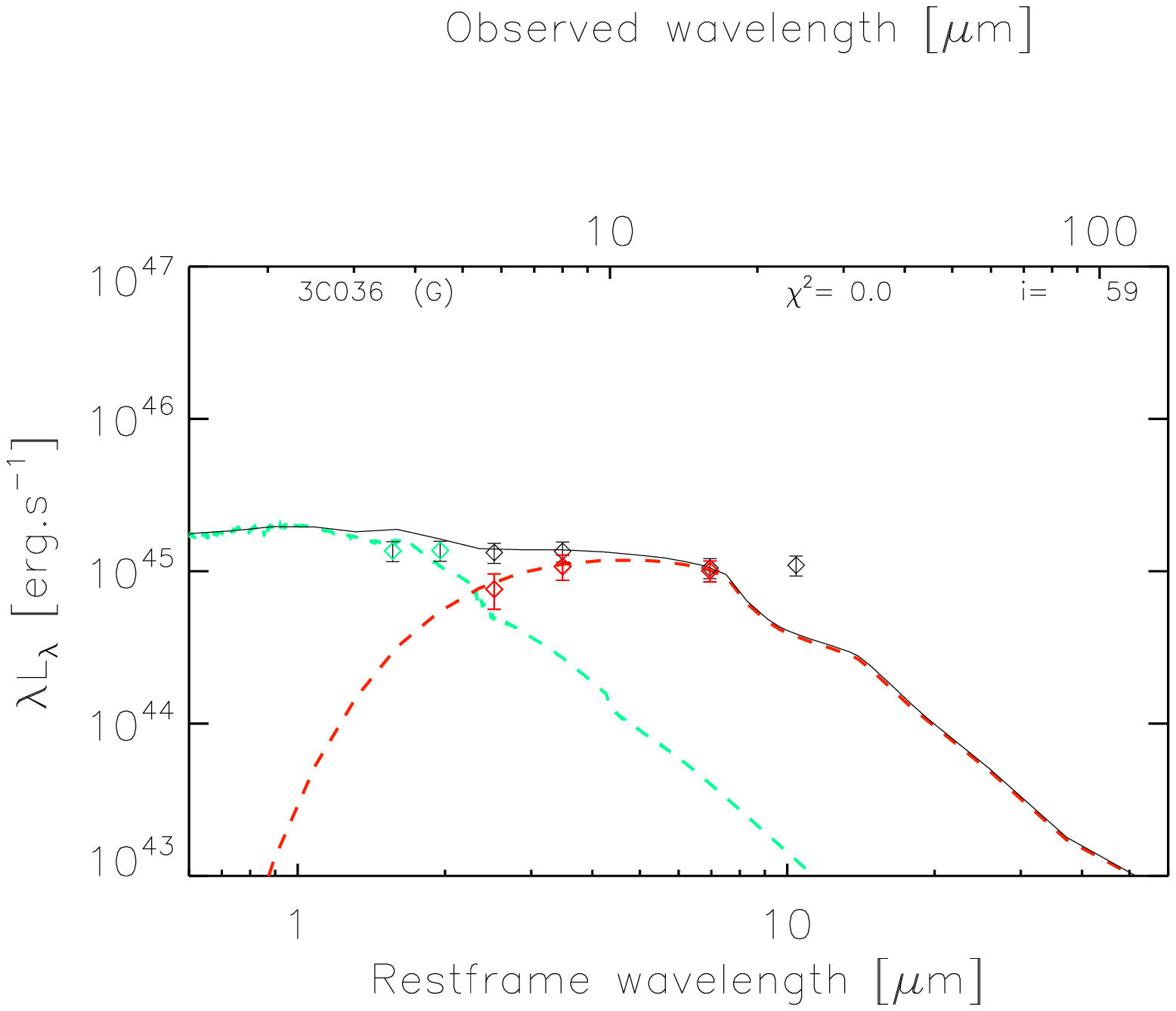} &
\includegraphics[height=37mm,trim=95 39 15 41,clip=true]{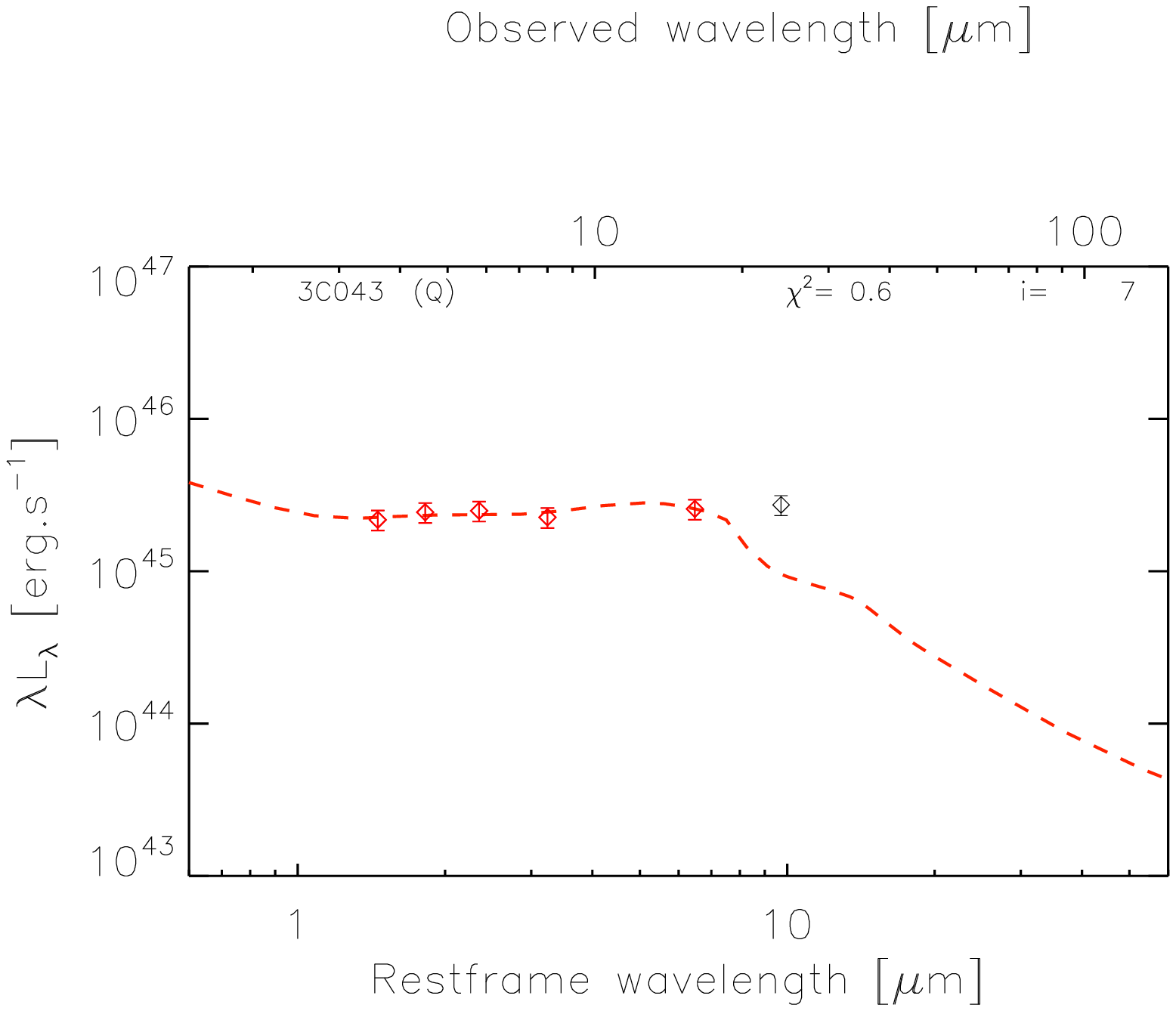} \\
\includegraphics[height=37mm,trim= 0 39 15 41,clip=true]{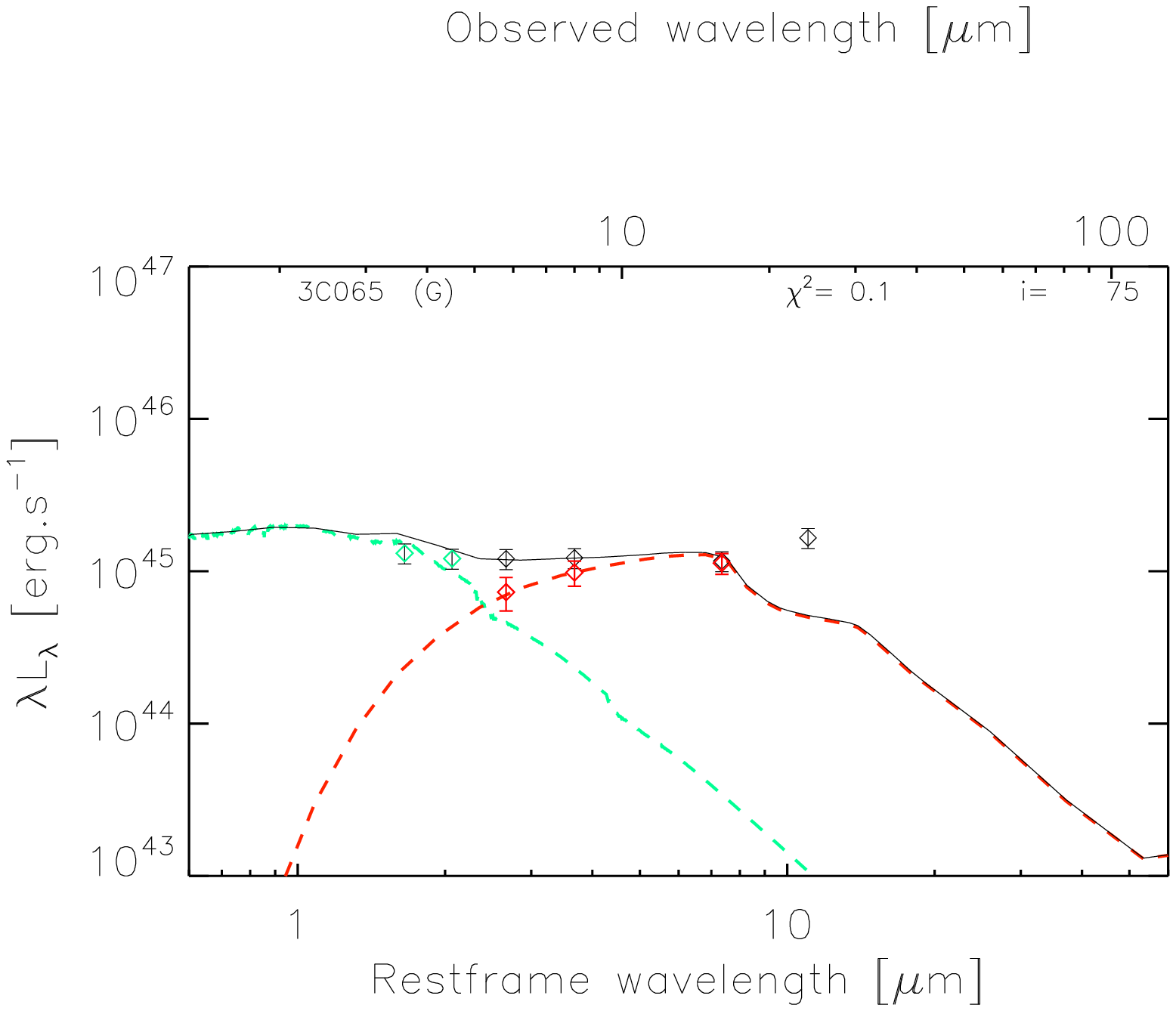} &
\includegraphics[height=37mm,trim=95 39 15 41,clip=true]{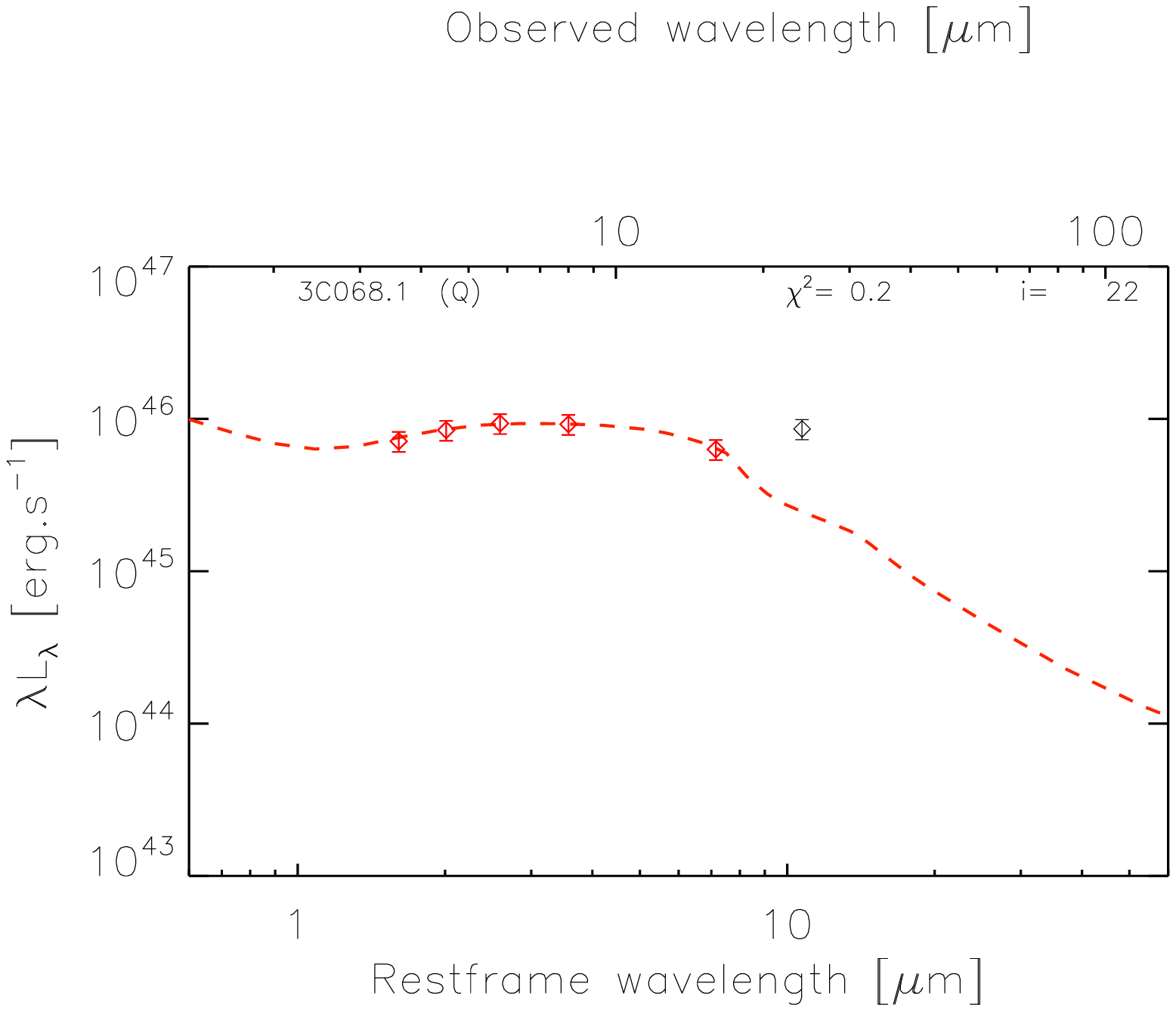} &
\includegraphics[height=37mm,trim=95 39 15 41,clip=true]{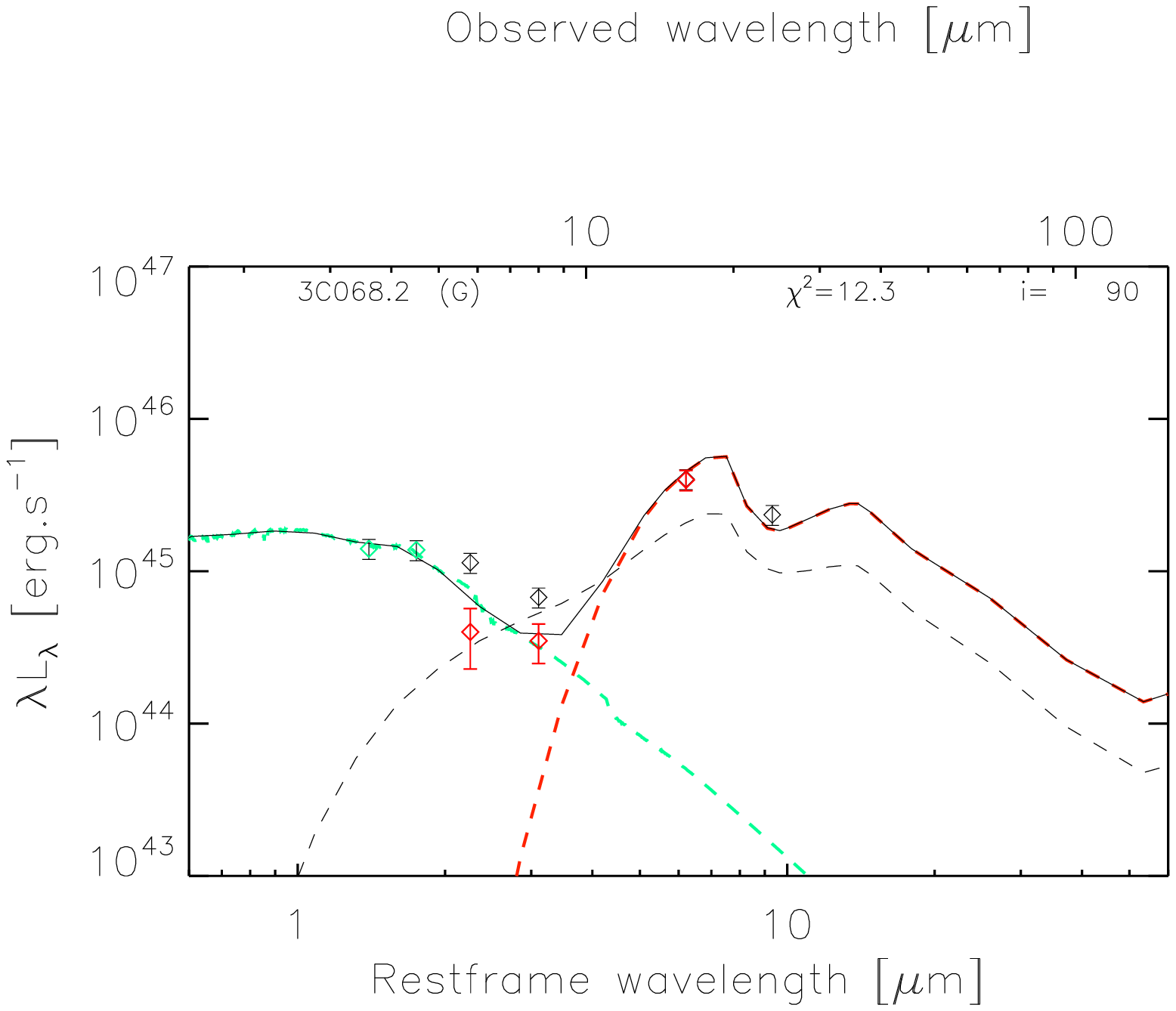} \\
\includegraphics[height=37mm,trim= 0 39 15 41,clip=true]{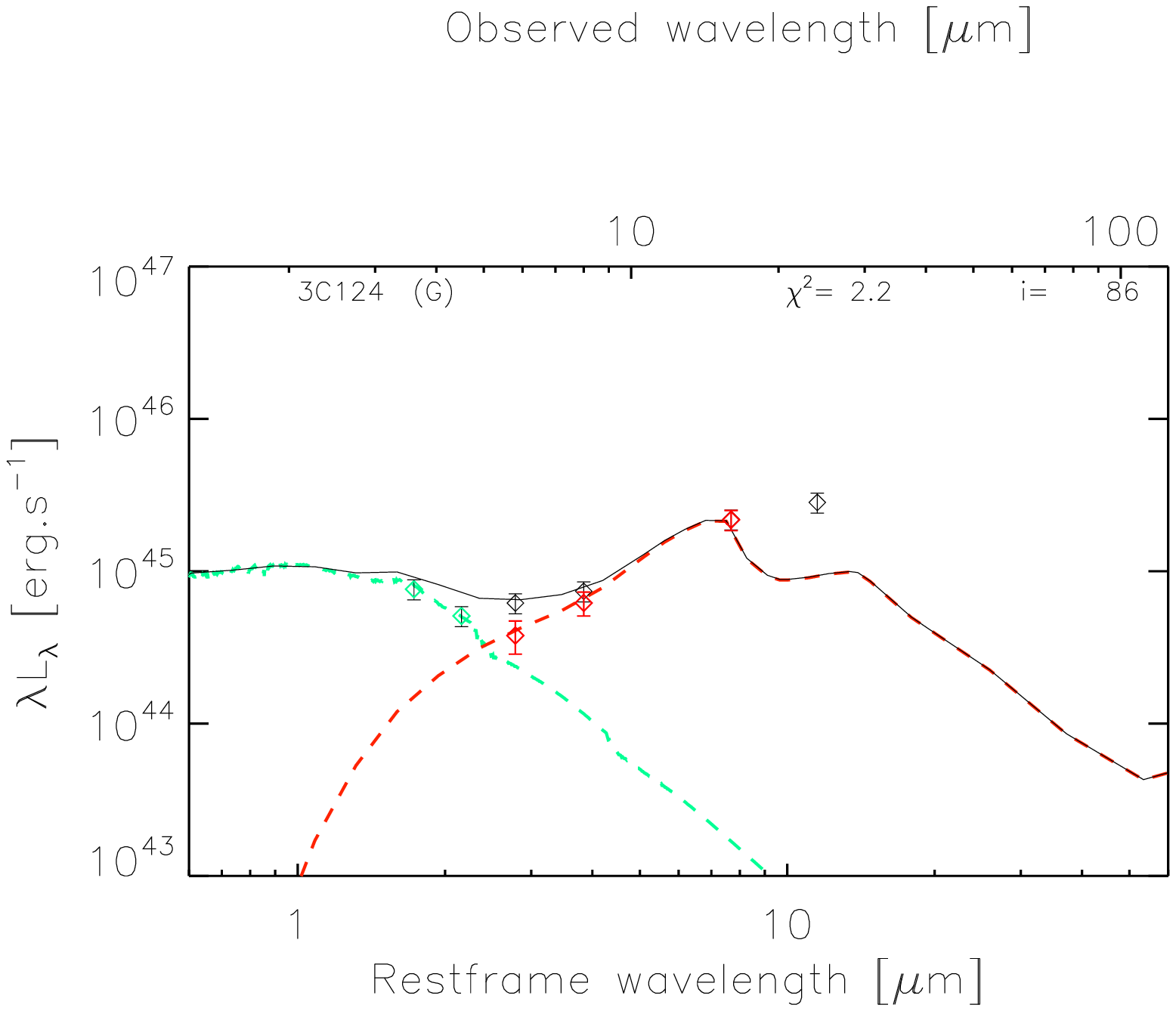} &
\includegraphics[height=37mm,trim=95 39 15 41,clip=true]{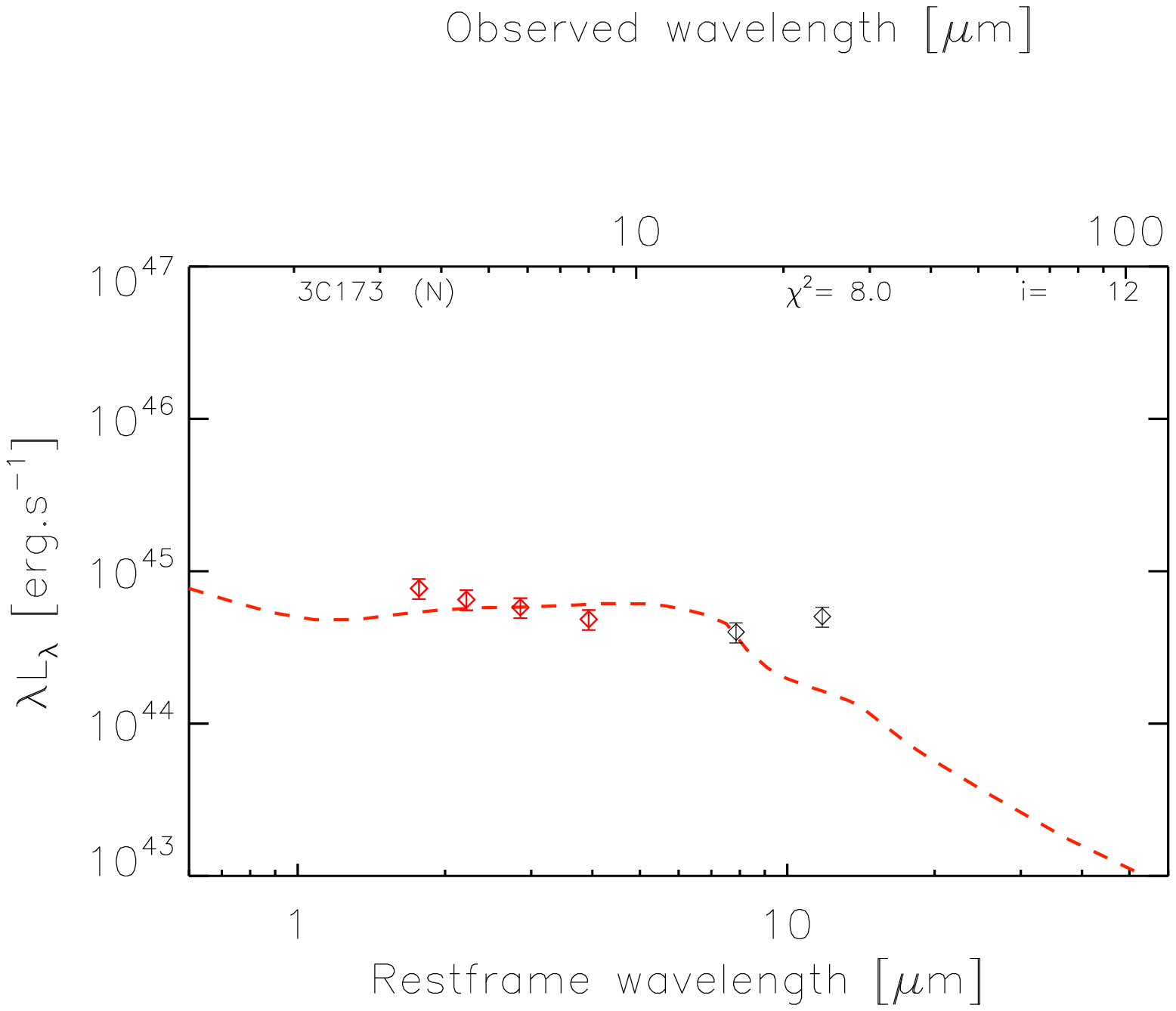} &
\includegraphics[height=37mm,trim=95 39 15 41,clip=true]{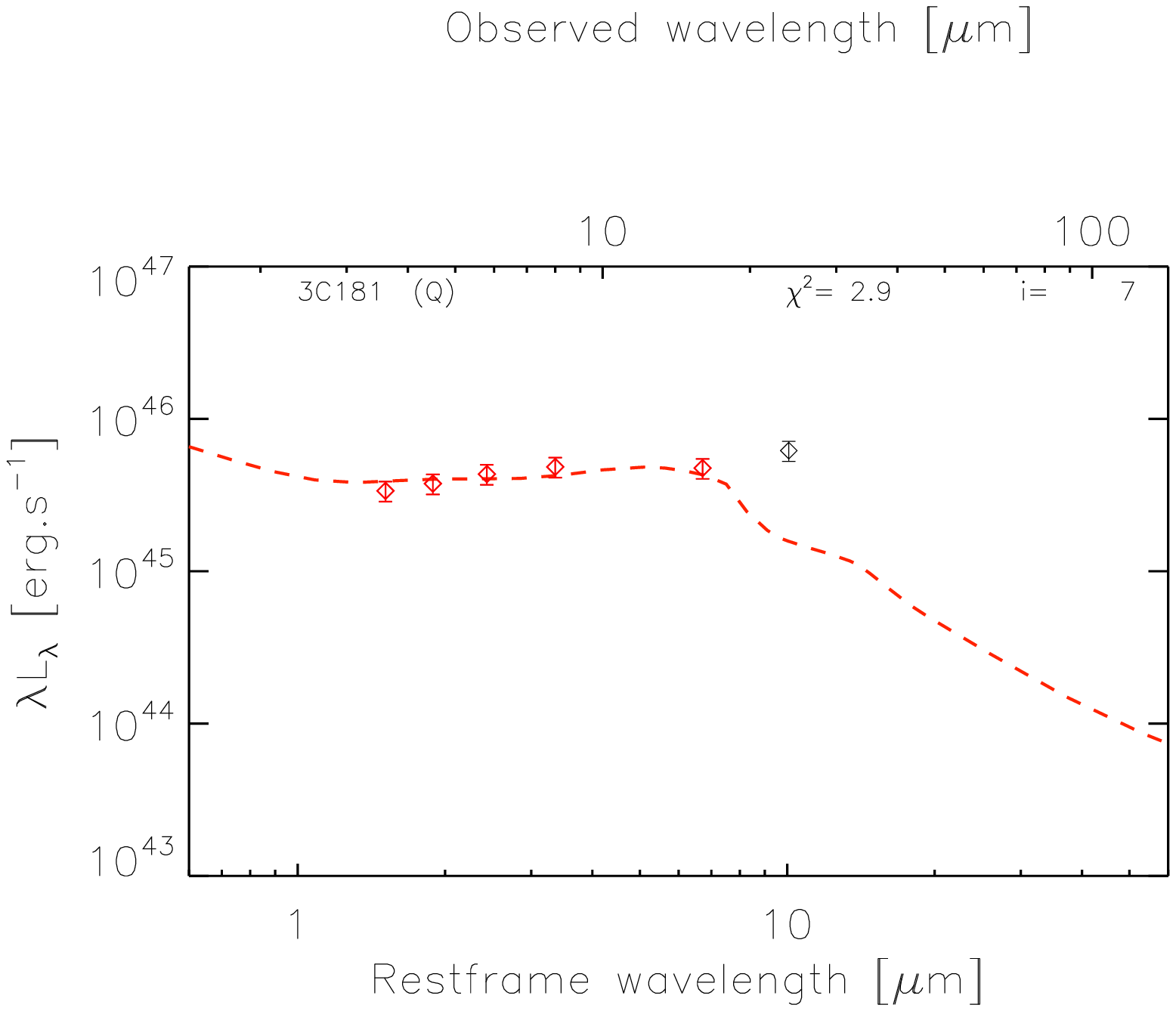} \\
\includegraphics[height=37mm,trim= 0 39 15 41,clip=true]{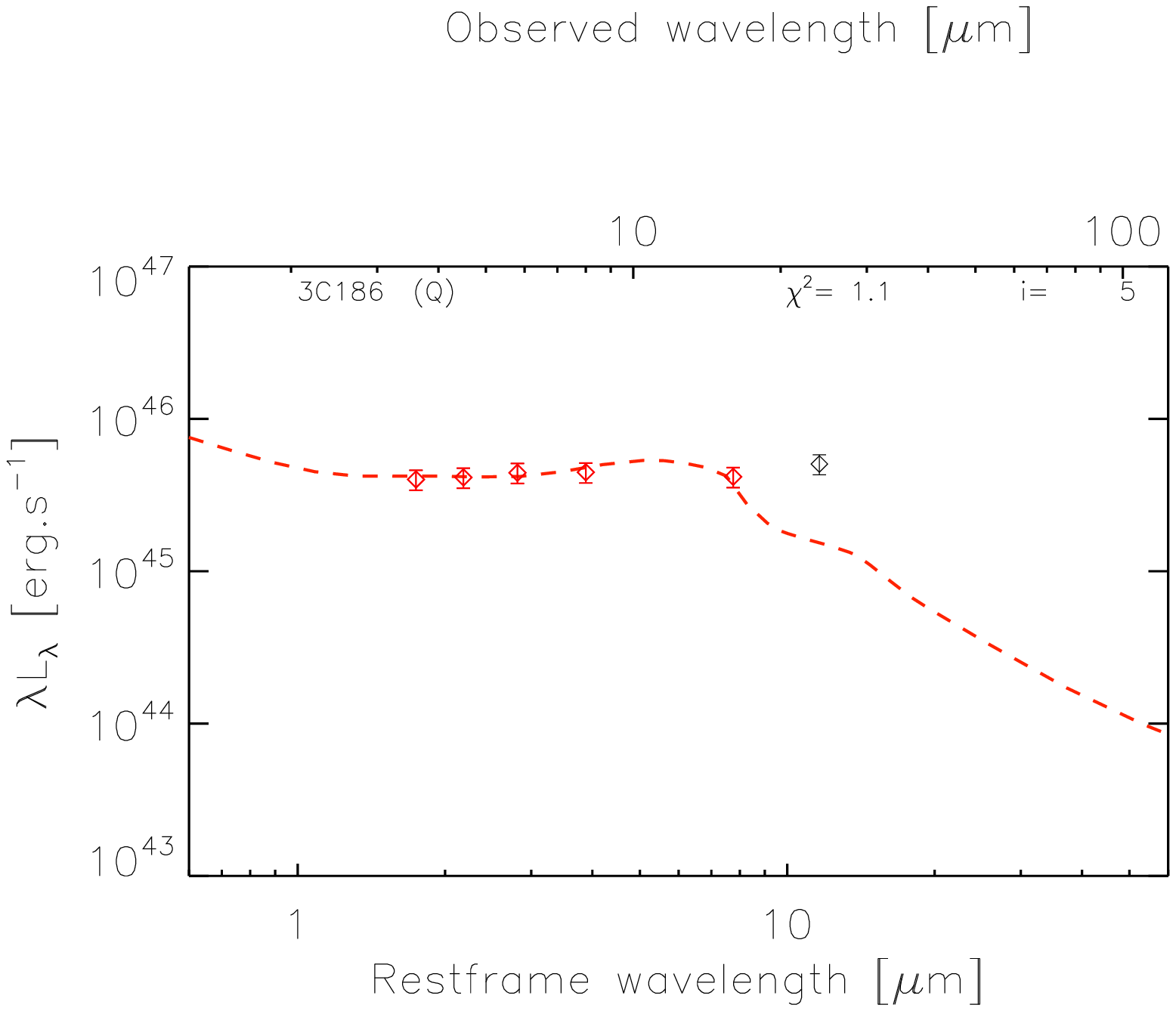} &
\includegraphics[height=37mm,trim=95 39 15 41,clip=true]{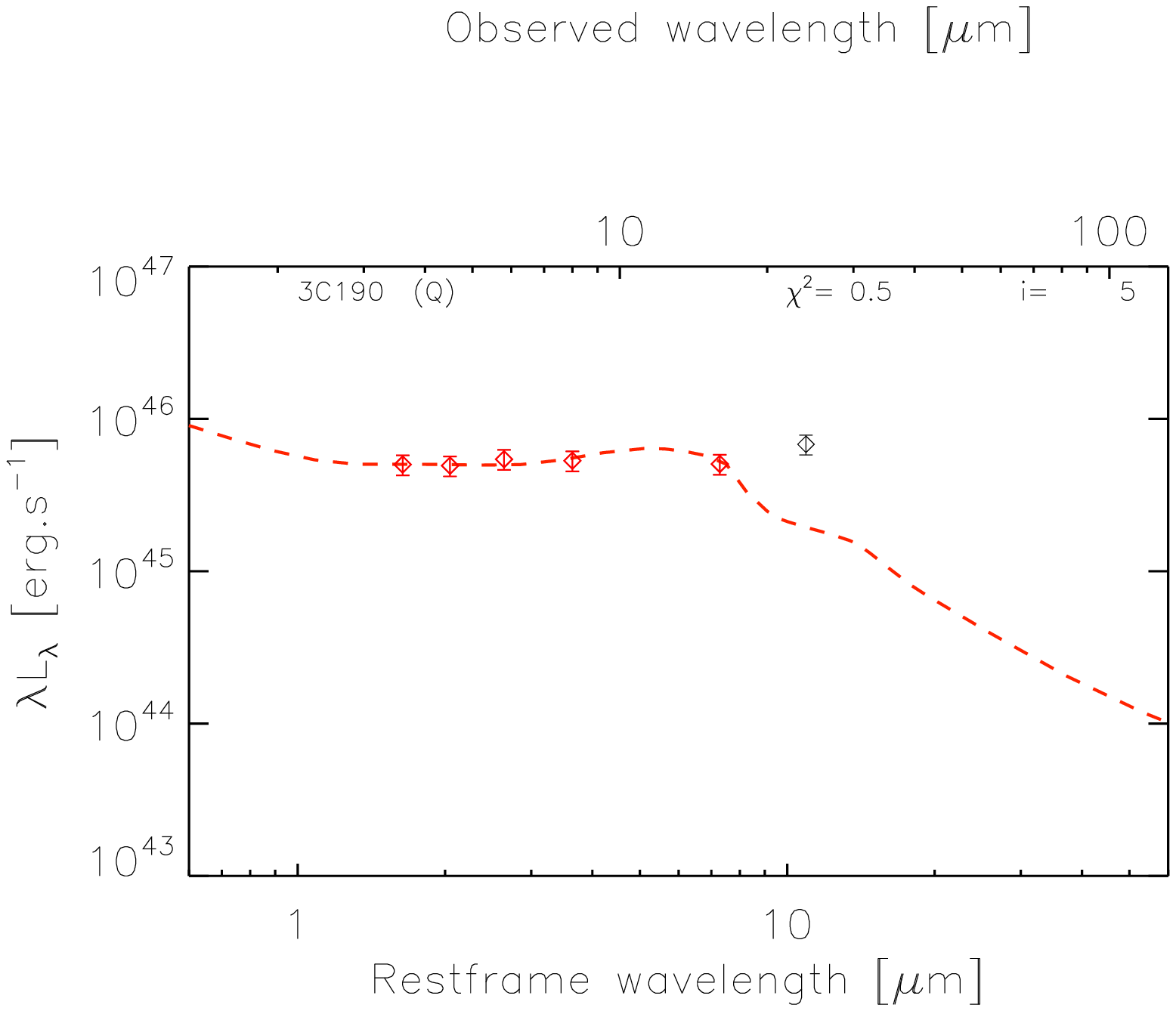} &
\includegraphics[height=37mm,trim=95 39 15 41,clip=true]{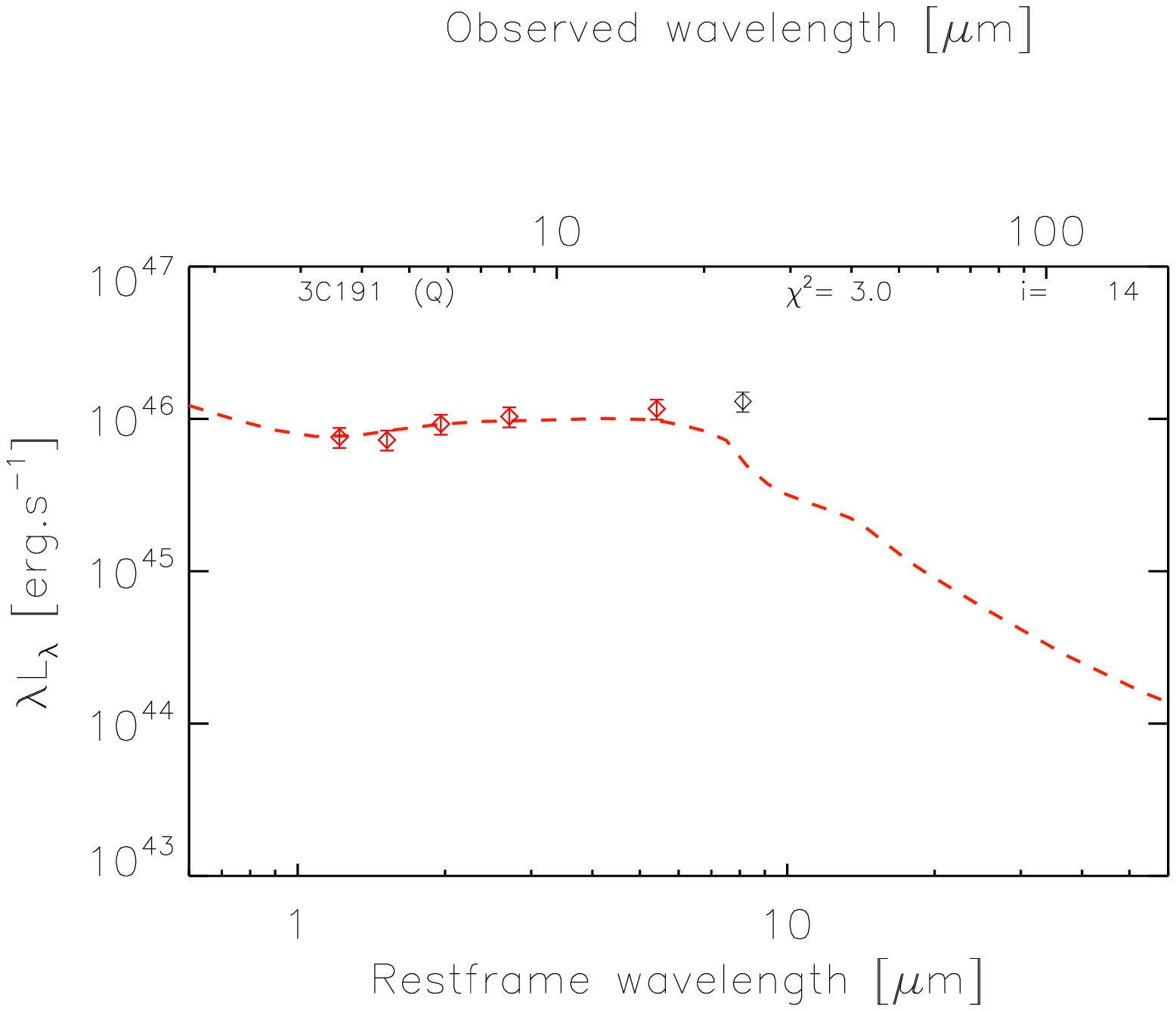} \\
\includegraphics[height=42mm,trim= 0  0 15 41,clip=true]{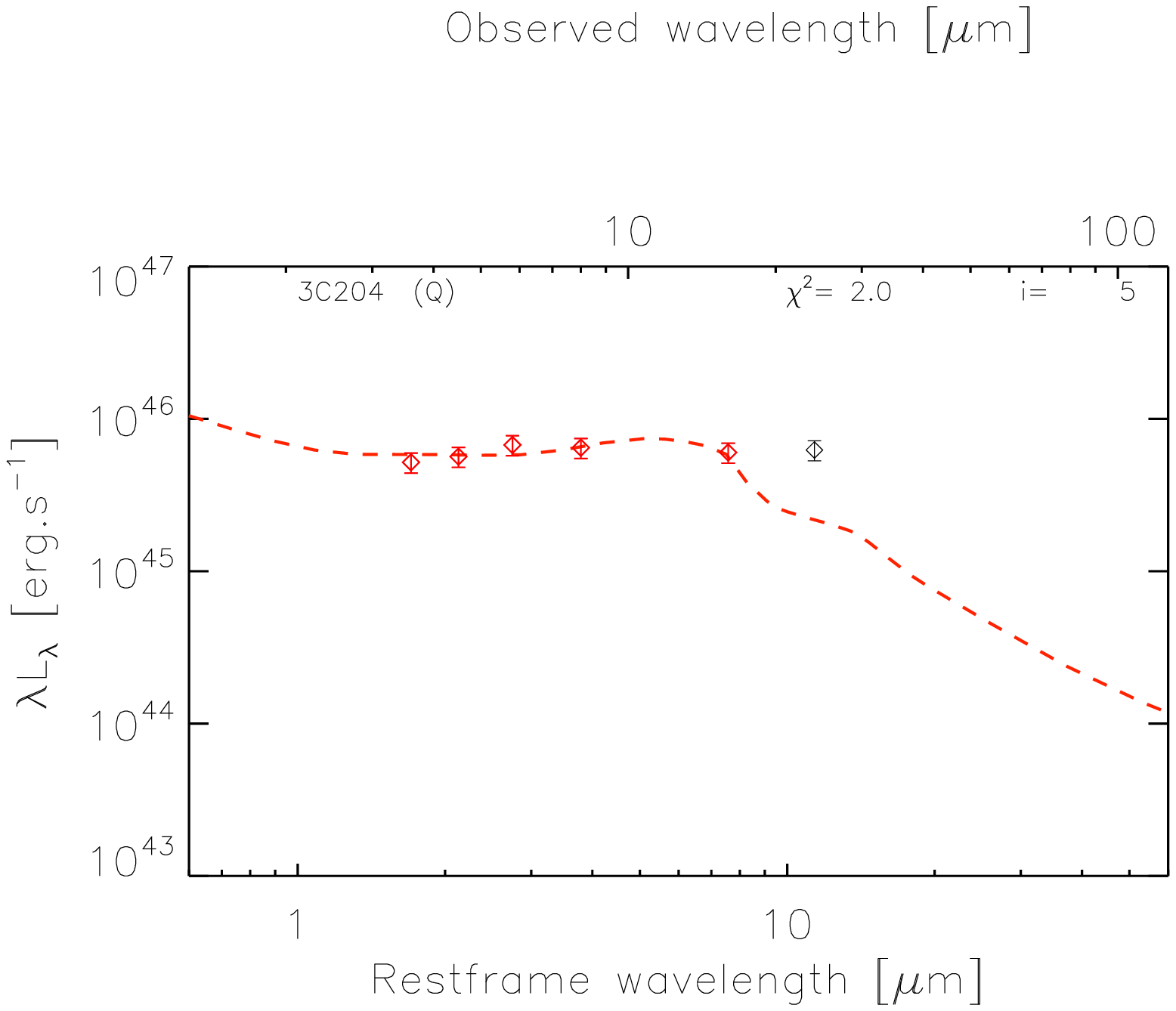} &
\includegraphics[height=42mm,trim=95  0 15 41,clip=true]{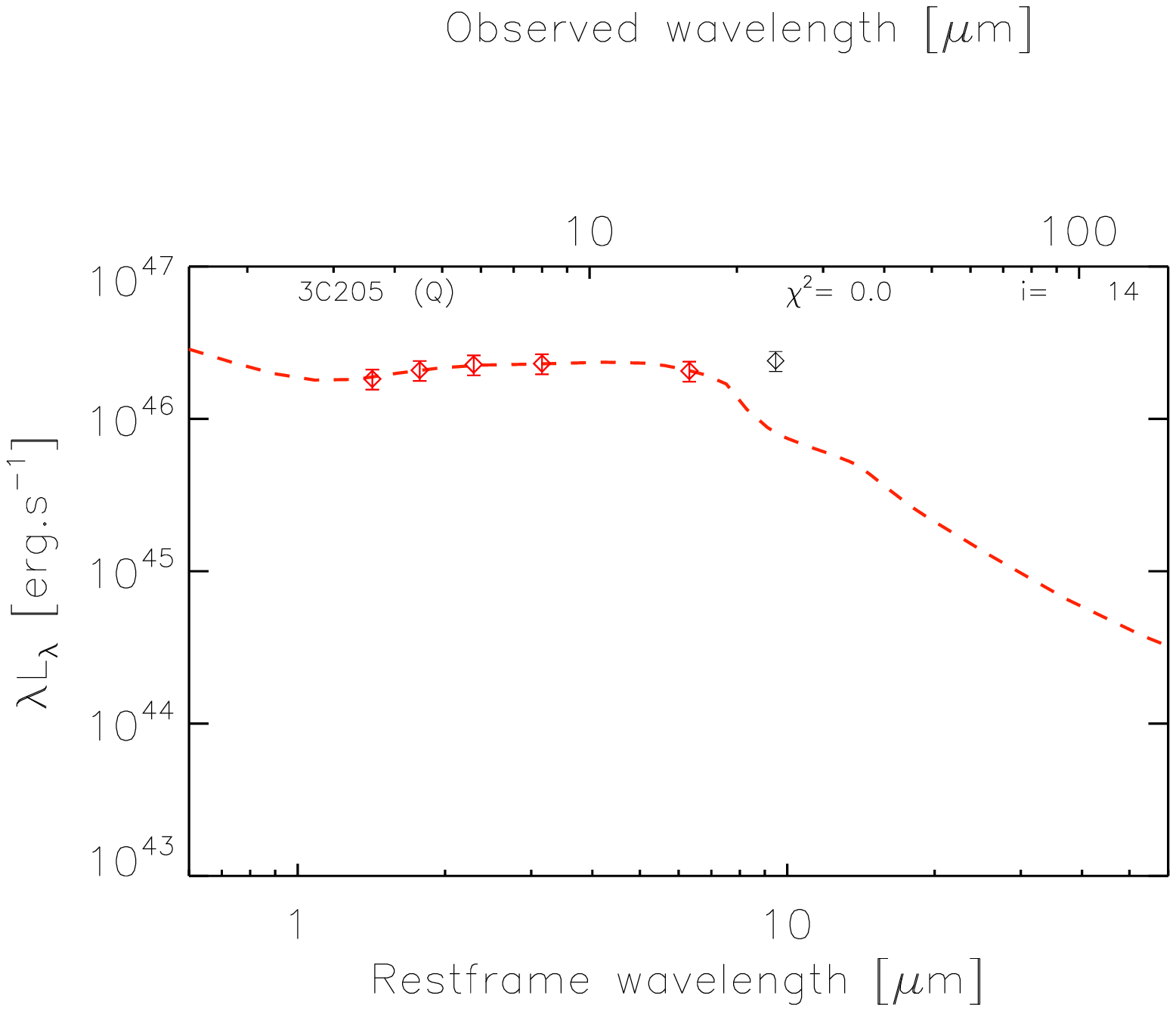} &
\includegraphics[height=42mm,trim=95  0 15 41,clip=true]{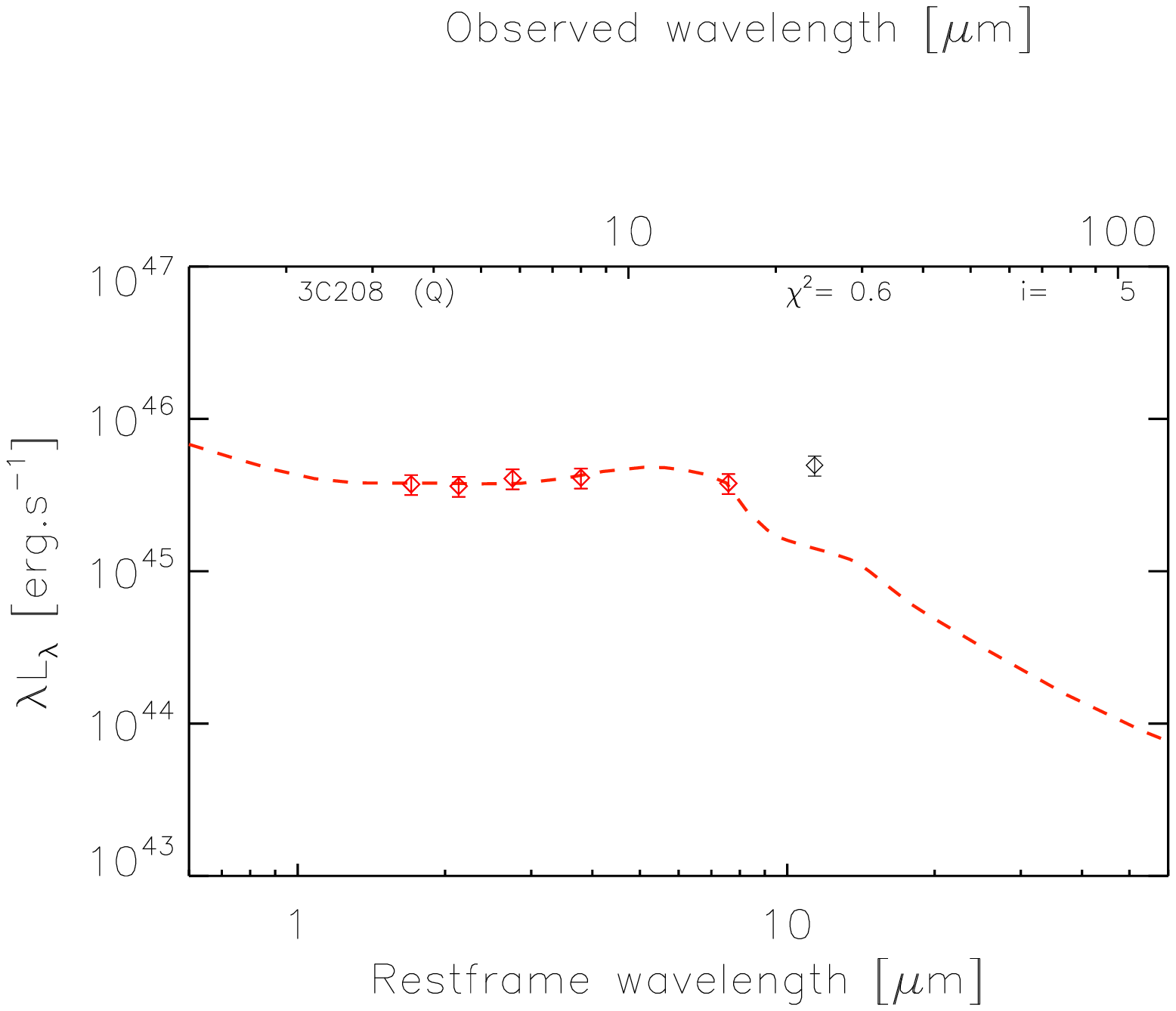} \\
  \end{tabular}
\end{center}
\end{figure*}

\begin{figure*}[ht]
  \begin{center}
  \begin{tabular}{r@{}c@{}l} 
\includegraphics[height=42mm,trim= 0 39 15  0,clip=true]{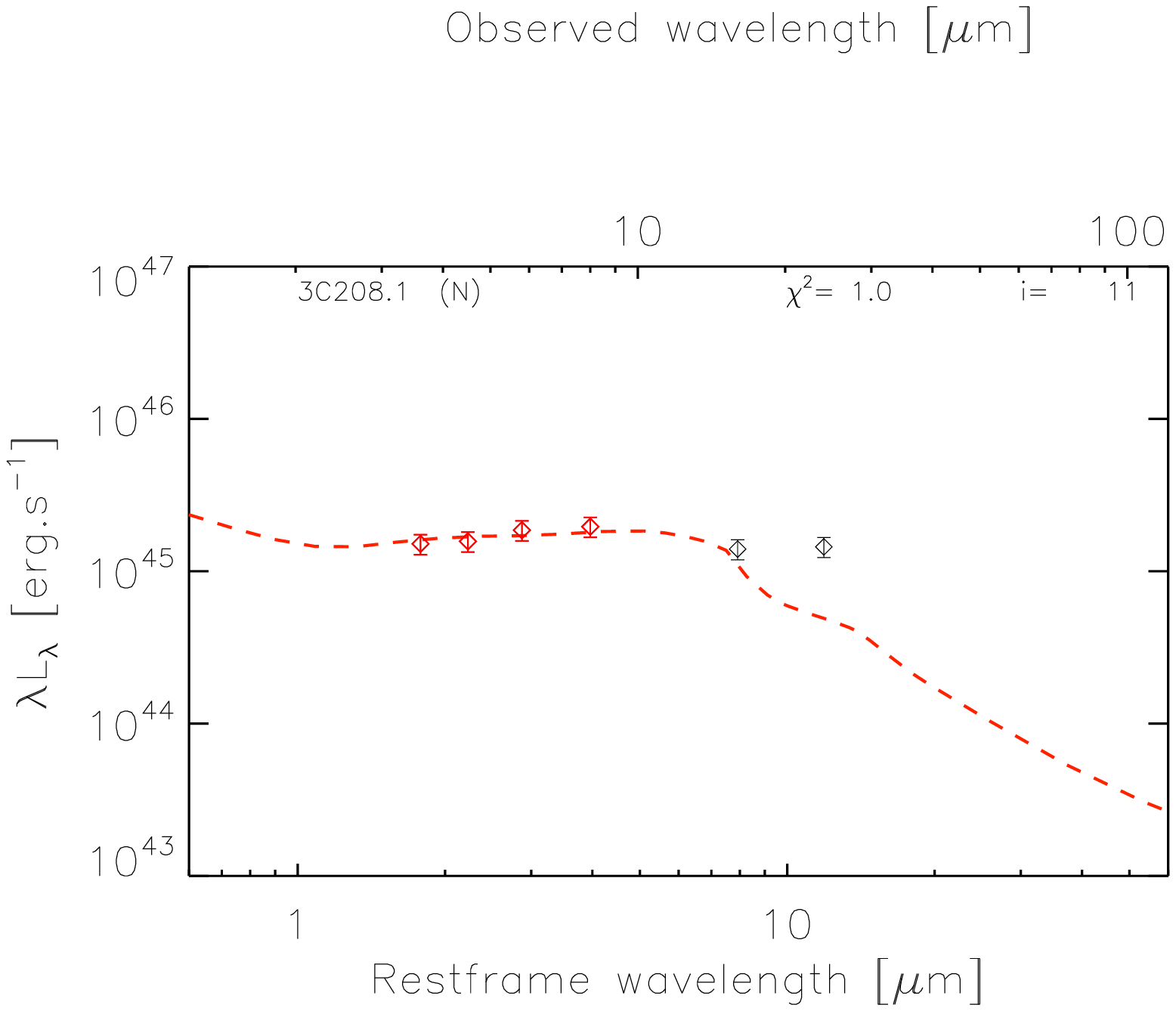} &
\includegraphics[height=42mm,trim=95 39 15  0,clip=true]{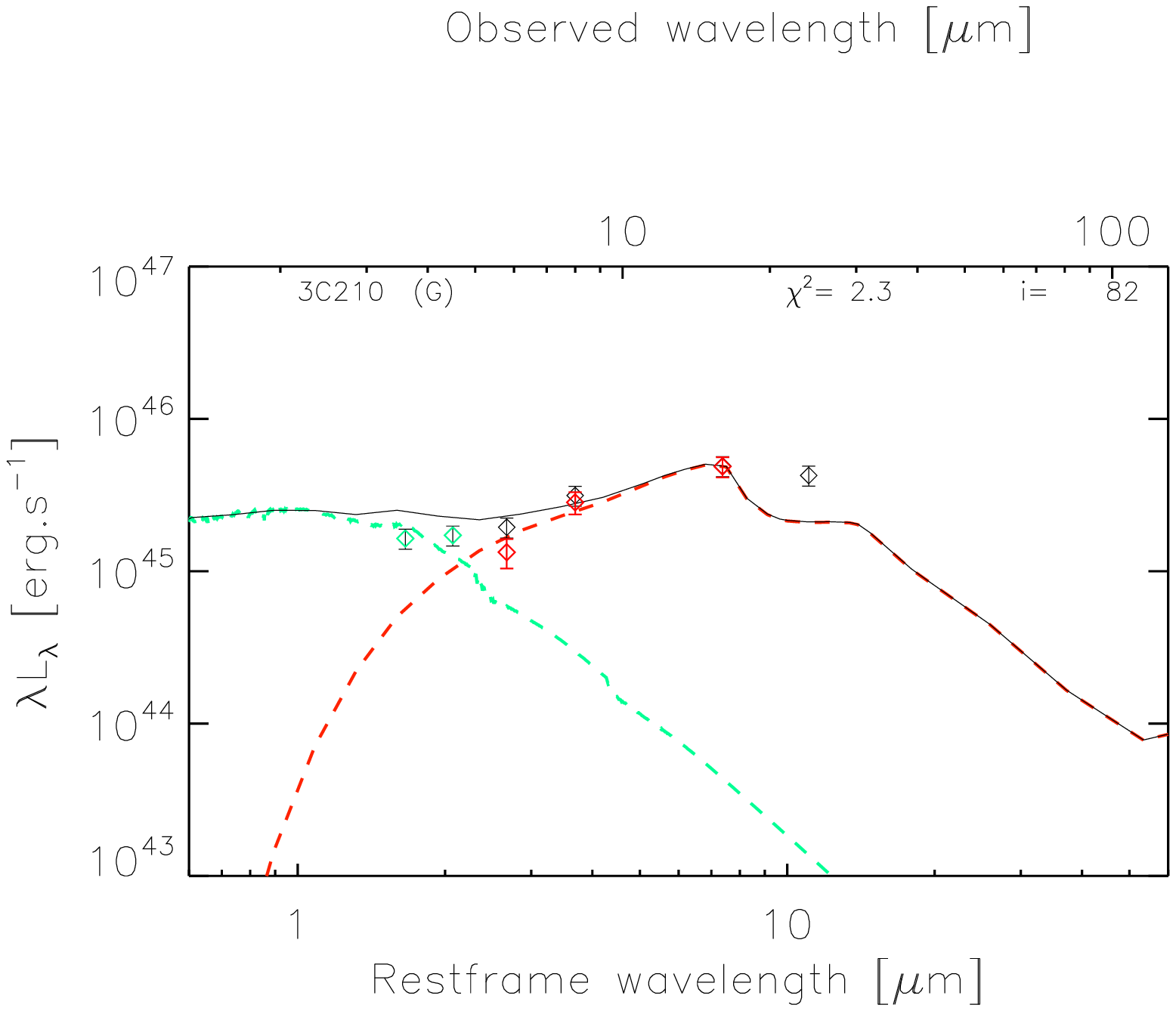} &
\includegraphics[height=42mm,trim=95 39 15  0,clip=true]{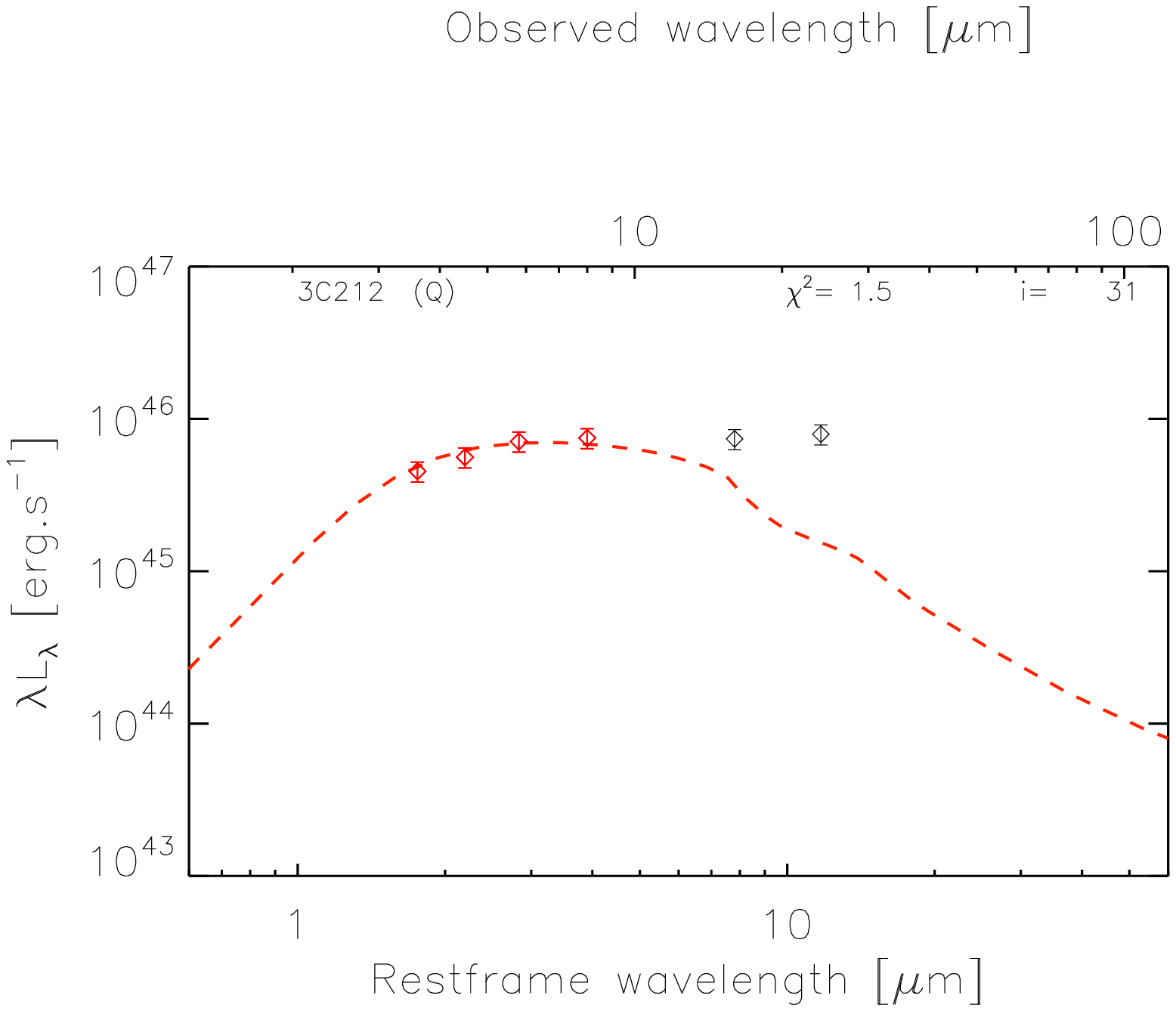} \\
\includegraphics[height=37mm,trim= 0 39 15 41,clip=true]{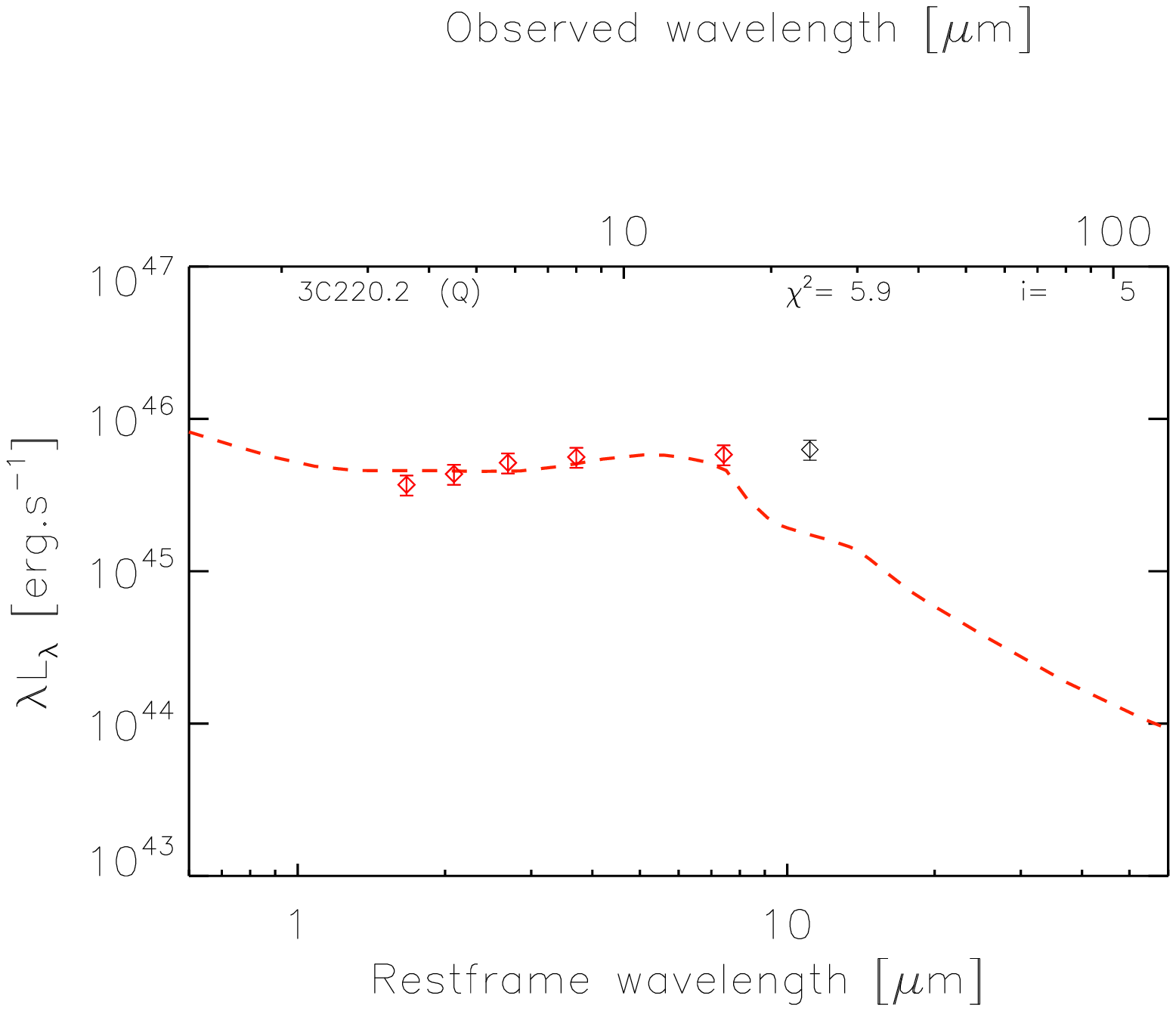} &
\includegraphics[height=37mm,trim=95 39 15 41,clip=true]{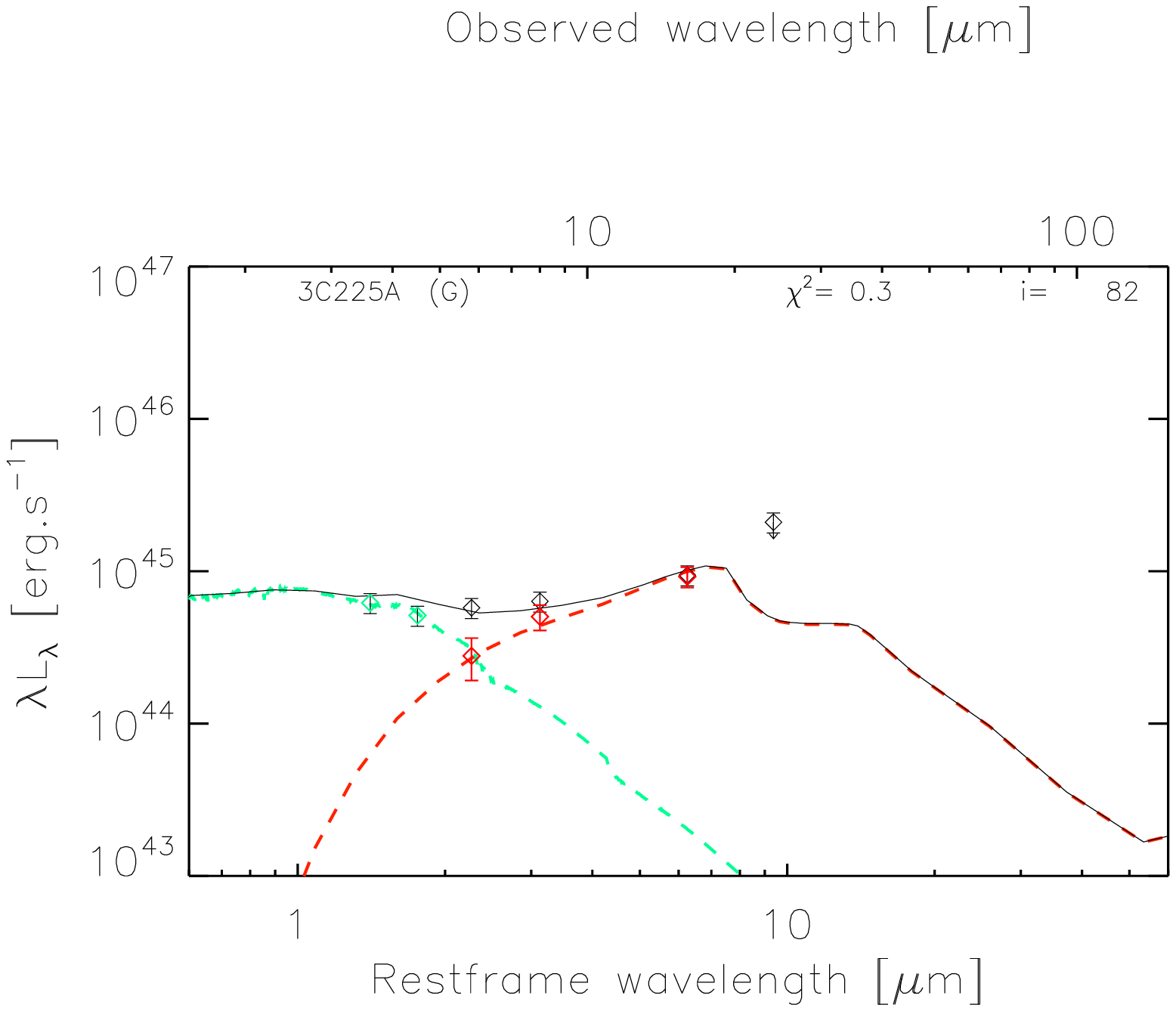} &
\includegraphics[height=37mm,trim=95 39 15 41,clip=true]{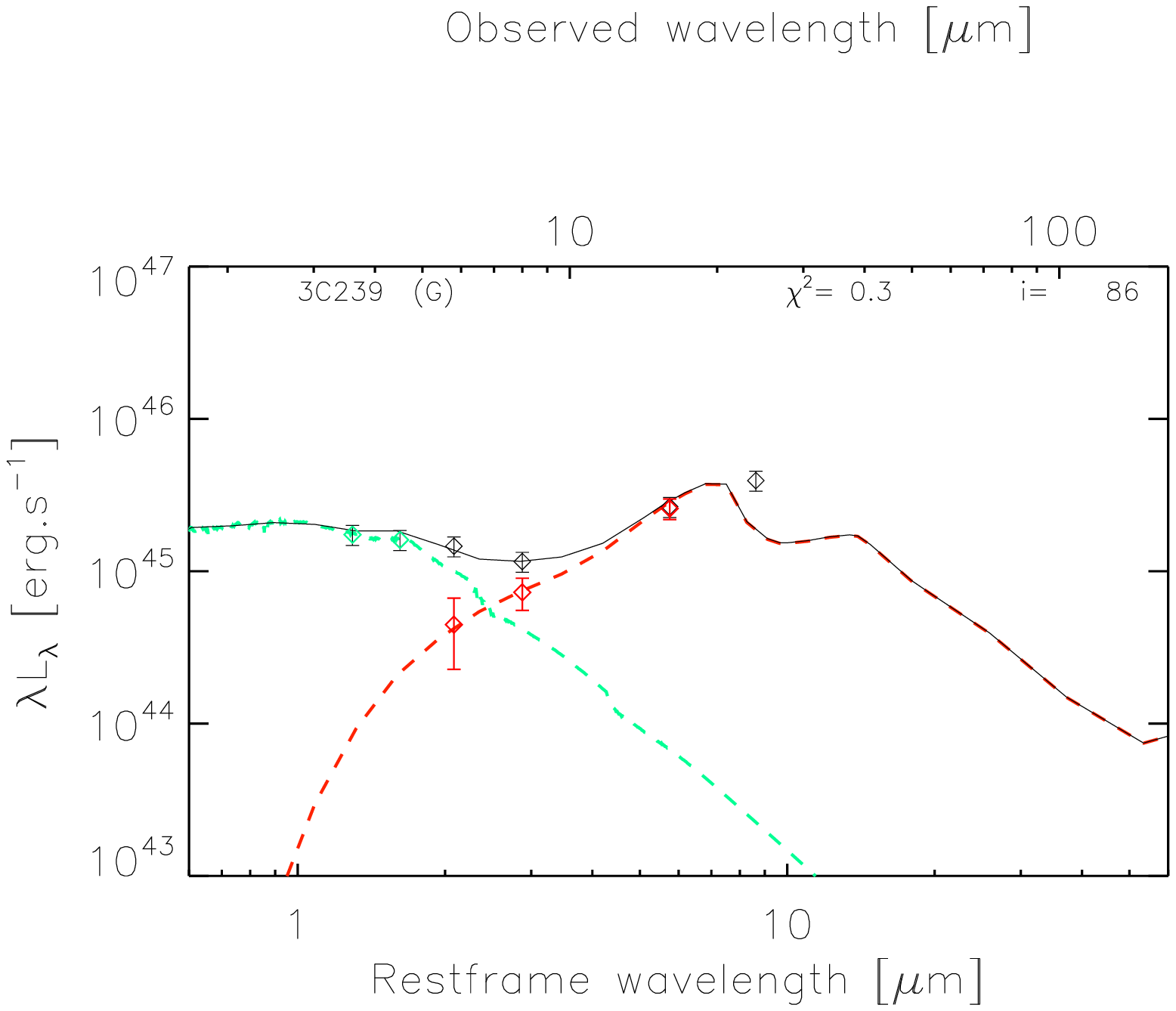} \\
\includegraphics[height=37mm,trim= 0 39 15 41,clip=true]{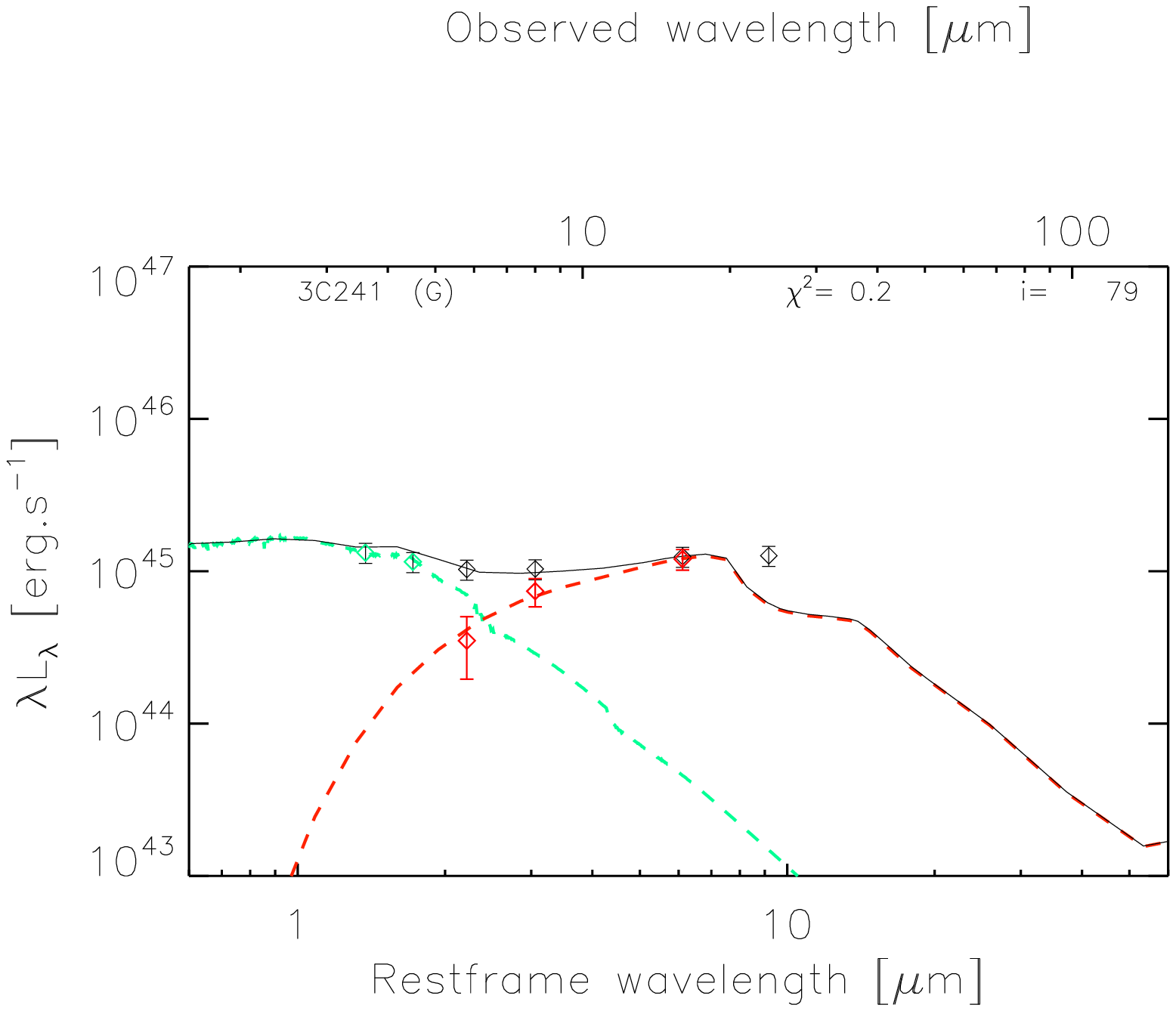} &
\includegraphics[height=37mm,trim=95 39 15 41,clip=true]{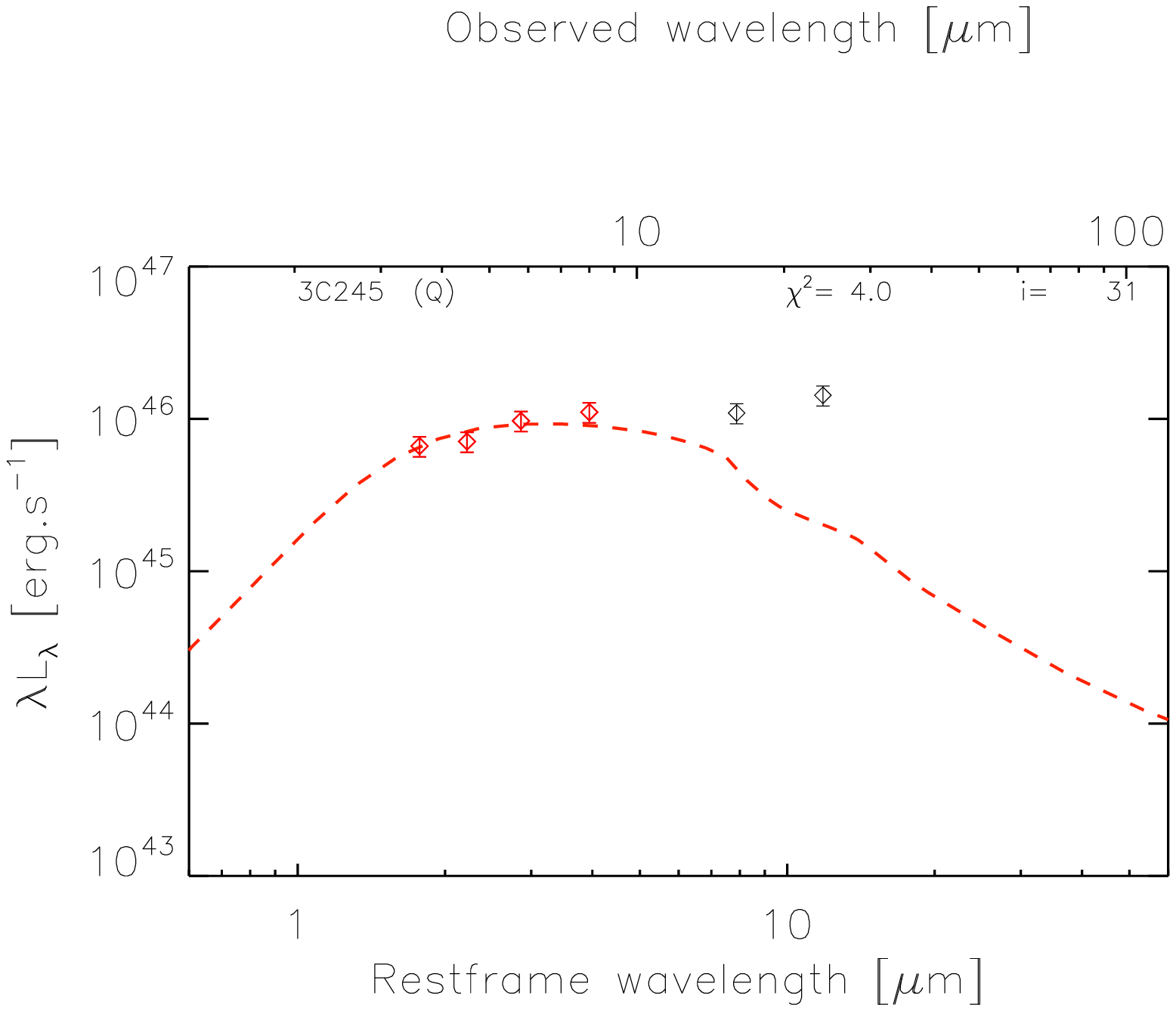} &
\includegraphics[height=37mm,trim=95 39 15 41,clip=true]{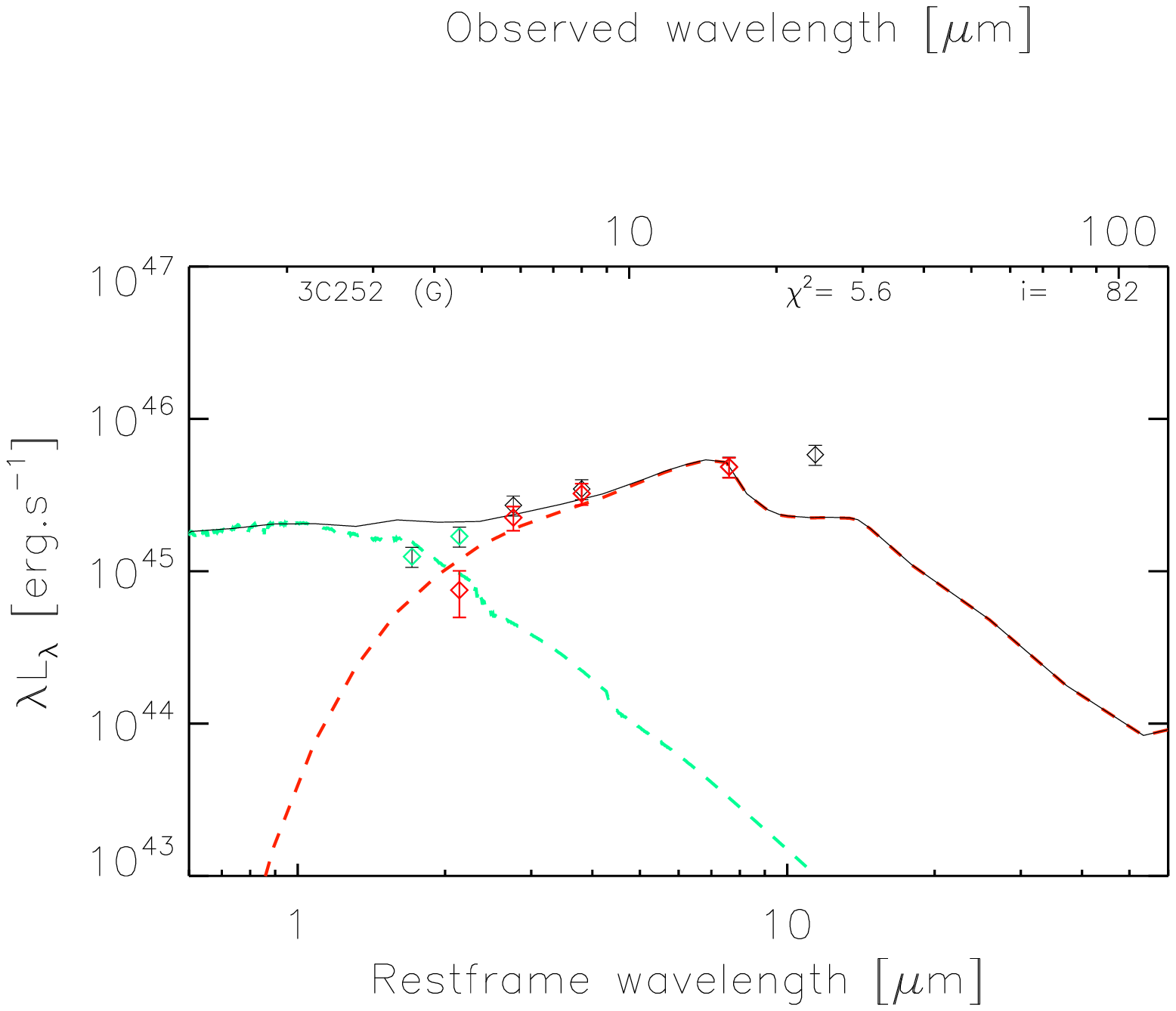} \\
\includegraphics[height=37mm,trim= 0 39 15 41,clip=true]{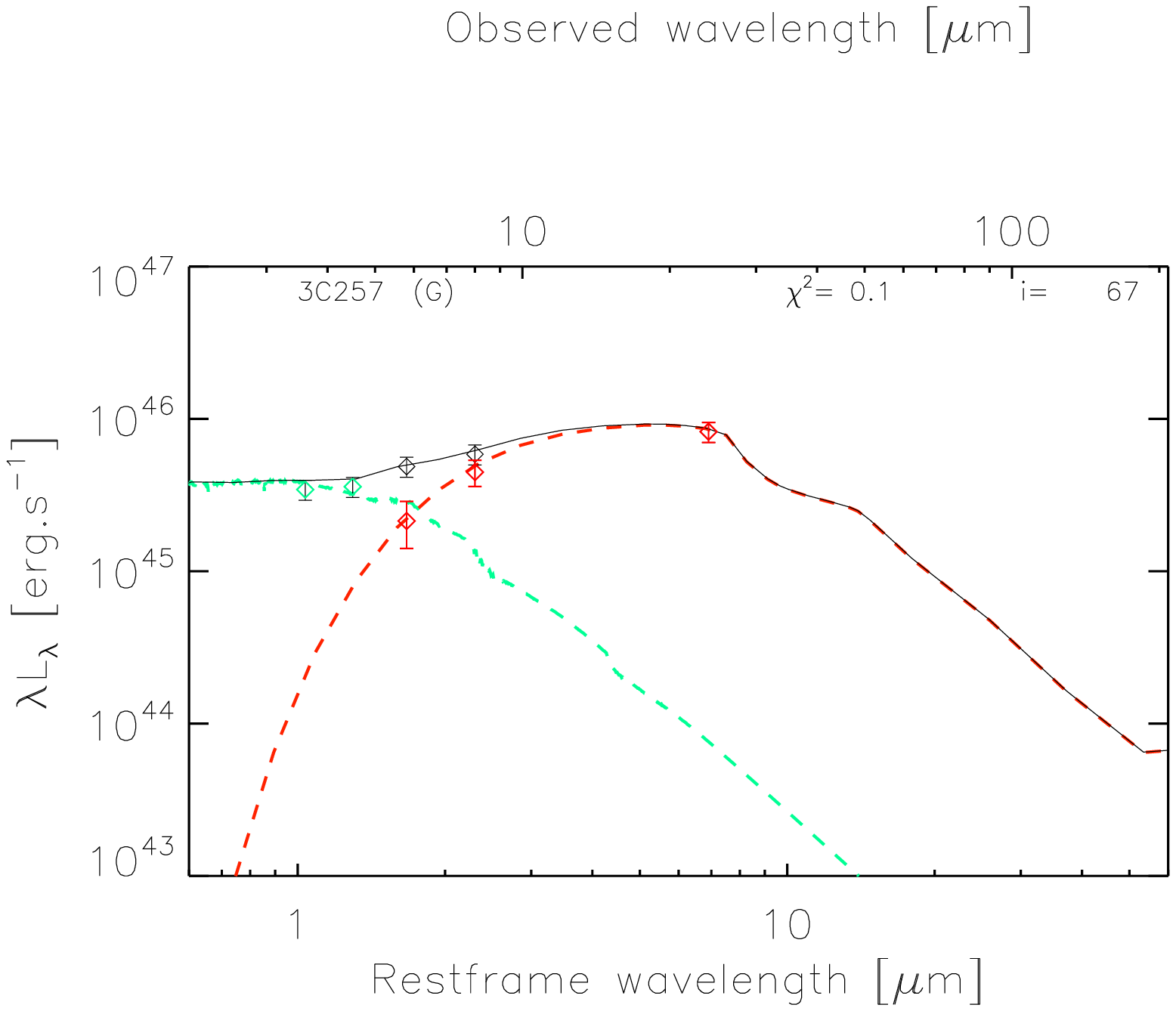} &
\includegraphics[height=37mm,trim=95 39 15 41,clip=true]{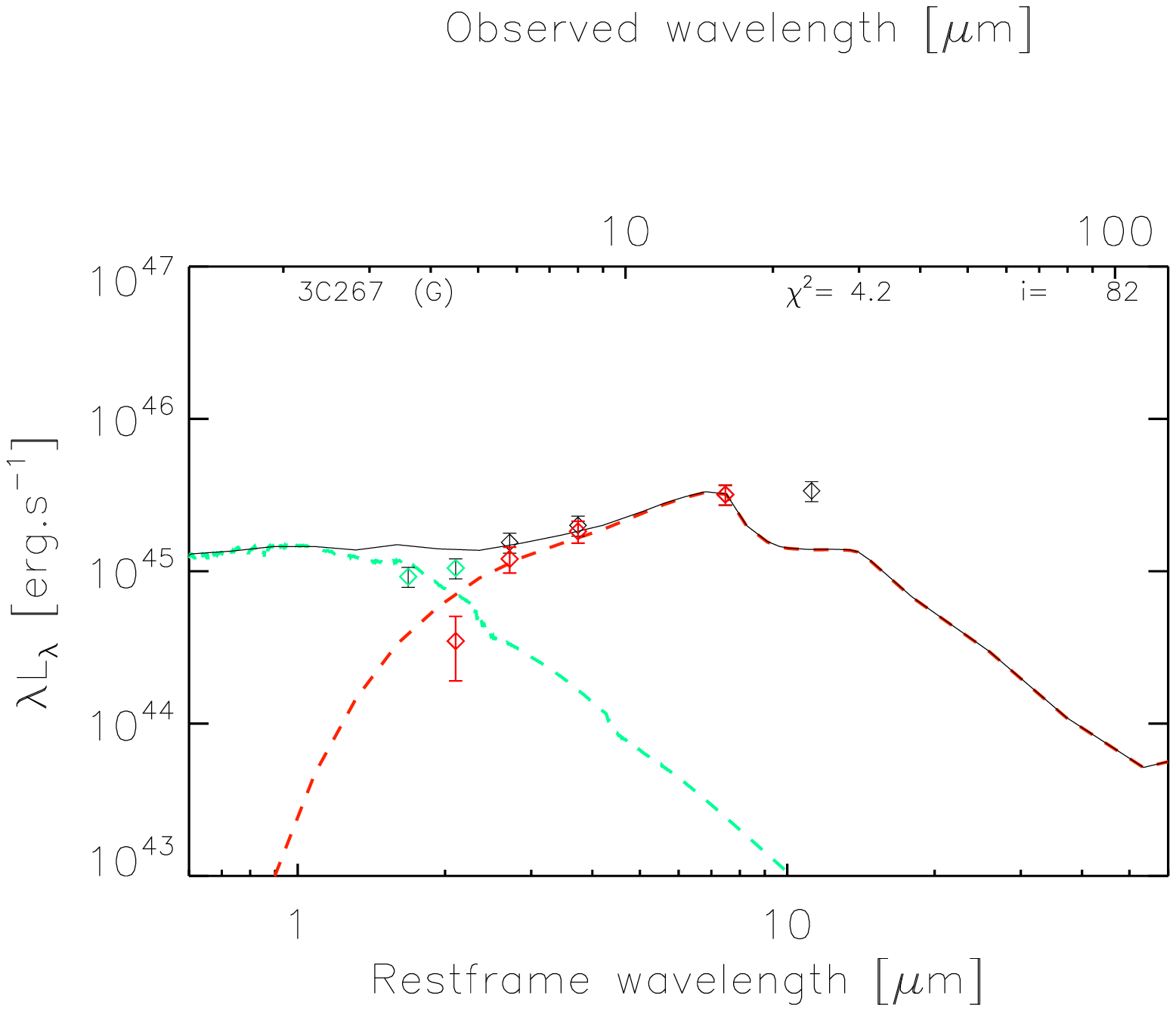} &
\includegraphics[height=37mm,trim=95 39 15 41,clip=true]{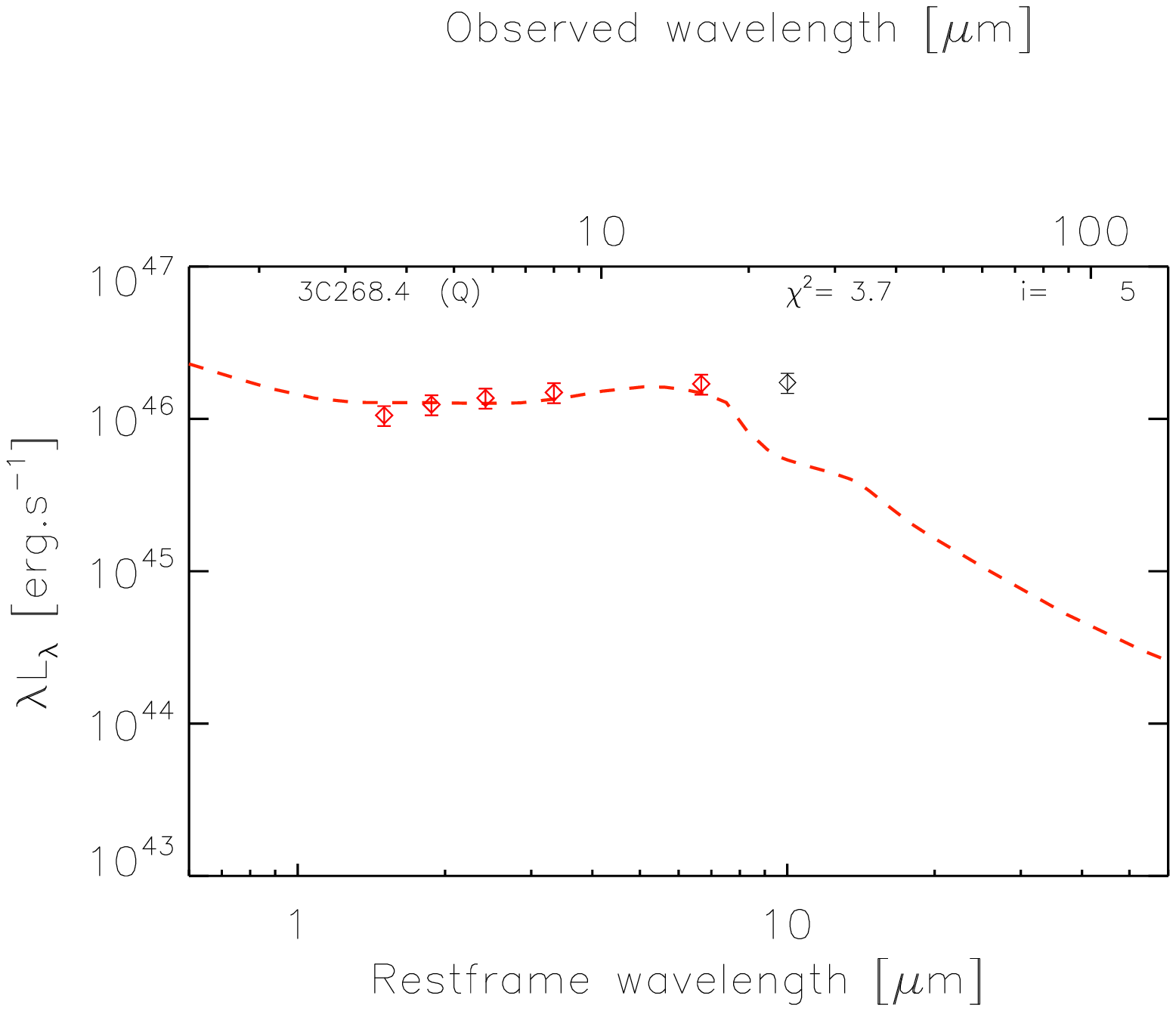} \\
\includegraphics[height=37mm,trim= 0 39 15 41,clip=true]{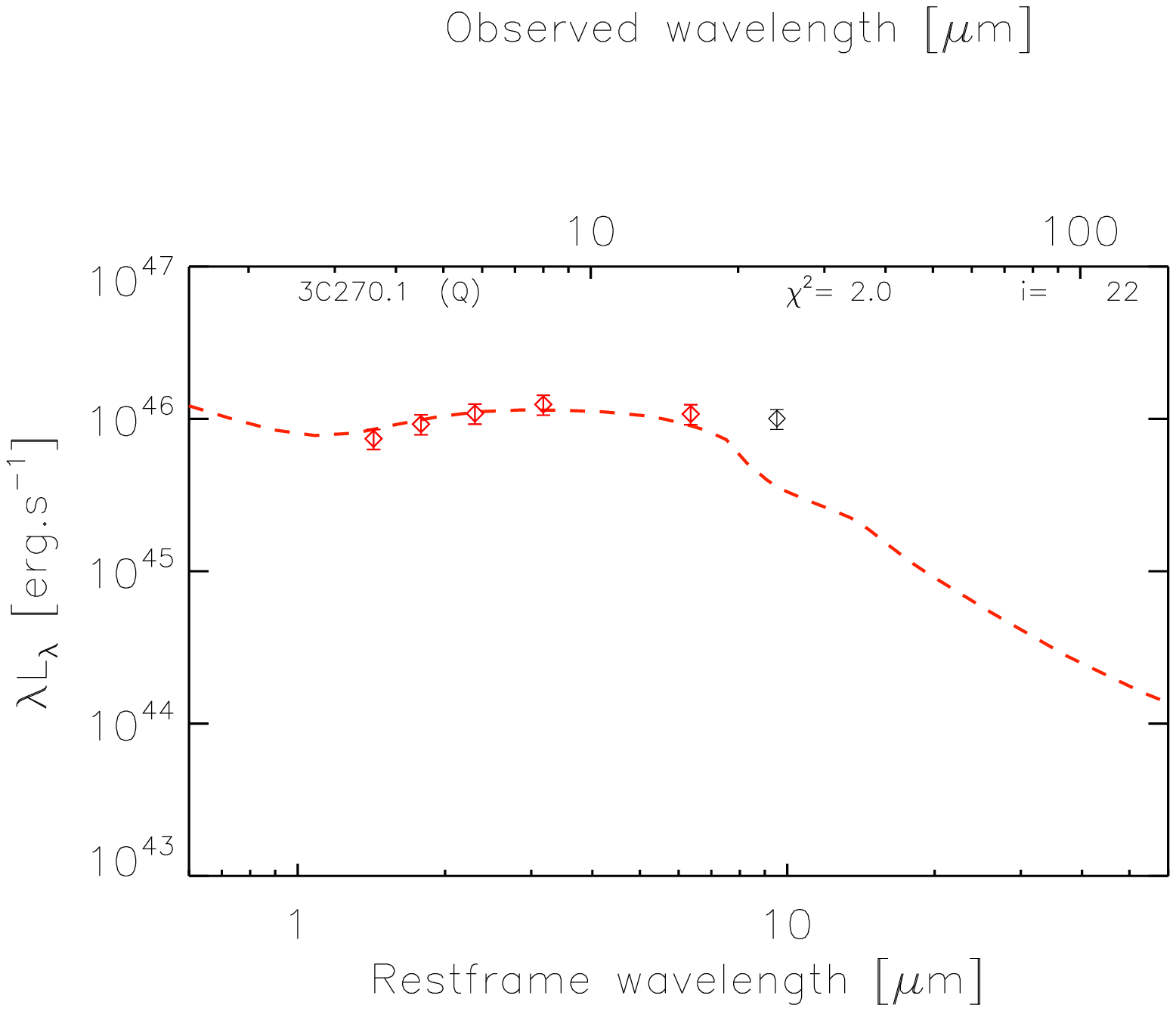} &
\includegraphics[height=37mm,trim=95 39 15 41,clip=true]{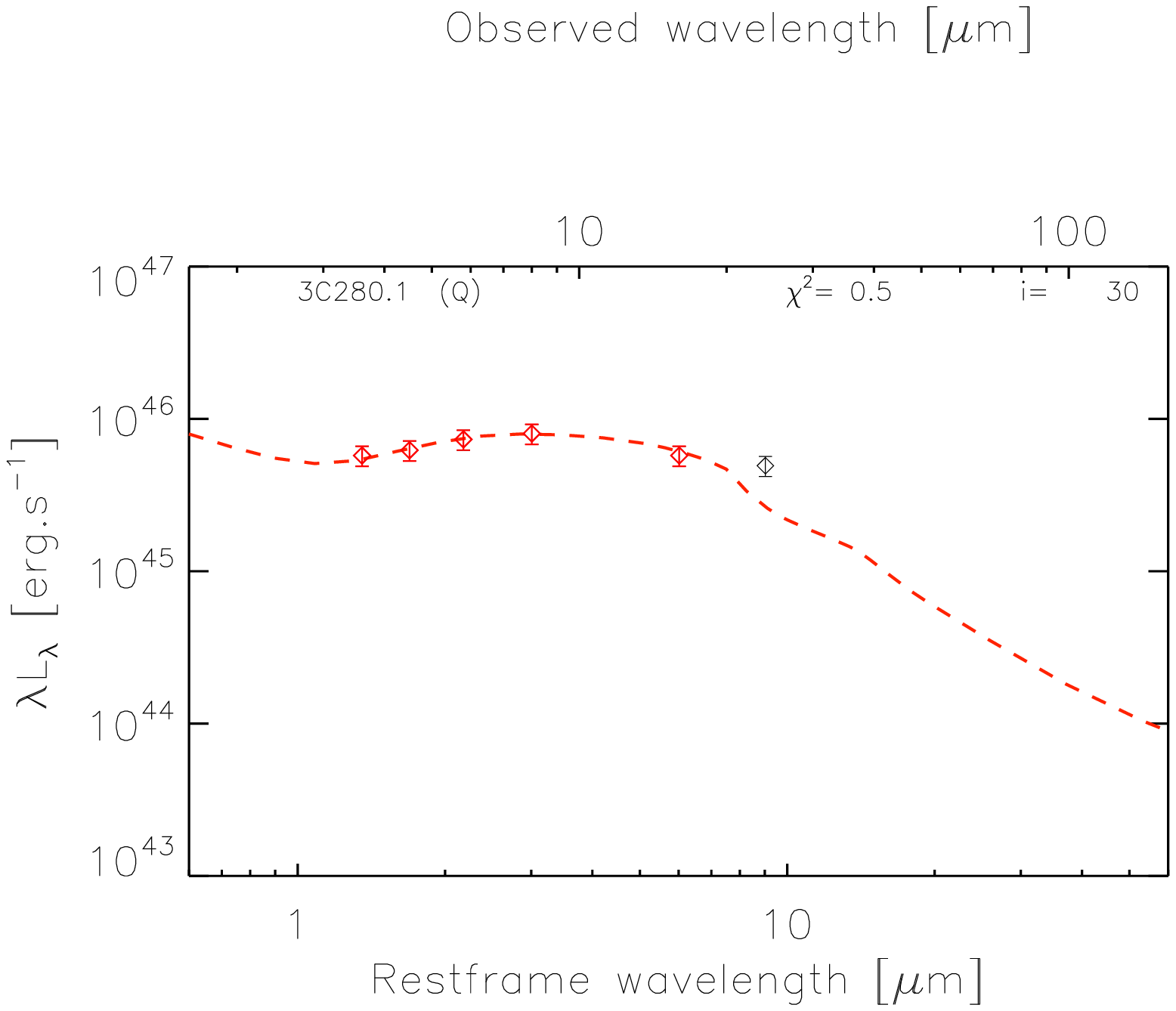} &
\includegraphics[height=37mm,trim=95 39 15 41,clip=true]{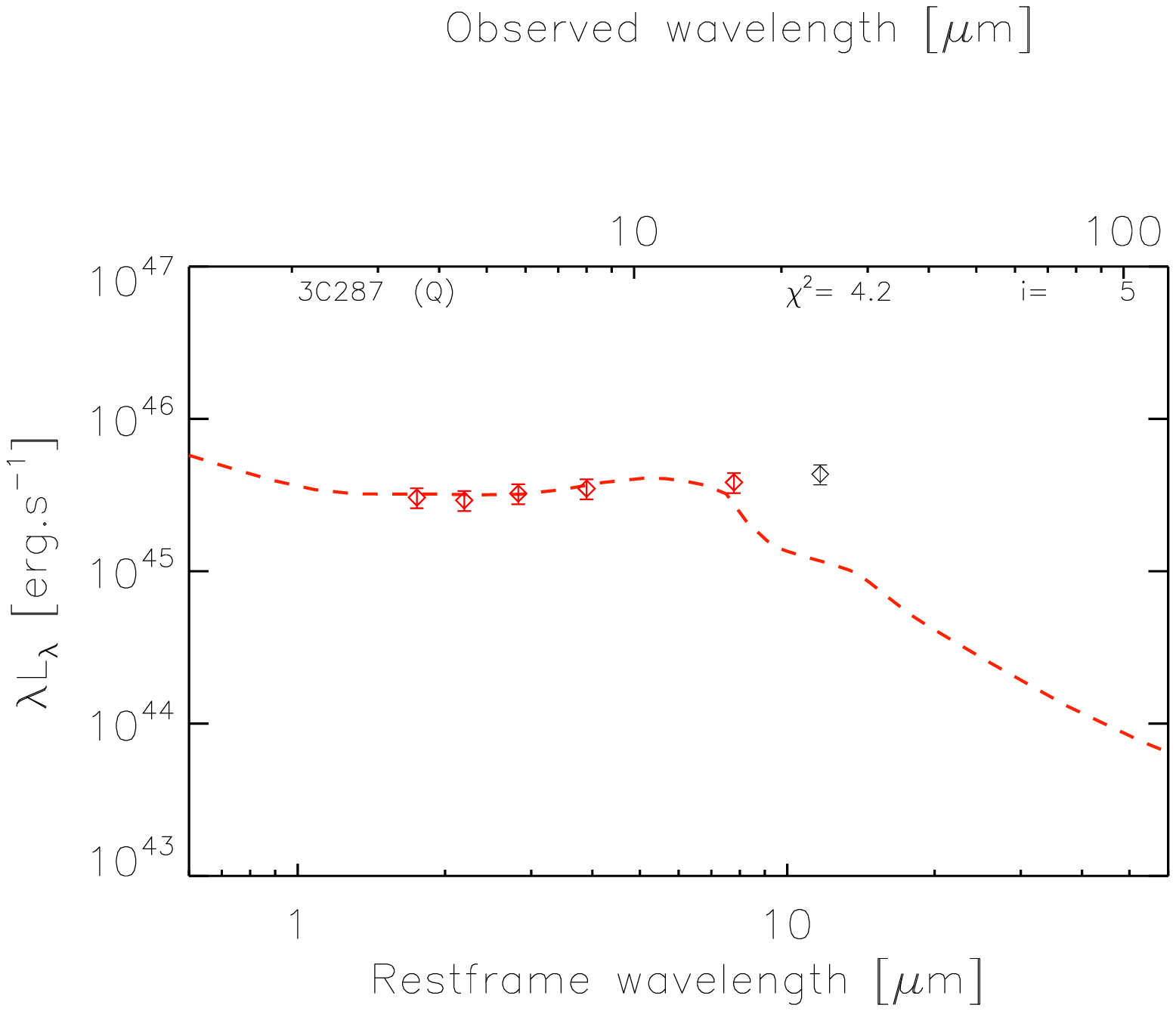} \\
\includegraphics[height=42mm,trim= 0  0 15 41,clip=true]{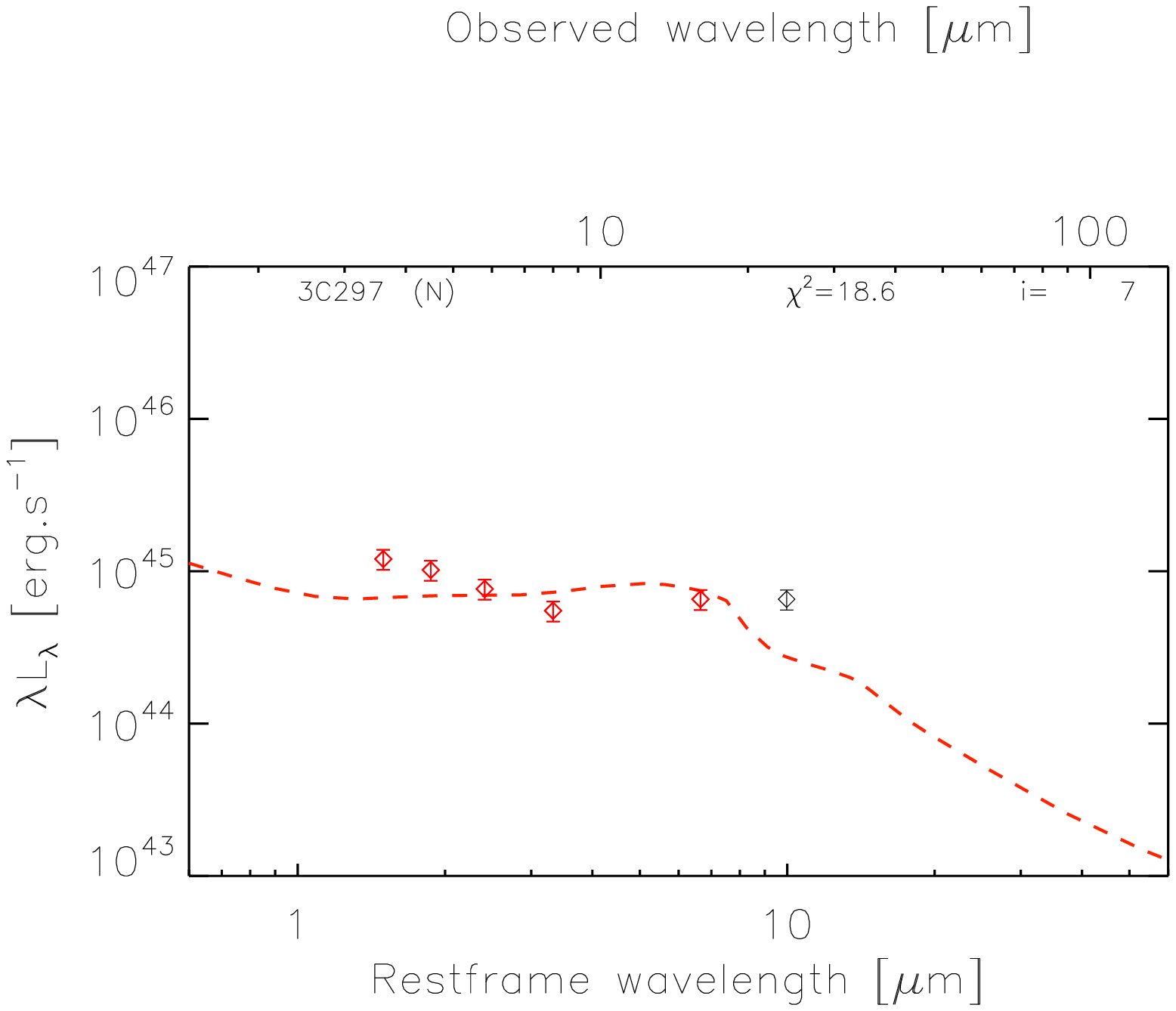} &
\includegraphics[height=42mm,trim=95  0 15 41,clip=true]{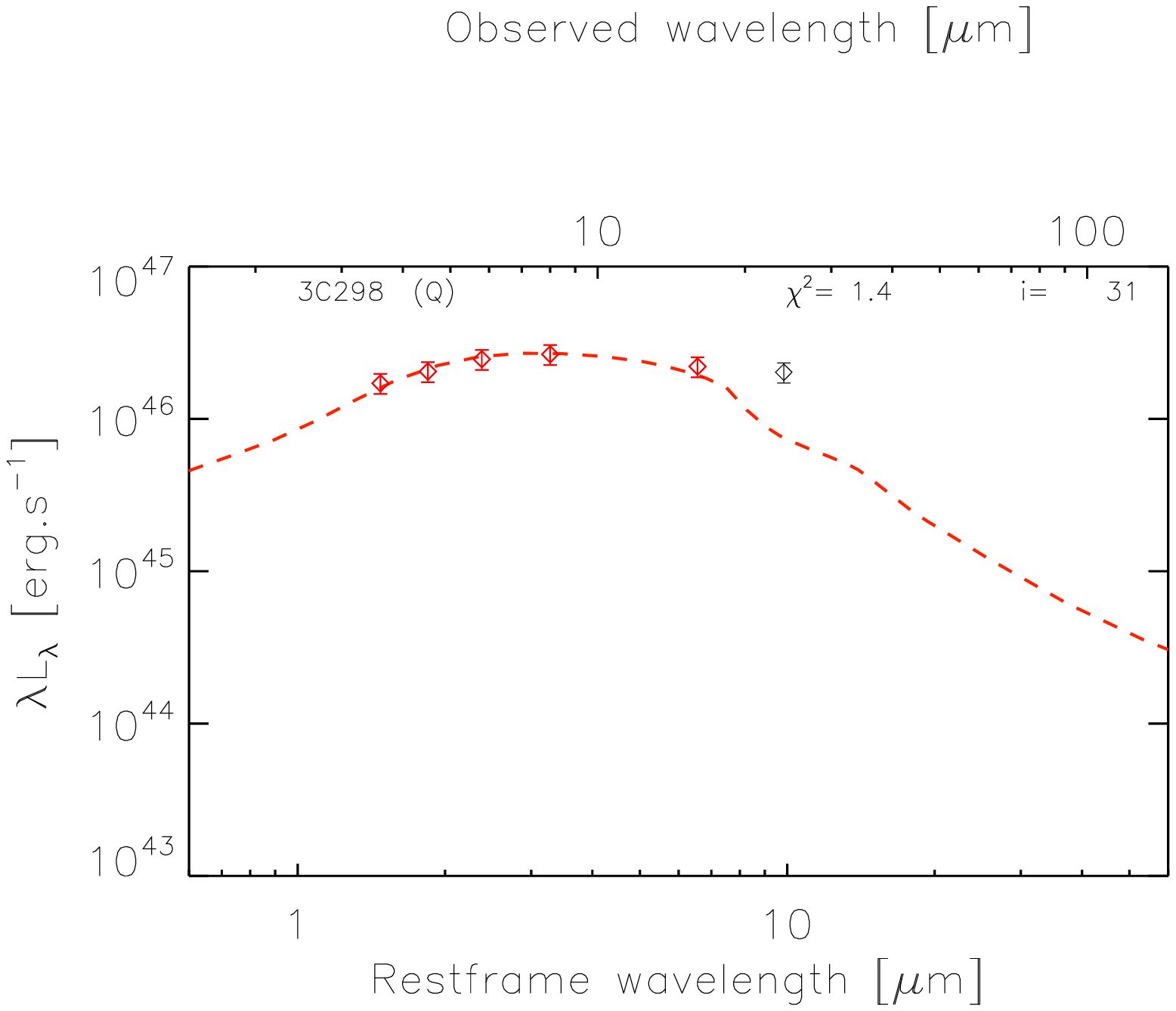} &
\includegraphics[height=42mm,trim=95  0 15 41,clip=true]{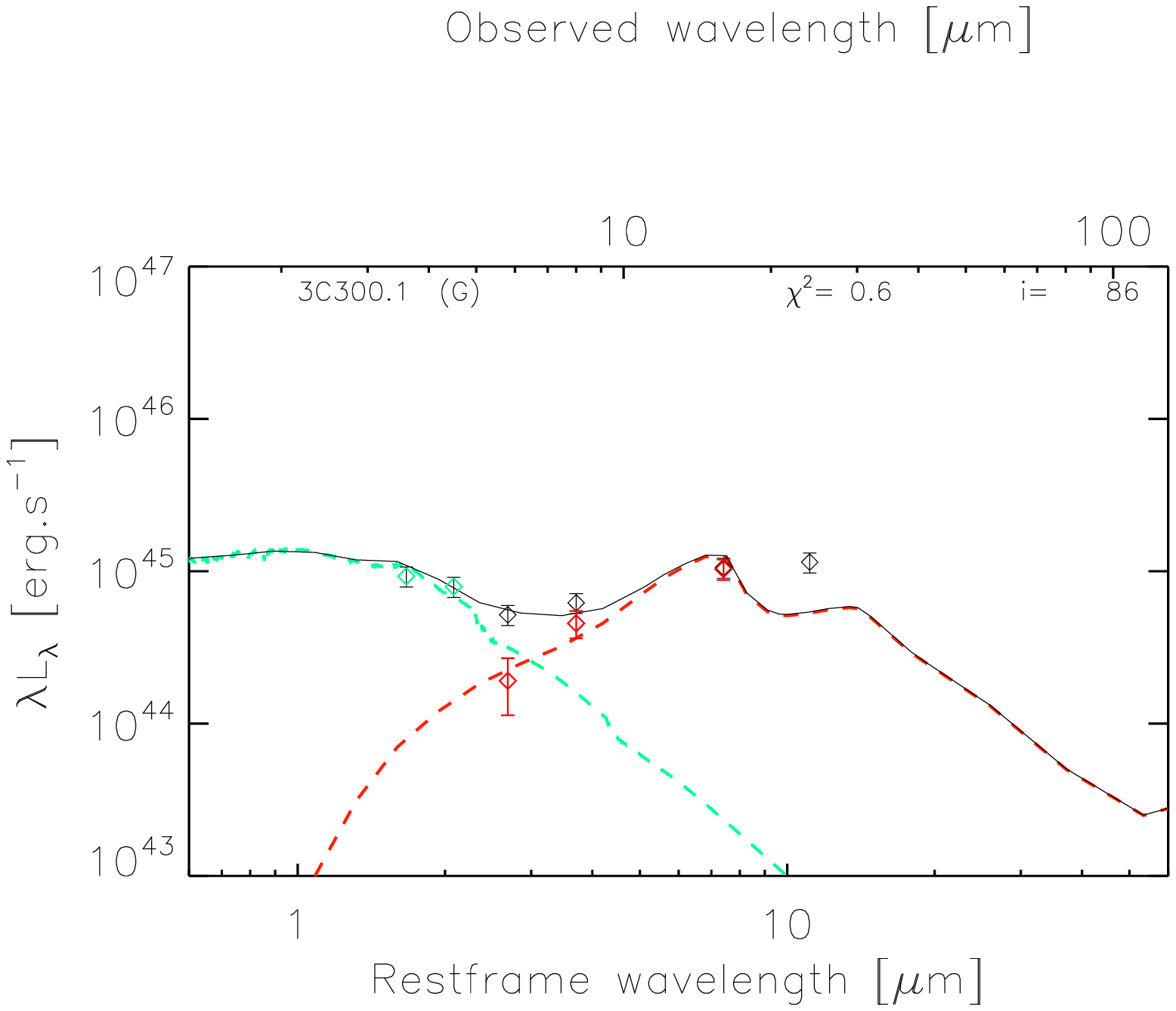} \\
  \end{tabular}
\end{center}
\end{figure*}

\begin{figure*}[ht]
  \begin{center}
  \begin{tabular}{r@{}c@{}l} 
\includegraphics[height=42mm,trim= 0 39 15  0,clip=true]{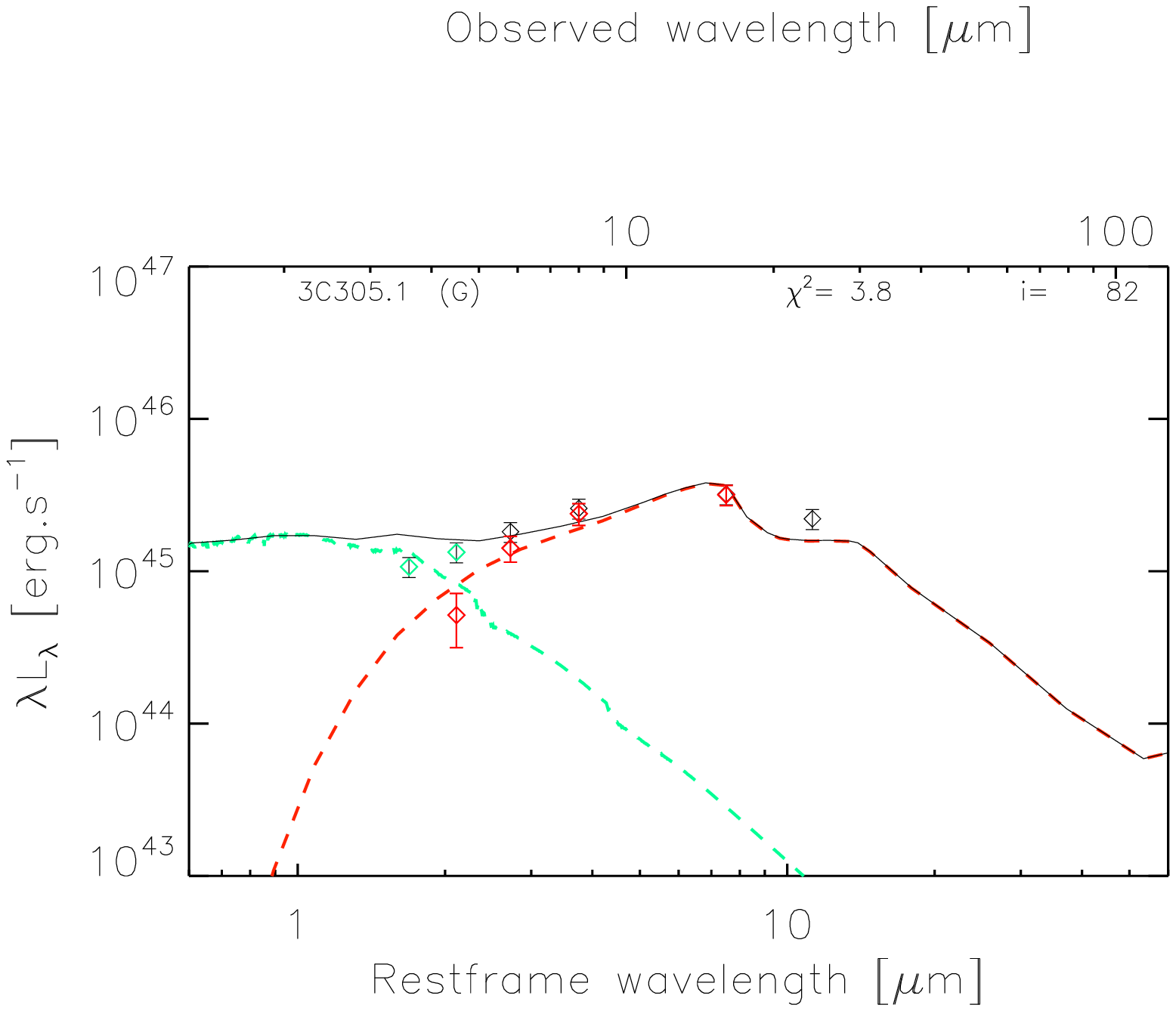} &
\includegraphics[height=42mm,trim=95 39 15  0,clip=true]{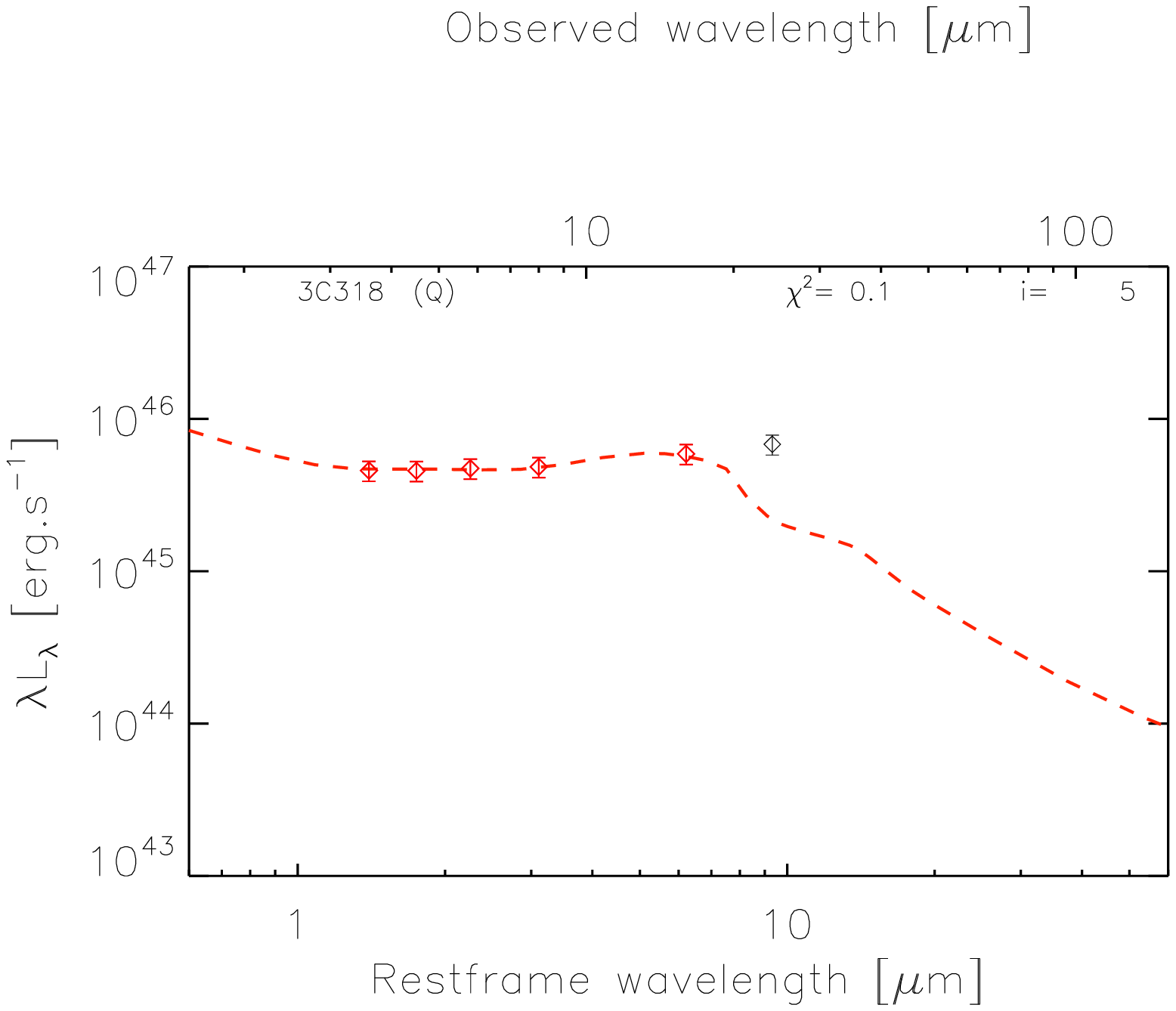} &
\includegraphics[height=42mm,trim=95 39 15  0,clip=true]{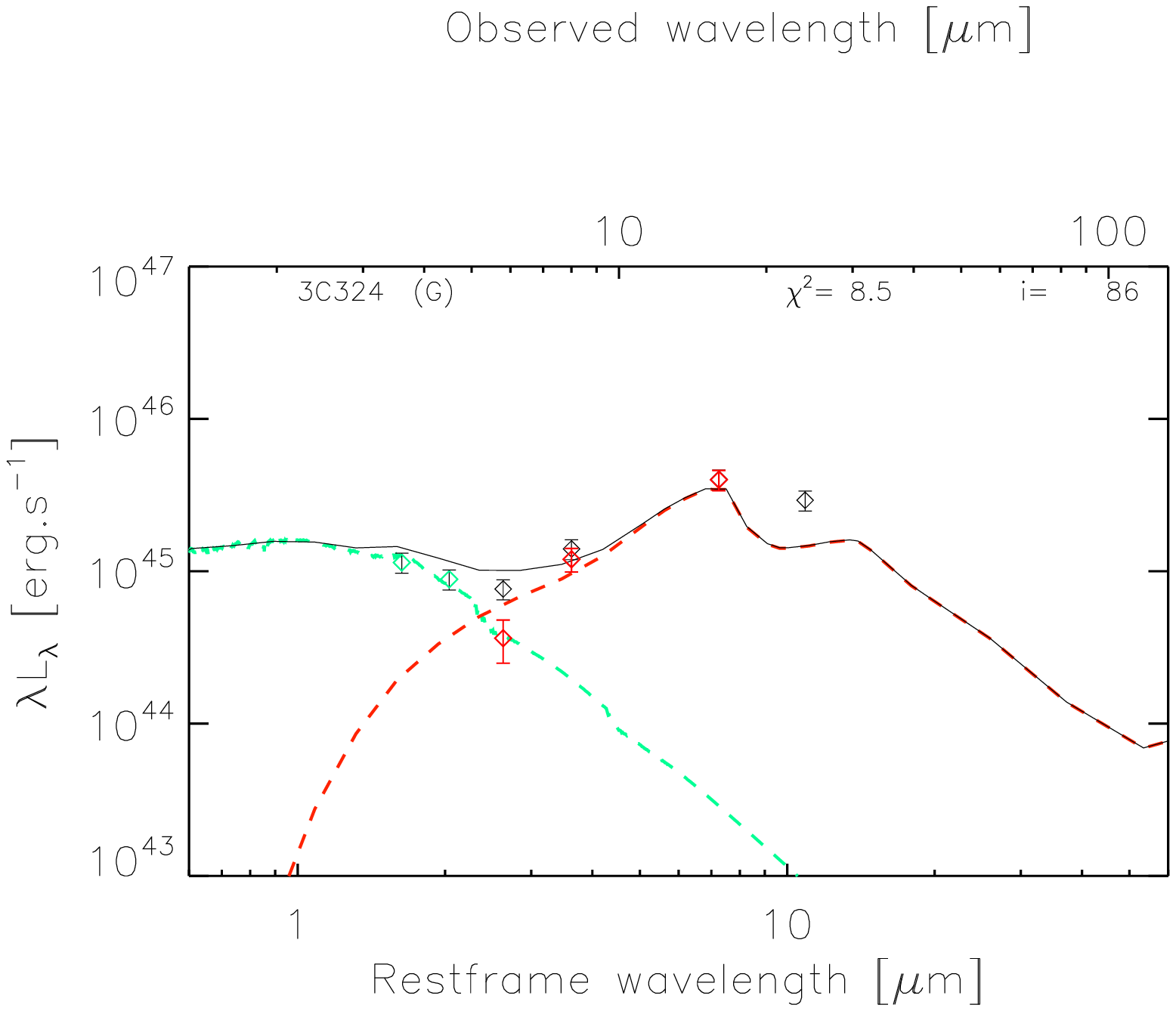} \\
\includegraphics[height=37mm,trim= 0 39 15 41,clip=true]{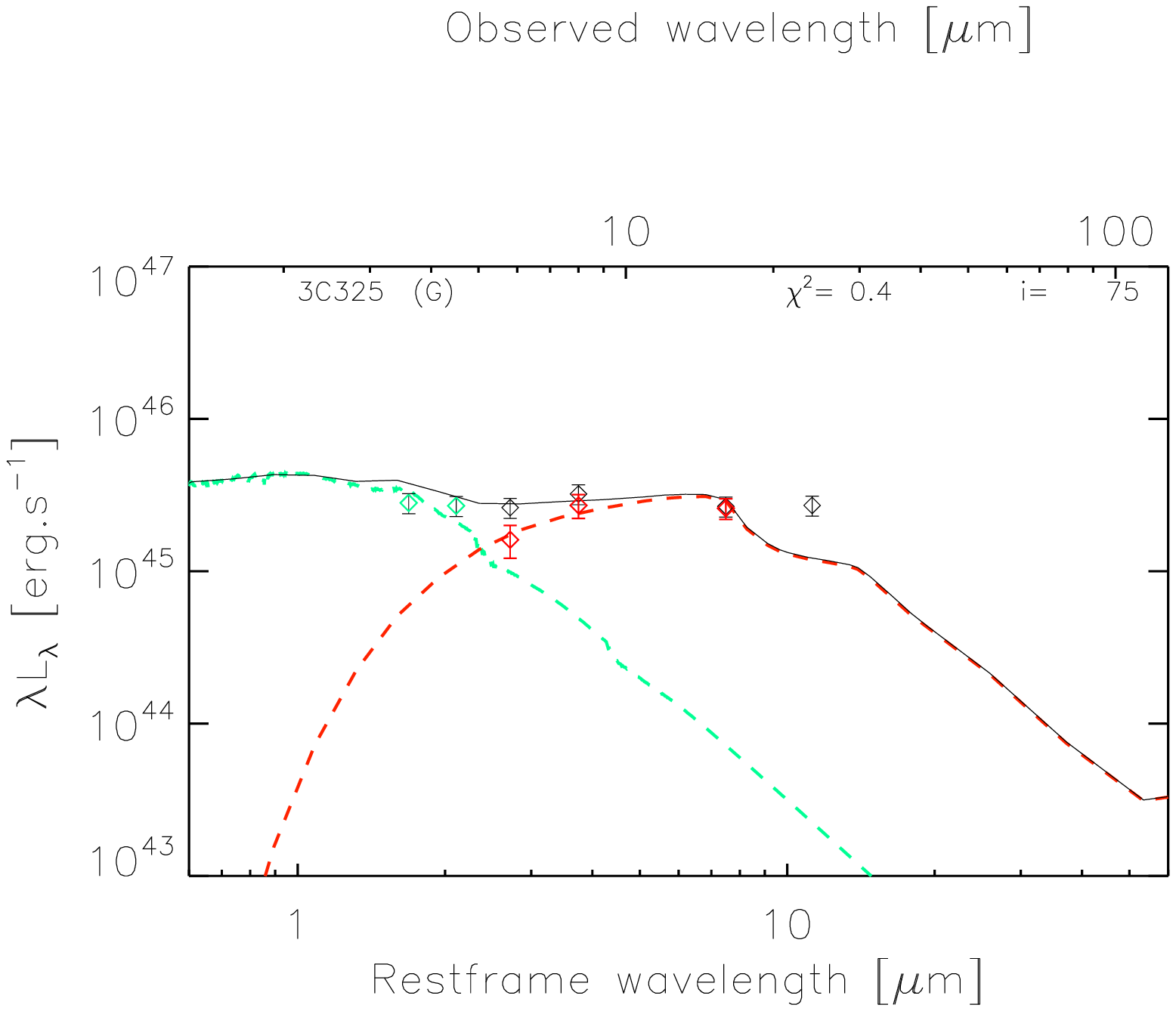} &
\includegraphics[height=37mm,trim=95 39 15 41,clip=true]{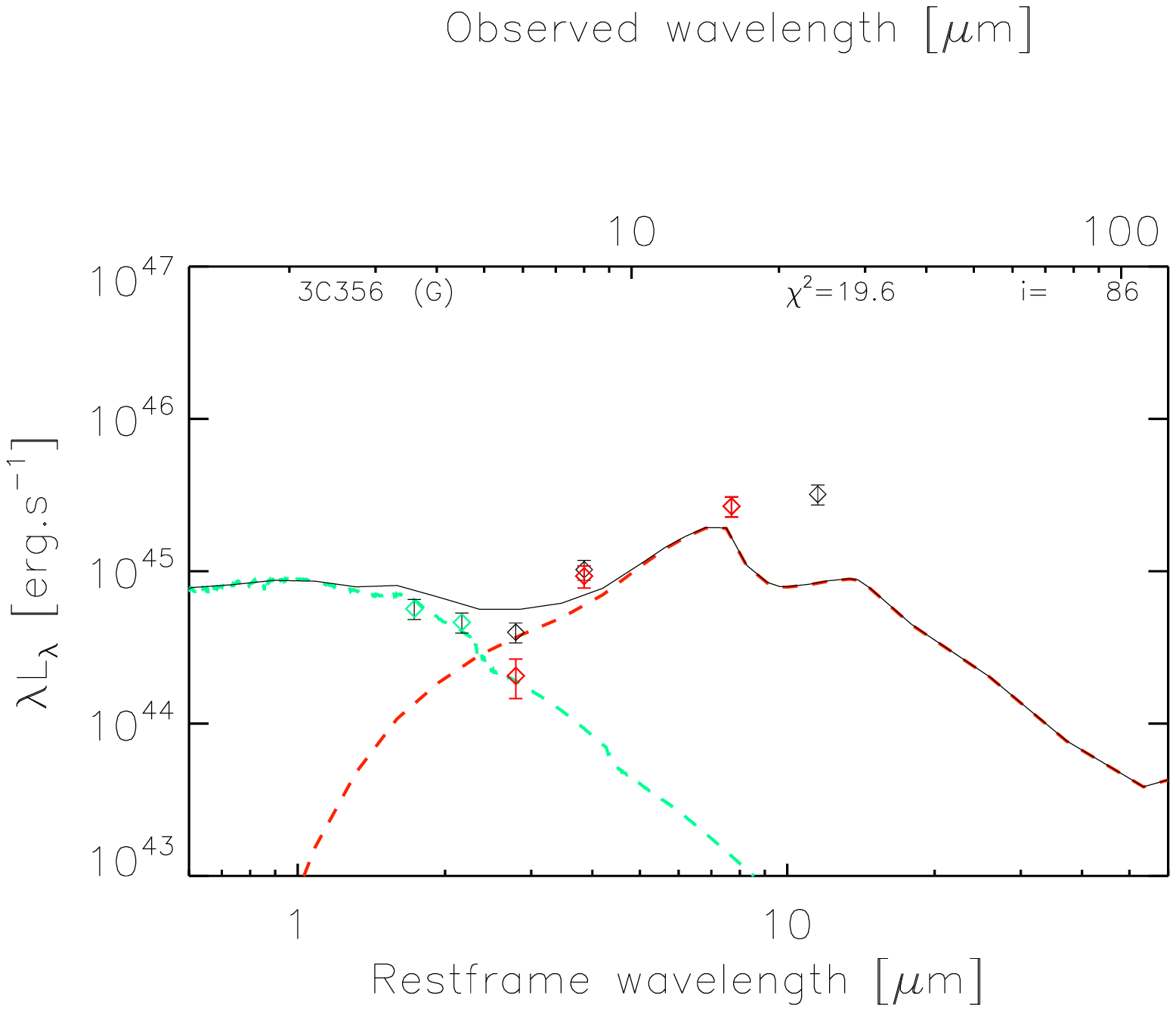} &
\includegraphics[height=37mm,trim=95 39 15 41,clip=true]{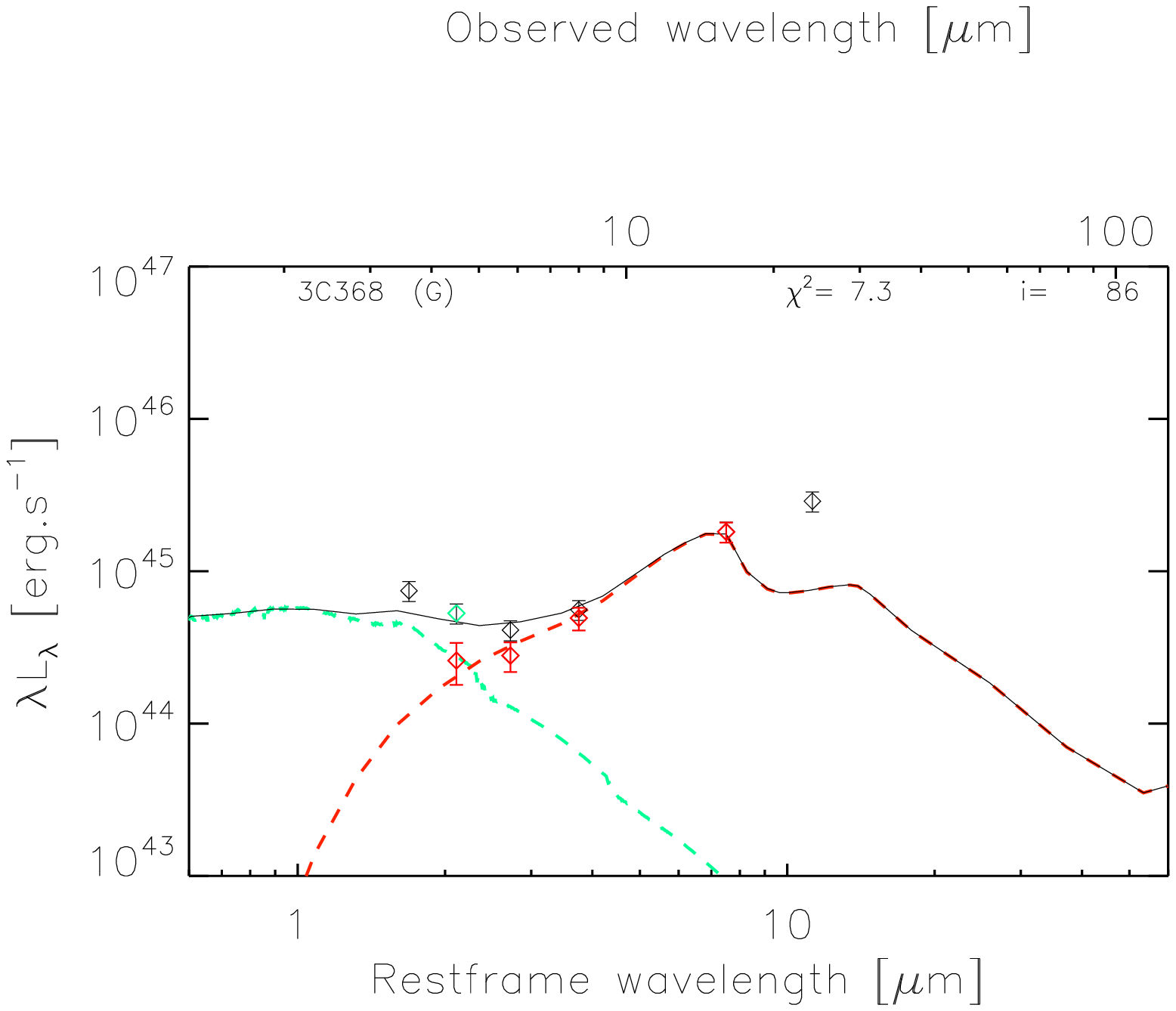} \\
\includegraphics[height=37mm,trim= 0 39 15 41,clip=true]{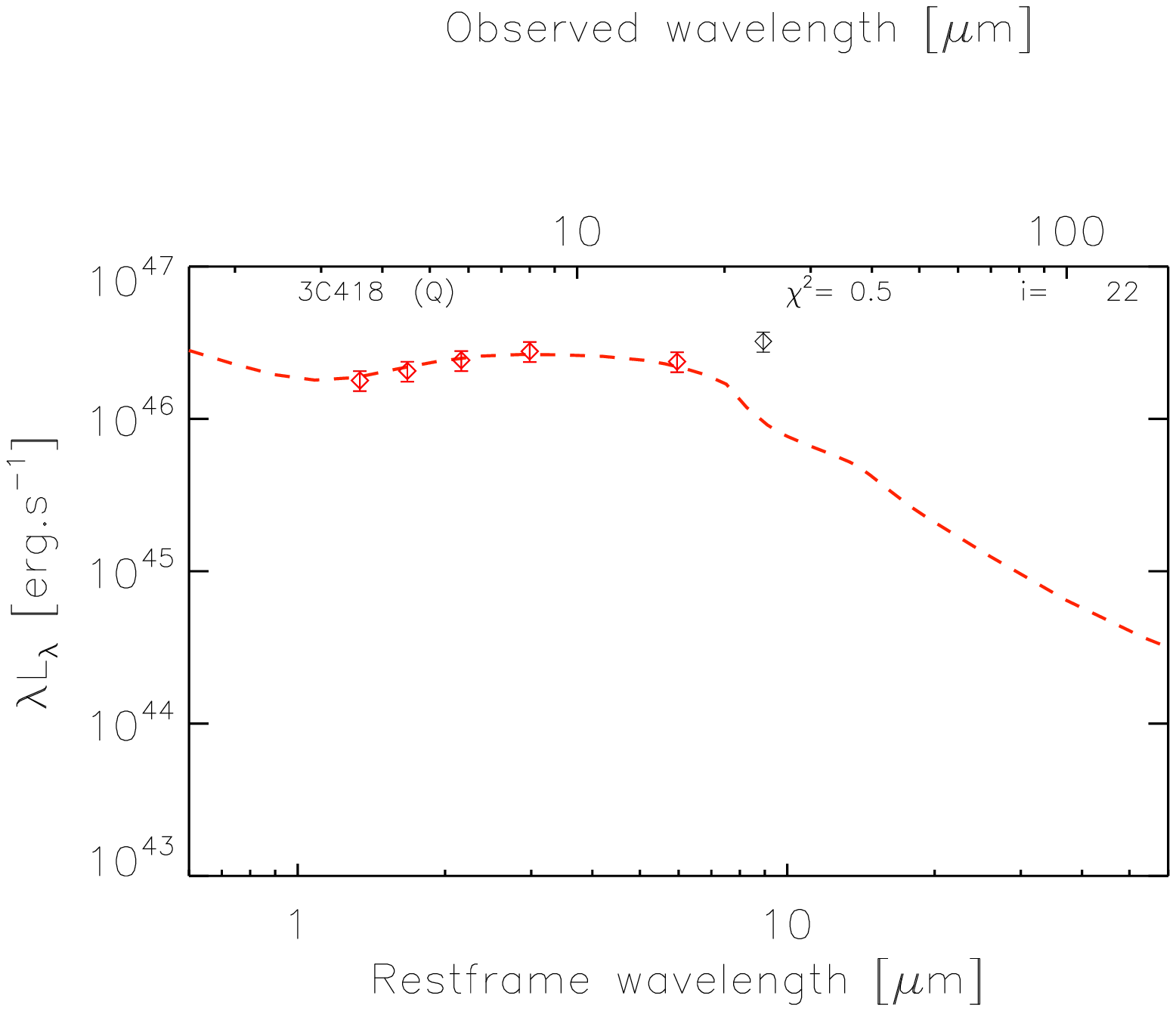} &
\includegraphics[height=37mm,trim=95 39 15 41,clip=true]{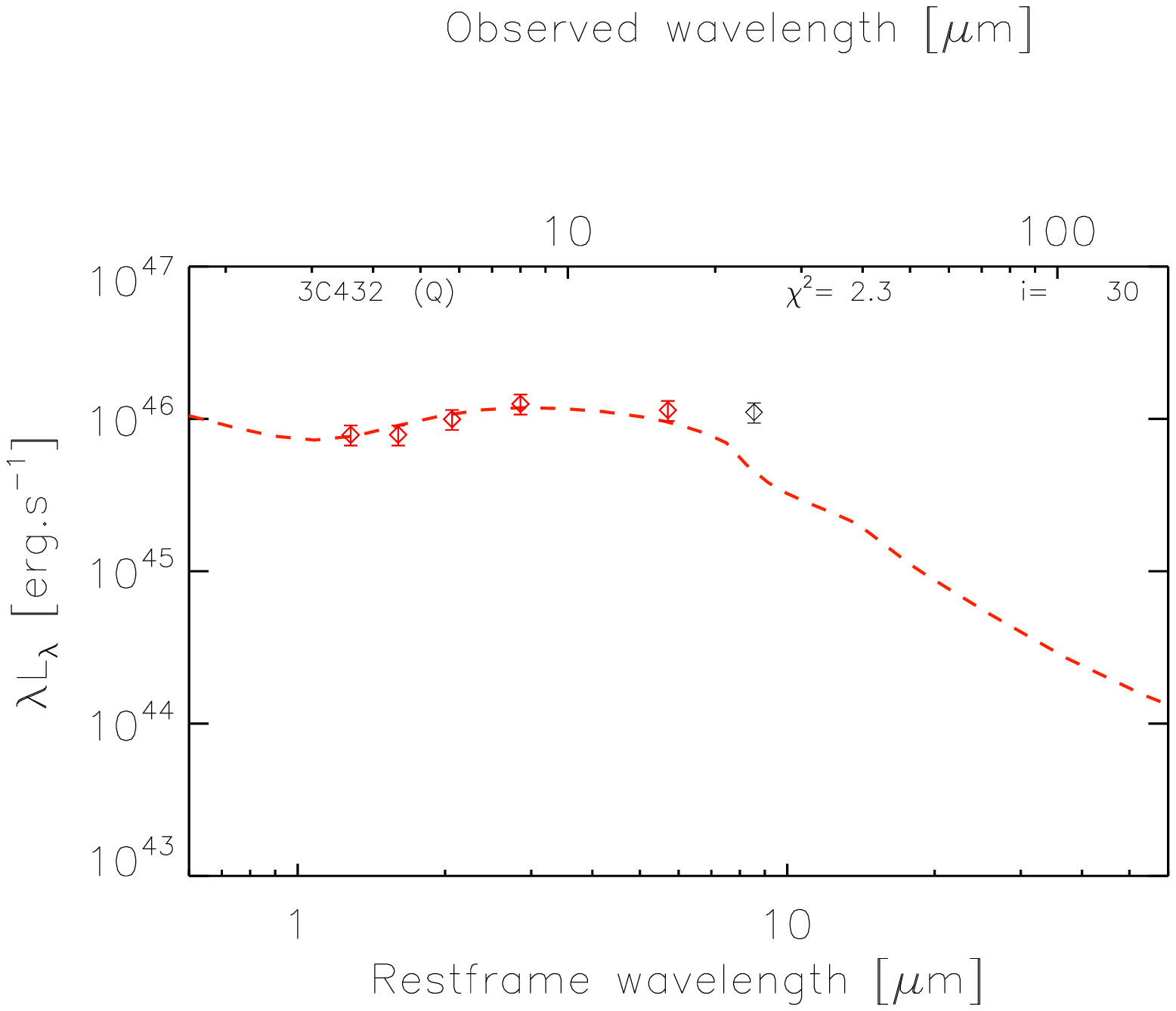} &
\includegraphics[height=37mm,trim=95 39 15 41,clip=true]{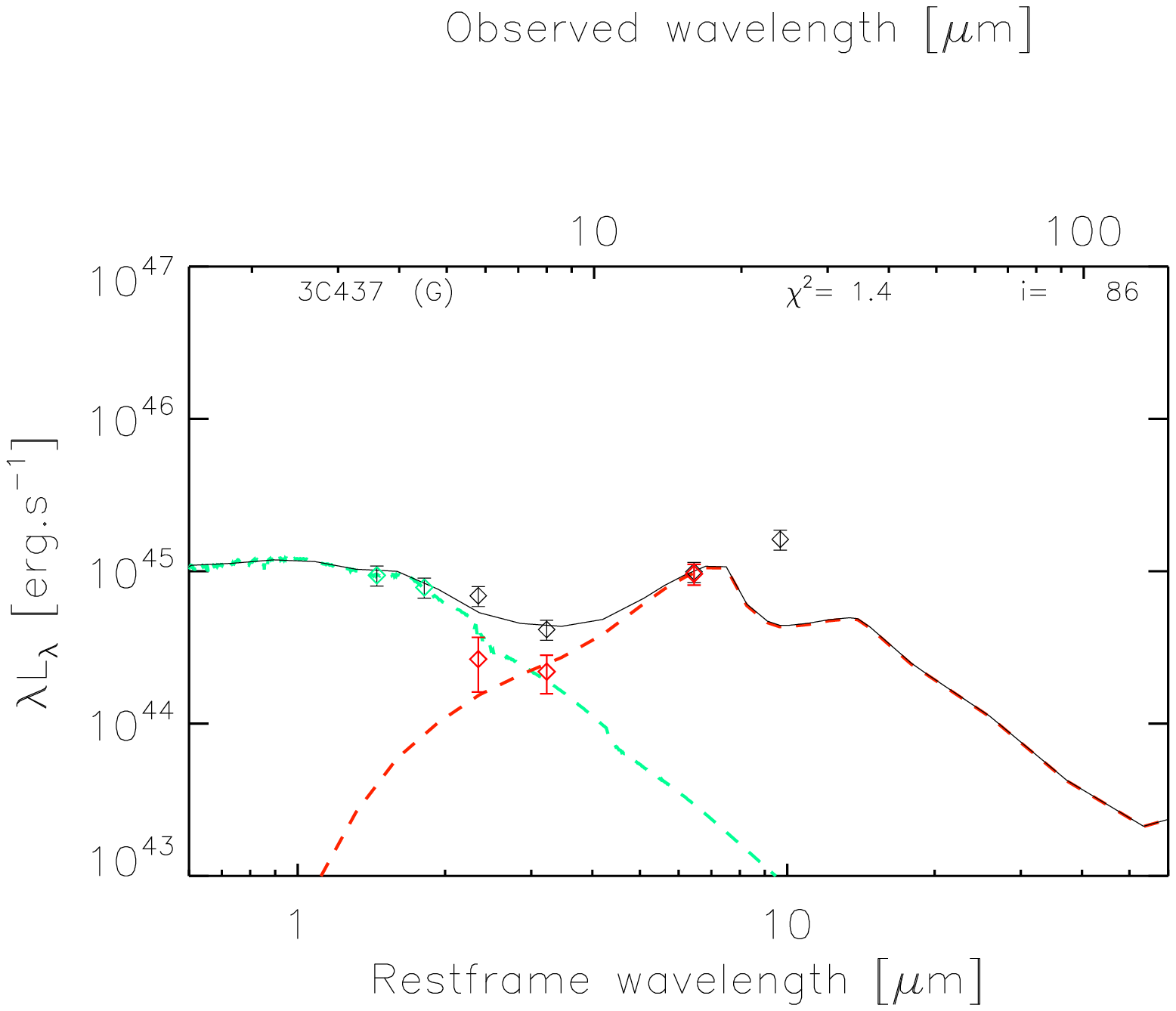} \\
\includegraphics[height=37mm,trim= 0 39 15 41,clip=true]{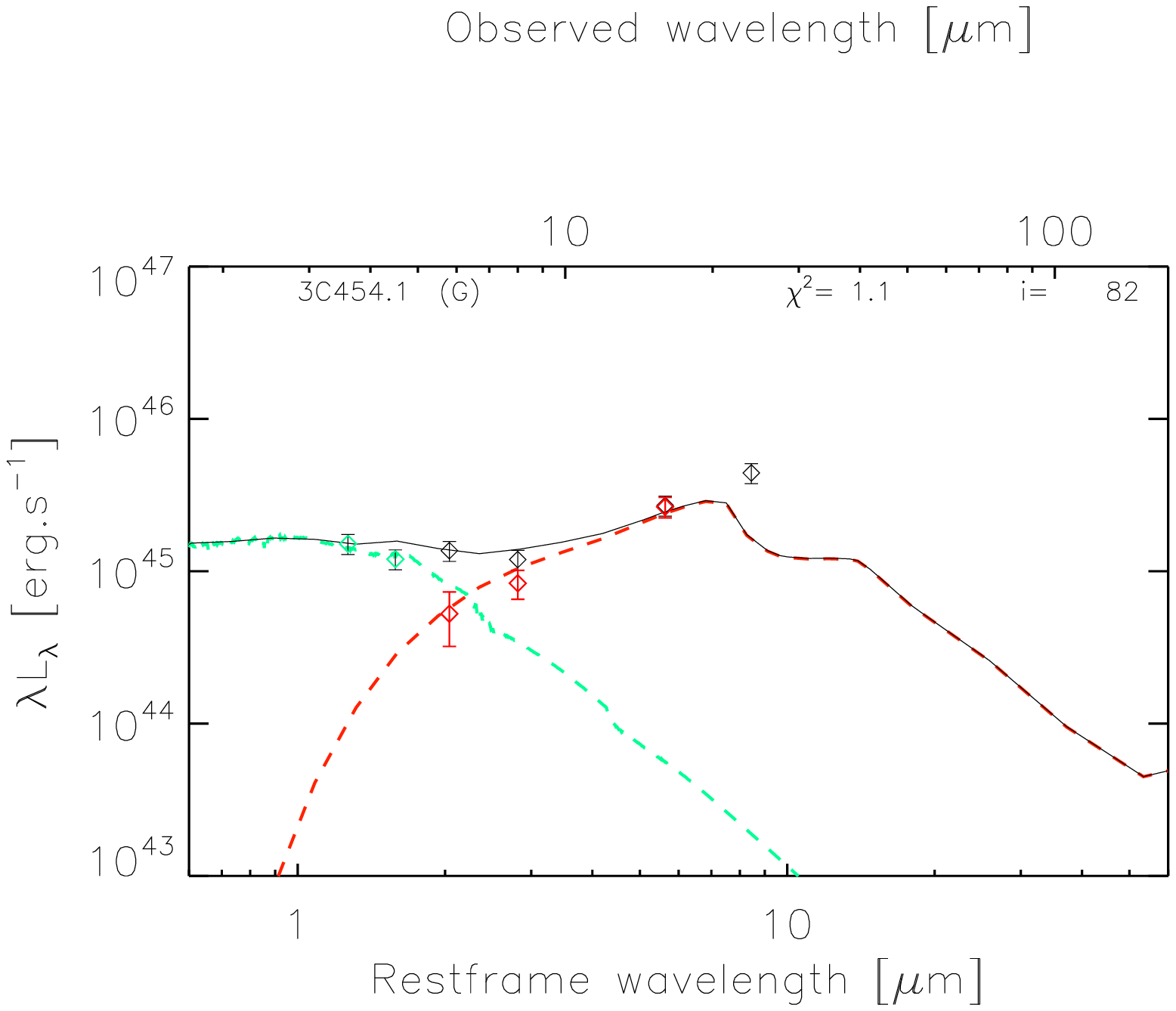} &
\includegraphics[height=37mm,trim=95 39 15 41,clip=true]{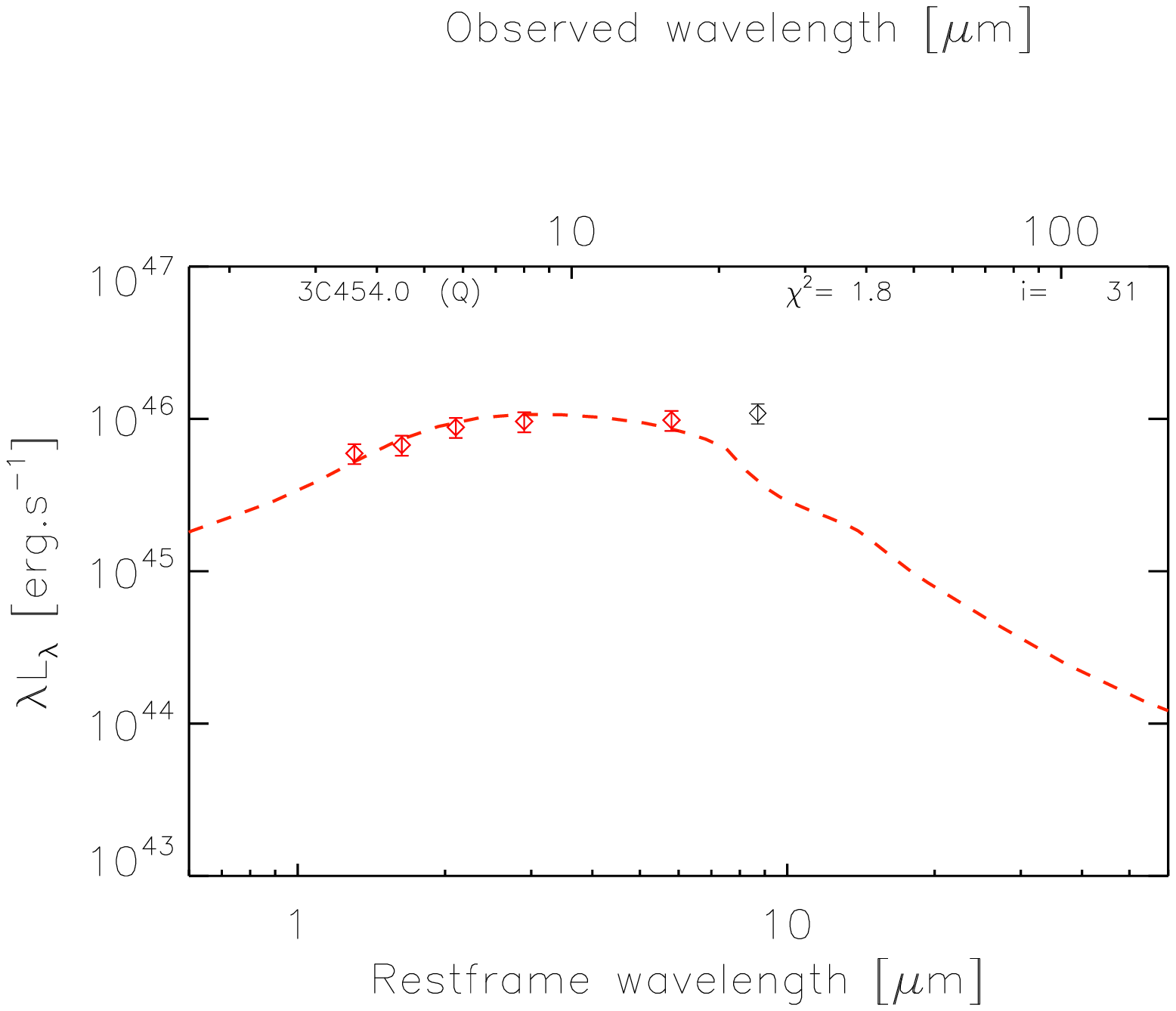} &
\includegraphics[height=37mm,trim=95 39 15 41,clip=true]{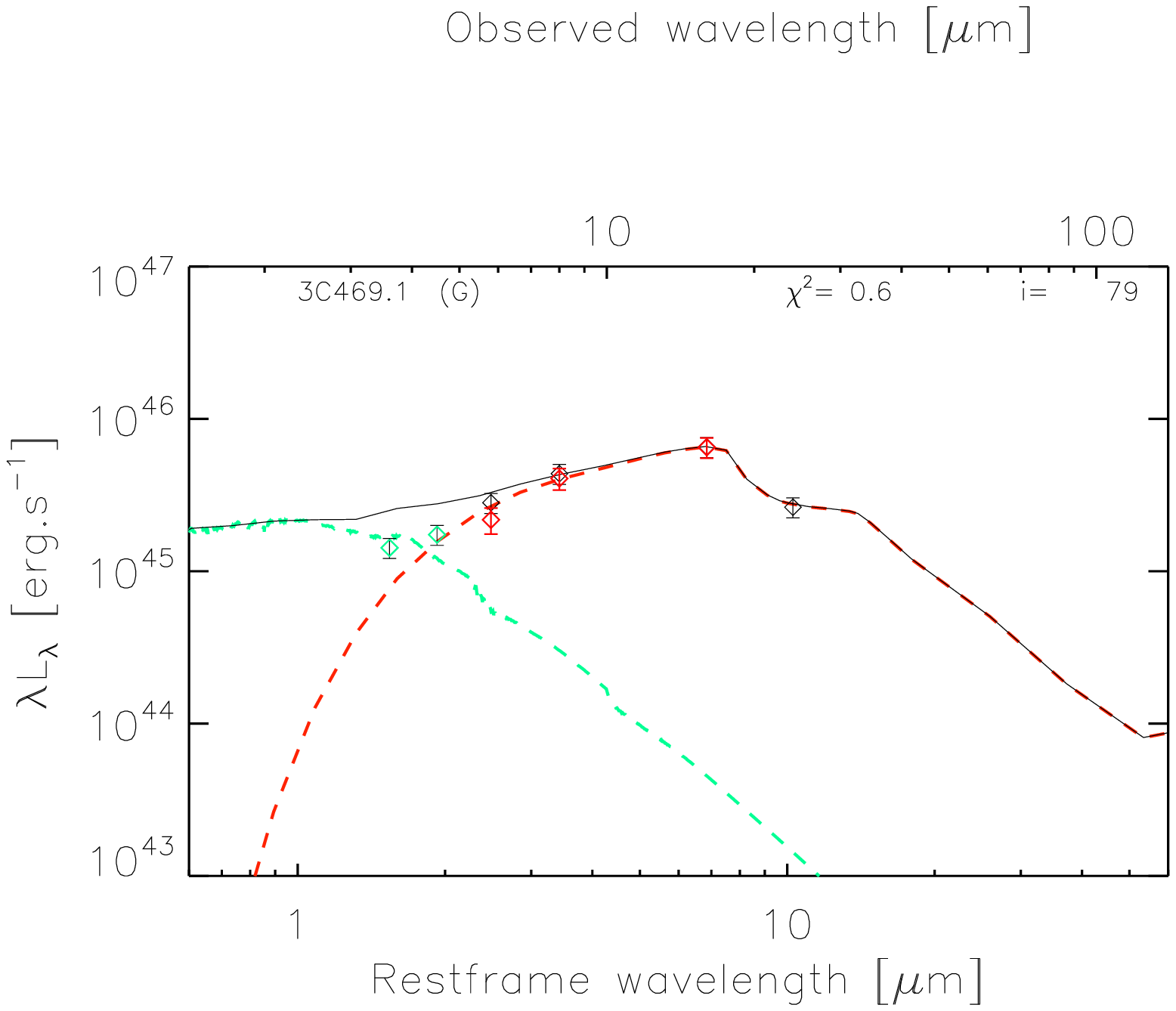} \\
\includegraphics[height=37mm,trim= 0 39 15 41,clip=true]{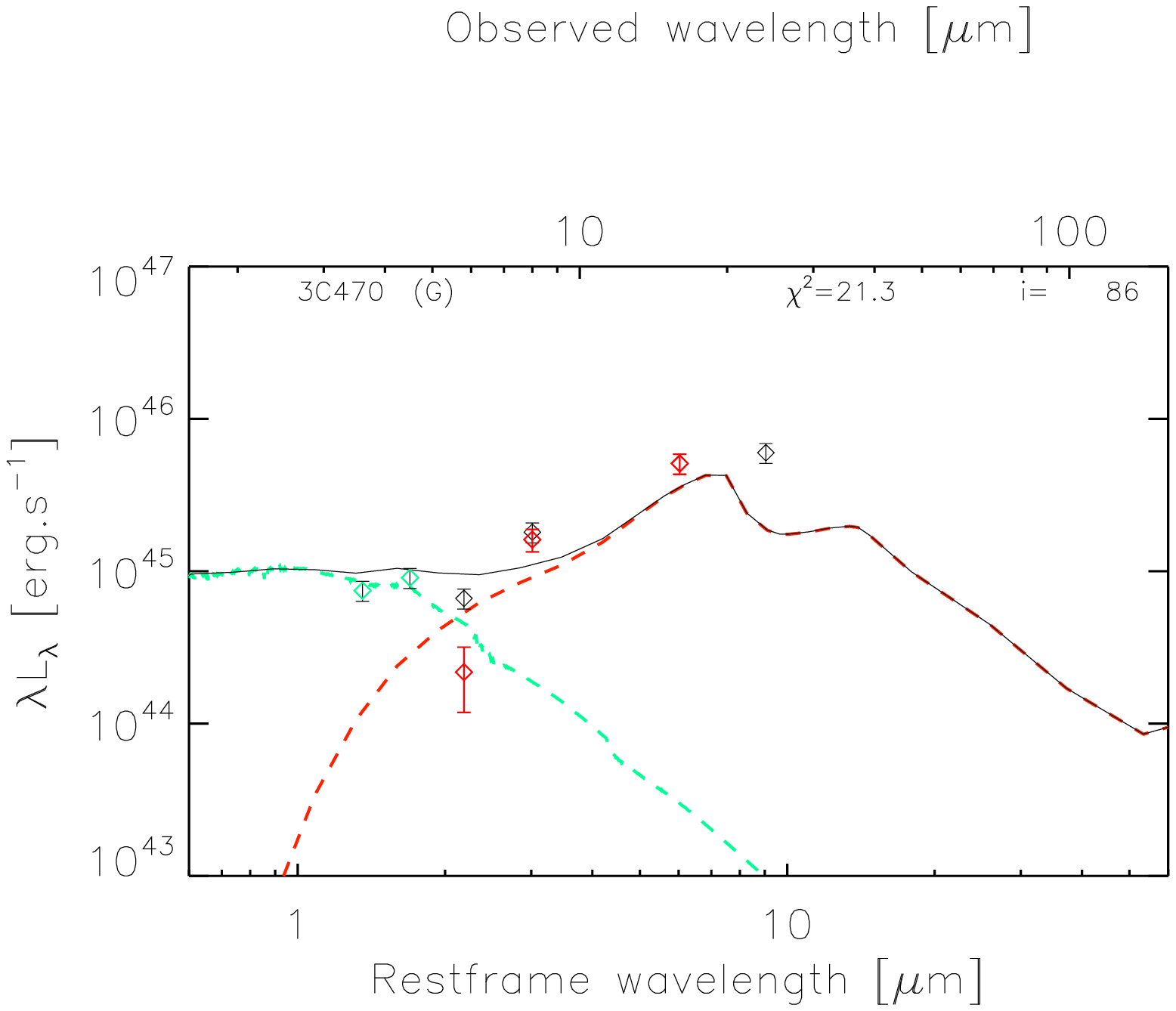} &
\includegraphics[height=37mm,trim=95 39 15 41,clip=true]{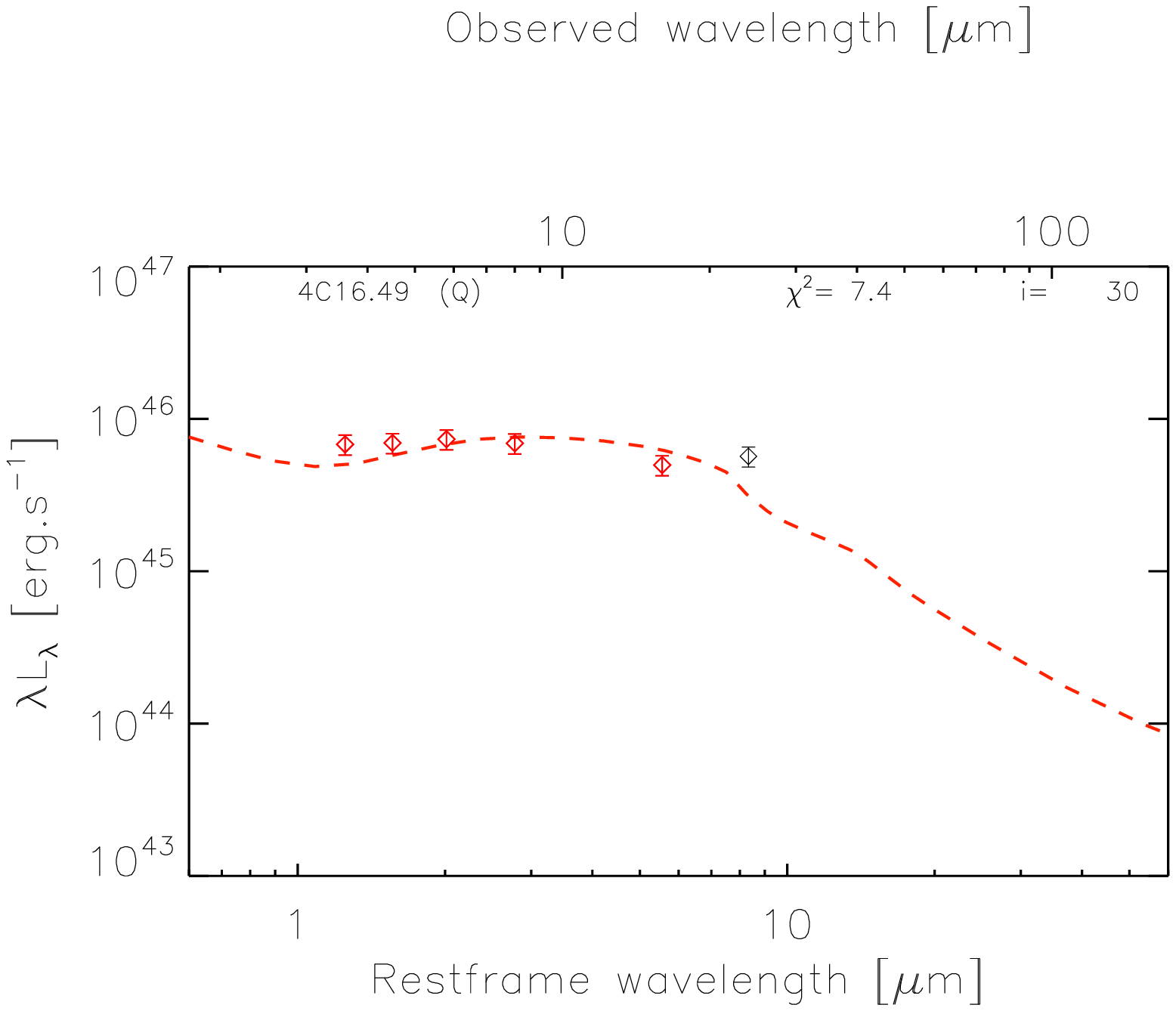} & \\

  \end{tabular}
  \caption{Mid-IR SEDs for the s3CR sample. The symbols are the same than 
    in Fig. \ref{fig:full_SED}.}
  \label{fig:full_SED_3C}
\end{center}
\end{figure*}

\appendix
\section{Notes on individual sources}
\label{notes}

\noindent B3\,J2330+3927 This galaxy has the highest core
dominance in the SHzRG sample ($R$=0.004). The radio 
emission is complex \citep{PerezTorres2005}.

\noindent 6C\,0140+326 This source (the second highest
redshift source in the SHzRG  sample) has been removed from this study, as a
foreground object contaminates the galaxy image in the IRAC bands.

\noindent 4C\,60.07, 3C\,356, MRC\,2048$-$272, 7C\,1756+5620.
These four objects have double components in the IRAC maps, but
only one of each pair coincides with a radio core.

\noindent MRC\,2025$-$218, MRC\,0156$-$252, MRC\,1017$-$220,
  TXS\,1113$-$178, MRC\,1138$-$262, MRC\,1558$-$003, MRC\,0251$-$273 These
galaxies show broad permitted lines 
\citep[][]{Nesvadba2011,Humphrey2008}.

\noindent 6C\,0032+412 This galaxy exhibits a very hot dust
component in its mid-IR SED \citep{DeBreuck2010}.

\noindent TNJ2007$-$1316 The previously quoted flux density at
5.6\,\mum has been replaced with a 3$\sigma$ upper limit of 
$F_{5.6\mu\rm{m}}<$146.0 $\mu$Jy.

\end{document}